\documentclass[]{aa}
\usepackage{graphicx}
\usepackage{txfonts}
\usepackage{lscape}
\usepackage{rotating}
\graphicspath{{.}}
\usepackage{natbib}
\usepackage{aalongtable}

\def\etal{et al. }

\def\Ia{SN~Ia}
\def\Iae{SNe~Ia}

\newcommand{\Yp}{{\tt Y}}
\newcommand{\Xp}{{\tt X}}

\begin{document}
\def\lesim{\stackrel{<}{{}_{\sim}}} 
\def\gesim{\stackrel{>}{{}_{\sim}}}
\title{The ESO/VLT 3rd year Type Ia supernova data set from the Supernova Legacy Survey \thanks{Based on observations obtained with FORS1 and FORS2 at the Very Large Telescope on Cerro Paranal, operated by the European Southern Observatory, Chile (ESO Large Programs 171.A-0486 and 176.A-0589)}$^,$\thanks{Figures A.1 to A.139 are only available in electronic form via http://www.edpsciences.org}}

\author{C. Balland\inst{1,2},  S. Baumont\inst{1}, S. Basa\inst{3}, M. Mouchet\inst{4,5}, 
D. A. Howell\inst{6,7}, P. Astier\inst{1}, 
R. G. Carlberg\inst{8}, A. Conley\inst{8}, D. Fouchez\inst{9}, J. Guy\inst{1}, 
D. Hardin\inst{1}, I. M. Hook\inst{10}, R. Pain\inst{1}, K. Perrett\inst{8},
C. J. Pritchet\inst{11}, N.  Regnault\inst{1}, J. Rich\inst{12}, M. Sullivan\inst{10}, P. Antilogus\inst{1},
V. Arsenijevic\inst{13,14}, J. Le Du\inst{9}, S. Fabbro\inst{13}, C. Lidman\inst{15}, 
A. Mour\~ao\inst{13}, N. Palanque-Delabrouille\inst{12}, E. P\'econtal\inst{16,17}, V. Ruhlmann-Kleider\inst{12}}

\institute{LPNHE, CNRS-IN2P3 and Universities of Paris 6 \& 7,F-75252
  Paris Cedex 05, France 
\and University Paris 11, Orsay, F-91405, France
\and LAM, CNRS, BP8, P\^{o}le de l'\'etoile, Site de Ch\^{a}teau-Gombert,
38 rue Fr\'ed\'eric Joliot-Curie, F-13388 Marseille Cedex 13, France
\and APC, UMR 7164 CNRS, 10 rue Alice Domon et L\'eonie Duquet, F-75205 Paris Cedex 13, France
\and LUTH, UMR 8102 CNRS, Observatoire de Paris, Section de Meudon, F-92195 Meudon Cedex, France
\and Las Cumbres Observatory Global Telescope Network, 6740 Cortona Dr., Suite 102, Goleta, CA 93117
\and Department of Physics, University of California, Santa Barbara, Broida Hall, Mail Code 9530, Santa Barbara, CA 93106-9530
\and Department of Astronomy and Astrophysics, 50 St. George Street, Toronto, ON M5S 3H4, Canada
\and CPPM, CNRS-Luminy, Case 907, F-13288 Marseille Cedex 9, France
\and University of Oxford, Astrophysics, Denys Wilkinson Building, Keble Road, Oxford OX1 3RH, UK
\and Department of Physics and Astronomy, University of Victoria, PO Box 3055, Victoria, BC V8W 3P6, Canada
\and CEA/Saclay, DSM/Irfu/Spp, F-91191 Gif-sur-Yvette Cedex, France
\and CENTRA-Centro M. de Astrofisica and Department of Physics, IST, Lisbon, Portugal
\and SIM/IDL, Faculdade de Ci\^encias da Universidade de Lisboa, Campo Grande, C8, 1749-016 Lisbon, Portugal
\and Oskar Klein Center, Roslagstullsbacken 21, 106 91 Stockholm, Sweden
\and CRAL, Observatoire de Lyon ; CNRS, UMR 5574 ; ENS de Lyon
\and Universit\'e de Lyon, F-69622, Lyon, France; Universit\'e Lyon 1 
}
 % end \institute

\offprints{balland@lpnhe.in2p3.fr}

\date{Received; accepted} \titlerunning{The ESO/VLT 3rd year Type Ia supernova data set from the SNLS}
\authorrunning{Balland \etal}

\abstract
  {}
  %Aim
  {We present 139 spectra of 124 Type Ia supernovae (SNe~Ia) that were
    observed at the ESO/VLT during the first three years of the
    Canada-France-Hawa\"{\i} Telescope (CFHT) Supernova Legacy Survey
    (SNLS). This homogeneous data set is used to test for redshift
    evolution of SN~Ia spectra, and will be used in the SNLS 3rd year
    cosmological analyses.}
  %Method
  {Spectra have been reduced and extracted with a dedicated pipeline
    that uses photometric information from deep CFHT Legacy Survey
    (CFHT-LS) reference images to trace, at sub-pixel accuracy, the
    position of the supernova on the spectrogram as a function of
    wavelength. It also separates the supernova and its host light in
    $\sim $60\% of cases. The identification of the supernova
    candidates is performed using a spectrophotometric SN~Ia model. }
  %Result 
  {A total of 124 SNe~Ia, roughly 50\% of the overall SNLS
spectroscopic sample, have been identified using the ESO/VLT during
the first three years of the survey. Their redshifts range from
$z=0.149$ to $z=1.031$. The average redshift of the sample is
$z=0.63\pm 0.02$. This constitutes the largest SN~Ia spectral set to
date in this redshift range. The spectra are presented along with
their best-fit spectral SN~Ia model and a host model where
relevant. In the latter case, a host subtracted spectrum is also
presented. We produce average spectra for pre-maximum, maximum and
post-maximum epochs for both $z<0.5$ and $z\geq0.5$ SNe~Ia. We find
that $z<0.5$ spectra have deeper intermediate mass element absorptions
than $z\geq 0.5$ spectra. The differences with redshift are consistent
with the selection of brighter and bluer supernovae at higher
redshift.

  } {} 

\maketitle

\section{Introduction}

Since the direct detection of the accelerated expansion of the
universe 10 years ago \citep{Riess98b, Perlmutter99}, constraining the
equation of state of the dark energy component responsible for this
acceleration has been a major goal of observational cosmology. Type Ia
supernovae (\Iae\ hereafter) samples have been gathered at low and
high redshift and extensively used for this purpose.  When combined
with other probes, the picture of a universe dominated by dark energy
emerges \citep{Tonry03, Astier06, Riess07, Wood-Vasey07, Kowalski08}.

Over the past five years\footnote{The SNLS started in June 2003.}, the
Supernova Legacy Survey (SNLS) has gathered more than 1000 light
curves of SN~Ia candidates on the Canada-France-Hawa\"{i} telescope
(CFHT) using {\sc megacam} \citep{Boulade03}, thanks to a rolling
search technique for discovery and photometric follow up of \Iae\ in
four 1 square degree fields \citep{Astier06}. Spectra of a subset
(about half) of these \Iae\ candidates have been observed on various 8
to 10-m class telescopes (VLT, Gemini N and S, Keck I and II).  About
50\% of spectroscopically observed SN candidates were observed at the
VLT.

In this paper, we present the VLT \Ia\ spectral set for the first
three years of operation of the SNLS. The non \Ia\ spectral set,
together with a description of the ``real-time'' operations and
procedures will be presented elsewhere \citep{Basa08}.  The spectra
shown here were obtained in the period running from June 1st, 2003 up
to July 31st, 2006, as part of two large VLT programs \footnote{These
  dates correspond to the first three years of the SNLS.}. The spectra
have been analysed using novel techniques for extraction and
identification which have been described in detail elsewhere
\citep{Baumont07,Baumont08}.  For each spectrum we provide a redshift
estimate.  The identification of \Iae\ relies in part on human
judgement, using the SALT2 spectral template of \citet{Guy07} as a
guide \citep[see][]{Baumont08}.

The SN spectra presented here will be used, along with \Ia\ spectra
obtained at Gemini and Keck telescopes, to build the 3rd year SNLS
Hubble diagram. Our primary goal is that the type and redshift of the
\Iae\ used for cosmological analysis are secure. In this paper, we
consider two classes of events: secure \Iae\ (``SN~Ia'') and probable
\Iae\ (``SN~Ia$\star$''). Studying the statistical properties of these
two classes, in order to assess the validity of using SN~Ia$\star$
events together with \Ia\ events, is therefore a goal of this paper.

\Ia\ spectra are a rich source of physical information about their
progenitor history and environment.  The possibility of evolution
among \Ia\ populations at low and high redshifts has been the subject
of considerable attention in recent years, as more and more \Ia\
spectral sets become available
\citep{Garavini07,Bronder08,Foley08,Blondin06,Ellis08}. Recently,
evidence has even been found for a demographic evolution among SN~Ia
populations, resulting in higher stretch, more luminous SNe~Ia at
higher redshift \citep{Howell07,Sullivan09}. Using \Ia\ spectra to
compare their physical properties at low- and high-redshift is
therefore a useful cross-check when using \Ia\ to constrain the
expansion history of the universe. In this paper, we take advantage of
the large number of high quality spectra obtained at the VLT to build
average composite spectra at various phases with respect to maximum
light for $z<0.5$ and $z\geq 0.5$. We also compare our average spectra
to composite spectra obtained in a similar way with different data
sets \citep{Ellis08,Foley08} and discuss the significance of the
differences found in terms of possible evolution or selection effects.

A plan of the paper follows. In Section \ref{obs}, we briefly describe
the SNLS photometric survey and the VLT spectral observation programs.
In Section \ref{datared}, we summarise the main steps of the data
reduction and spectrum extraction. We detail our identification
procedure and classification scheme in Section \ref{snid}. In Section
\ref{results}, the spectra are individually presented. Composite
spectra at $z<0.5$ and $z\geq 0.5$ are built in Section \ref{sec:composit}.
In Section \ref{discussion}, we discuss our sample in the light of
other existing \Ia\ spectroscopic data. Concluding remarks are made in
Section \ref{conclusion}.

\section{SNLS Observations}
\label{obs}

\subsection{The SNLS imaging survey}

The SNLS is composed of an imaging survey devoted to the detection and
the photometric follow up of SN candidates, and a spectroscopic survey
of a sample of the detected candidates, prioritised for spectroscopy
on various telescopes. The imaging survey ran from June 2003, after a
period of pre-survey, until June 2008. It was based on the Deep survey
of the Canada-France-Hawaii Telescope Legacy Survey (CFHT-LS)
(amounting to half of the 474 nights allocated to the CFHT-LS). Full
details of the survey can be found elsewhere \citep{Astier06}. In
brief, SNLS observed 4 fields (D1-4) every 3--5 nights during
dark/grey time in the $griz$ filters, each field followed for 5-6
lunations per year.  Around 1000 well sampled multi-colour light
curves of SN~Ia candidates have been obtained up to $z\sim 1.1$.

\subsection{Spectroscopic follow-up}

Spectroscopy of SNLS SN candidates was performed on several 8 to 10-m
class telescopes in both hemispheres, namely the VLT, Gemini-N and S,
Keck I and II.  Almost 50\% of SNLS candidates identified as certain
or probable \Ia\ were spectroscopically observed at the VLT.
\citet{Howell05} and \citet{Bronder08} describe the SNLS first three
years of Gemini spectral data (up to May 2006), while \citet{Ellis08}
present 36 high signal-to-noise ratio (S/N) \Ia\ spectra obtained at
Keck. In this paper, we focus on spectra taken at the VLT on Cerro
Paranal.

Candidate selection for spectroscopic follow-up was based on the
multi-band photometry procedure of \citet{Sullivan06a}.  This
selection was performed on the rising part of the light curves,
routinely available thanks to the rolling search strategy (see e.g.,
\citealt{Perrett09}).  Candidates were generally sent for spectroscopy
at, or slightly after, maximum light, which optimised the time budget
allocated for spectroscopy. We triggered a target on VLT every three
to four days during dark and grey time.

During the first large program (2003-2005), we performed long slit
spectroscopy (LSS) of SN candidates on FORS1 for a total of 60 hours
of dark/grey time per semester. During the second large program
(2005-2007), we observed using both FORS1 and FORS2 with the standard
collimator in LSS and multi object spectroscopy (MOS)
mode\footnote{Only eight candidates from SNLS3 were observed in MOS
  mode and have been identified as \Ia\ in real-time. As the MOS mode
  is currently not supported by our new extraction pipeline (see
  \citealt{Baumont08}), we do not include them in here. Note also that
  only 3 SNe~Ia were observed on FORS2 in the period covered in this
  study.}.  Most observations were carried out with the 300V grism,
along with the GG435 order-sorting filter.  Grism 300V was chosen to
optimise spectral resolution, spectral coverage and high enough S/N
for an unambiguous identification, even for the faintest candidates of
our survey ($i'_{AB}\sim 24$).  Moreover, using the 300V grism for
high redshift SNe allows us to study the interesting rest frame UV
region of the spectra. The pixel scale is $0.2''$ along the spatial
axis and 2.65 $\AA$ along the dispersion axis. At 5000 $\AA$, the
resolution limit with this setup is $\approx$ 11 $\AA$.  The
efficiency of the 300V grism peaks around 4700 $\AA$ at a level of
$\approx 80\%$.  The 300I grism along with the OG590 order-sorting
filter was sometimes used for the faintest ($z> 0.8$) SNe.  Being
typically more distant, and due to strong sky emission and fringing
beyond 6000 $\AA$, spectra obtained in this way have a much lower S/N
than those acquired with the 300V grism.

The slit width was chosen according to the following rule of thumb:
``slit width $\approx$ seeing + 0.2$''$", as a compromise between
observing most of the flux from the targeted candidate and limiting
the sky background flux. An air mass $\chi<1.6$ was required for each
spectroscopic observation. A ``blind offset" technique was used to
target the candidate, using a bright star located within $<1'$ of the
target and then offsetting the telescope to position the slit onto the
candidate. When possible, the slit position was chosen to observe both
the SN and its host. Differential slit losses were corrected by a
Longitudinal Atmospheric Dispersion Corrector (LADC). Residual losses
are taken into account with the recalibration procedure described in
Section \ref{sec:idsalt2}.

All spectra were acquired in Service Observing mode.  With a limiting
magnitude of $i'_{AB}\sim 24$, 3-4 exposures of 750 or 900s were taken
for each candidate, with small offsets along the spatial axis ($Y$;
the dispersion axis $X$ is horizontal with our setup). Thanks to the
regular time sampling of the rolling search, it was possible to
acquire most of candidates around or slightly past maximum light.

\section{Data processing}
\label{datared}

Data reduction and spectral extraction were performed in two separate
ways. A quick, ``real-time'' analysis (within a day of acquisition)
was used to assess the type and redshift of the candidate
\citep{Basa08}, an essential task for efficiently allocating other
candidates to the various telescopes.  In parallel, we developed tools
``off-line'' to cleanly extract the SN from its host.  A dedicated
pipeline, PHASE (PHotometry Assisted Spectral Extraction), was used
for the final reduction and extraction \citep{Baumont07,Baumont08}.
Full details can be found in these papers; we give only a brief
description here. All extractions presented in this paper used the
PHASE technique.

The PHASE {\em reduction} technique improves over the real-time
reductions in refinements of the master flat-fields, the dispersion
relation, and in the sky estimation. As an example, the 2D dispersion
relation is modeled by a fourth order polynomial in $X_i$ (and of 2nd
order in $Y_i$ and $XY_i$) to further reduce the residuals. For flux
calibration, we build a single average response curve (one for UT1 and
one for UT2 as FORS1 moved from UT1 to UT2 in June 2005) from previous
individual standard star observations.  We prefer using a well
controlled average response for the whole set, rather than using a
response built from standard star observations of a different night.
This is at the expense of absolute flux calibration, as we average out
sky transmission variations from night to night, but permits a more
robust estimation of the sensitivity function near the blue edge of
the order sorting filter.

PHASE {\em extraction} uses photometric information on the SN location
and host brightness from the deep reference images ($i'_{AB}\sim
27-28$) used for building the SN light curves\footnote{We use $g_M$,
  $r_M$ and $i_M$ for spectra obtained with grism 300V
  \citep{Baumont08}}. This allows us to accurately trace the SN
position along the dispersion axis on the two-dimensional spectrogram.
Moreover, we build a multi-component model of the galaxies present in
the slit (including the host, if resolved) by measuring the spatial
photometric profiles of these galaxies on the deep stacked reference
images projected along the slit direction. We then add a SN component,
modeled as a Gaussian of width equal to the seeing of the
spectroscopic observation. The location of the SN is accurately known
from the light curves. The flux of each component (SN + host and
potentially other galaxies in the slit) is a free parameter, the sum
of the profile fluxes being normalised to unity. Such a model is built
for each column (the dispersion direction is horizontal with our
setup) and fluxes assigned to each component in each column $i$ are
estimated by a $\chi^2$ minimisation where:
\begin{equation}
\label{eq2} \chi^2=\sum_j\Big(\frac{F_i(\Yp_j)-{\mathcal
M}_i(\Yp_j)}{\sigma_j}\Big)^2.
\end{equation} Here, $\sigma_j$ is the error associated with the
column pixel data, ${\mathcal M}_i$ is the multi-component model in
column $i$ and $F_i(\Yp_j)=F(\Xp_i,\Yp)$ is the column flux on the 2D
spectrogram. We distinguish three types of galaxies:
\begin{itemize}
\item PSF: unresolved, point-like galaxies
\item EXT: extended, but regularly shaped profiles, e.g. ellipticals
\item Mix: extended, but irregularly shaped profiles, e.g. galaxies
  with spiral arms.
\end{itemize}

We note that the latter two cases are not an accurate indication of
the physical morphology of the host galaxy, but instead just model the
spatial profile of the host that enters the slit.

In the first case, the galactic component of the model is a Gaussian
of width equal to the seeing of the deep reference observation. The
seeing variation with wavelength is estimated from standard star
spectra as a power law of index $\sim -0.3$, in good agreement with
\citet{Blondin05} measurements on FORS1 spectra. In the second case,
the model is the ``bolometric'' spatial profile (the sum of the
galactic profile in all observed filters) as measured on the {\sc
  megacam} combined deep reference images. For the third case, we use
a mixture of a Gaussian with width equal to the spectroscopic seeing
to model the core and the photometric spatial profile to model the
extended arms. From a pure algorithmic point of view, this latter case
is equivalent to have two distinct galaxies, a point-like source and
an extended one.

Host models used to extract our VLT spectra are about equally divided
into EXT and PSF models ($\sim 30$\% each), with only a few percent of
cases being Mix models. In the remaining $\sim 40\%$ of cases, no
separate extraction of the SN and the host was possible.

PHASE uses a set of default parameters to select the correct host
model and make the extraction as automatic as possible. These
parameters include cuts on flux, galactic compactness, extension
minimum level, colour variation between the centre and the extended
part of the host (identifying ``Mix'' host types), and a minimum value
of the SN to host centre distance to perform a separate extraction
(usually 0.15$''$, a bit less than one pixel). These default
parameters allow an automatic extraction of most spectra, though the
parameters can be adjusted for specific cases. PHASE performance and
limitations have been discussed in \citet{Baumont08} and we refer the
reader to that paper.

The main hypotheses in using PHASE are that 1) the PSF is a Gaussian
of width equal to the seeing; 2) the coordinates of the SN are
accurate; 3) CFHT-LS reference images and VLT spectrograms have
comparable seeings; 4) no flux of the SN is present in the reference
image. Any deviation from these assumptions result in weak flux
losses, increased noise, and contamination of a SN spectrum by its
host. PHASE performs well for most spectra encountered. In particular,
it succeeds at recovering the SN from the host, even in the case of a
SN located close to the host centre (typically $\lesim 1$ pixel):
strong correlations between the host and the SN are often unavoidable
in standard extractions. Even for ``non favourable'' cases, such as
sub-pixel SN/host separations, both component spectra are recovered
and are essentially non correlated.  This is a remarkable feature of
our pipeline, as most SN spectral extractions are hampered by host
contamination in these cases.  Nevertheless, if the SN is too close to
the host centre (separation less than one fifth of the seeing), no
separate extraction is possible. In that case, a host spectral
template is used at the stage of identification to estimate the host
contamination (see Section \ref{sec:hostsub}).

A comparison between PHASE and standard extractions has been done and
illustrated on a few examples in \citet{Baumont08}.  A major
difference is that, in PHASE, we do not re-sample the data until the
final step of flux calibration. This avoids introducing correlations
across pixels and allows us to trace the statistical noise along the
reduction and extraction procedure. For the same reason, we also avoid
rebinning the data at a constant wavelength step, as is done in most
standard spectroscopic reduction pipelines. As a consequence, the
final PHASE calibrated spectrum has unequal steps. We have checked
that the statistical noise is properly propagated with PHASE by
computing $<S/N>$, the S/N per pixel averaged over the whole spectral
range, and $<S/\sigma>$, the signal-to-r.m.s. ratio. Here, $\sigma$ is
the standard deviation of a group of measurements around a low-order
polynomial fit. We find relatively good agreement between these two
quantities for PHASE extracted spectra.

\section{Spectral analysis}
\label{snid}

\subsection{Redshift determination}

Where possible, a redshift is obtained from the host galaxy spectrum
using characteristic emission or absorption lines, yielding a typical
uncertainty of 0.001 on the redshift \citep{Baumont08}. When no lines
are detected, the redshift is determined from the fit of a model to
the SN spectrum (see Section \ref{sec:idsalt2}). The typical redshift
uncertainty is then $\delta z \approx 0.01$, due to the diversity of
ejecta velocities among SNe~Ia, (e.g. \citealt{Hachinger06}).  In
about 80\% of cases, the redshift is obtained from host emission
and/or absorption lines. In SNLS we have two independent
identification pipelines. The redshifts (as well as identifications,
see below) have been carefully cross-checked using the two-dimensional
data to match host lines.  Checking the corresponding noise map,
``bright'' spots visible on the 2D frames are easily identified as
true emission lines or cosmic/sky subtraction residuals.

\subsection{Identifying SNe~Ia}
\label{sec:idsalt2}

To identify SNe~Ia, we use a minimisation procedure using the SALT2
spectral template of \citet{Guy07} with a combined fit of the light
curves and the spectrum consistently performed \citep{Baumont08}. As
the training sample of the SALT2 model only contains SN~Ia spectra and
light-curves, it does not allow a direct identification of non-SN~Ia
objects. However, the best-fitting parameter values obtained when
fitting a non-SN~Ia with SALT2 are in themselves an indirect
indication of the SN type (see Section \ref{sec:nonIa}).  In addition,
a $\chi^2$ template-fitting code \citep{Howell05} has been used for
cross-check.  Both techniques use varying levels of human judgement.

The main output parameters of the SALT2 fit are 1) the light curve fit
parameters $x_0$ (overall normalisation), $x_1$ (light curve shape),
and colour $c$, and 2) spectroscopic fit parameters: the host fraction
$f_{gal}$ in the model when relevant, and recalibration parameters.
The latter enter a recalibration function applied to the photometric
model in order to fit the spectrum and account for possible errors in
flux calibration \citep{Guy07}.  This function is a polynomial of
order $n$, with coefficients $\gamma_n$ inside an exponential (to
ensure positivity).  We usually only use two recalibration parameters:
an overall normalisation $\gamma_0$ and a first order coefficient
$\gamma_1$ (tilt applied at rest frame 4400 $\AA$), adding higher
order corrections for a few objects.  The $x_1$ parameter is a light
curve shape parameter that can be converted into a stretch factor $s$
or $\Delta m_{15}$ parameter of \cite{Phillips93}; see \citet{Guy07}.
The colour $c$ parameter is defined as the difference between
$(B-V)_{SN}$ and the average $<B-V>$ value at maximum light for the
whole training sample of SALT2.

There are some advantages in using SALT2 as a tool for identification.
First, the fit of the light curve yields the date of maximum light.
The phase $\phi$ of the spectrum (the number of rest frame days
between the date of $B$-band maximum and the date of acquisition of
the spectrum) is accurately known, usually within a fraction of day.
This alleviates possible degeneracies between SN types, such as
between pre-maximum Type Ic supernovae (SNe~Ic) and post-maximum
SNe~Ia, whose spectra show similarities. Second, the training set of
the SALT2 model is built from a large collection of spectra and light
curves from local and distant SNe~Ia. These latter include SNLS SNe
themselves, added to the training sample once identified.  Third,
using a model instead of a set of spectrum templates (as is typical of
standard identification techniques) alleviates the problems due to
unavoidable incompleteness in phase and wavelength coverage of the
template libraries: all SNe are treated on an equal footing. Note also
that only SNe typed as SN~Ia (as opposed to SNIa$\star$) are included
in the training set of SALT2. As an erroneous typing of a SN~Ic as a
SN~Ia is very unlikely, this limits the chance of polluting the
training sample. Even in such unlikely case, SALT2 being a model, it
is robust against the inclusion of a non-SN~Ia object.

Once the SALT2 fit is performed, the identification is guided by the
best-fitting spectral parameter values. As the phase is fixed by the
light curve fit, the spectral fit implicitly uses the photometry.
However, we will not classify a SN candidate as a certain \Ia\ if we
do not get an adequate fit of the spectrum, even if the light curve
fit is good. On the contrary, a convincing spectral fit is sufficient
for an \Ia\ identification, even if the light curve fit is poor.  Our
goal is to obtain a clean, spectroscopically confirmed, SN~Ia sample.

\subsection{Host galaxy subtraction}
\label{sec:hostsub}

Spectroscopic identification of distant SNe is challenging. One of the
key issues discussed in the literature is host contamination. Low S/N
spectra are common, either due to high redshift or because sometimes
candidates are observed at a late phase (up to a few weeks past
maximum) due to telescope scheduling.  Several techniques have been
developed to improve host--SN separation.  Standard techniques involve
template $\chi^2$ fitting (e.g., \citealt{Howell05,Lidman05}) once the
spectrum has been extracted, and/or cross-correlation methods, such as
the SuperNova IDentification (SNID) algorithm \citep{Blondin07}, used
for the first two years of the ESSENCE project
\citep{Matheson05,Foley09}. \citet{Blondin05} proposed a PSF
deconvolution technique that separates the two components, provided
that the spatial extension of the Gaussian profile is very different
for the SN and for its host.  \citet{Zheng08} use a principal
component analysis decomposition, combined with template fitting to
assess the level of host contamination.

SNLS has a key advantage in that deep photometric $ugriz$ data are
available for the host galaxy that can assist with SN--host
separation. \citet{Bronder08} used deep $i_M$-band photometry to
estimate host contamination.  More recently, \citet{Ellis08} used the
multi-colour SNLS photometry to do a more sophisticated estimate of
host contamination in extracting Keck \Iae\ spectra.  This approach is
very efficient at removing the host contribution.  Though a wide range
of host templates are used and are colour-matched to the host galaxy
photometric data, it nevertheless has the drawback of using synthetic
templates. By contrast, PHASE, while relying on photometric priors,
does not use fixed host templates \citep{Baumont08}.

PHASE measures the photometric profile of the host and, once combined
with a PSF model for the SN placed at the SN position, estimates the
flux of each component in each pixel along the SN trace on the
spectrogram using a $\chi^2$ minimisation procedure.  Inspection of
the residual image after extraction shows that this technique is
efficient provided the centre of the two components is separated by
more than $\sim 0.15 ''$ and the seeings of the reference and
spectroscopic images are similar.  \citet{Baumont08} have shown that
fitting a SN spectrum extracted separately from its host, with a
SN+host SALT2 model, yields a small amount of residual galactic
contribution, i.e.  less than 10\% on average. This is a good hint
that our component separation performs well.  Note that our technique
is designed within the framework of a rolling search (very deep
reference images are required) and may not be well suited for other
search techniques.

When it is not possible to extract the two components separately, a
$\chi^2$ fit of the full (SN+host) spectrum using SALT2 is done. SALT2
has been adapted to fit a galaxy template in addition to the SN model
whenever a separate extraction of the SN from its host was not
possible with PHASE \citep{Baumont08}.  These galaxy templates include
\citet{Kinney96} types, as well as a series of template spectra
synthesised using PEGASE2 \citep{Fioc97,Fioc99} ranging from
ellipticals to late-type spirals.  Templates are ordered in a regular
sequence from red to blue spectra and the best-fit model is
interpolated between two contiguous templates in the sequence. We do
not add emission lines to the PEGASE templates: whenever PEGASE
templates are used to model the host contribution, residual emission
lines might be found in the host subtracted spectrum. This technique
is then essentially comparable to the PCA+$\chi^2$ fitting technique
used by \citet{Zheng08} to evaluate the host contamination in the
first season of the SDSS-II survey.

At high redshift and/or for SNe close to their host centre, host
subtraction remains a difficulty. We find a higher average host
fraction for probable SNe~Ia (SN~Ia$\star$ type) than for certain
SNe~Ia (SN~Ia type) spectra (see Section \ref{sec:nonIa}), which shows
that, as expected, significant host contamination can alter the
quality of the identification. A major improvement would be to have a
spectrum of the host. We are currently in the process of obtaining
``SN free'' spectra for those cases in which the subtraction of the
host failed. We plan to use those to improve the efficiency of the
host subtraction in the final 5-year SNLS sample.

\subsection{SN Classification}
\label{sec:classify}

We classify candidate spectra into six categories adapted from the
classification of \citet{Lidman05}. We define \Ia\ (certain SNe~Ia),
SN~Ia$\star$ (probable \Iae\ but other types, in particular SNe~Ic,
can not be excluded given the S/N or the phase of the spectrum),
SN~Ia\_pec (peculiar \Iae), SN? (possible SN of unclear type), SN~Ib/c
and SN~II.  In this paper, we only present SNe from the first 3
categories:

$\bullet$ A SN~Ia classification requires the presence of at least one
of the following features: \ion{Si}{\sc ii} $\lambda$ 4000 or
\ion{Si}{\sc ii} $\lambda$ 6150, or the \ion{S}{\sc ii} W-shaped
feature around 5600 $\AA$.  Where these features are not clearly
visible, a \Ia\ classification is still possible provided that the
following criteria are met:
\begin{itemize}
\item 1) The overall fit is visually good over the entire spectral
  range,
\item 2) The spectral phase is earlier than about +5 days. At about
  one week past maximum, it has been noted that SNe~Ia and SNe~Ic show
  strong similarities and confusion between types is possible
  \citep[e.g.,][]{Hook05,Howell05},
\item 3) No strong recalibration is necessary to obtain a
  good fit (typical flux correction less than 20\%). 
A large recalibration usually indicates that the candidate is {\em
    not} a SN~Ia, a fact that would also be reflected in unusual
  photometric parameters (very red or blue colour, i.e.  positive or
  negative $c$ respectively, very high or low $x_1$ value).
\end{itemize}

$\bullet$ Classifying a candidate spectrum as a SN~Ia$\star$ implies
that no typical \Ia\ absorption (Si or S) can be found but that the
overall fit is acceptable over a large spectral range and broad
features are well reproduced.  Spectra of low S/N, or spectra one week
(or more) past maximum fall into the SN~Ia$\star$ category unless
\ion{Si}{\sc ii} is clearly seen.

$\bullet$ A spectrum is classified as a SN~Ia\_pec when spectral
features characteristic of under- or over-luminous objects (e.g.
\citealt{Li01b}) are present.

While the SN~Ia category represents two thirds of our total sample,
the number of SNe identified as ``peculiar'' is very small, which may
reflect selection biases against such events \citep{Bronder08}.
Low-stretch, underluminous \Iae\ are very hard to identify
spectroscopically because of their dimness compared to their usually
bright host.  Moreover, low S/N spectra of peculiar \Iae\ at high
redshift may make them harder to identify than their low-redshift
counterparts. Finally, our SALT2-based identification is by essence
not well suited for handling peculiar SNe~Ia \citep{Guy07}.

As described above, all types presented in this paper have been
cross-checked independently using two techniques: PHASE/SALT2, and the
code developed by \citet{Howell05}. Difficult spectra were discussed,
on a case-by-case basis, until agreement was reached. In case of
disagreement, the most conservative typing was chosen. This procedure
makes the identification of the SNe~Ia presented in this paper
homogeneous to the identification of Gemini spectra of
\citet{Howell05} and \citet{Bronder08}. Note however that we do not
use here the confidence index (CI) classification of \citet{Howell05}.
As a guide, the SNe~Ia of the present paper correspond to CI=4 and 5
while our SNe~Ia$\star$ correspond to CI=3.

\section{Results}
\label{results}

\subsection{Individual spectra}

In this section, we present the spectra of the 124 identified \Iae\ of
the SNLS 3rd year VLT spectroscopic survey, along with their
identification as SN~Ia, SN~Ia$\star$ or SN~Ia\_pec. Only two objects
have been identified as SN~Ia\_pec (SNLS-03D4ag and SNLS-05D1hk).
Their properties are described individually in Section
\ref{subsec:pec}.  In the following, we include them into the SN~Ia
subsample.

Table \ref{tab:tab2} lists the SN~Ia and SN~Ia$\star$ spectra and
their observing conditions. In some cases, several spectra of the same
candidate were taken due to poor conditions at the telescope, or an
insufficient S/N for a secure identification. The asterisk in the
column $D_{sep}$ denotes cases for which a separate extraction of the
SN from the host component was not possible (including, for some
candidates, the absence of a detectable host).

Table \ref{tab:tab1} summarises the results of our identification for
each of the 139 spectra corresponding to the whole set of 124 SNe~Ia.
Name, type, redshift, redshift source, i.e. host (H) or SN (S), phase,
host type, fraction of host used in the best-fitting model, and
average S/N per 5 $\AA$ bin are given. This latter quantity is
computed on the host subtracted spectrum (when relevant) for each SN.
PEGASE best-fitting host model is given as a letter for the morphology
(E, S0, Sa, Sb, Sc, Sd) followed by a figure between parentheses
indicating the age (in Gyrs). When the best-fit is obtained for a
\citet{Kinney96} template, we indicate the two contiguous Hubble types
between which the best-fitting galaxy model is interpolated. Note that
a host component is always allowed (this includes a null template),
even when the SN spectrum has been extracted separately from its host.
Where the best-fit is obtained for a model with no host contribution,
the label 'NoGalaxy' is given.

\begin{figure*}[ht]
                    \begin{center}
\begin{tabular}[b]{cc}
                    \hspace{-5mm}
                    \vspace*{1cm}
                    \begin{minipage}[b]{10cm}
                    \centering
                    \includegraphics[width=10cm]{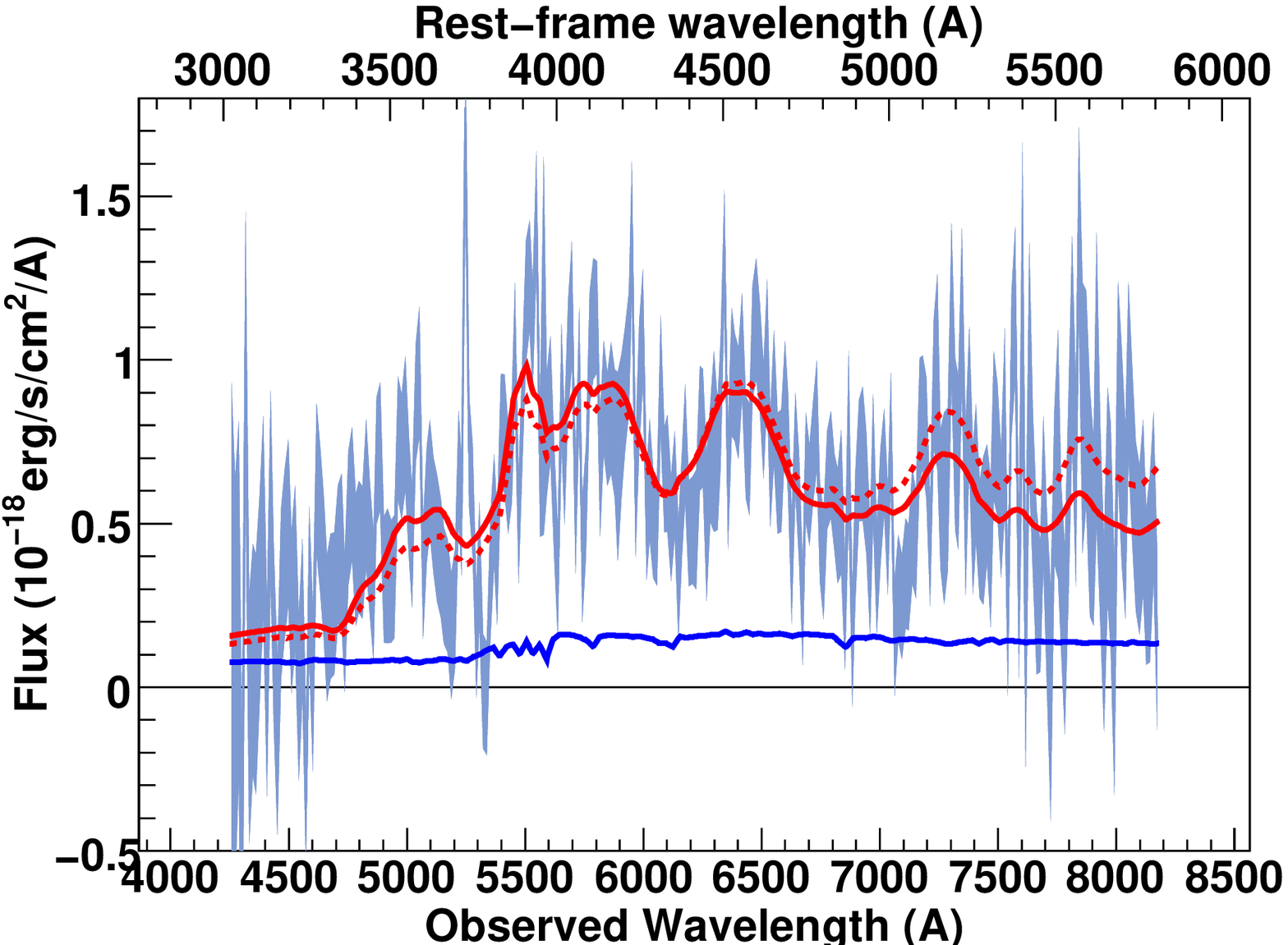}
                    \end{minipage}
                    &
                    \begin{minipage}[b]{10cm}
                    \hspace{-5mm}\includegraphics[width=10cm]{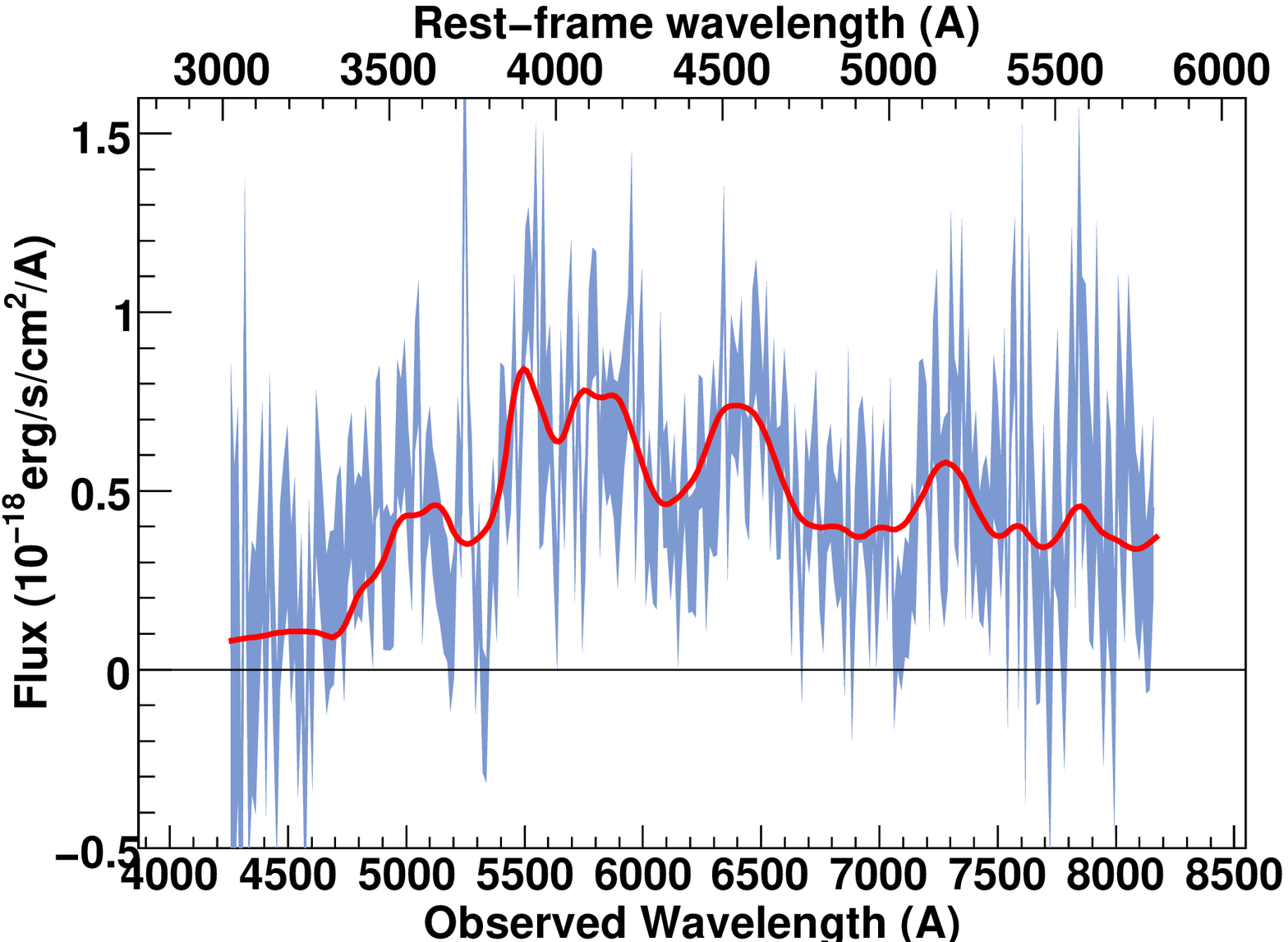}
                    \end{minipage}
                    \\
                    \end{tabular}
                    \caption{\label{fig:03D1ar270} SNLS-  03D1ar270: a SN~Ia$\star$ supernova at z=0.408. The spectrum phase is 5.3. A         E(7) host model has been subtracted.}
                    \end{center}
                    \end{figure*}

\begin{figure*}[ht]
                    \begin{center}
                    \begin{tabular}[b]{cc}
                    \hspace{-5mm}
                    \vspace*{1cm}
                    \begin{minipage}[b]{10cm}
                    \centering
                    \includegraphics[width=10cm]{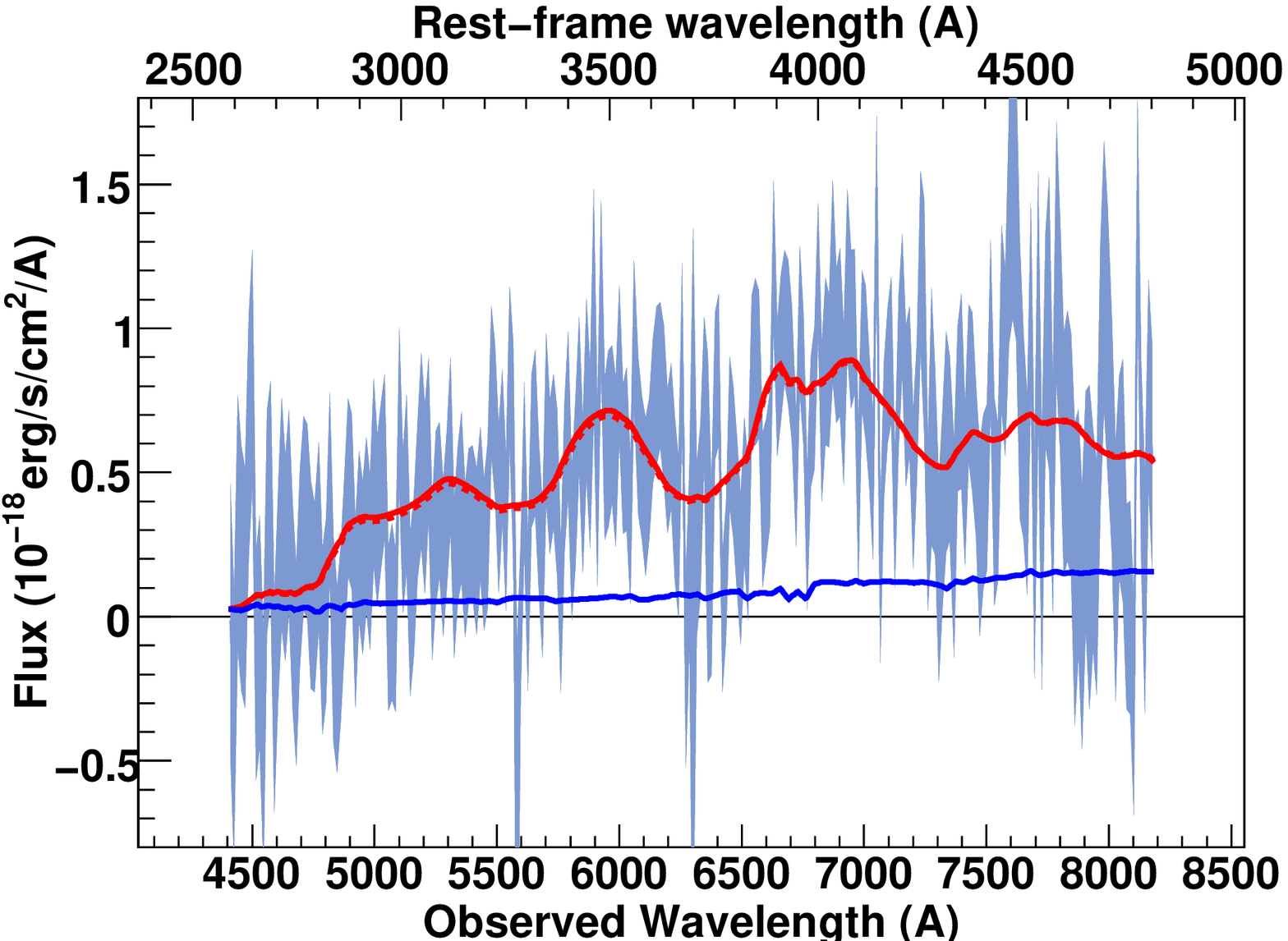}
                    \end{minipage}
                    &
                    \begin{minipage}[b]{10cm}
                    \hspace{-5mm}\includegraphics[width=10cm]{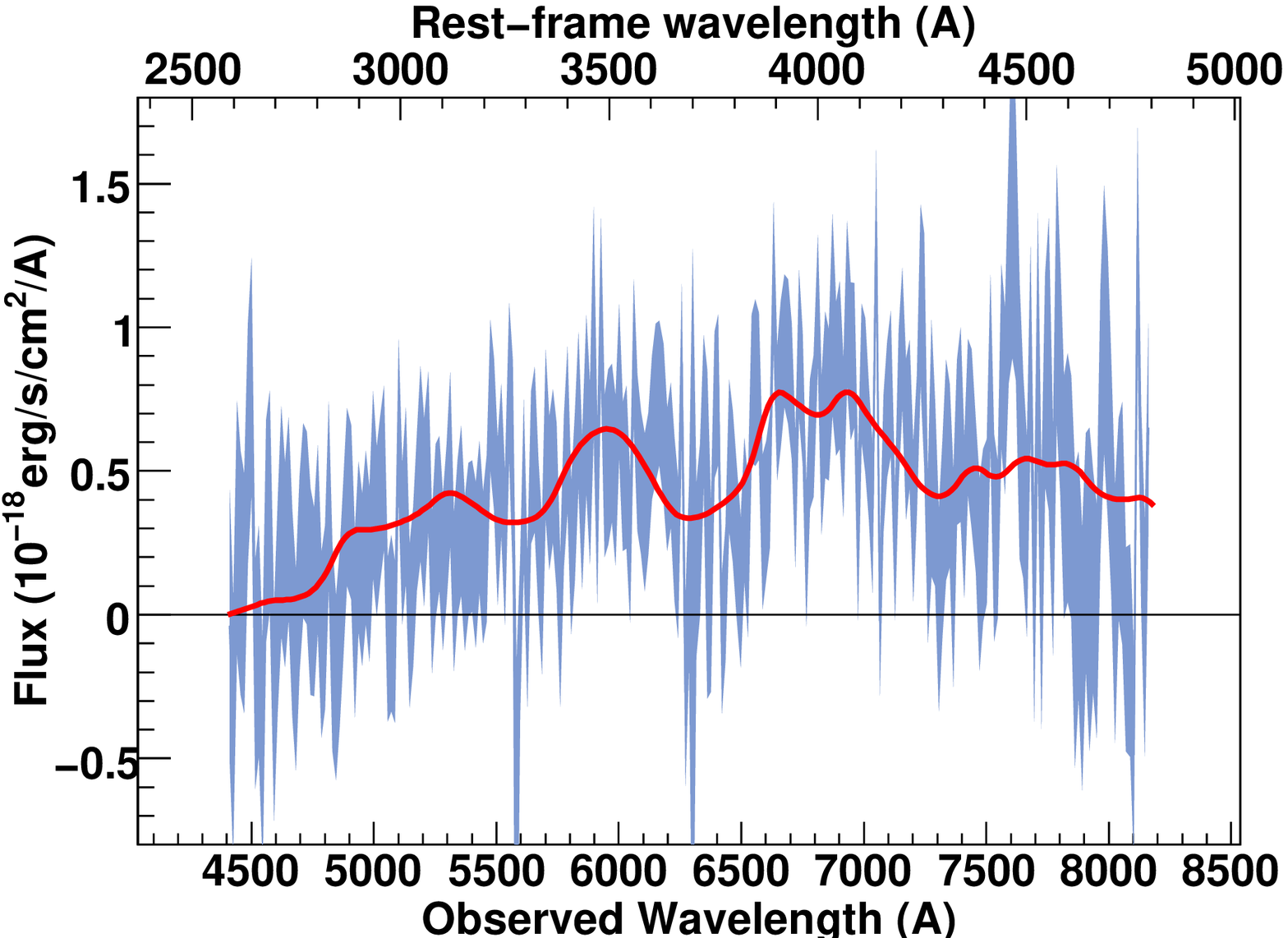}
                    \end{minipage}
                    \\
                    \end{tabular}
                    \caption{\label{fig:03D1bf269} SNLS-  03D1bf269: a SN~Ia$\star$ supernova at z=0.703. The spectrum phase is -2.7. A         E(3) host model has been subtracted.}
                    \end{center}
                    \end{figure*}

\begin{figure*}[ht]
                    \begin{center}
                    \begin{tabular}[b]{cc}
                    \hspace{-5mm}
                    \vspace*{1cm}
                    \begin{minipage}[b]{10cm}
                    \centering
                    \includegraphics[width=10cm]{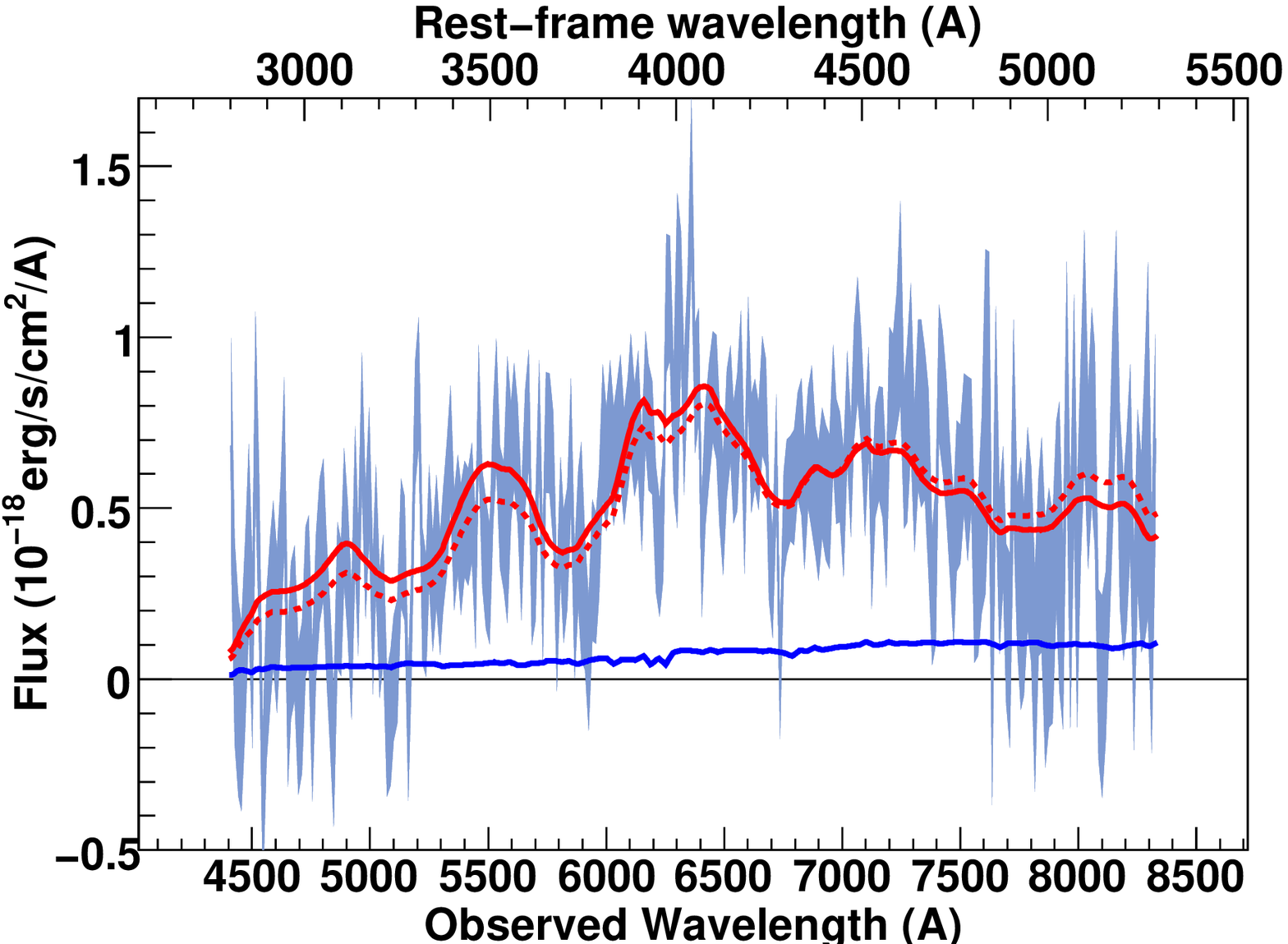}
                    \end{minipage}
                    &
                    \begin{minipage}[b]{10cm}
                    \hspace{-5mm}\includegraphics[width=10cm]{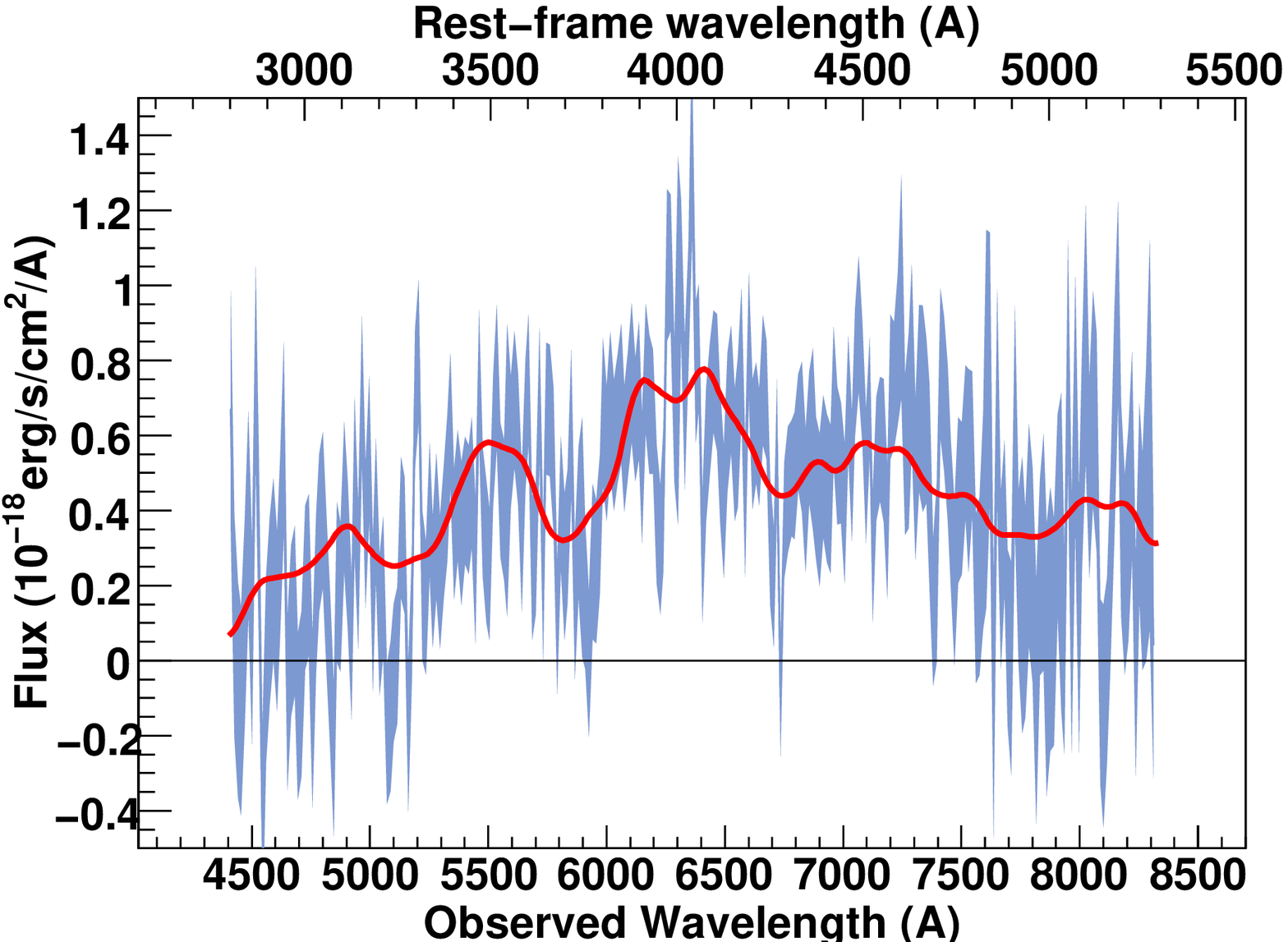}
                    \end{minipage}
                    \\
                    \end{tabular}
                    \caption{\label{fig:03D1bm272} SNLS-  03D1bm272: a SN~Ia$\star$ supernova at z=0.575. The spectrum phase is -5.1. A         E(3) host model has been subtracted.}
                    \end{center}
                    \end{figure*}

\begin{figure*}[ht]
                    \begin{center}
                    \begin{tabular}[b]{cc}
                    \hspace{-5mm}
                    \vspace*{1cm}
                    \begin{minipage}[b]{10cm}
                    \centering
                    \includegraphics[width=10cm]{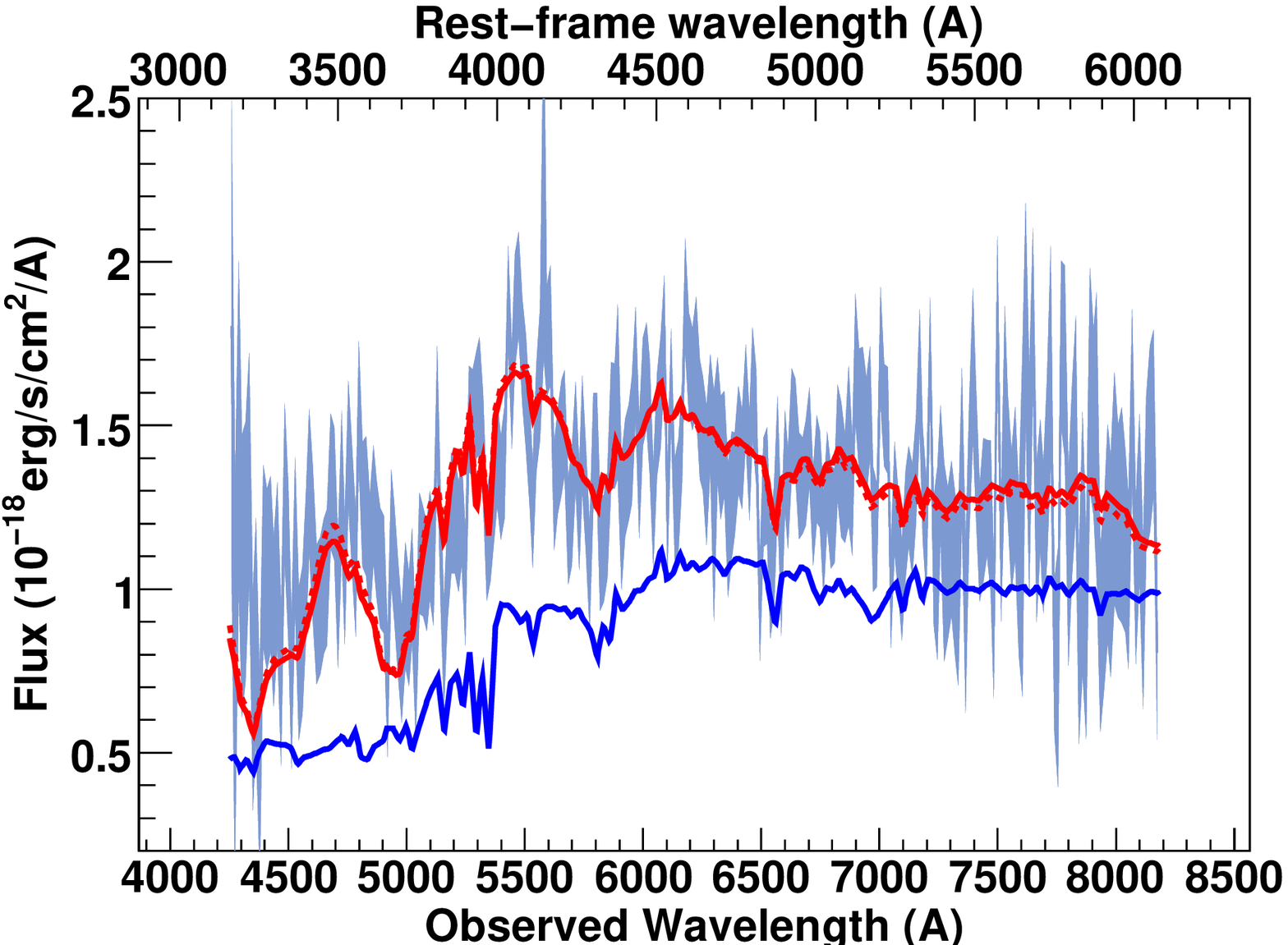}
                    \end{minipage}
                    &
                    \begin{minipage}[b]{10cm}
                    \hspace{-5mm}\includegraphics[width=10cm]{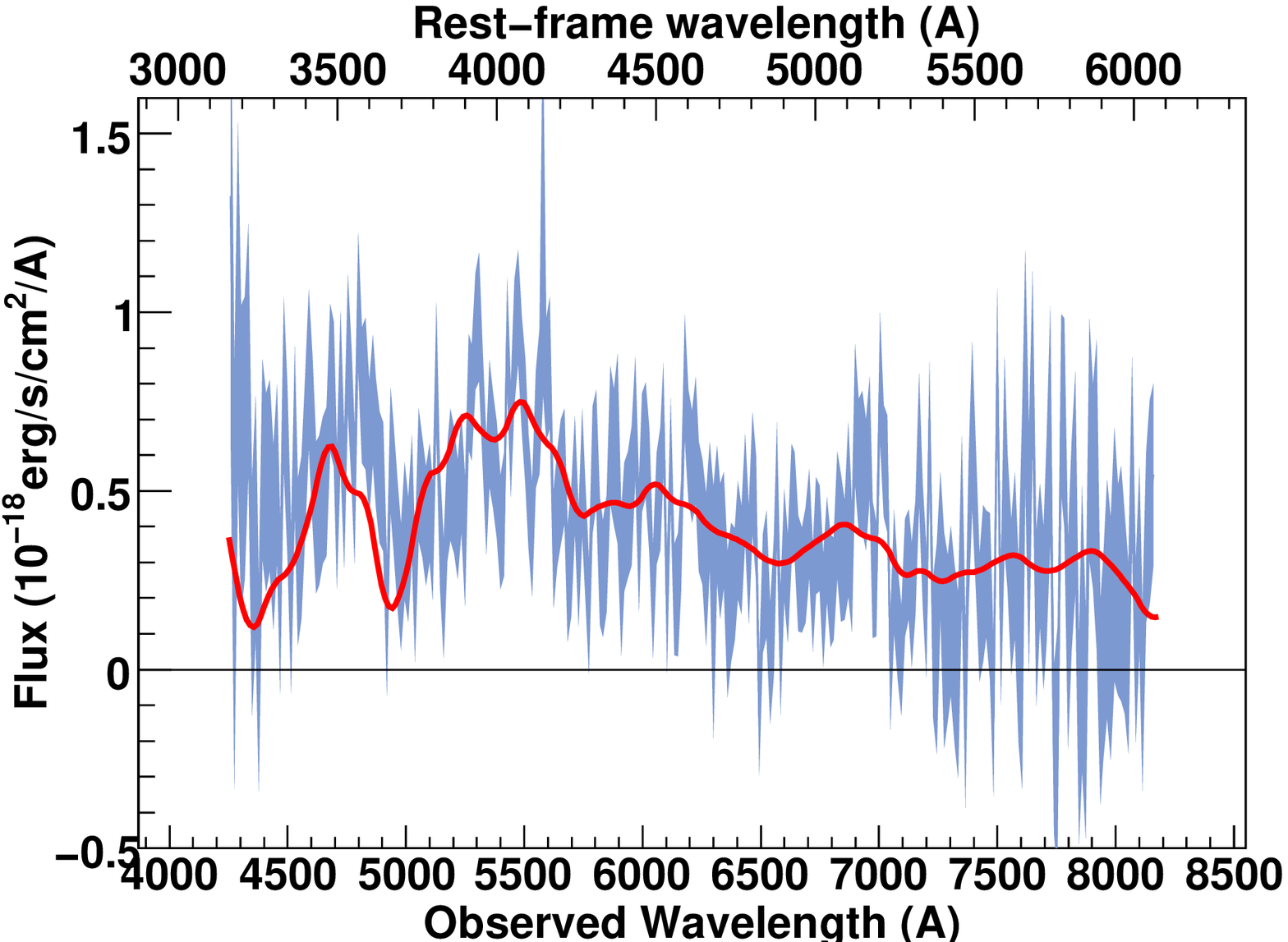}
                    \end{minipage}
                    \\
                    \end{tabular}
                    \caption{\label{fig:03D1bp270} SNLS-  03D1bp270: a SN~Ia$\star$ supernova at z=0.347. The spectrum phase is -7.5. A         E(2) host model has been subtracted.}
                    \end{center}
                    \end{figure*}

\begin{figure*}[ht]
                    \begin{center}
                    \begin{tabular}[b]{cc}
                    \hspace{-5mm}
                    \vspace*{1cm}
                    \begin{minipage}[b]{10cm}
                    \centering
                    \includegraphics[width=10cm]{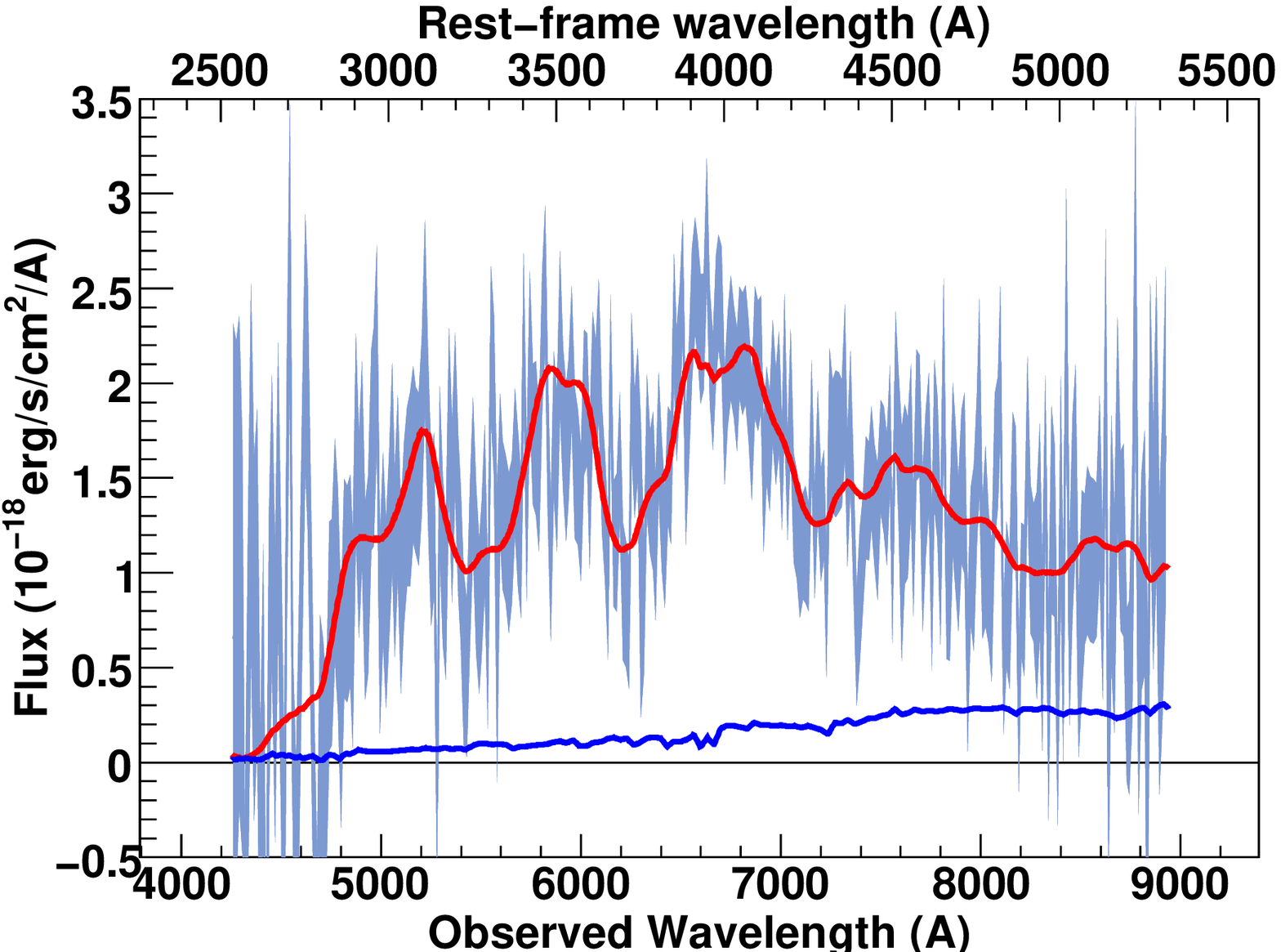}
                    \end{minipage}
                    &
                    \begin{minipage}[b]{10cm}
                    \hspace{-5mm}\includegraphics[width=10cm]{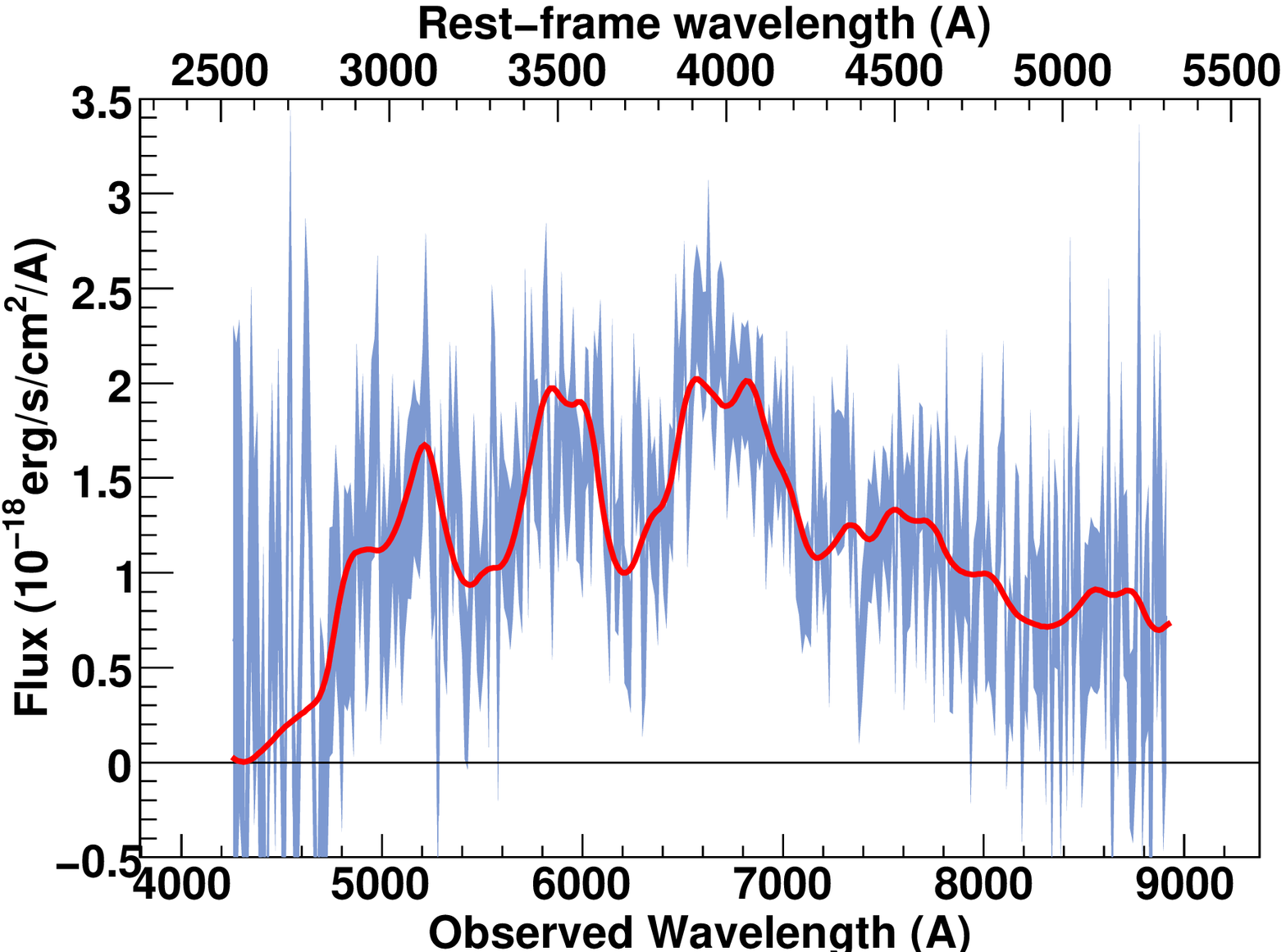}
                    \end{minipage}
                    \\
                    \end{tabular}
                    \caption{\label{fig:03D1co307} SNLS-  03D1co307: a SN~Ia supernova at z=0.679. The spectrum phase is -4.1. A         E(8) host model has been subtracted.}
                    \end{center}
                    \end{figure*}

\begin{figure*}[ht]
                    \begin{center}
                    \begin{tabular}[b]{cc}
                    \hspace{-5mm}
                    \vspace*{1cm}
                    \begin{minipage}[b]{10cm}
                    \centering
                    \includegraphics[width=10cm]{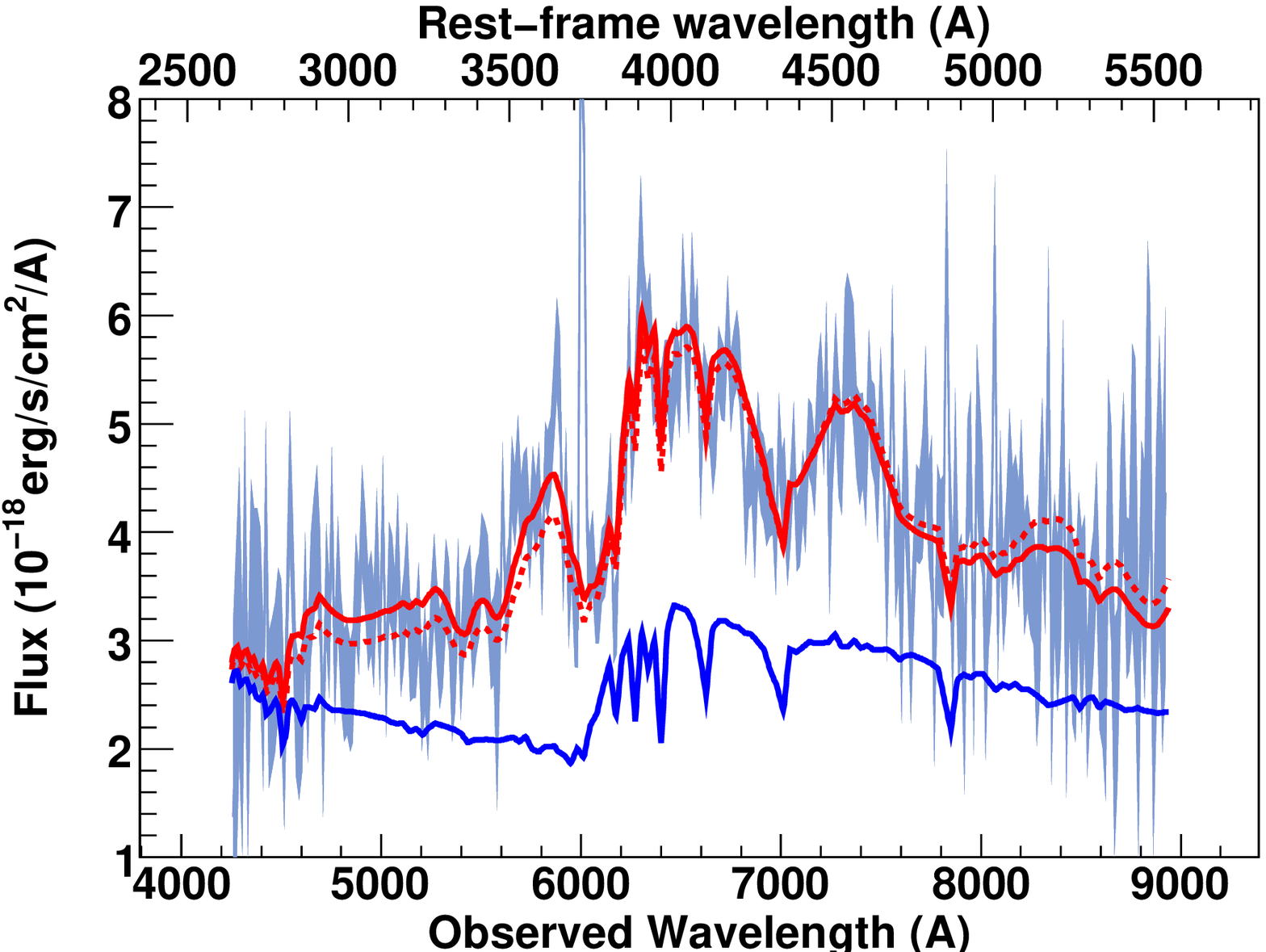}
                    \end{minipage}
                    &
                    \begin{minipage}[b]{10cm}
                    \hspace{-5mm}\includegraphics[width=10cm]{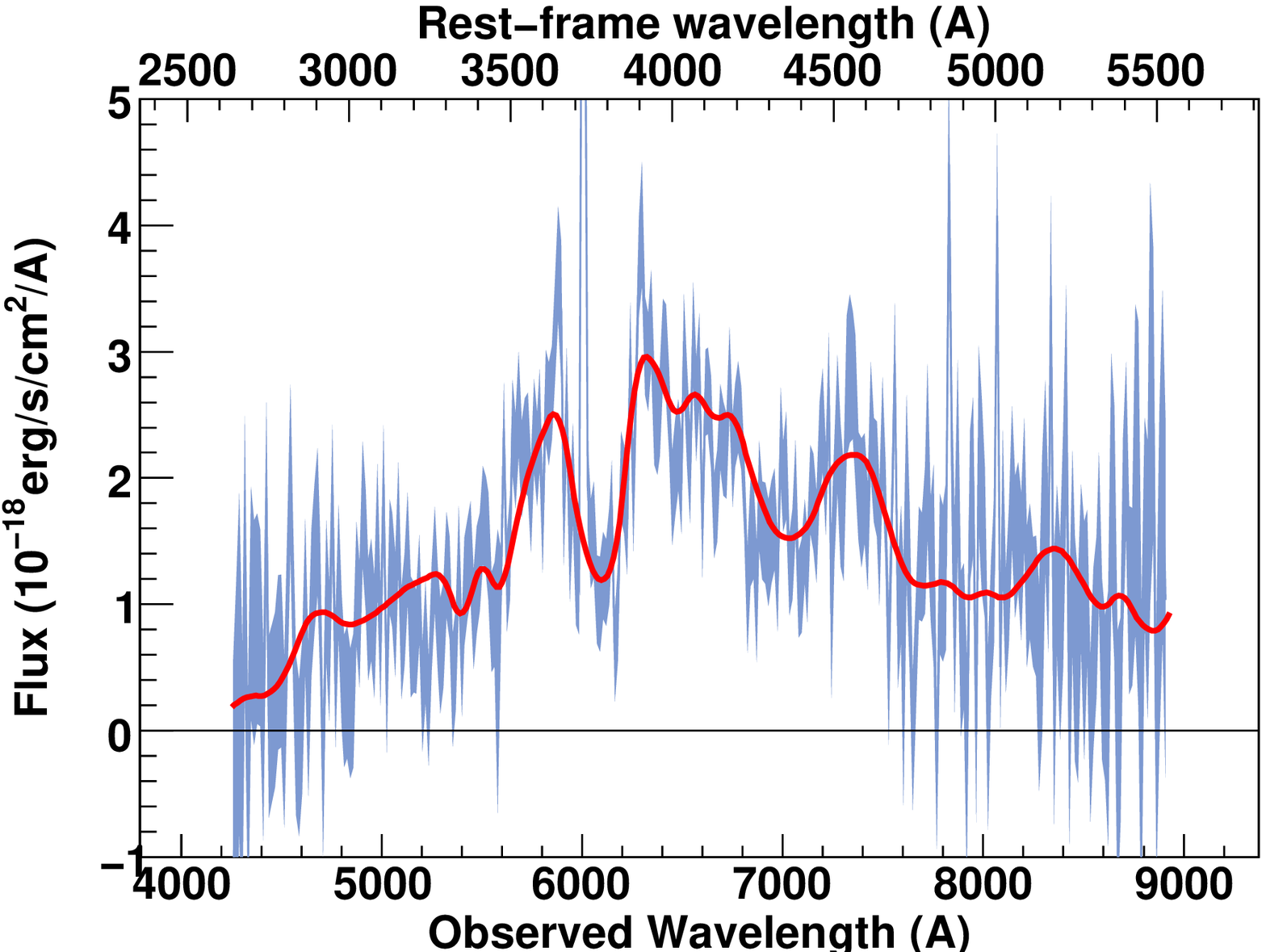}
                    \end{minipage}
                    \\
                    \end{tabular}
                    \caption{\label{fig:03D1dt333} SNLS-  03D1dt333: a SN~Ia supernova at z=0.612. The spectrum phase is 5.1. A        Sc(4) host model has been subtracted.}
                    \end{center}
                    \end{figure*}

\begin{figure*}[ht]
                    \begin{center}
                    \begin{tabular}[b]{cc}
                    \hspace{-5mm}
                    \vspace*{1cm}
                    \begin{minipage}[b]{10cm}
                    \centering
                    \includegraphics[width=10cm]{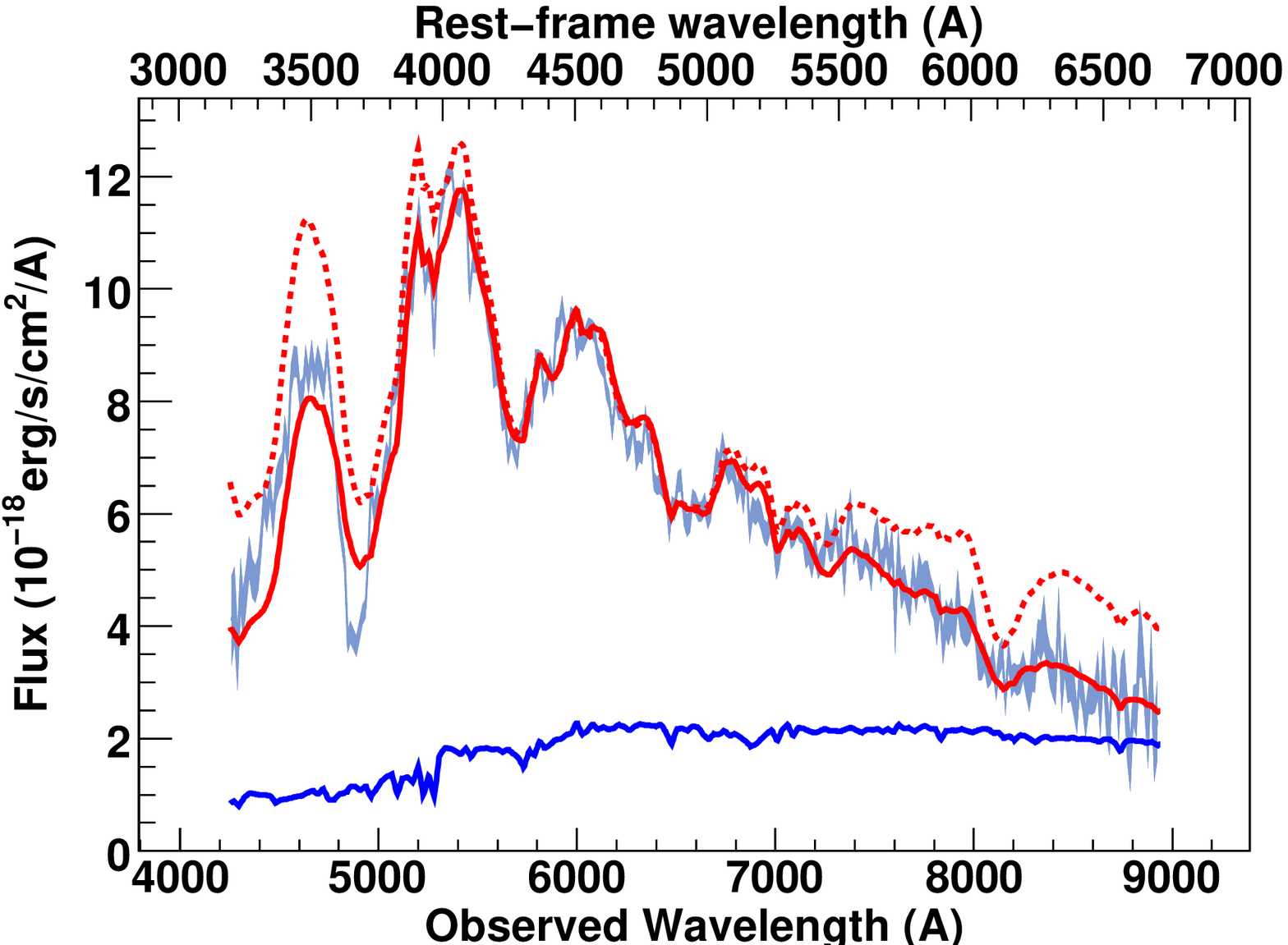}
                    \end{minipage}
                    &
                    \begin{minipage}[b]{10cm}
                    \hspace{-5mm}\includegraphics[width=10cm]{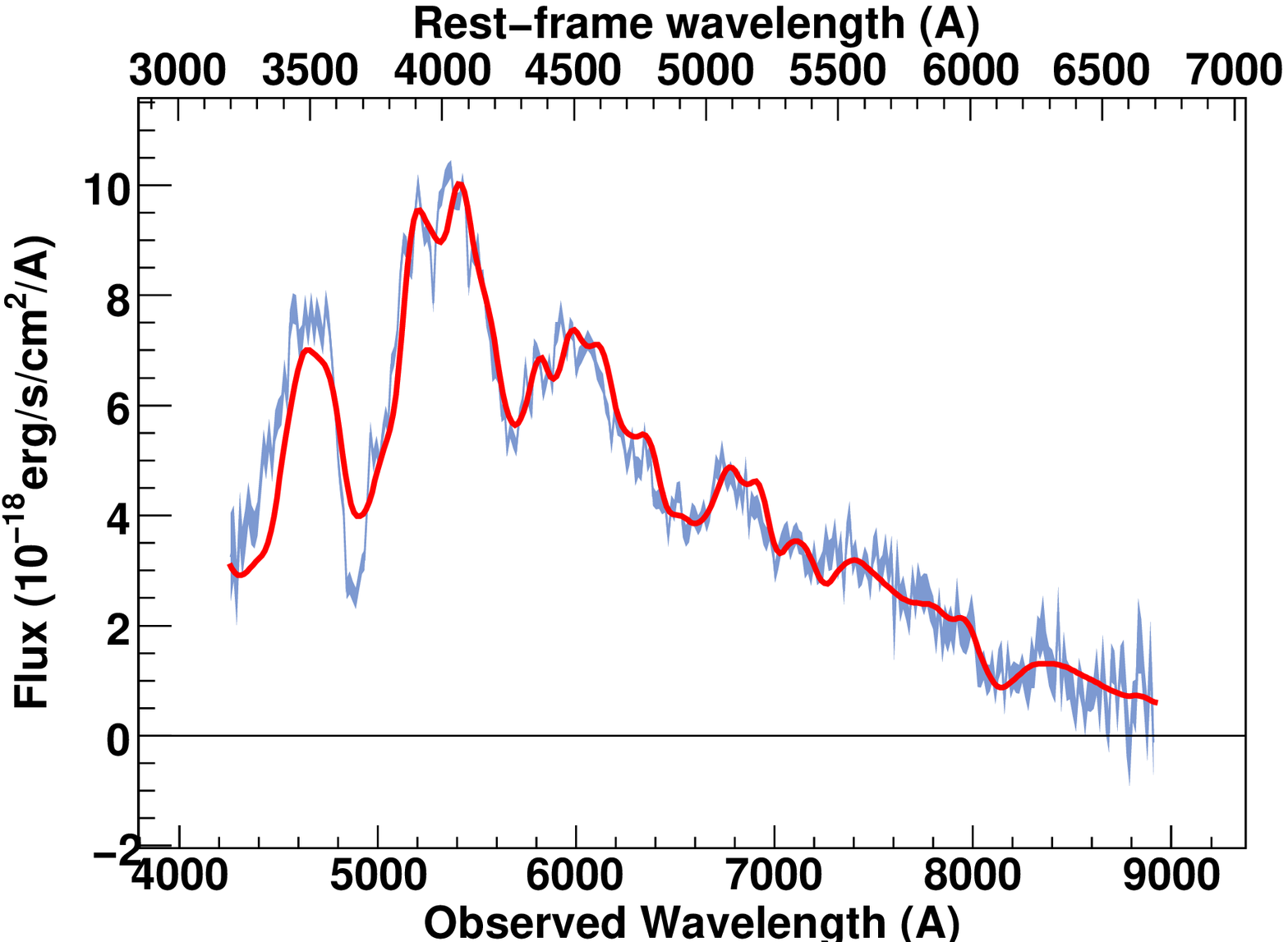}
                    \end{minipage}
                    \\
                    \end{tabular}
                    \caption{\label{fig:03D1fc358} SNLS-  03D1fc358: a SN~Ia supernova at z=0.332. The spectrum phase is -4.4. A         E(3) host model has been subtracted. Three re-calibration parameters have been used to fit this spectrum.}
                    \end{center}
                    \end{figure*}

\begin{figure*}[ht]
                    \begin{center}
                    \begin{tabular}[b]{cc}
                    \hspace{-5mm}
                    \vspace*{1cm}
                    \begin{minipage}[b]{10cm}
                    \centering
                    \includegraphics[width=10cm]{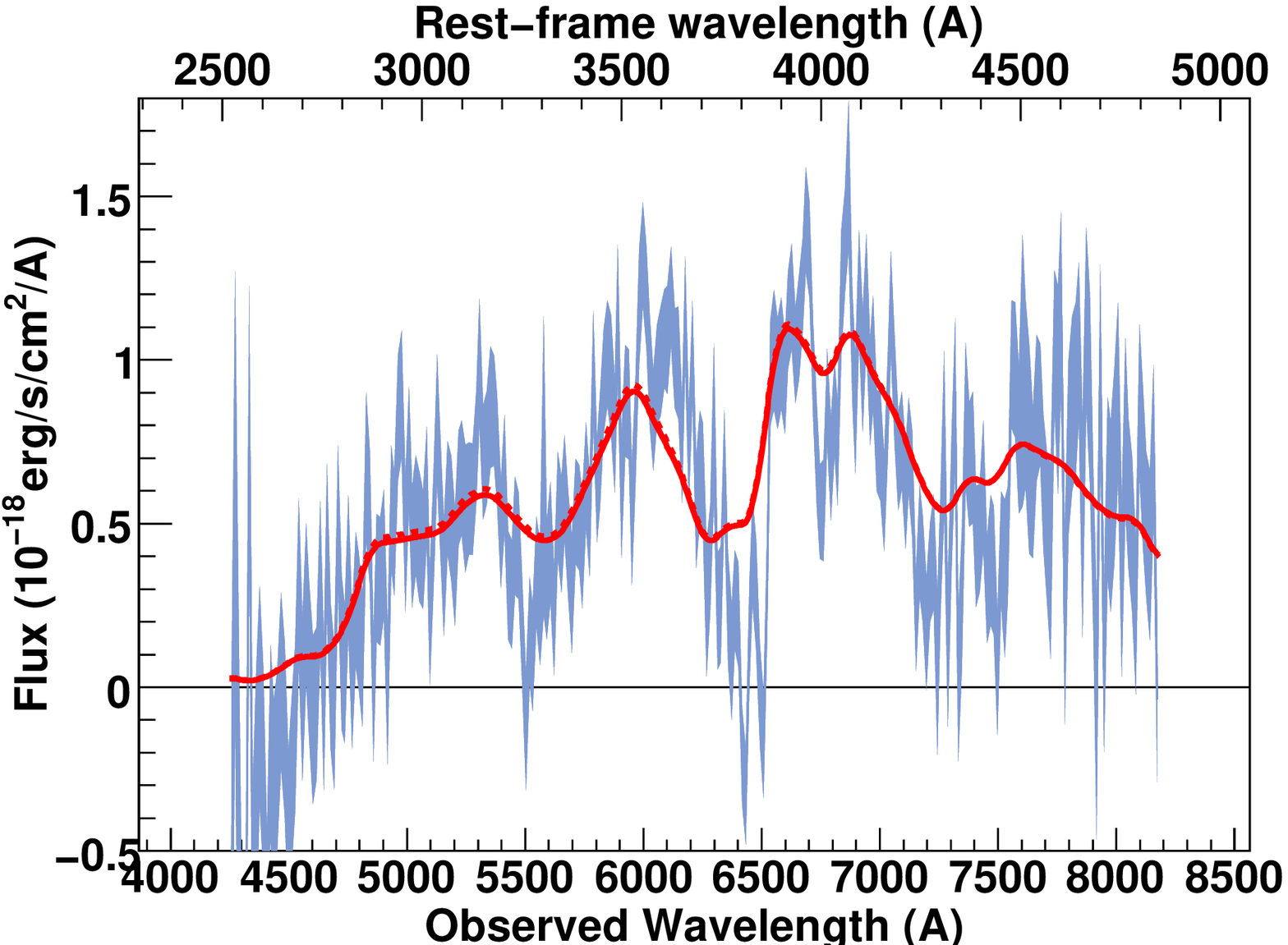}
                    \end{minipage}
                    &
                    \begin{minipage}[b]{10cm}
                    \end{minipage}
                    \\
                    \end{tabular}
                    \caption{\label{fig:03D1fl355} SNLS-  03D1fl355: a SN~Ia supernova at z=0.688. The spectrum phase is 0.5. Best-fit obtained for a model with no host galaxy component.}
                    \end{center}
                    \end{figure*}

\begin{figure*}[ht]
                    \begin{center}
                    \begin{tabular}[b]{cc}
                    \hspace{-5mm}
                    \vspace*{1cm}
                    \begin{minipage}[b]{10cm}
                    \centering
                    \includegraphics[width=10cm]{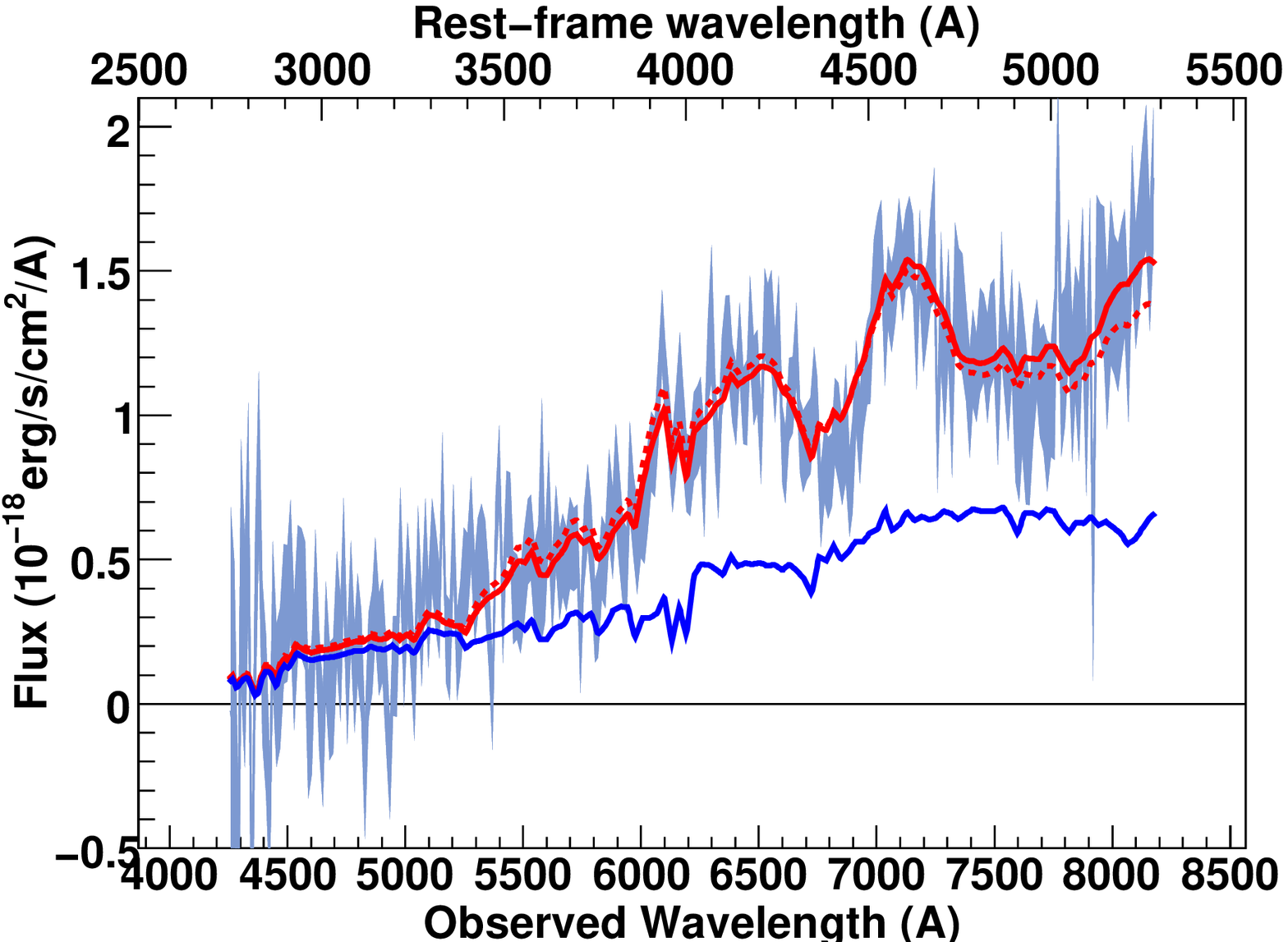}
                    \end{minipage}
                    &
                    \begin{minipage}[b]{10cm}
                    \hspace{-5mm}\includegraphics[width=10cm]{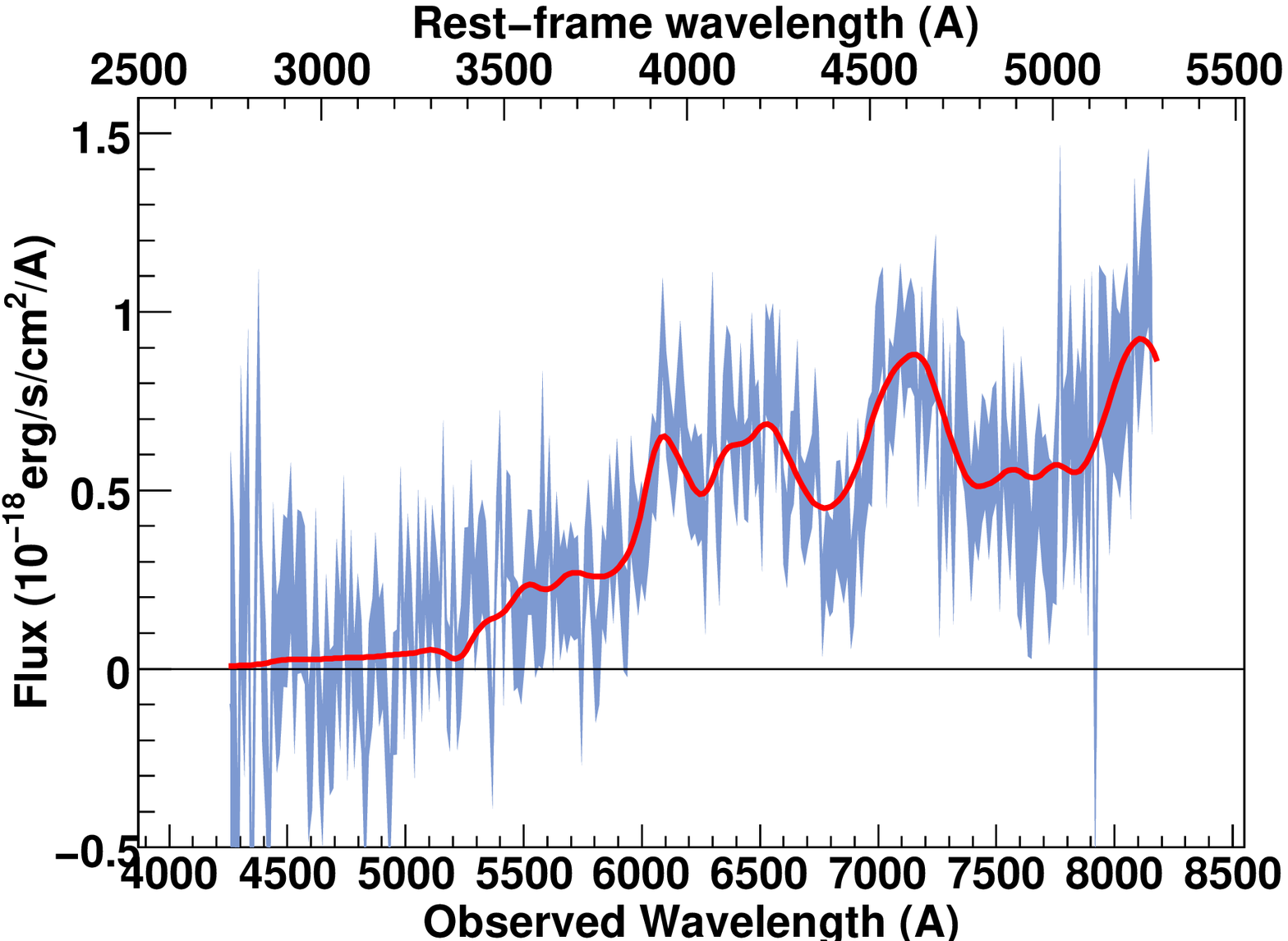}
                    \end{minipage}
                    \\
                    \end{tabular}
                    \caption{\label{fig:03D1gt385} SNLS-  03D1gt385: a SN~Ia supernova at z=0.560. The spectrum phase is 7.0. A         E(4) host model has been subtracted.}
                    \end{center}
                    \end{figure*}

\begin{figure*}[ht]
                    \begin{center}
                    \begin{tabular}[b]{cc}
                    \hspace{-5mm}
                    \vspace*{1cm}
                    \begin{minipage}[b]{10cm}
                    \centering
                    \includegraphics[width=10cm]{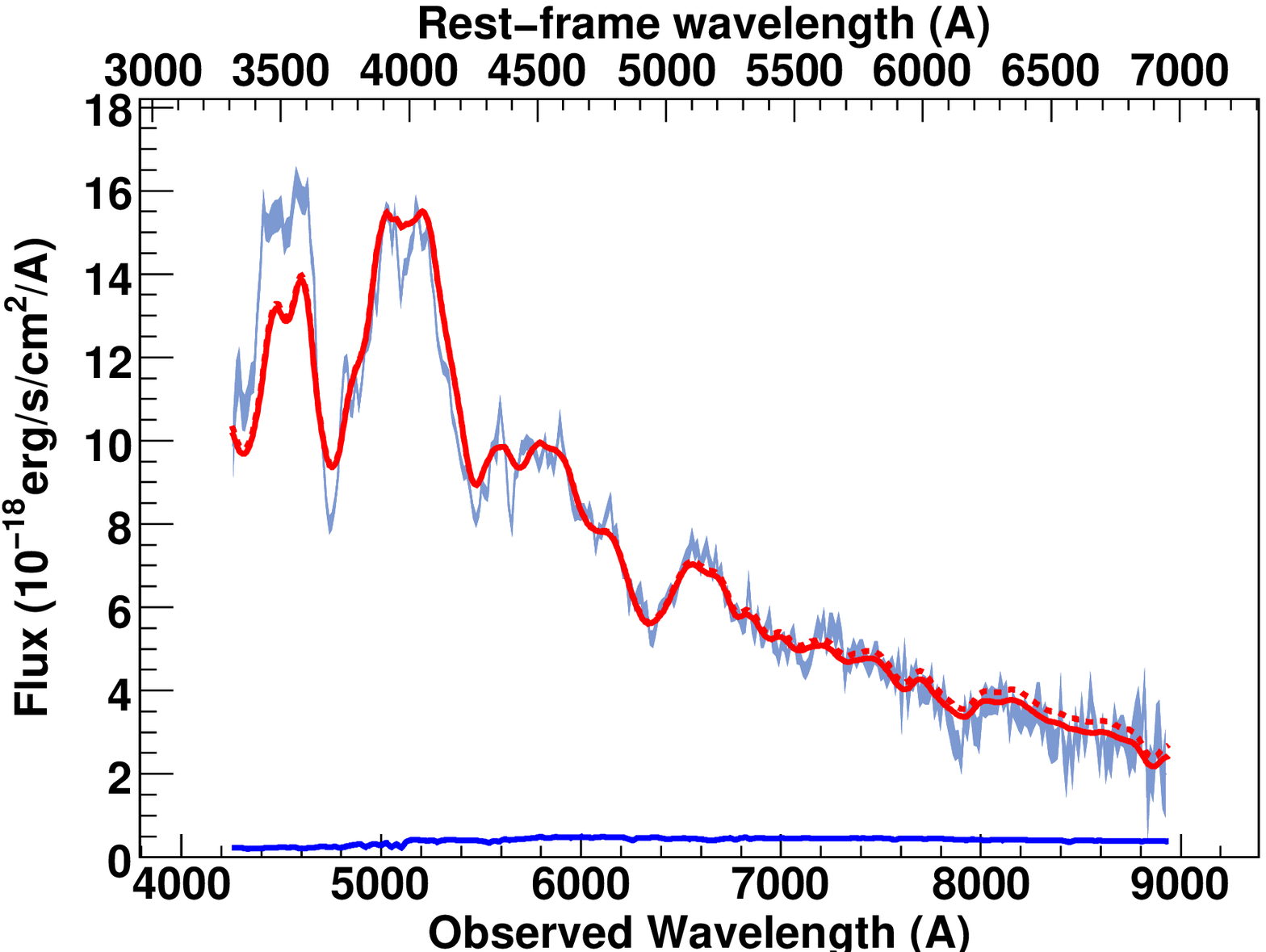}
                    \end{minipage}
                    &
                    \begin{minipage}[b]{10cm}
                    \hspace{-5mm}\includegraphics[width=10cm]{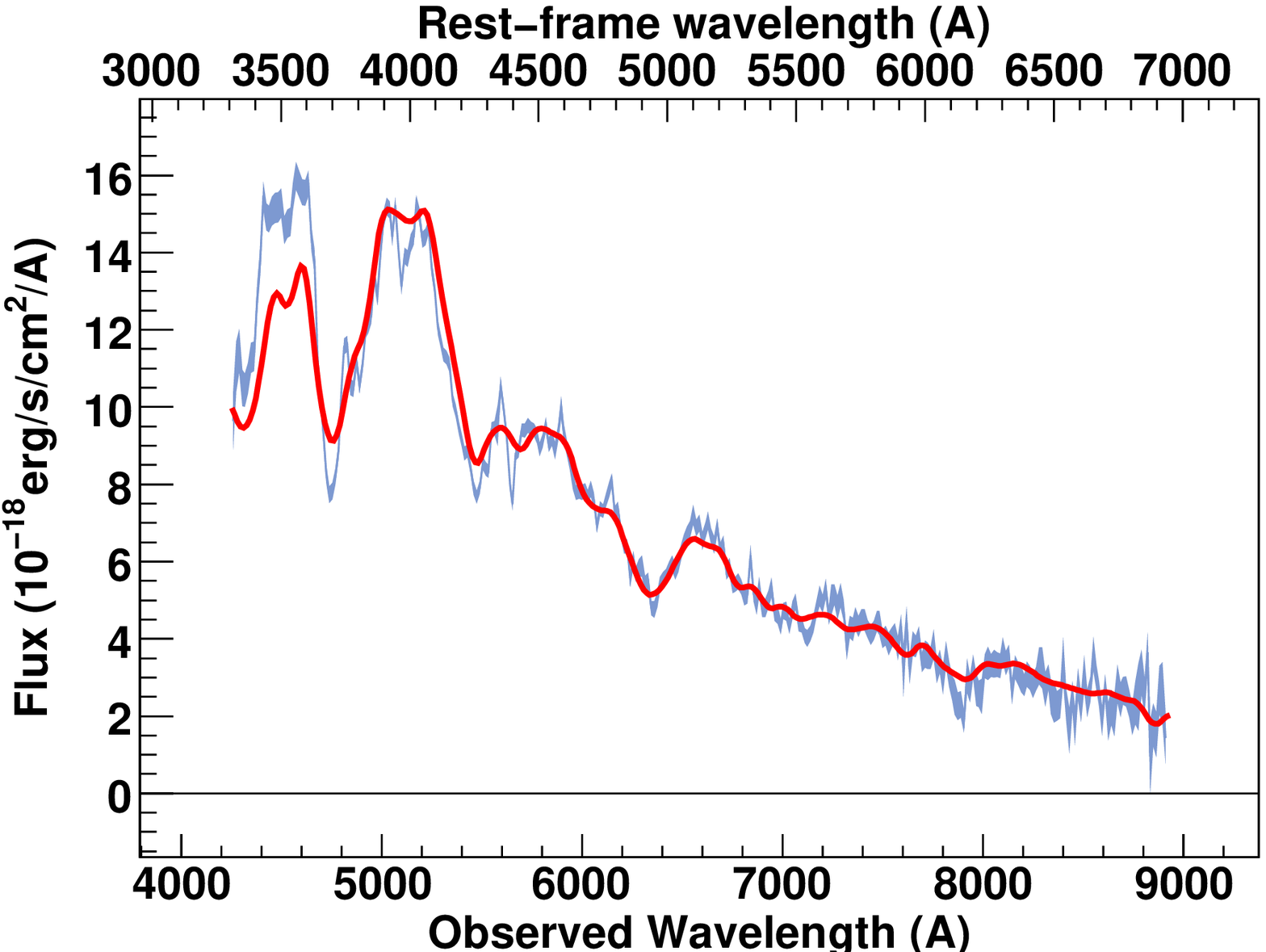}
                    \end{minipage}
                    \\
                    \end{tabular}
                    \caption{\label{fig:03D4ag180} SNLS-  03D4ag180: a SN~Ia\_pec supernova at z=0.285. The spectrum phase is -8.6. A         E(2) host model has been subtracted. Note the poor fit below rest-frame 4000 $\AA$ of this SN~1999aa-like supernova.}
                    \end{center}
                    \end{figure*}

\begin{figure*}[ht]
                    \begin{center}
                    \begin{tabular}[b]{cc}
                    \hspace{-5mm}
                    \vspace*{1cm}
                    \begin{minipage}[b]{10cm}
                    \centering
                    \includegraphics[width=10cm]{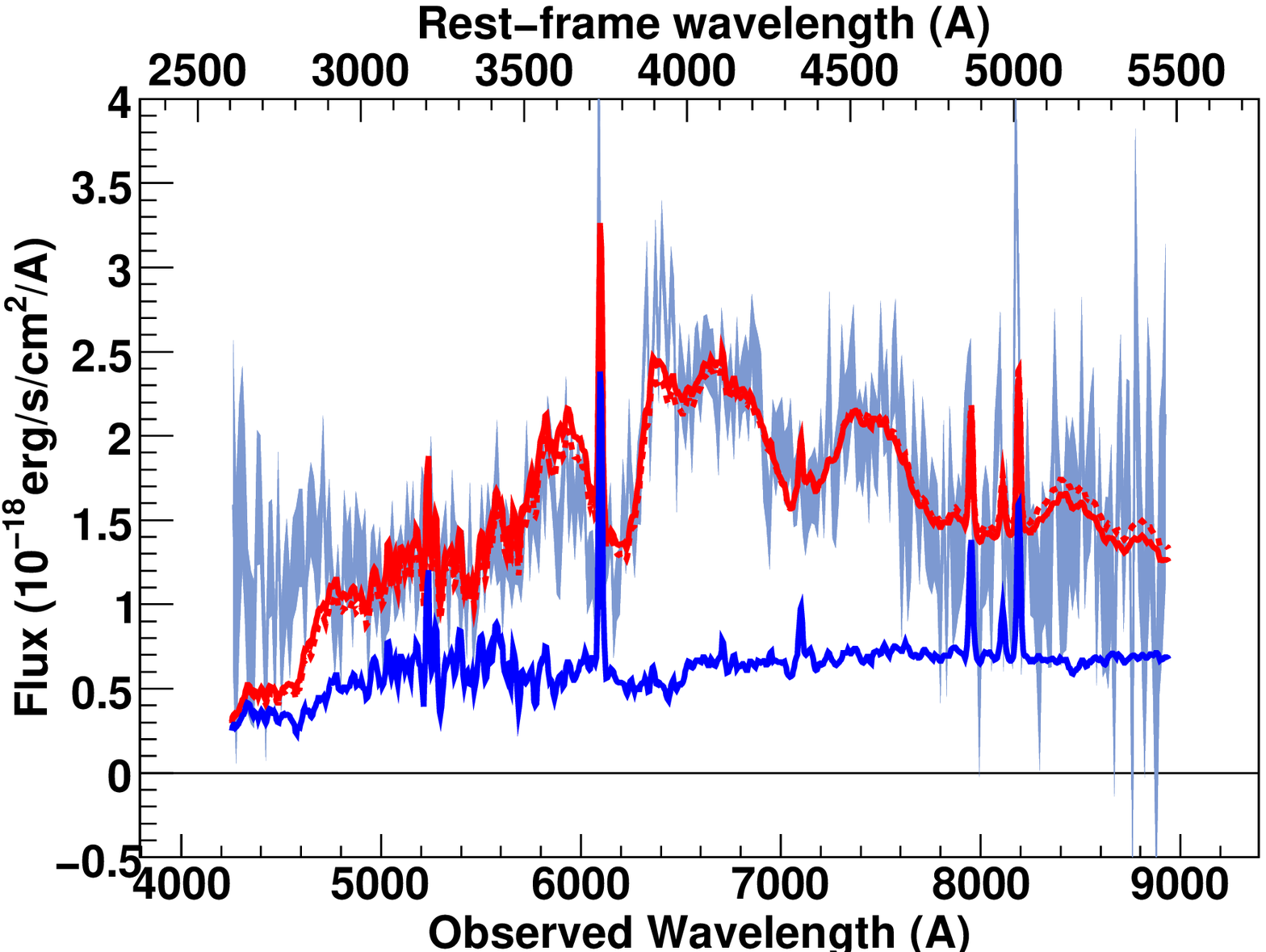}
                    \end{minipage}
                    &
                    \begin{minipage}[b]{10cm}
                    \hspace{-5mm}\includegraphics[width=10cm]{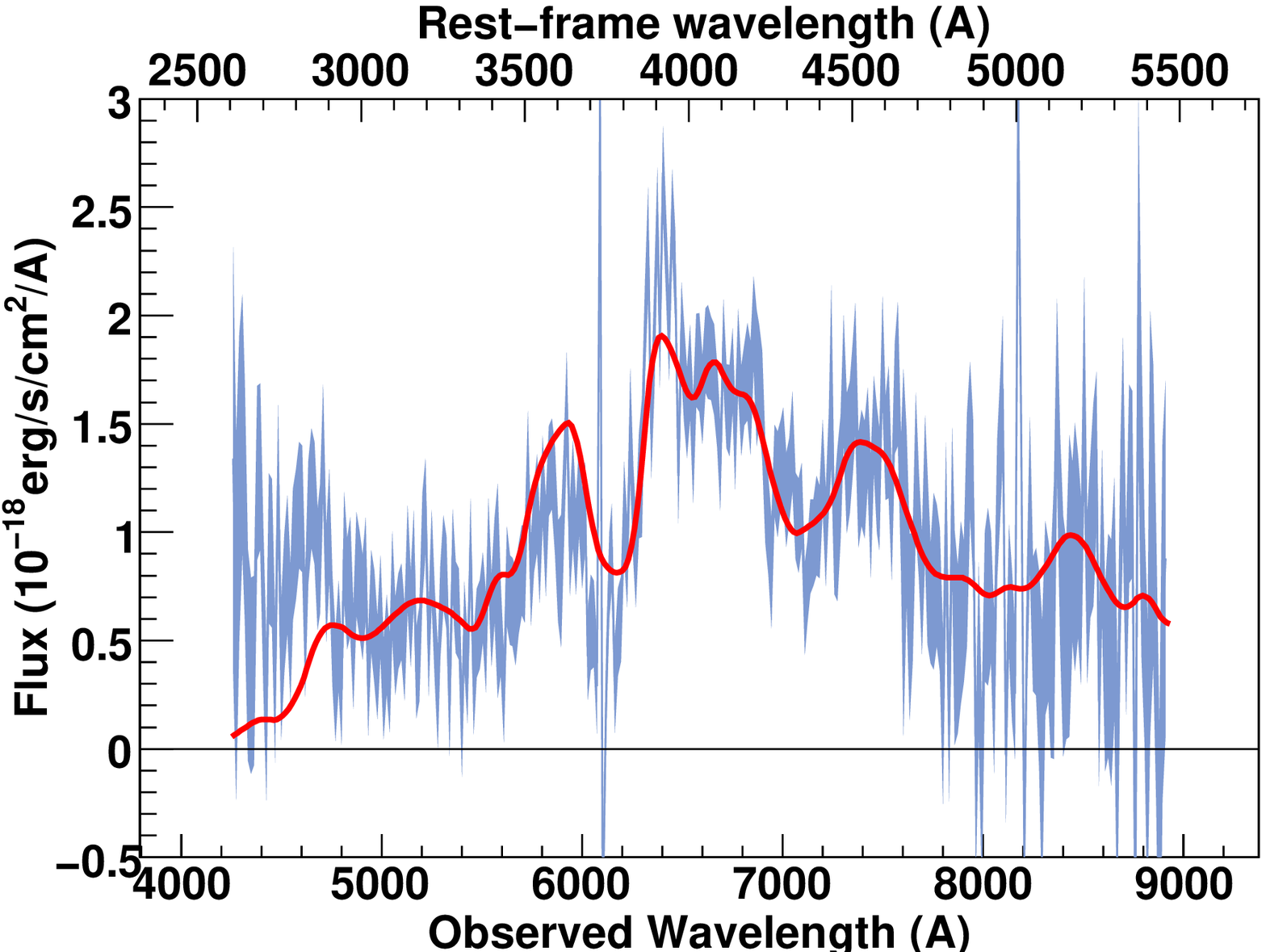}
                    \end{minipage}
                    \\
                    \end{tabular}
                    \caption{\label{fig:03D4at186} SNLS-  03D4at186: a SN~Ia supernova at z=0.634. The spectrum phase is 5.5. A Sb-Sc host model has been subtracted.}
                    \end{center}
                    \end{figure*}

\begin{figure*}[ht]
                    \begin{center}
                    \begin{tabular}[b]{cc}
                    \hspace{-5mm}
                    \vspace*{1cm}
                    \begin{minipage}[b]{10cm}
                    \centering
                    \includegraphics[width=10cm]{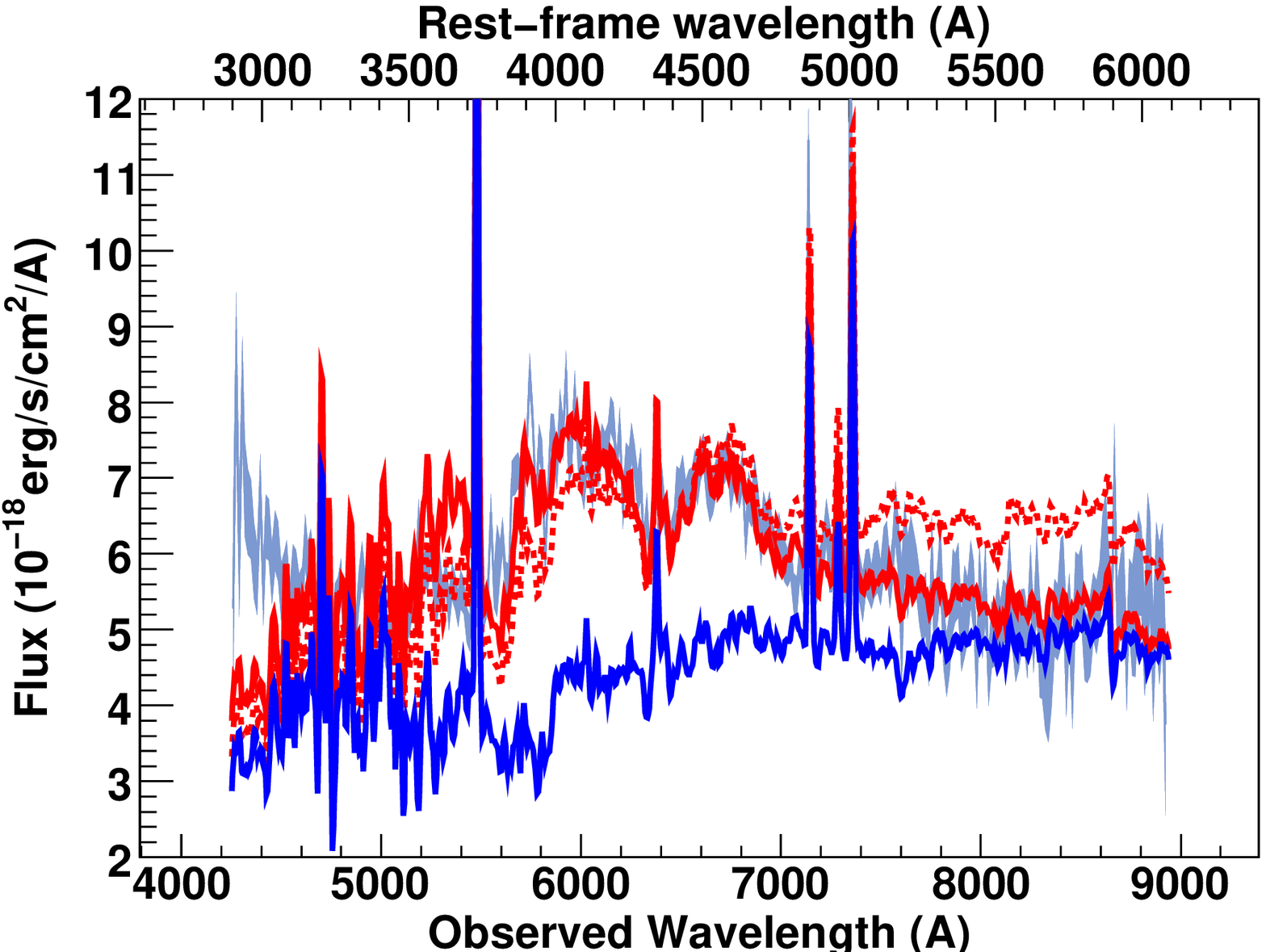}
                    \end{minipage}
                    &
                    \begin{minipage}[b]{10cm}
                    \hspace{-5mm}\includegraphics[width=10cm]{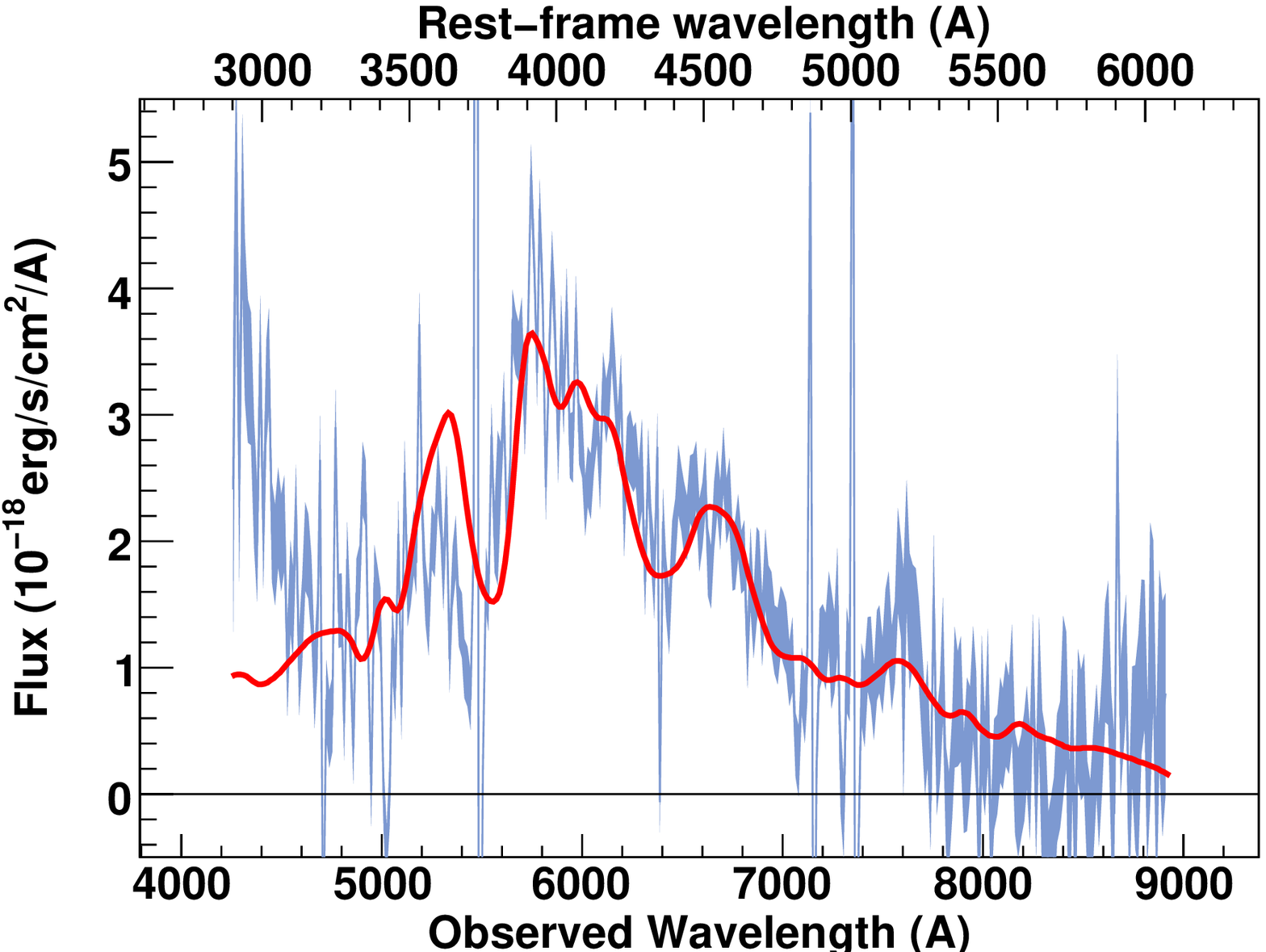}
                    \end{minipage}
                    \\
                    \end{tabular}
                    \caption{\label{fig:03D4au186} SNLS-  03D4au186: a SN~Ia$\star$ supernova at z=0.468. The spectrum phase is 6.5. A   Sb-Sc host model has been subtracted. Note the UV flux excess in the spectrum compared to the model. This is due to a host galaxy residual in this region of the spectrum.}
                    \end{center}
                    \end{figure*}

\begin{figure*}[ht]
                    \begin{center}
                    \begin{tabular}[b]{cc}
                    \hspace{-5mm}
                    \vspace*{1cm}
                    \begin{minipage}[b]{10cm}
                    \centering
                    \includegraphics[width=10cm]{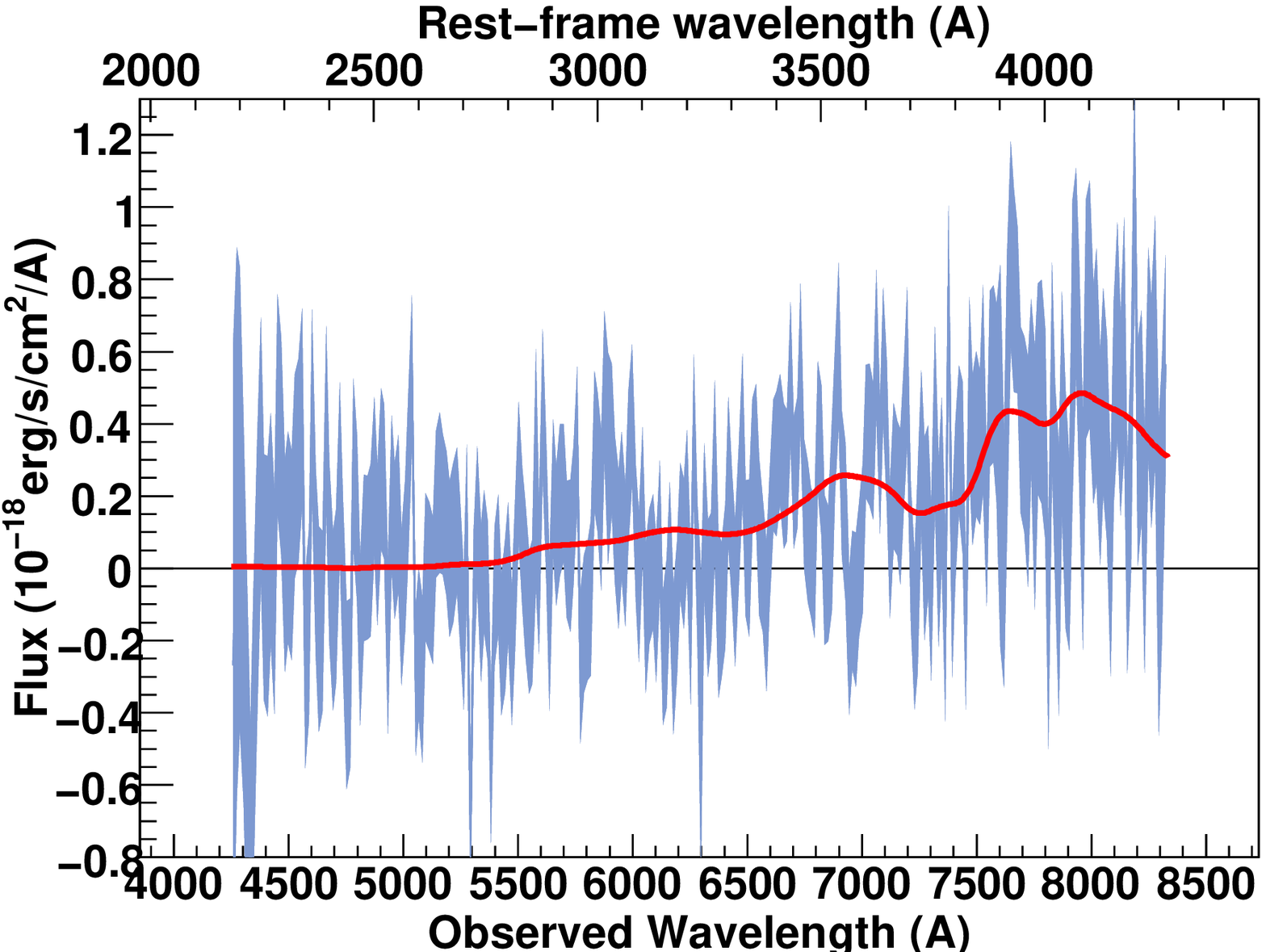}
                    \end{minipage}
                    &
                    \begin{minipage}[b]{10cm}
                    \end{minipage}
                    \\
                    \end{tabular}
                    \caption{\label{fig:03D4cx245} SNLS-  03D4cx245: a SN~Ia supernova at z=0.949. The spectrum phase is 1.2. Best-fit obtained for a model with no host galaxy component.}
                    \end{center}
                    \end{figure*}

\begin{figure*}[ht]
                    \begin{center}
                    \begin{tabular}[b]{cc}
                    \hspace{-5mm}
                    \vspace*{1cm}
                    \begin{minipage}[b]{10cm}
                    \centering
                    \includegraphics[width=10cm]{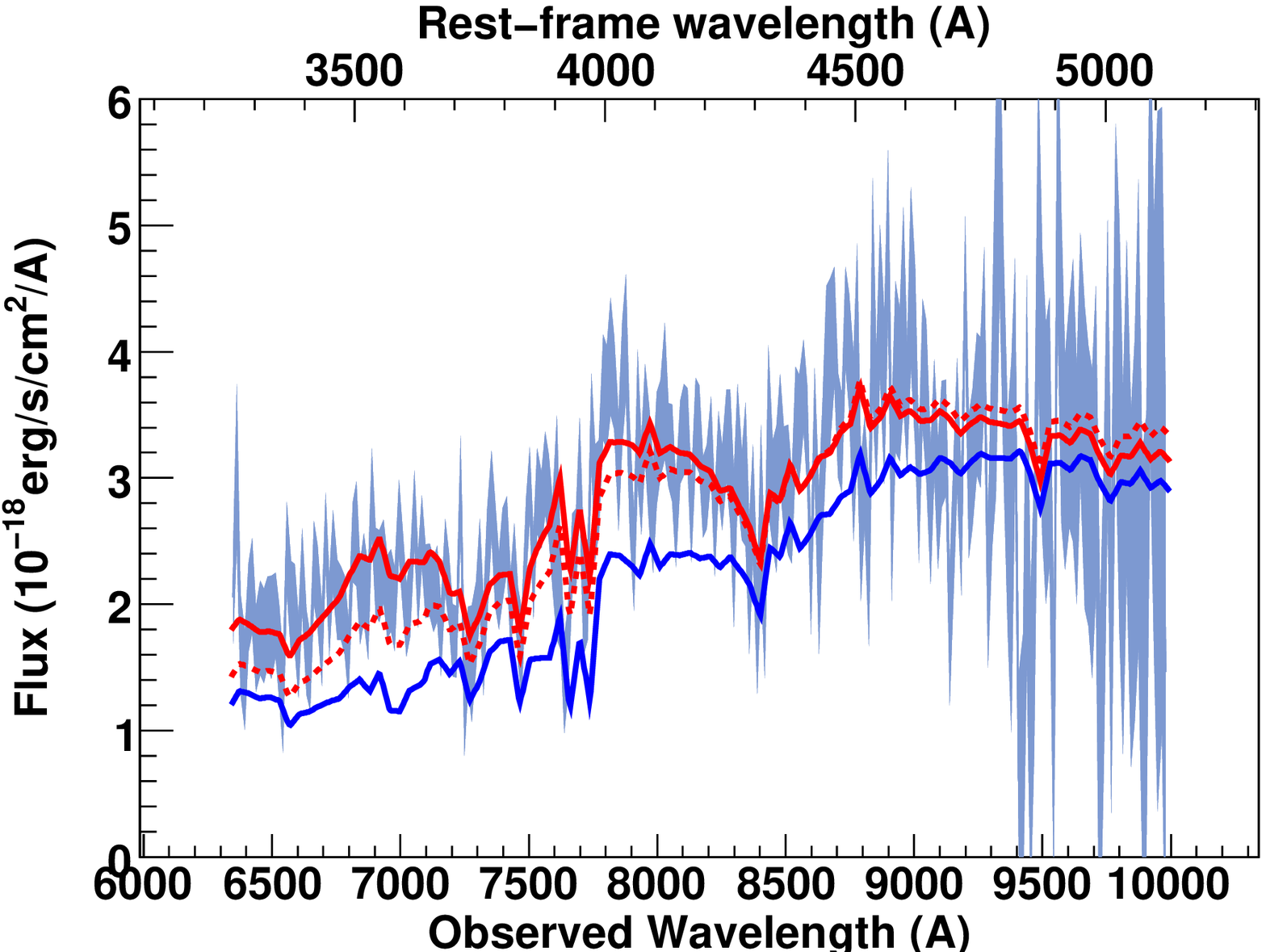}
                    \end{minipage}
                    &
                    \begin{minipage}[b]{10cm}
                    \hspace{-5mm}\includegraphics[width=10cm]{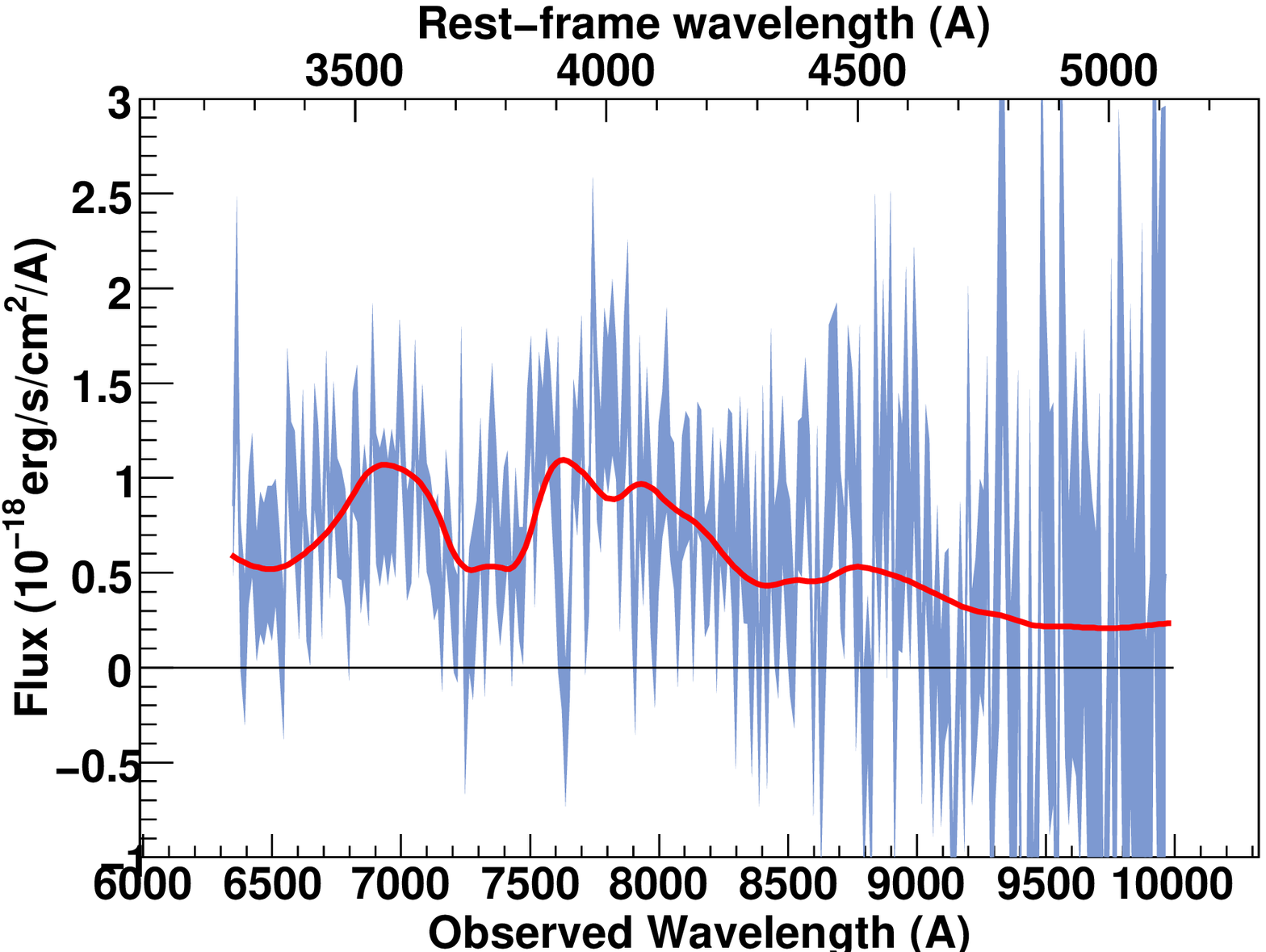}
                    \end{minipage}
                    \\
                    \end{tabular}
                    \caption{\label{fig:03D4cx248} SNLS-  03D4cx248: a SN~Ia supernova at z=0.949. Spectrum obtained with the 300I Grism. The spectrum phase is 2.7. A        S0(4) host model has been subtracted.}
                    \end{center}
                    \end{figure*}

\begin{figure*}[ht]
                    \begin{center}
                    \begin{tabular}[b]{cc}
                    \hspace{-5mm}
                    \vspace*{1cm}
                    \begin{minipage}[b]{10cm}
                    \centering
                    \includegraphics[width=10cm]{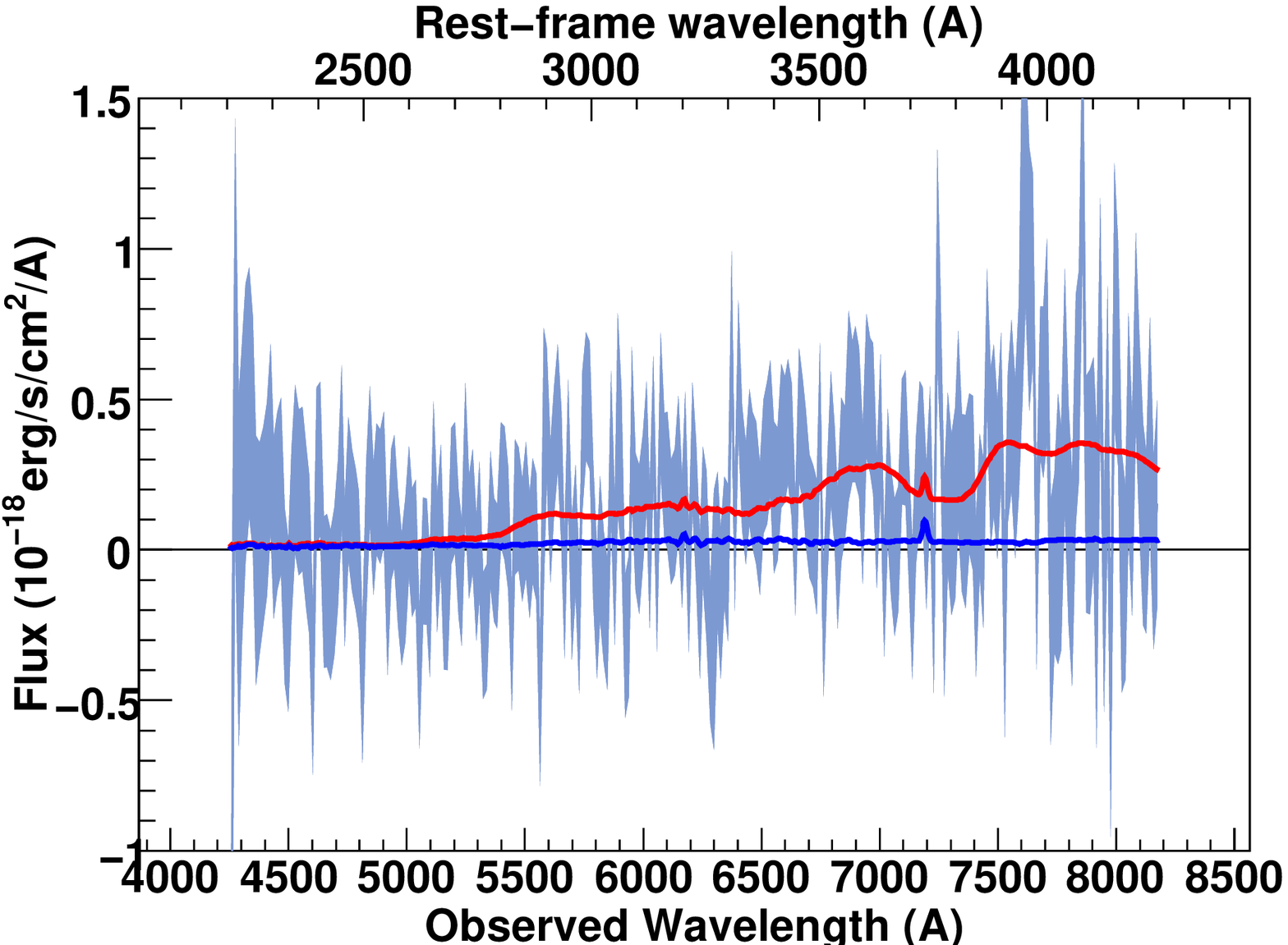}
                    \end{minipage}
                    &
                    \begin{minipage}[b]{10cm}
                    \hspace{-5mm}\includegraphics[width=10cm]{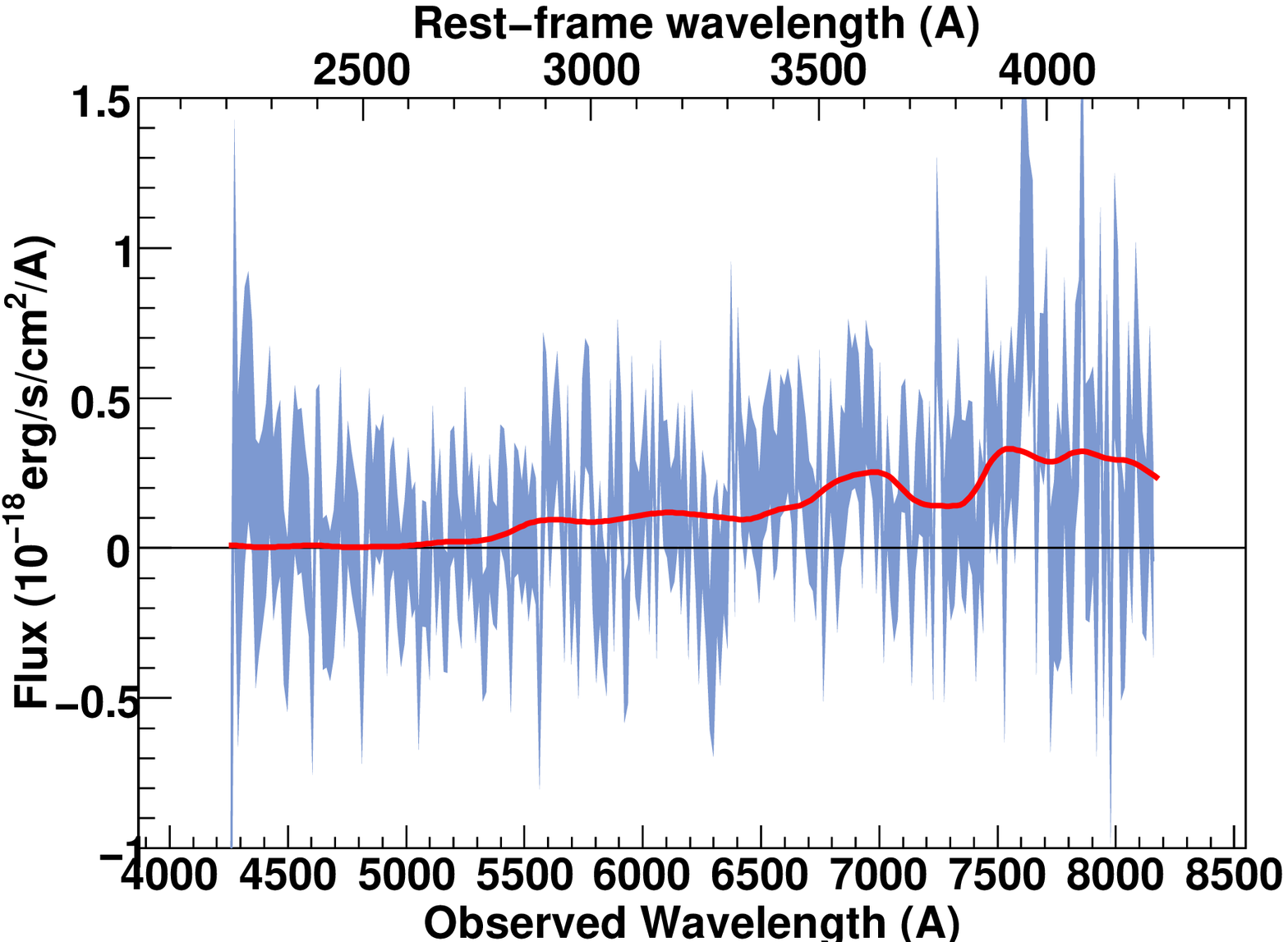}
                    \end{minipage}
                    \\
                    \end{tabular}
                    \caption{\label{fig:03D4cy269} SNLS-  03D4cy269: a SN~Ia supernova at z=0.927. The spectrum phase is 4.6. A         Sb-Sc host model has been subtracted. The identification of this supernova as a SN~Ia is based on a Gemini spectrum of this same SN \citep{Howell05}}
                    \end{center}
                    \end{figure*}

\begin{figure*}[ht]
                    \begin{center}
                    \begin{tabular}[b]{cc}
                    \hspace{-5mm}
                    \vspace*{1cm}
                    \begin{minipage}[b]{10cm}
                    \centering
                    \includegraphics[width=10cm]{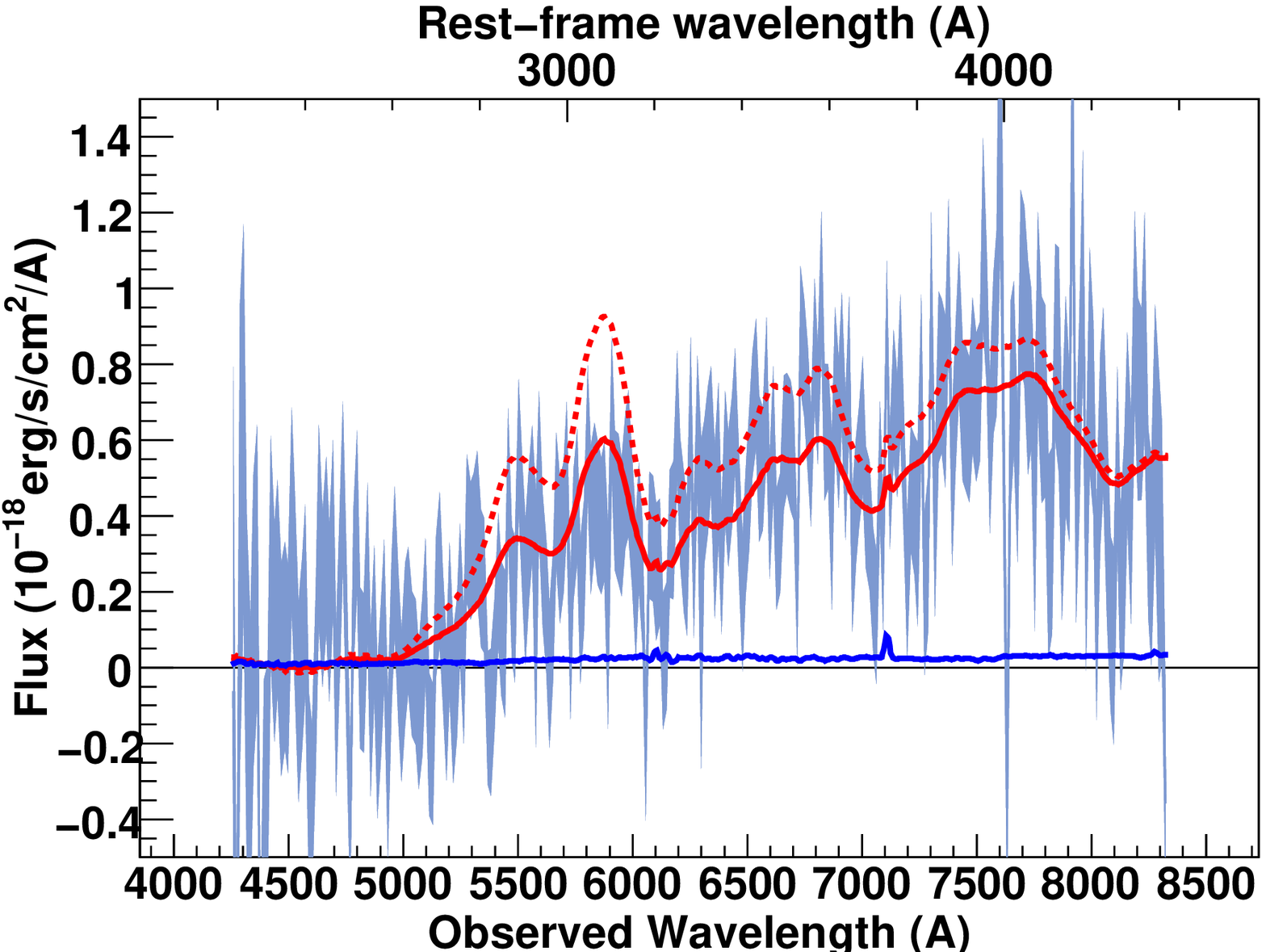}
                    \end{minipage}
                    &
                    \begin{minipage}[b]{10cm}
                    \hspace{-5mm}\includegraphics[width=10cm]{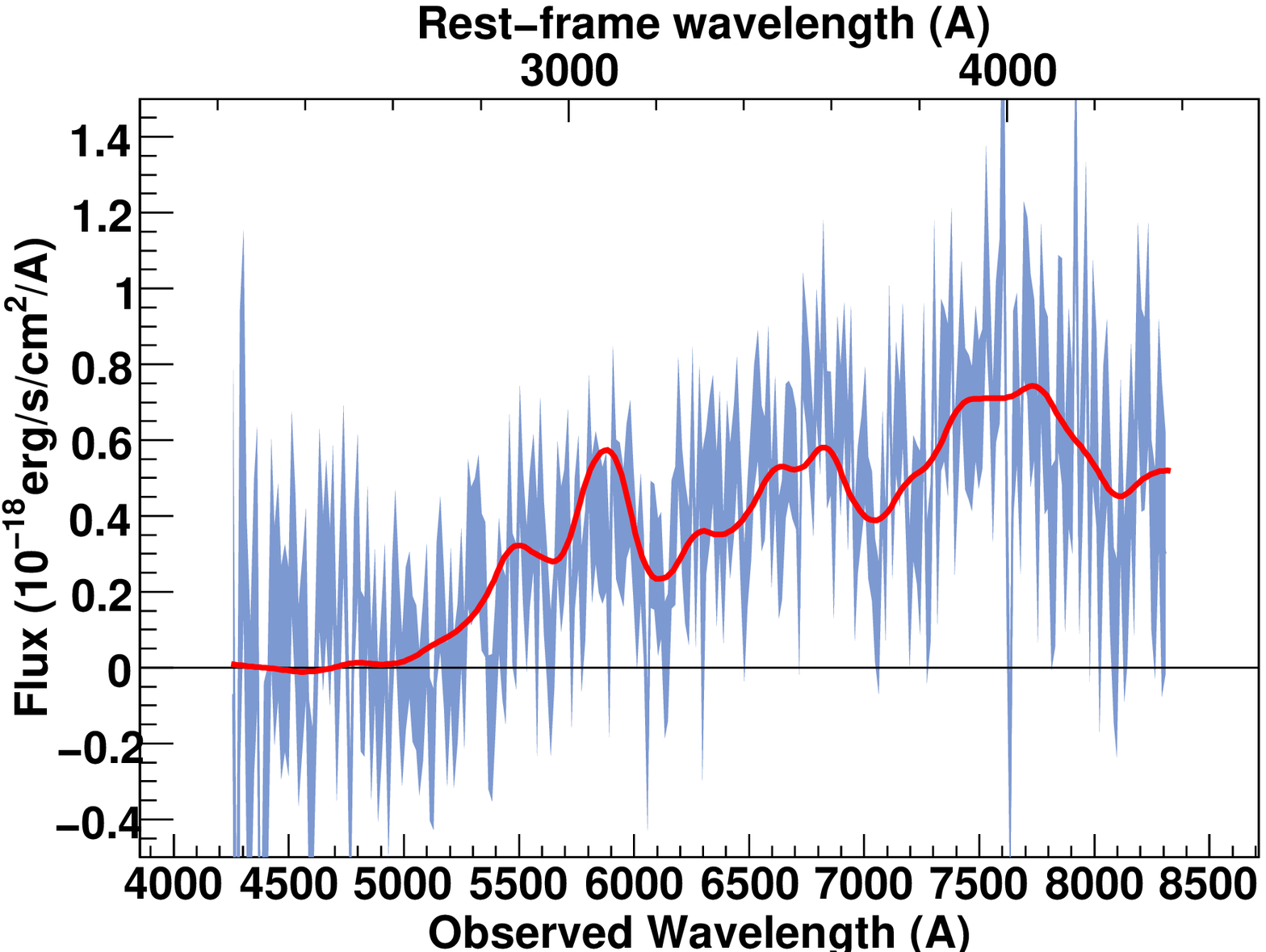}
                    \end{minipage}
                    \\
                    \end{tabular}
                    \caption{\label{fig:03D4di245} SNLS-  03D4di245: a SN~Ia$\star$ supernova at z=0.899. The spectrum phase is -8.6. A         Sb-Sc host model has been subtracted.}
                    \end{center}
                    \end{figure*}

\begin{figure*}[ht]
                    \begin{center}
                    \begin{tabular}[b]{cc}
                    \hspace{-5mm}
                    \vspace*{1cm}
                    \begin{minipage}[b]{10cm}
                    \centering
                    \includegraphics[width=10cm]{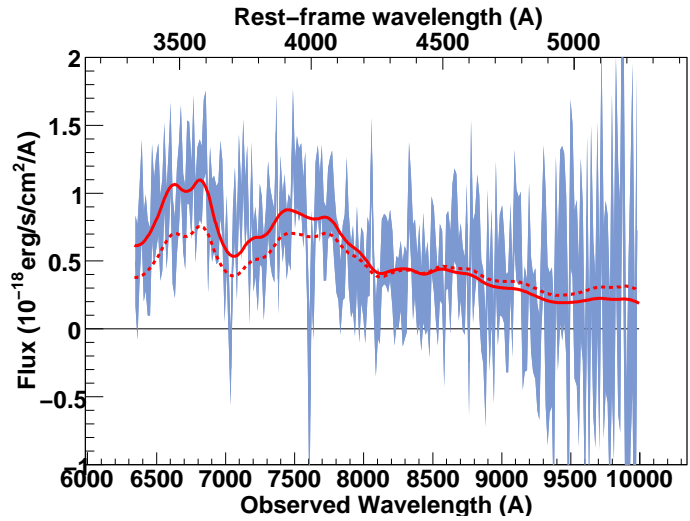}
                    \end{minipage}
                    &
                    \begin{minipage}[b]{10cm}
                    \end{minipage}
                    \\
                    \end{tabular}
                    \caption{\label{fig:03D4di249} SNLS-  03D4di249: a SN~Ia$\star$ supernova at z=0.899. Spectrum obtained with the 300I Grism. The spectrum phase is -6.5. Best-fit obtained for a model with no host galaxy component.}
                    \end{center}
                    \end{figure*}

\begin{figure*}[ht]
                    \begin{center}
                    \begin{tabular}[b]{cc}
                    \hspace{-5mm}
                    \vspace*{1cm}
                    \begin{minipage}[b]{10cm}
                    \centering
                    \includegraphics[width=10cm]{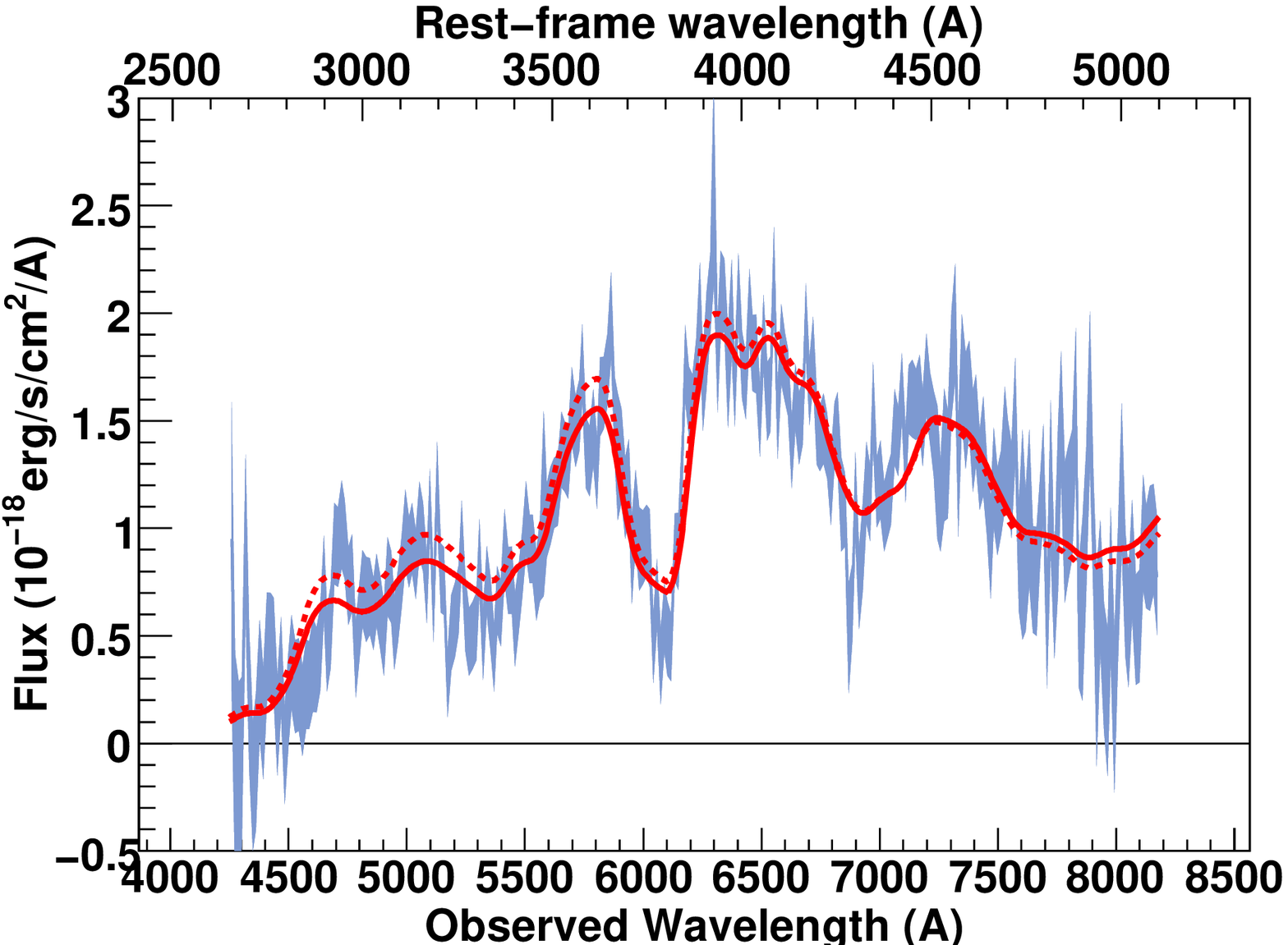}
                    \end{minipage}
                    &
                    \begin{minipage}[b]{10cm}
                    \end{minipage}
                    \\
                    \end{tabular}
                    \caption{\label{fig:03D4dy272} SNLS-  03D4dy272: a SN~Ia supernova at z=0.61. The spectrum phase is 4.8. Best-fit obtained for a model with no host galaxy component.}
                    \end{center}
                    \end{figure*}

\begin{figure*}[ht]
                    \begin{center}
                    \begin{tabular}[b]{cc}
                    \hspace{-5mm}
                    \vspace*{1cm}
                    \begin{minipage}[b]{10cm}
                    \centering
                    \includegraphics[width=10cm]{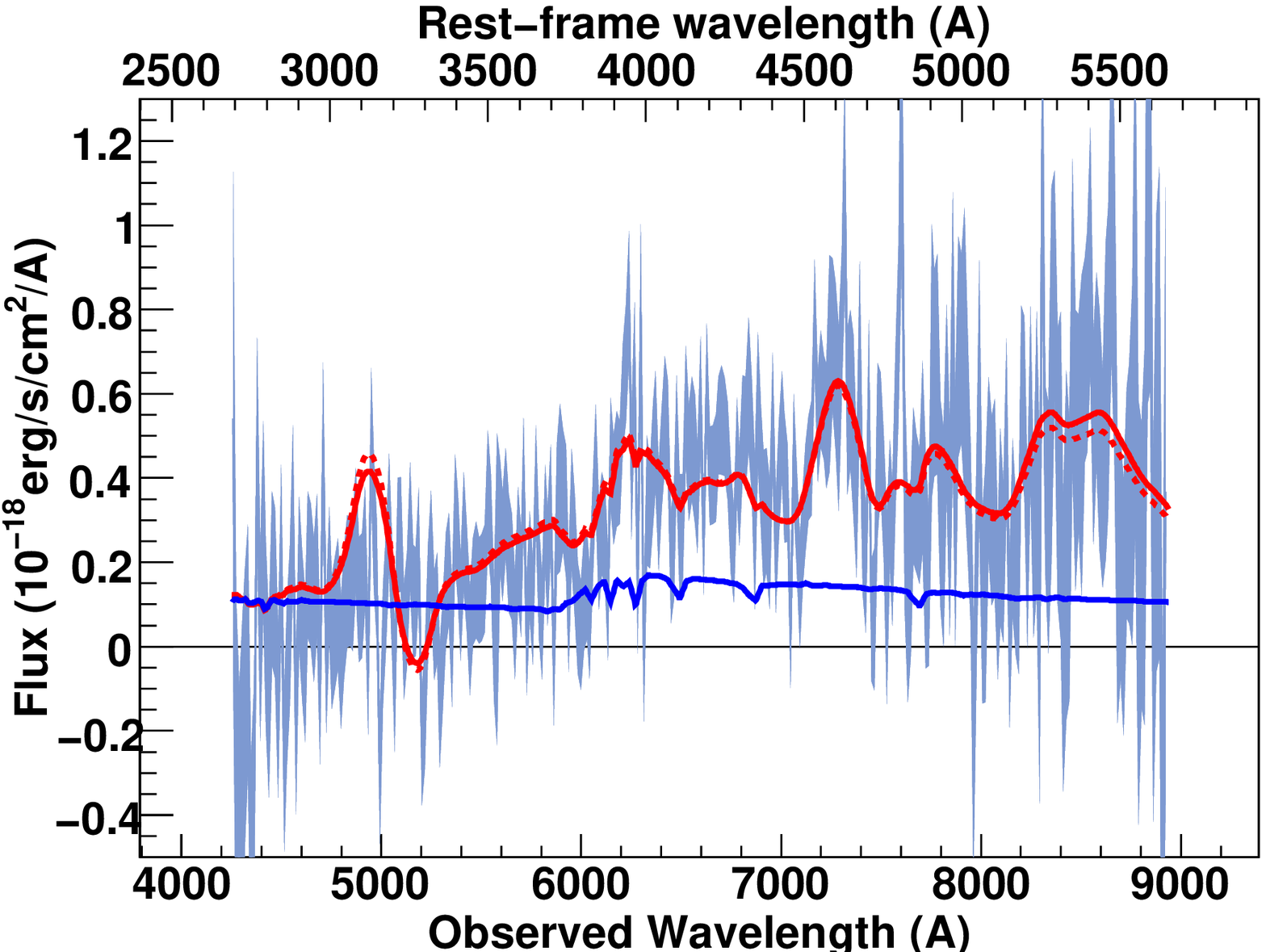}
                    \end{minipage}
                    &
                    \begin{minipage}[b]{10cm}
                    \hspace{-5mm}\includegraphics[width=10cm]{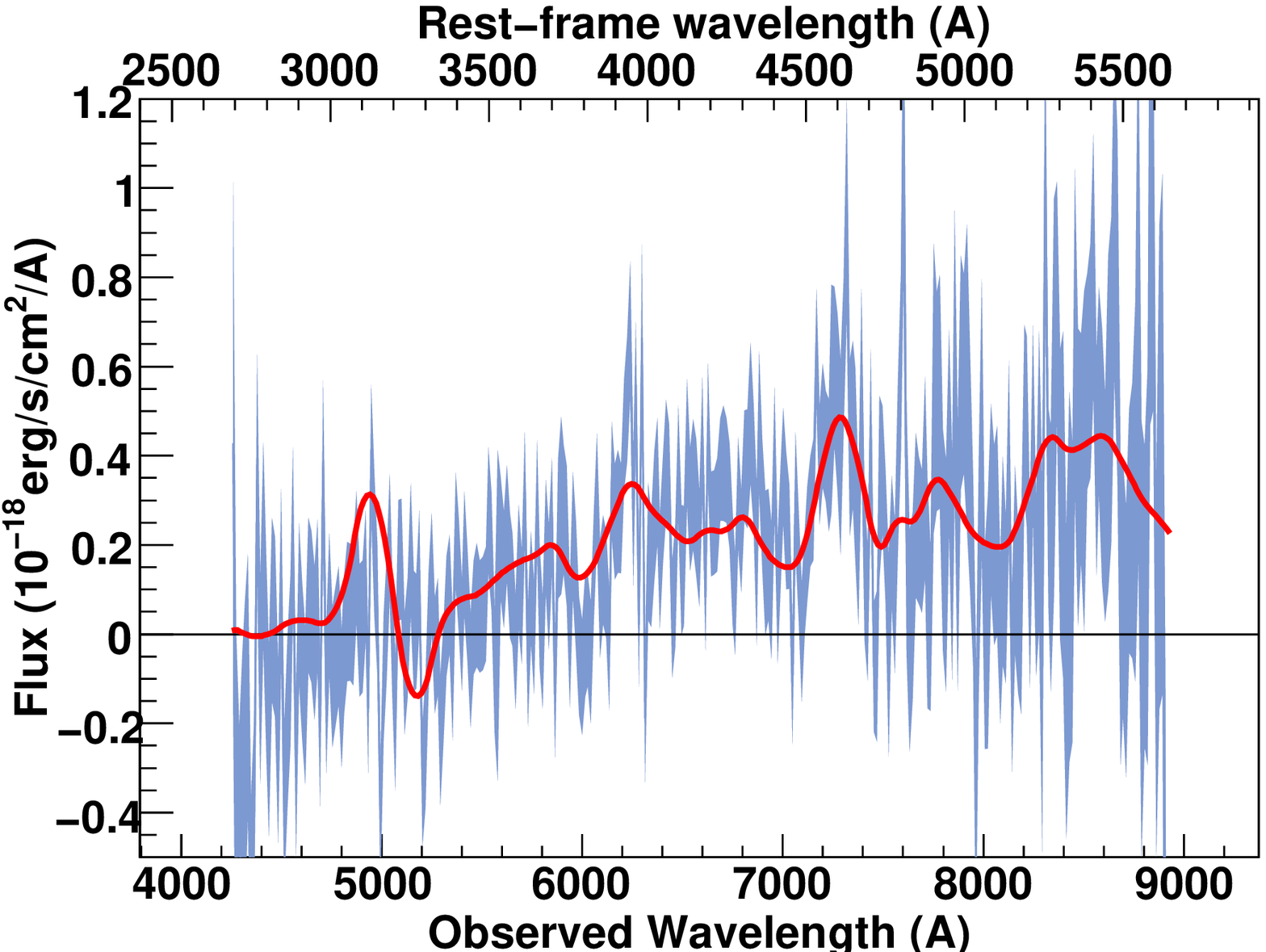}
                    \end{minipage}
                    \\
                    \end{tabular}
                    \caption{\label{fig:03D4gf329} SNLS-  03D4gf329: a SN~Ia$\star$ supernova at z=0.580. The spectrum phase is 20.4. A        S0(1) host model has been subtracted. The UV part of the model is poorly constrained due to the lack of UV templates at late phases in the version of the SALT2 training set used in this work.}
                    \end{center}
                    \end{figure*}

\begin{figure*}[ht]
                    \begin{center}
                    \begin{tabular}[b]{cc}
                    \hspace{-5mm}
                    \vspace*{1cm}
                    \begin{minipage}[b]{10cm}
                    \centering
                    \includegraphics[width=10cm]{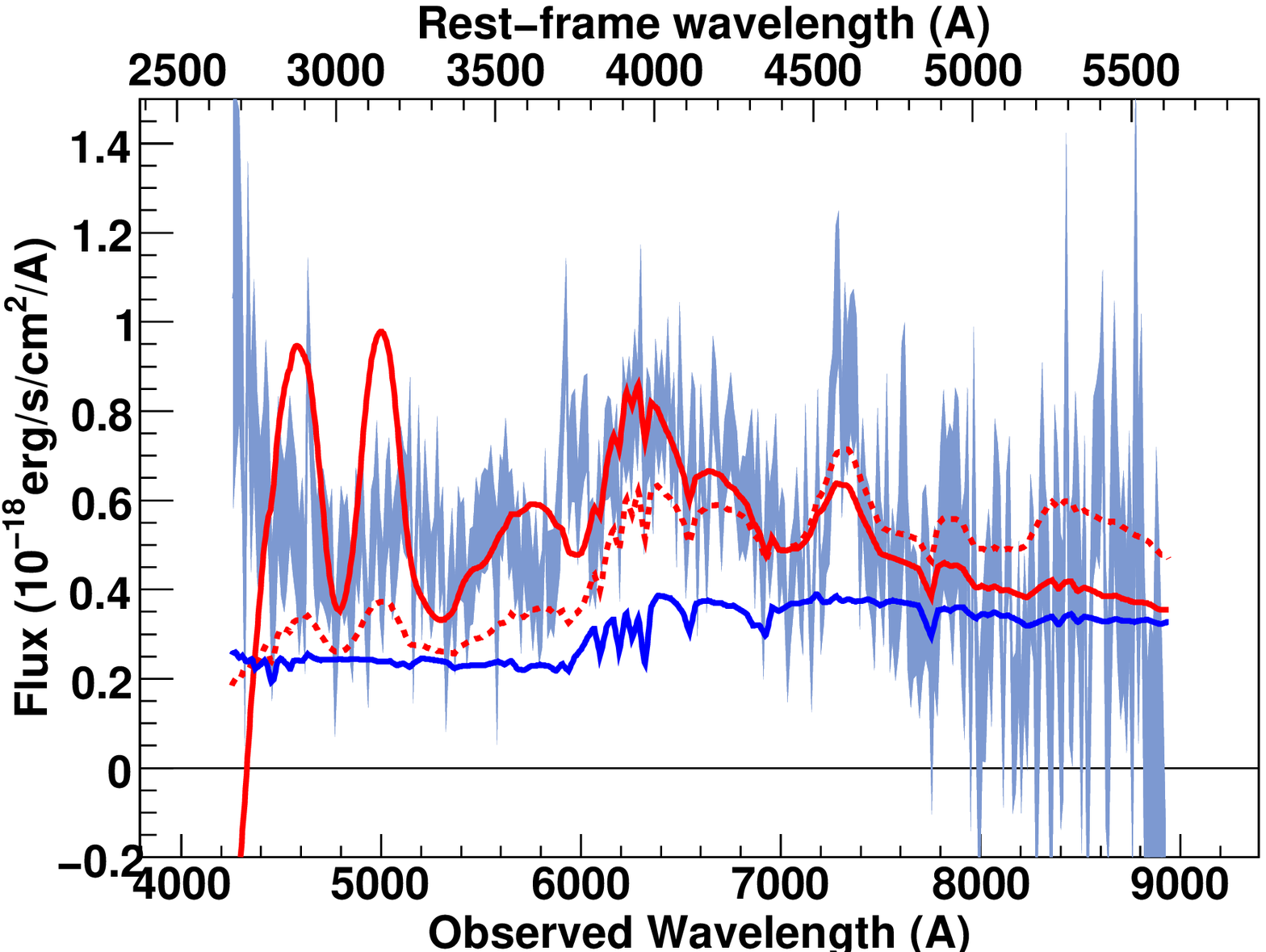}
                    \end{minipage}
                    &
                    \begin{minipage}[b]{10cm}
                    \hspace{-5mm}\includegraphics[width=10cm]{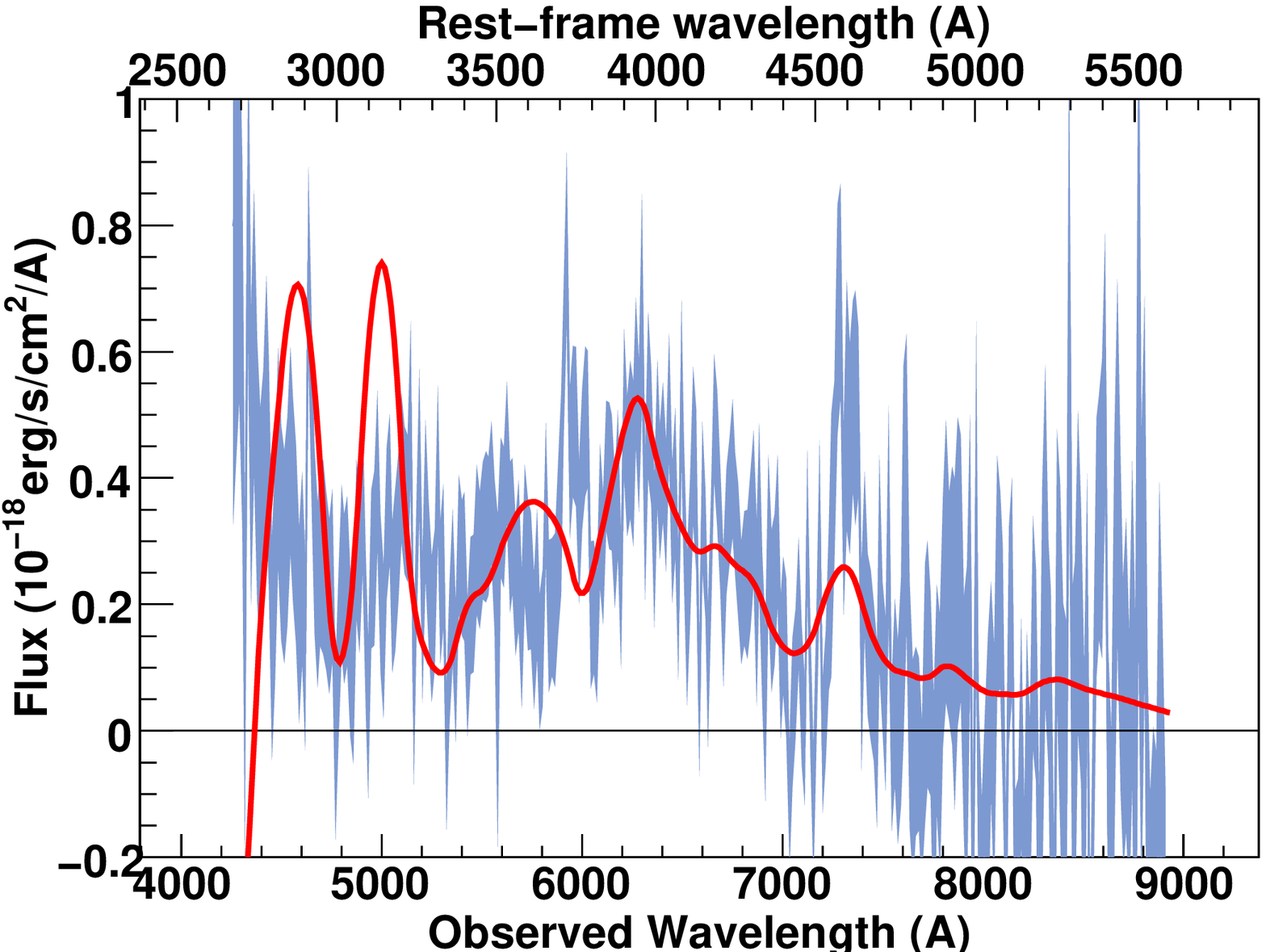}
                    \end{minipage}
                    \\
                    \end{tabular}
                    \caption{\label{fig:03D4gg330} SNLS-  03D4gg330: a SN~Ia$\star$ supernova at z=0.592. The spectrum phase is 16.4. A       Sc(10) host model has been subtracted. The UV part of the model is poorly constrained due to the lack of UV templates at late phases in the version of the SALT2 training set used in this work.}
                    \end{center}
                    \end{figure*}

\begin{figure*}[ht]
                    \begin{center}
                    \begin{tabular}[b]{cc}
                    \hspace{-5mm}
                    \vspace*{1cm}
                    \begin{minipage}[b]{10cm}
                    \centering
                    \includegraphics[width=10cm]{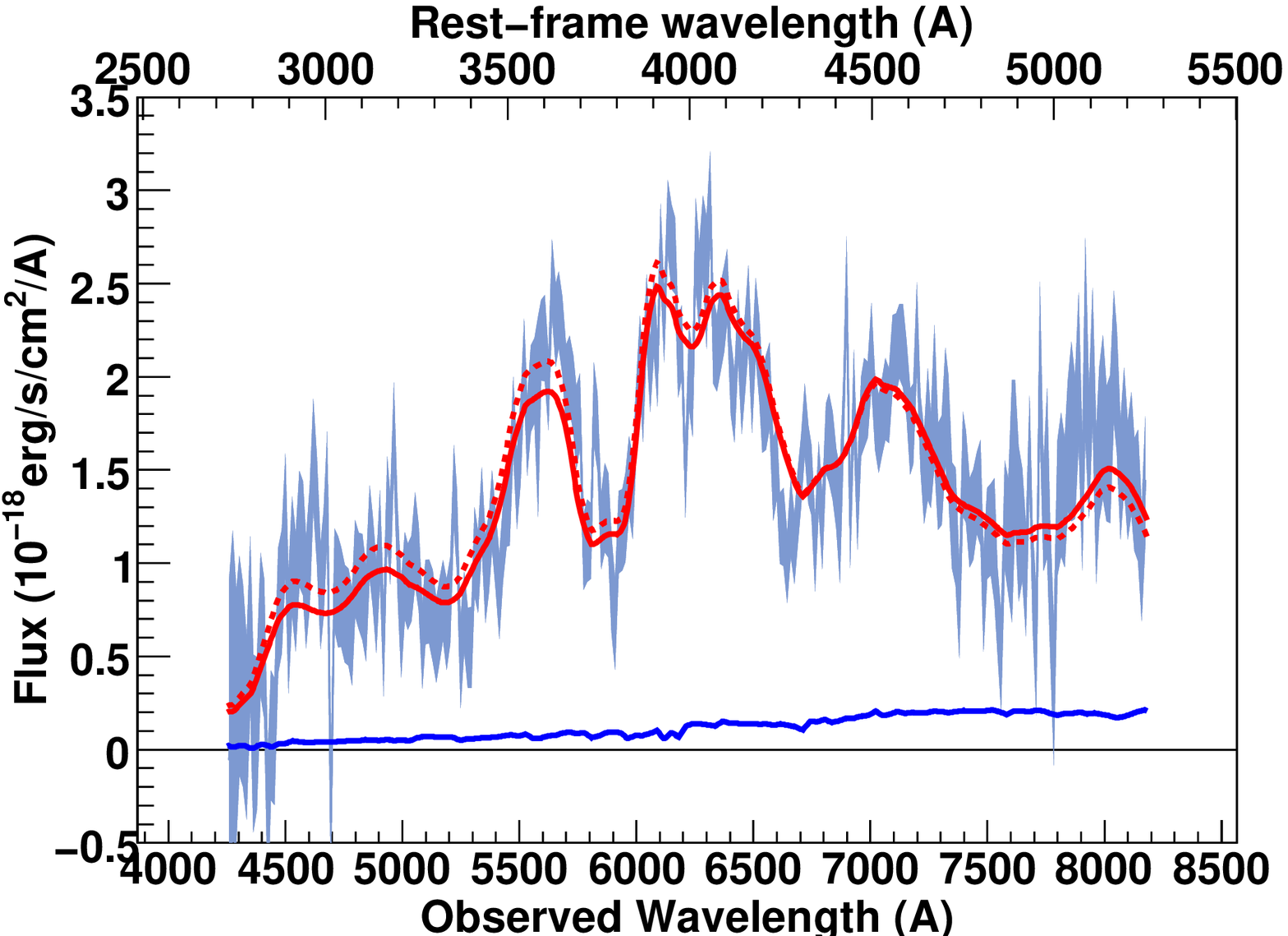}
                    \end{minipage}
                    &
                    \begin{minipage}[b]{10cm}
                    \hspace{-5mm}\includegraphics[width=10cm]{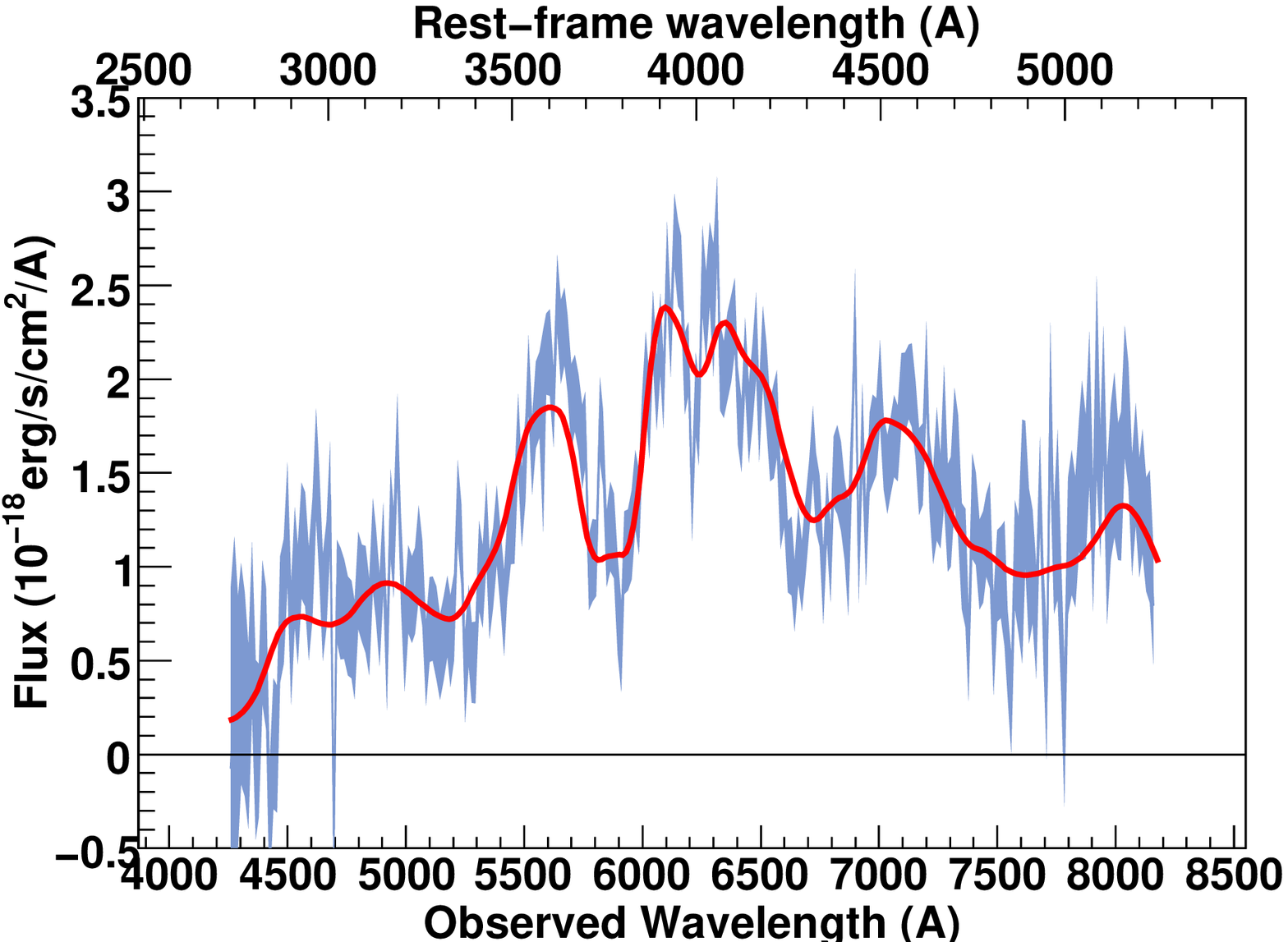}
                    \end{minipage}
                    \\
                    \end{tabular}
                    \caption{\label{fig:04D1ag383} SNLS-  04D1ag383: a SN~Ia supernova at z=0.557. The spectrum phase is 4.3. A        E(12) host model has been subtracted.}
                    \end{center}
                    \end{figure*}

\begin{figure*}[ht]
                    \begin{center}
                    \begin{tabular}[b]{cc}
                    \hspace{-5mm}
                    \vspace*{1cm}
                    \begin{minipage}[b]{10cm}
                    \centering
                    \includegraphics[width=10cm]{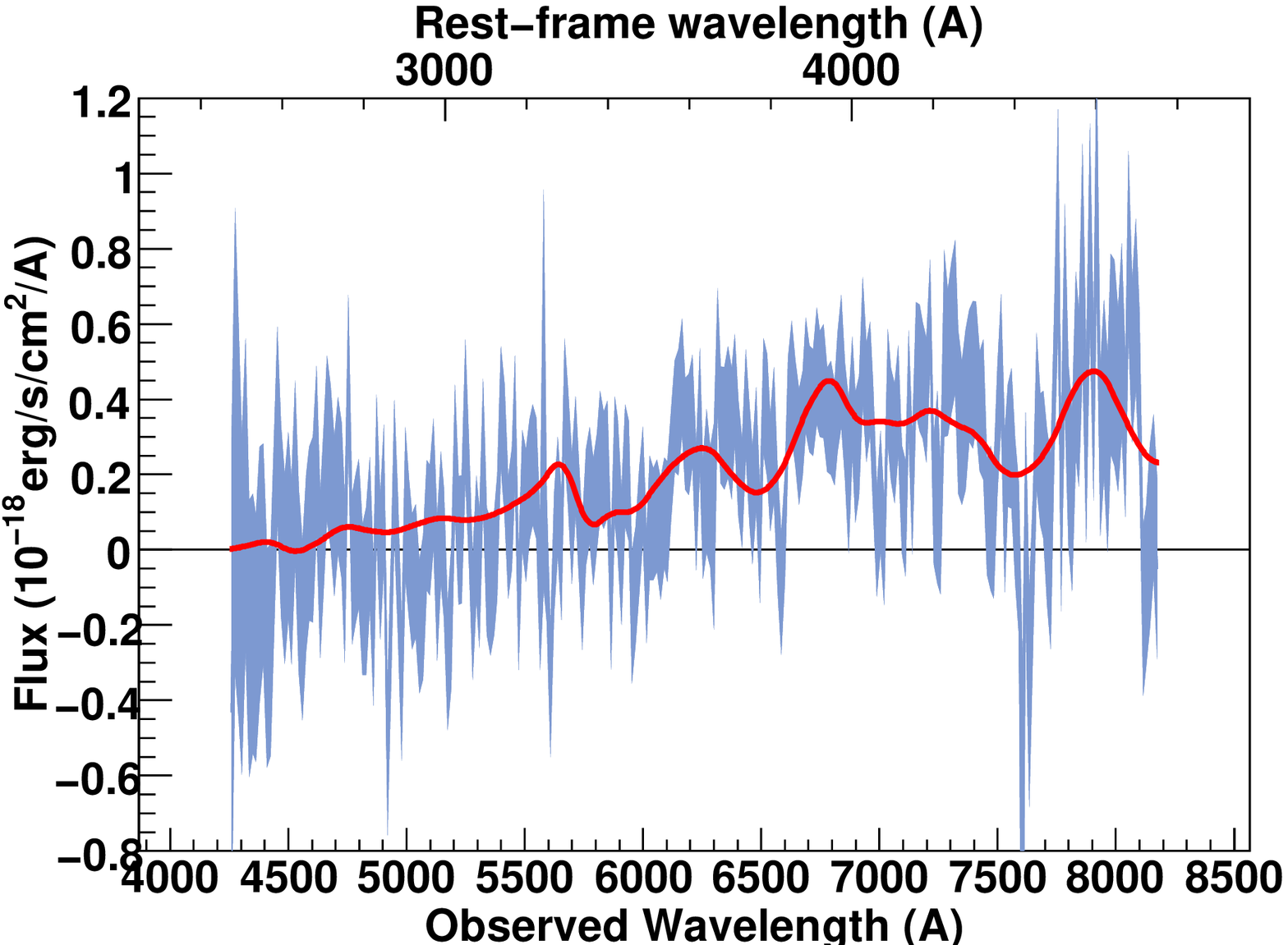}
                    \end{minipage}
                    &
                    \begin{minipage}[b]{10cm}
                    \end{minipage}
                    \\
                    \end{tabular}
                    \caption{\label{fig:04D1aj381} SNLS-  04D1aj381: a SN~Ia$\star$ supernova at z=0.721. The spectrum phase is 11.6. Best-fit obtained for a model with no host galaxy component.}
                    \end{center}
                    \end{figure*}

\begin{figure*}[ht]
                    \begin{center}
                    \begin{tabular}[b]{cc}
                    \hspace{-5mm}
                    \vspace*{1cm}
                    \begin{minipage}[b]{10cm}
                    \centering
                    \includegraphics[width=10cm]{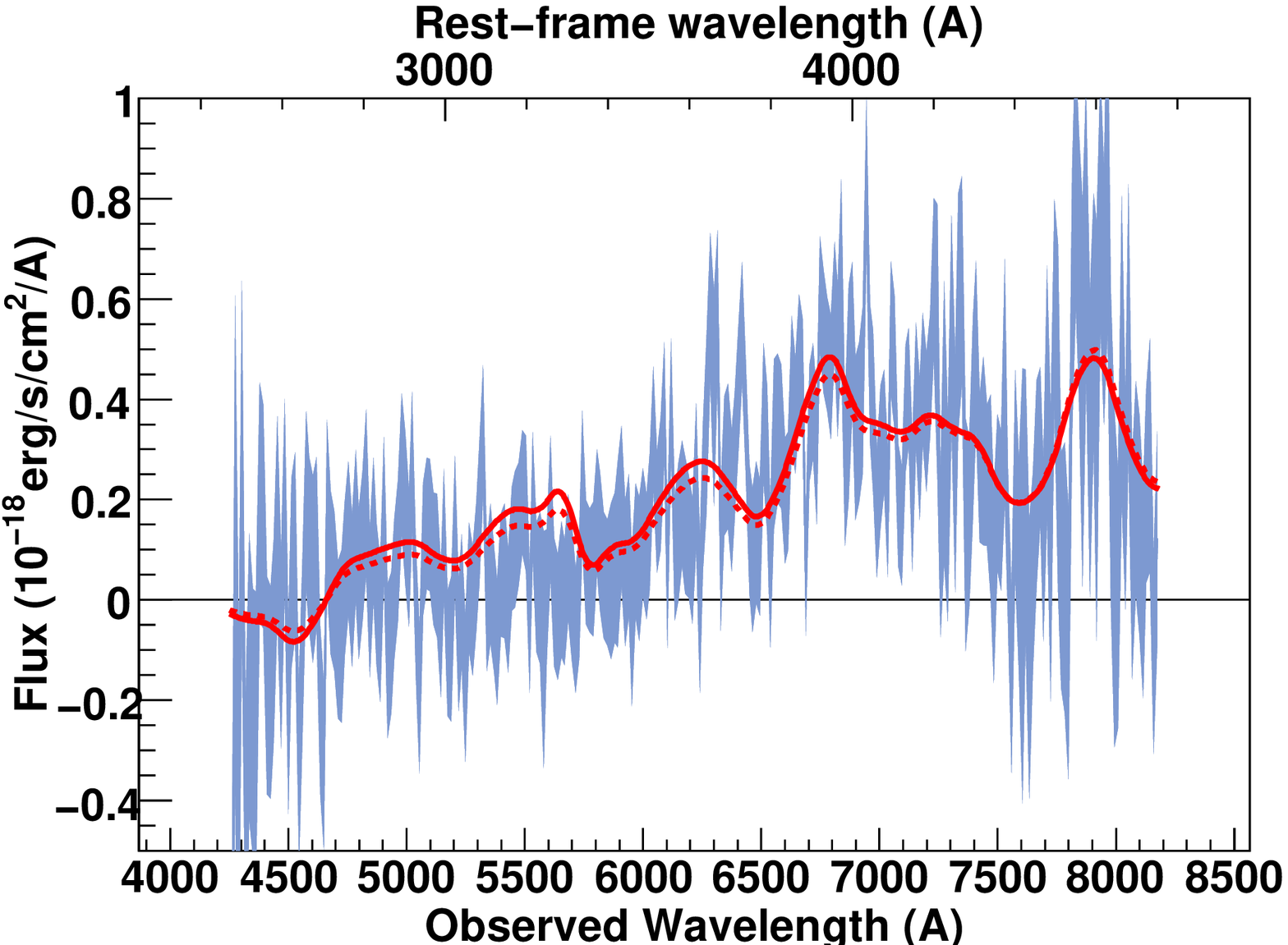}
                    \end{minipage}
                    &
                    \begin{minipage}[b]{10cm}
                    \end{minipage}
                    \\
                    \end{tabular}
                    \caption{\label{fig:04D1aj384} SNLS-  04D1aj384: a SN~Ia$\star$ supernova at z=0.721. The spectrum phase is 13.4.  Best-fit obtained for a model with no host galaxy component.}
                    \end{center}
                    \end{figure*}

\begin{figure*}[ht]
                    \begin{center}
                    \begin{tabular}[b]{cc}
                    \hspace{-5mm}
                    \vspace*{1cm}
                    \begin{minipage}[b]{10cm}
                    \centering
                    \includegraphics[width=10cm]{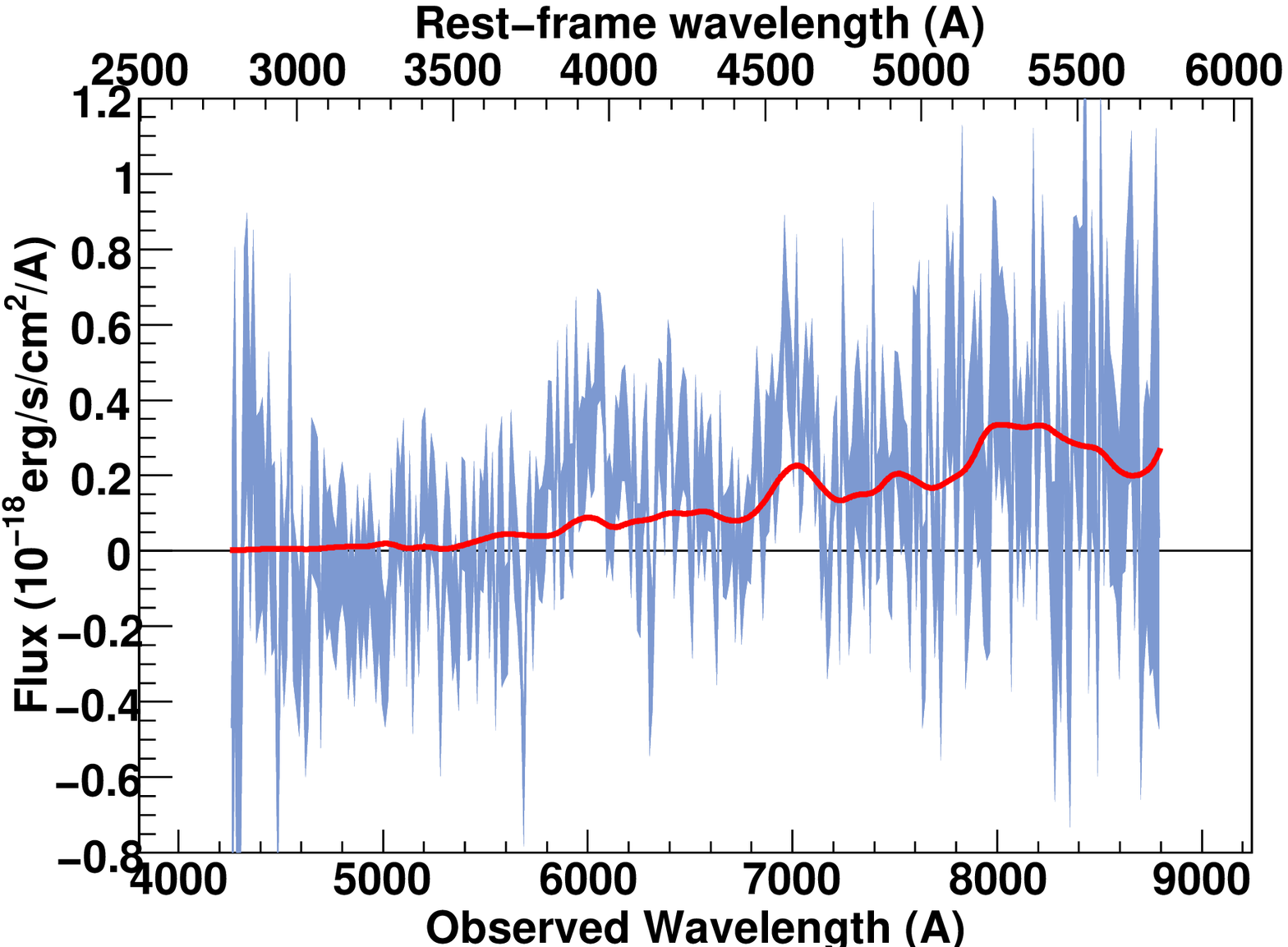}
                    \end{minipage}
                    &
                    \begin{minipage}[b]{10cm}
                    \end{minipage}
                    \\
                    \end{tabular}
                    \caption{\label{fig:04D1ak388} SNLS-  04D1ak388: a SN~Ia$\star$ supernova at z=0.526. The spectrum phase is 11.8. Best-fit obtained for a model with no host galaxy component.}
                    \end{center}
                    \end{figure*}

\begin{figure*}[ht]
                    \begin{center}
                    \begin{tabular}[b]{cc}
                    \hspace{-5mm}
                    \vspace*{1cm}
                    \begin{minipage}[b]{10cm}
                    \centering
                    \includegraphics[width=10cm]{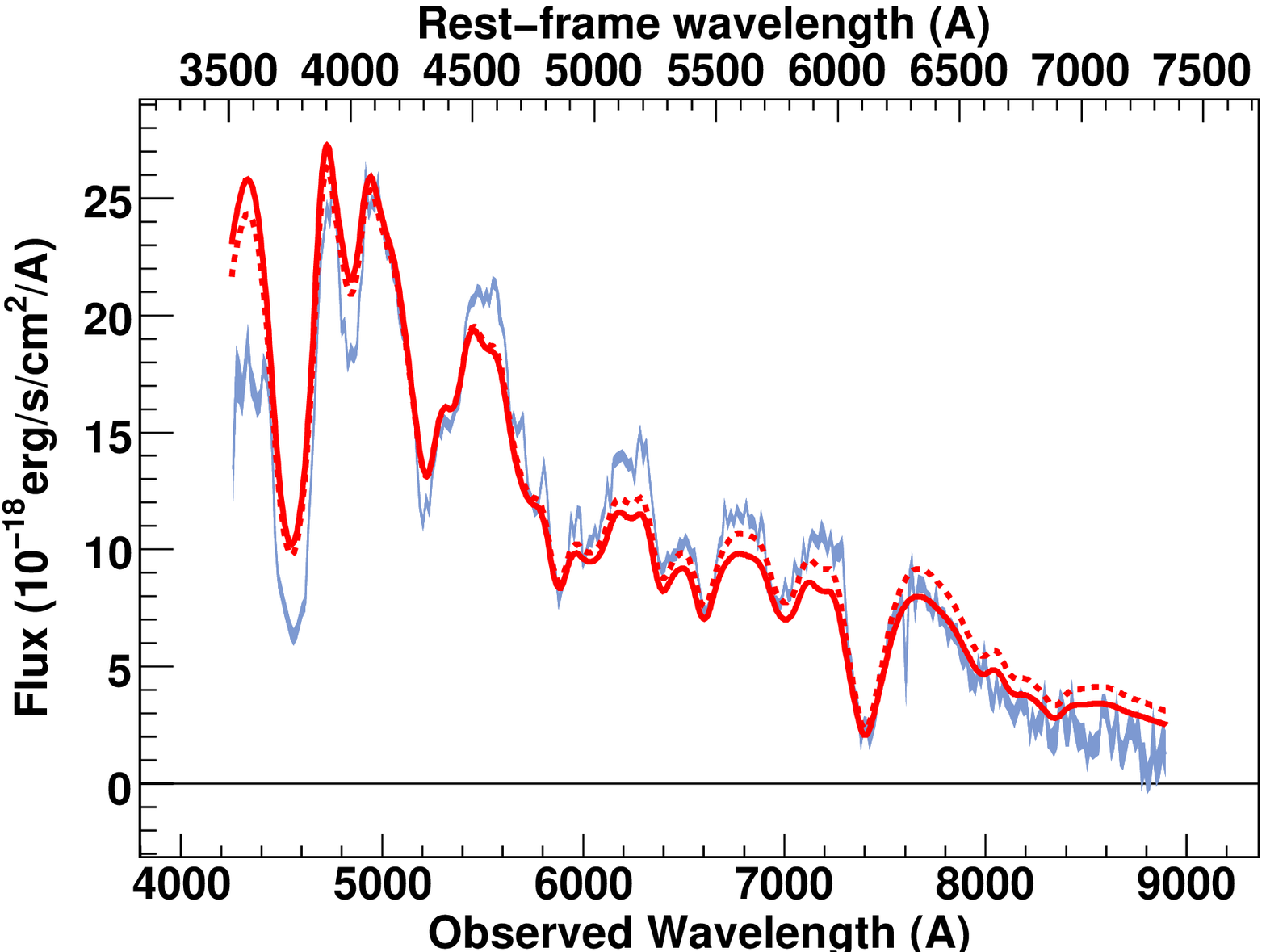}
                    \end{minipage}
                    &
                    \begin{minipage}[b]{10cm}
                    \end{minipage}
                    \\
                    \end{tabular}
                    \caption{\label{fig:04D1dc589} SNLS-  04D1dc589: a SN~Ia supernova at z=0.211. The spectrum phase is -0.4.  Best-fit obtained for a model with no host galaxy component.}
                    \end{center}
                    \end{figure*}

\begin{figure*}[ht]
                    \begin{center}
                    \begin{tabular}[b]{cc}
                    \hspace{-5mm}
                    \vspace*{1cm}
                    \begin{minipage}[b]{10cm}
                    \centering
                    \includegraphics[width=10cm]{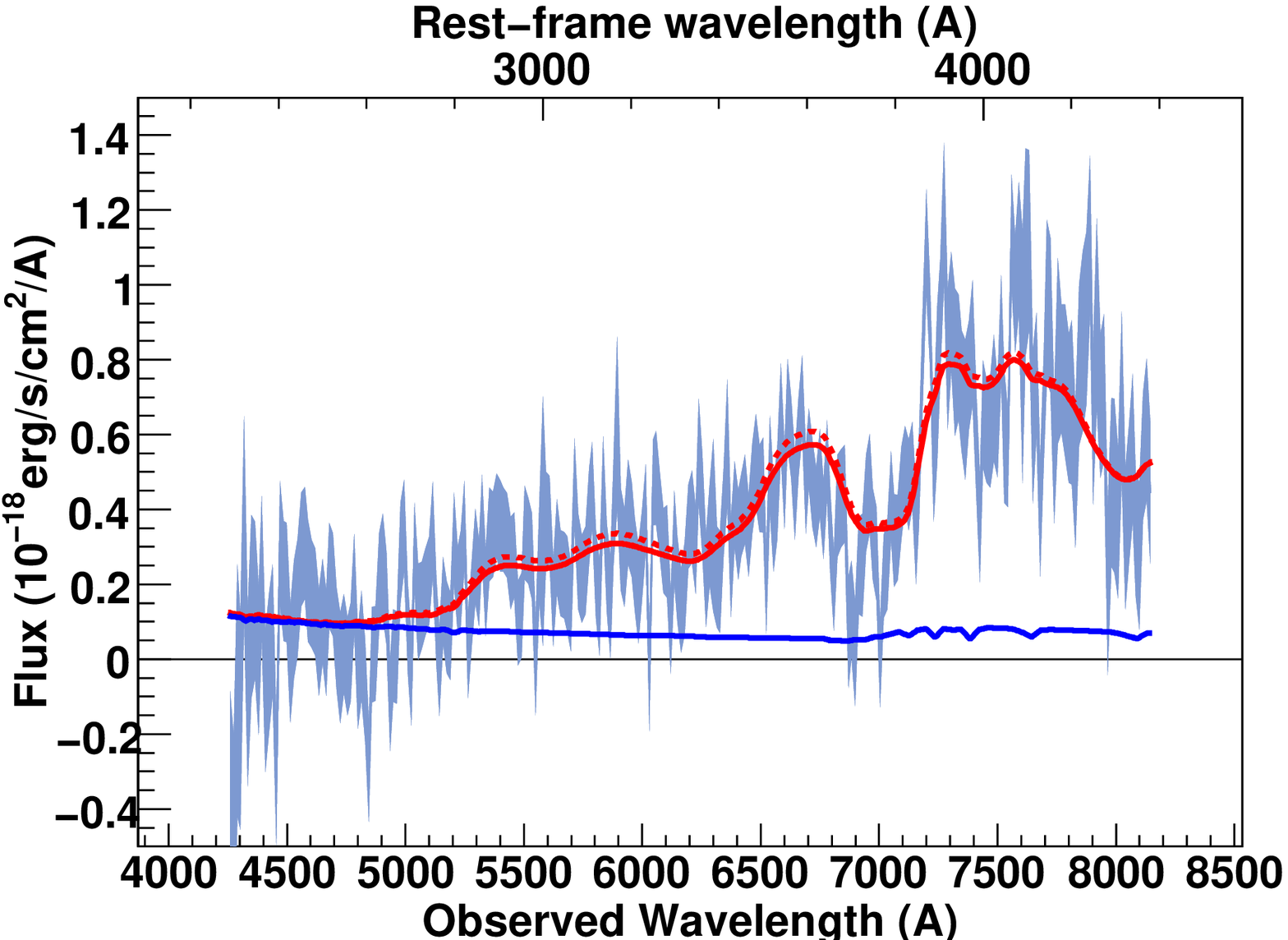}
                    \end{minipage}
                    &
                    \begin{minipage}[b]{10cm}
                    \hspace{-5mm}\includegraphics[width=10cm]{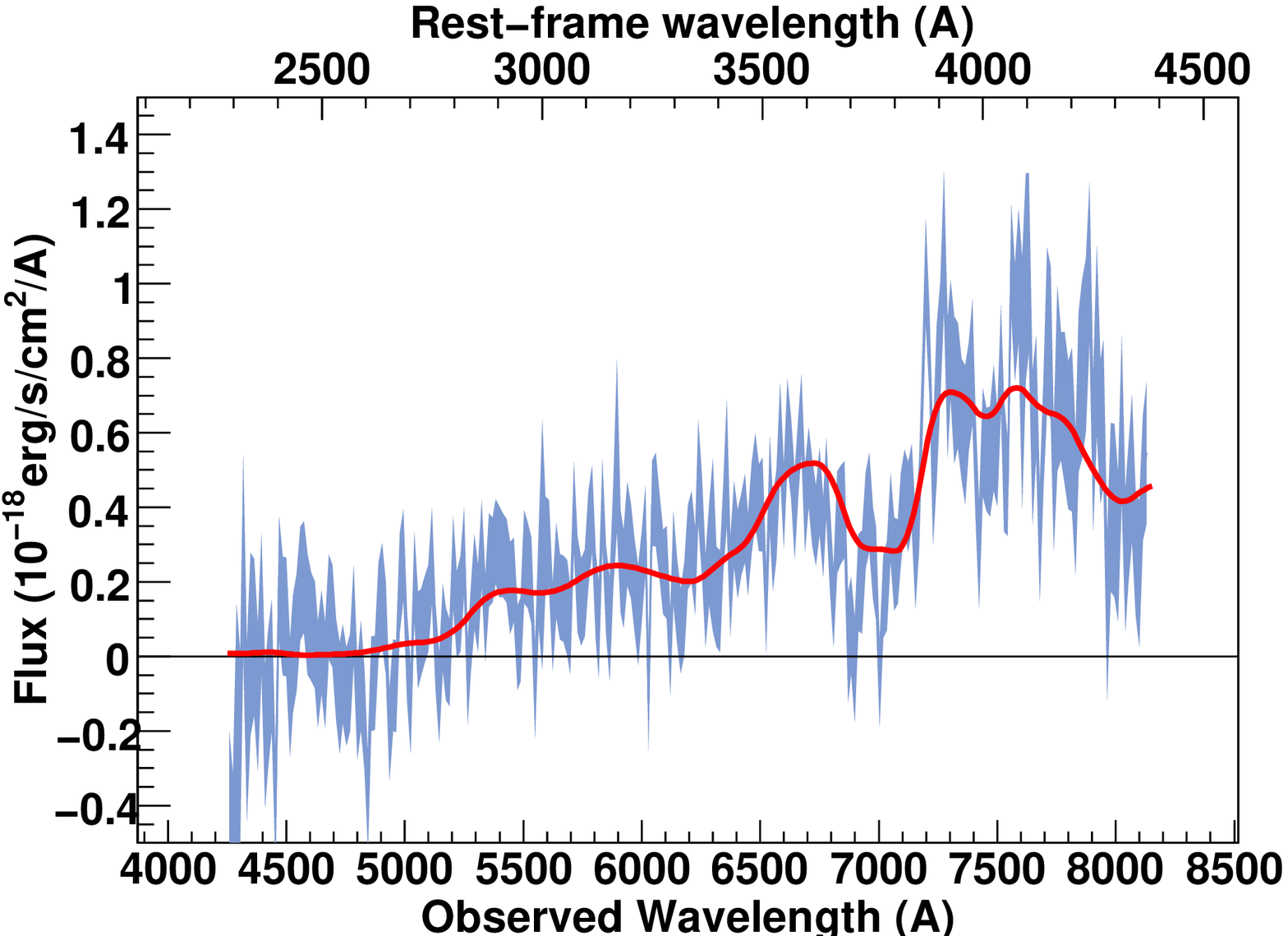}
                    \end{minipage}
                    \\
                    \end{tabular}
                    \caption{\label{fig:04D1ff619} SNLS-  04D1ff619: a SN~Ia supernova at z=0.86. The spectrum phase is 4.6. A        Sc(2) host model has been subtracted.}
                    \end{center}
                    \end{figure*}

\begin{figure*}[ht]
                    \begin{center}
                    \begin{tabular}[b]{cc}
                    \hspace{-5mm}
                    \vspace*{1cm}
                    \begin{minipage}[b]{10cm}
                    \centering
                    \includegraphics[width=10cm]{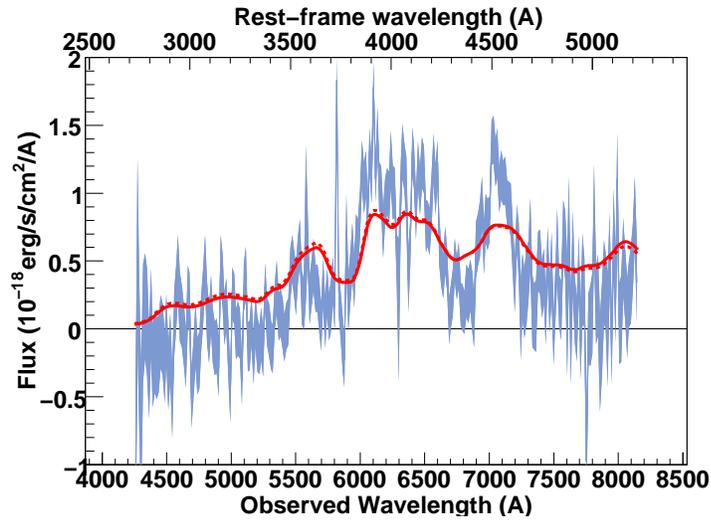}
                    \end{minipage}
                    &
                    \begin{minipage}[b]{10cm}
                    \end{minipage}
                    \\
                    \end{tabular}
                    \caption{\label{fig:04D1hx628} SNLS-  04D1hx628: a SN~Ia supernova at z=0.560. The spectrum phase is 5.9.  Best-fit obtained for a model with no host galaxy component. Note the model is unable to reproduce the large feature at rest-frame 4500 $\AA$.}
                    \end{center}
                    \end{figure*}

\begin{figure*}[ht]
                    \begin{center}
                    \begin{tabular}[b]{cc}
                    \hspace{-5mm}
                    \vspace*{1cm}
                    \begin{minipage}[b]{10cm}
                    \centering
                    \includegraphics[width=10cm]{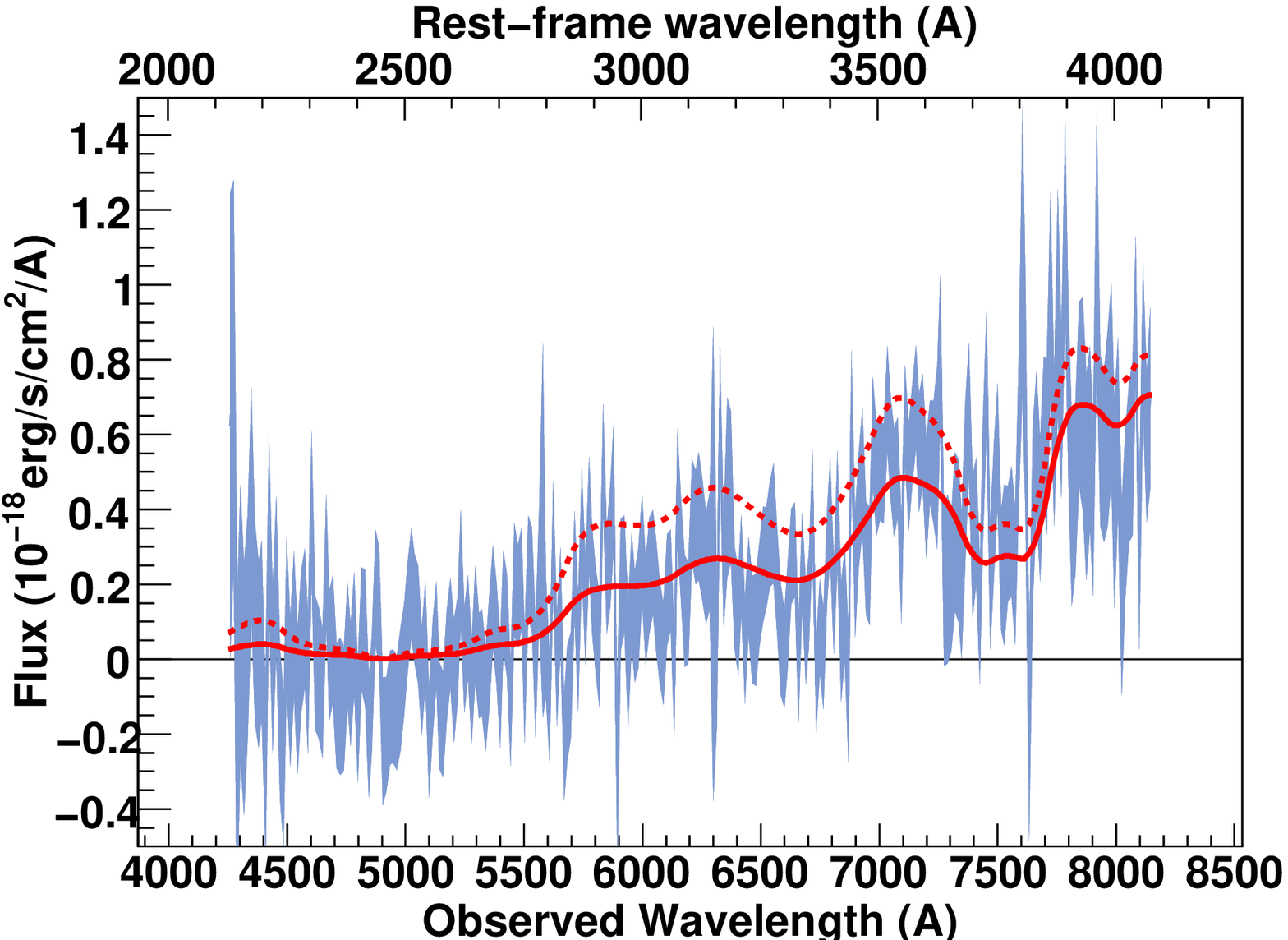}
                    \end{minipage}
                    &
                    \begin{minipage}[b]{10cm}
                    \end{minipage}
                    \\
                    \end{tabular}
                    \caption{\label{fig:04D1iv630} SNLS-  04D1iv630: a SN~Ia supernova at z=0.998. The spectrum phase is 3.0.  Best-fit obtained for a model with no host galaxy component.}
                    \end{center}
                    \end{figure*}

\begin{figure*}[ht]
                    \begin{center}
                    \begin{tabular}[b]{cc}
                    \hspace{-5mm}
                    \vspace*{1cm}
                    \begin{minipage}[b]{10cm}
                    \centering
                    \includegraphics[width=10cm]{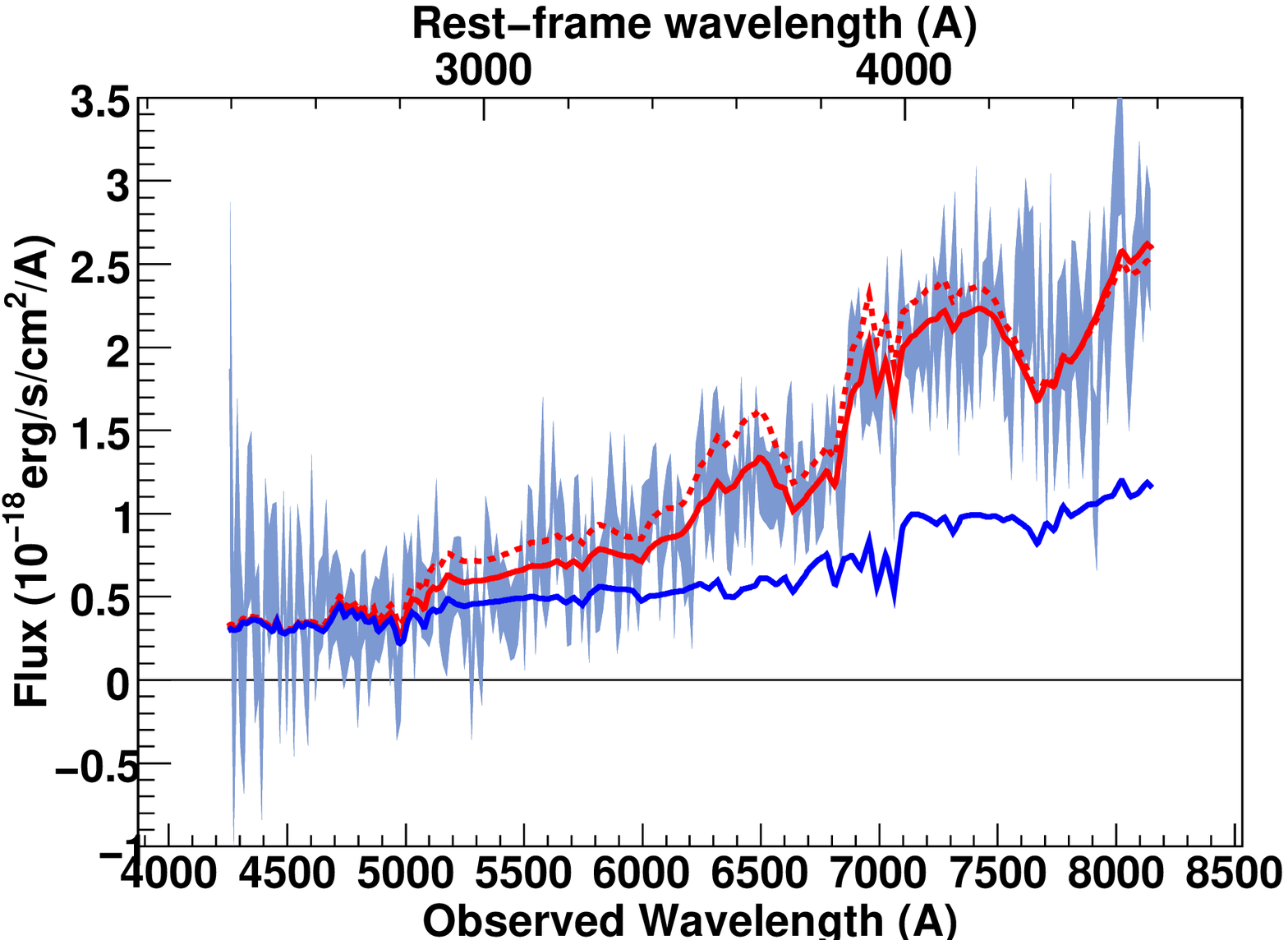}
                    \end{minipage}
                    &
                    \begin{minipage}[b]{10cm}
                    \hspace{-5mm}\includegraphics[width=10cm]{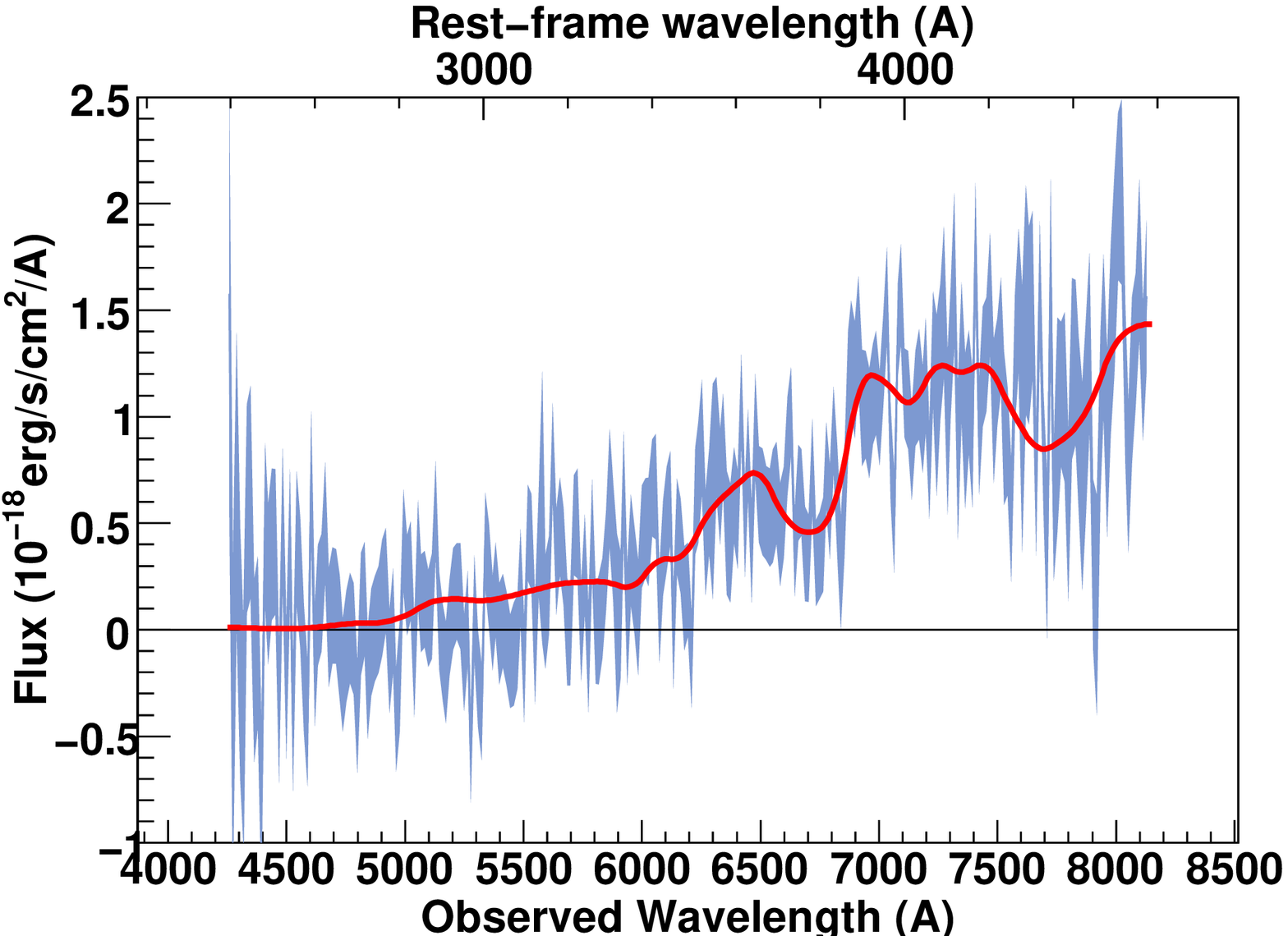}
                    \end{minipage}
                    \\
                    \end{tabular}
                    \caption{\label{fig:04D1jd630} SNLS-  04D1jd630: a SN~Ia$\star$ supernova at z=0.778. The spectrum phase is 6.8. A         E(2) host model has been subtracted.}
                    \end{center}
                    \end{figure*}

\begin{figure*}[ht]
                    \begin{center}
                    \begin{tabular}[b]{cc}
                    \hspace{-5mm}
                    \vspace*{1cm}
                    \begin{minipage}[b]{10cm}
                    \centering
                    \includegraphics[width=10cm]{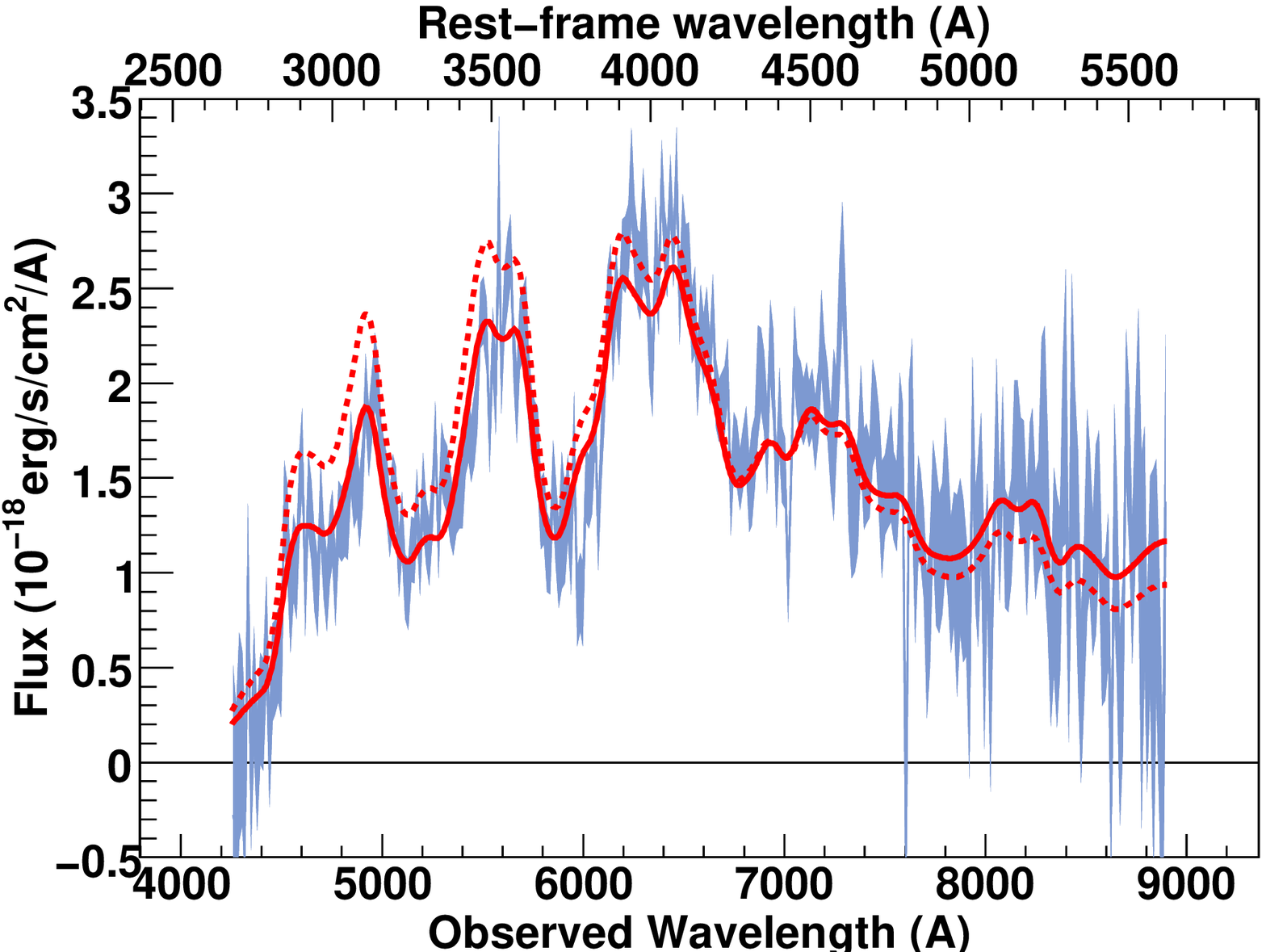}
                    \end{minipage}
                    &
                    \begin{minipage}[b]{10cm}
                    \end{minipage}
                    \\
                    \end{tabular}
                    \caption{\label{fig:04D1kj650} SNLS-  04D1kj650: a SN~Ia supernova at z=0.585. The spectrum phase is -3.7. Best-fit obtained for a model with no host galaxy component}
                    \end{center}
                    \end{figure*}

\begin{figure*}[ht]
                    \begin{center}
                    \begin{tabular}[b]{cc}
                    \hspace{-5mm}
                    \vspace*{1cm}
                    \begin{minipage}[b]{10cm}
                    \centering
                    \includegraphics[width=10cm]{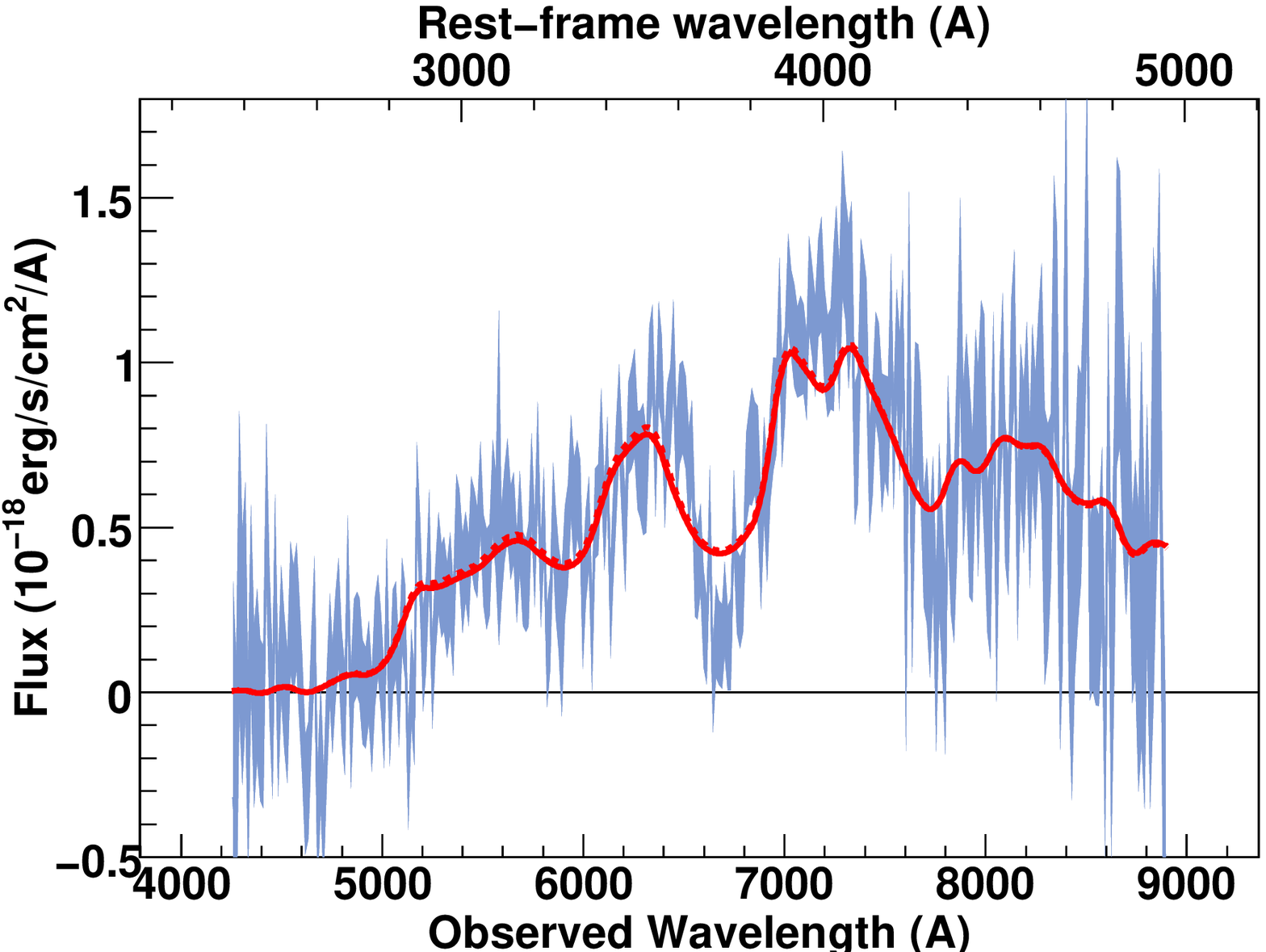}
                    \end{minipage}
                    &
                    \begin{minipage}[b]{10cm}
                    \end{minipage}
                    \\
                    \end{tabular}
                    \caption{\label{fig:04D1ks649} SNLS-  04D1ks649: a SN~Ia supernova at z=0.80. The spectrum phase is -1.0. Best-fit obtained for a model with no host galaxy component.}
                    \end{center}
                    \end{figure*}

\begin{figure*}[ht]
                    \begin{center}
                    \begin{tabular}[b]{cc}
                    \hspace{-5mm}
                    \vspace*{1cm}
                    \begin{minipage}[b]{10cm}
                    \centering
                    \includegraphics[width=10cm]{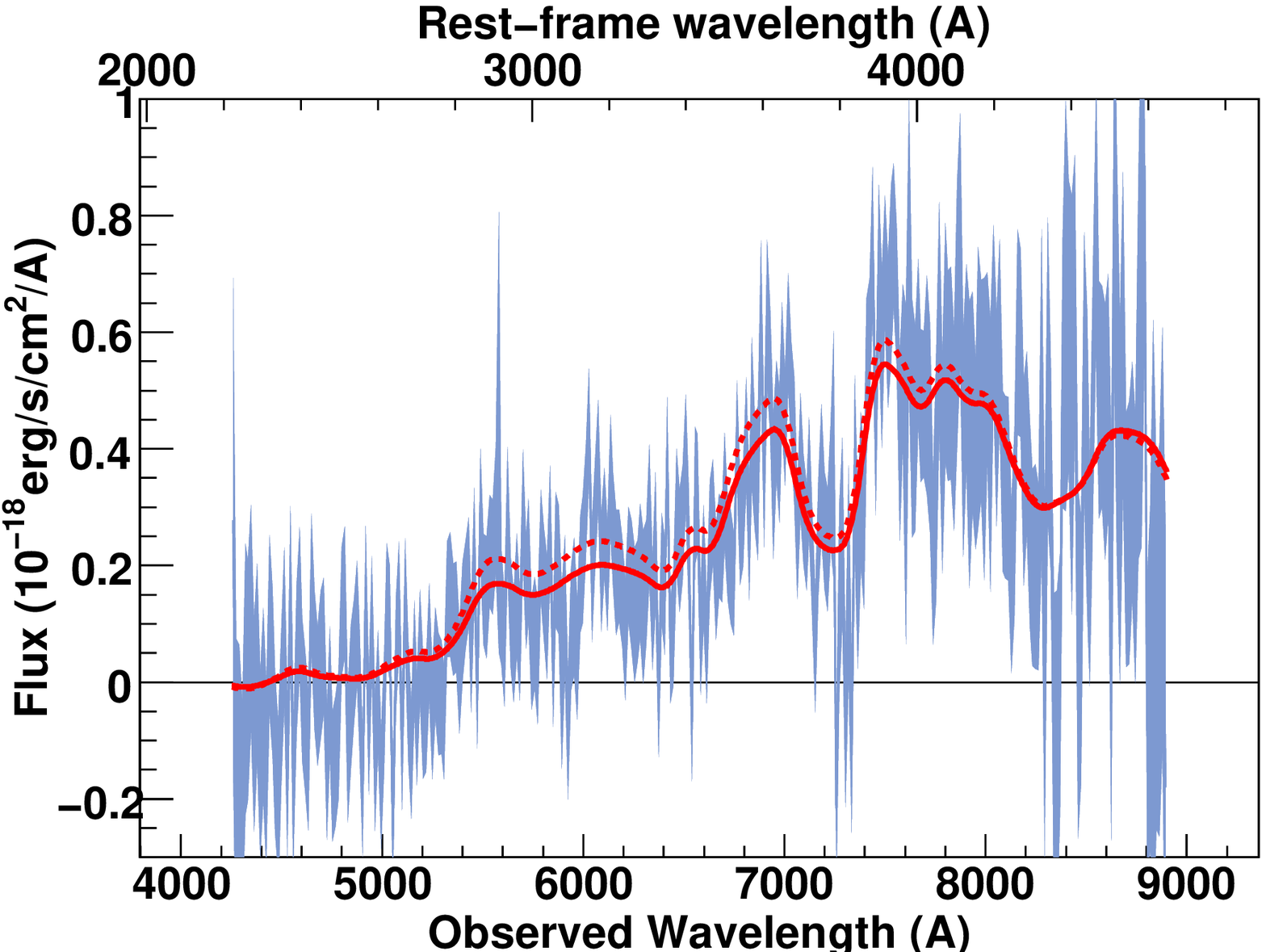}
                    \end{minipage}
                    &
                    \begin{minipage}[b]{10cm}
                    \end{minipage}
                    \\
                    \end{tabular}
                    \caption{\label{fig:04D1ow683} SNLS-  04D1ow683: a SN~Ia supernova at z=0.915. The spectrum phase is 6.4.  Best-fit obtained for a model with no host galaxy component.}
                    \end{center}
                    \end{figure*}

\begin{figure*}[ht]
                    \begin{center}
                    \begin{tabular}[b]{cc}
                    \hspace{-5mm}
                    \vspace*{1cm}
                    \begin{minipage}[b]{10cm}
                    \centering
                    \includegraphics[width=10cm]{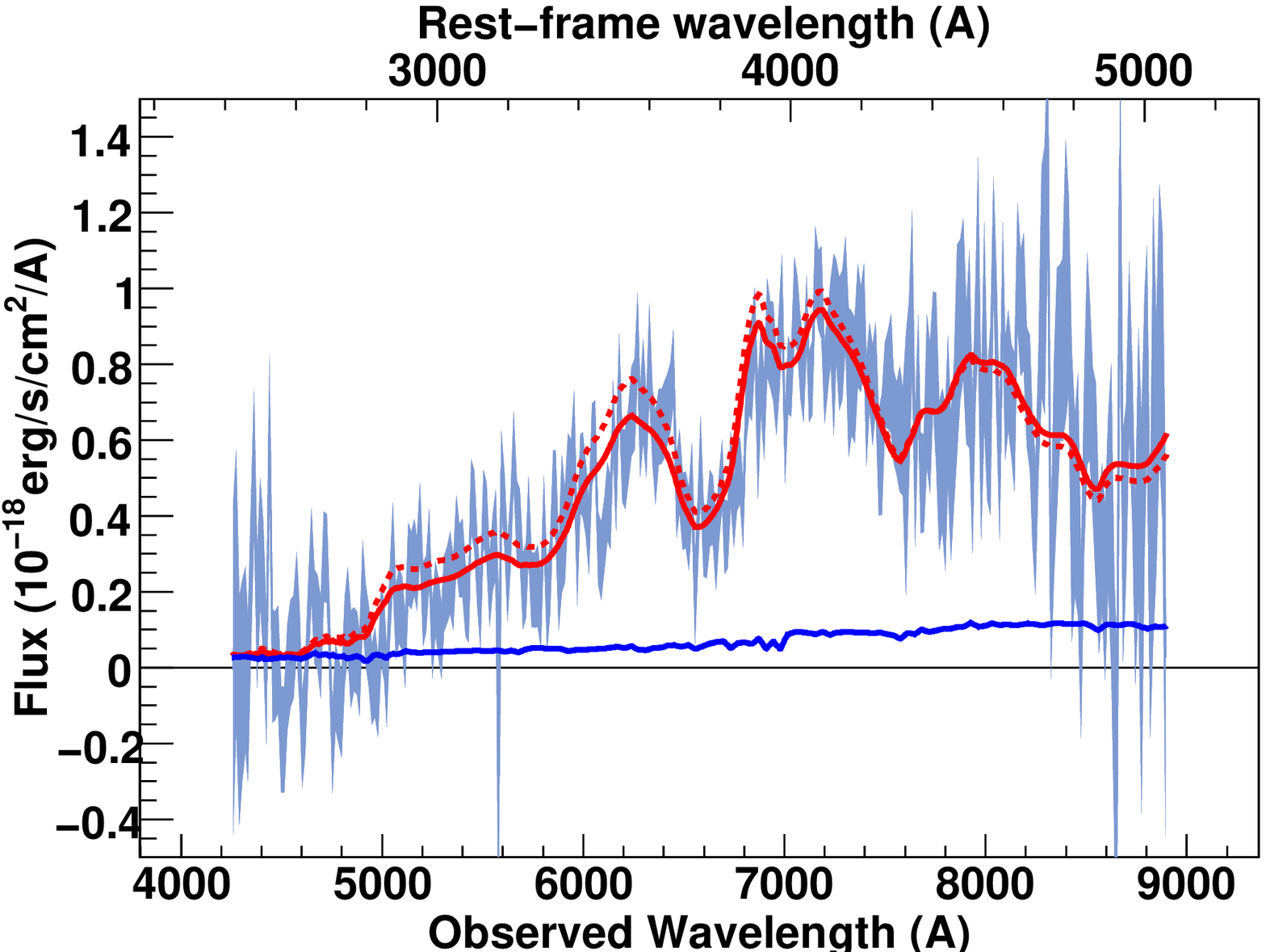}
                    \end{minipage}
                    &
                    \begin{minipage}[b]{10cm}
                    \hspace{-5mm}\includegraphics[width=10cm]{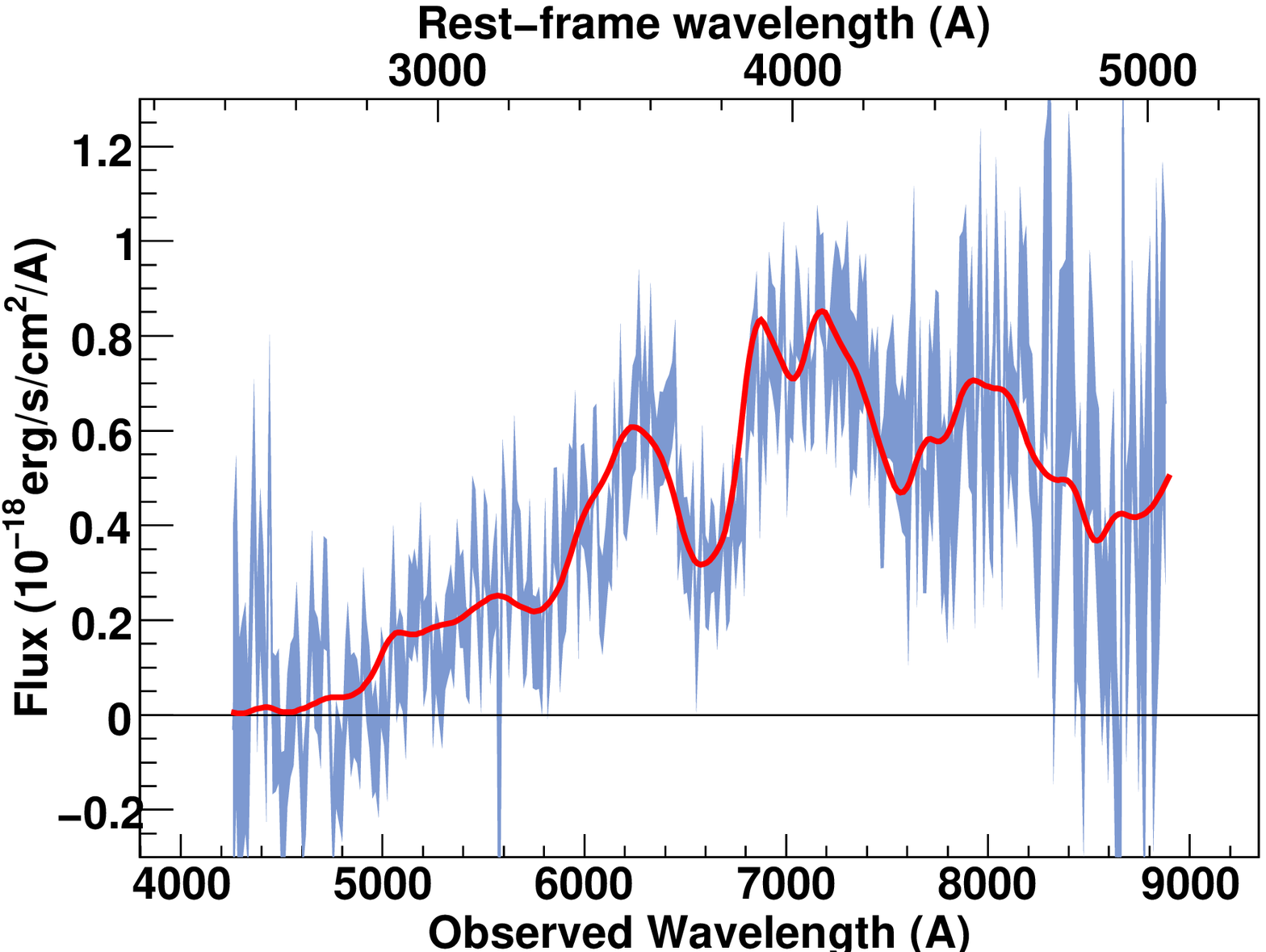}
                    \end{minipage}
                    \\
                    \end{tabular}
                    \caption{\label{fig:04D1pc684} SNLS-  04D1pc684: a SN~Ia supernova at z=0.770. The spectrum phase is 0.1. A         E(3) host model has been subtracted.}
                    \end{center}
                    \end{figure*}

\begin{figure*}[ht]
                    \begin{center}
                    \begin{tabular}[b]{cc}
                    \hspace{-5mm}
                    \vspace*{1cm}
                    \begin{minipage}[b]{10cm}
                    \centering
                    \includegraphics[width=10cm]{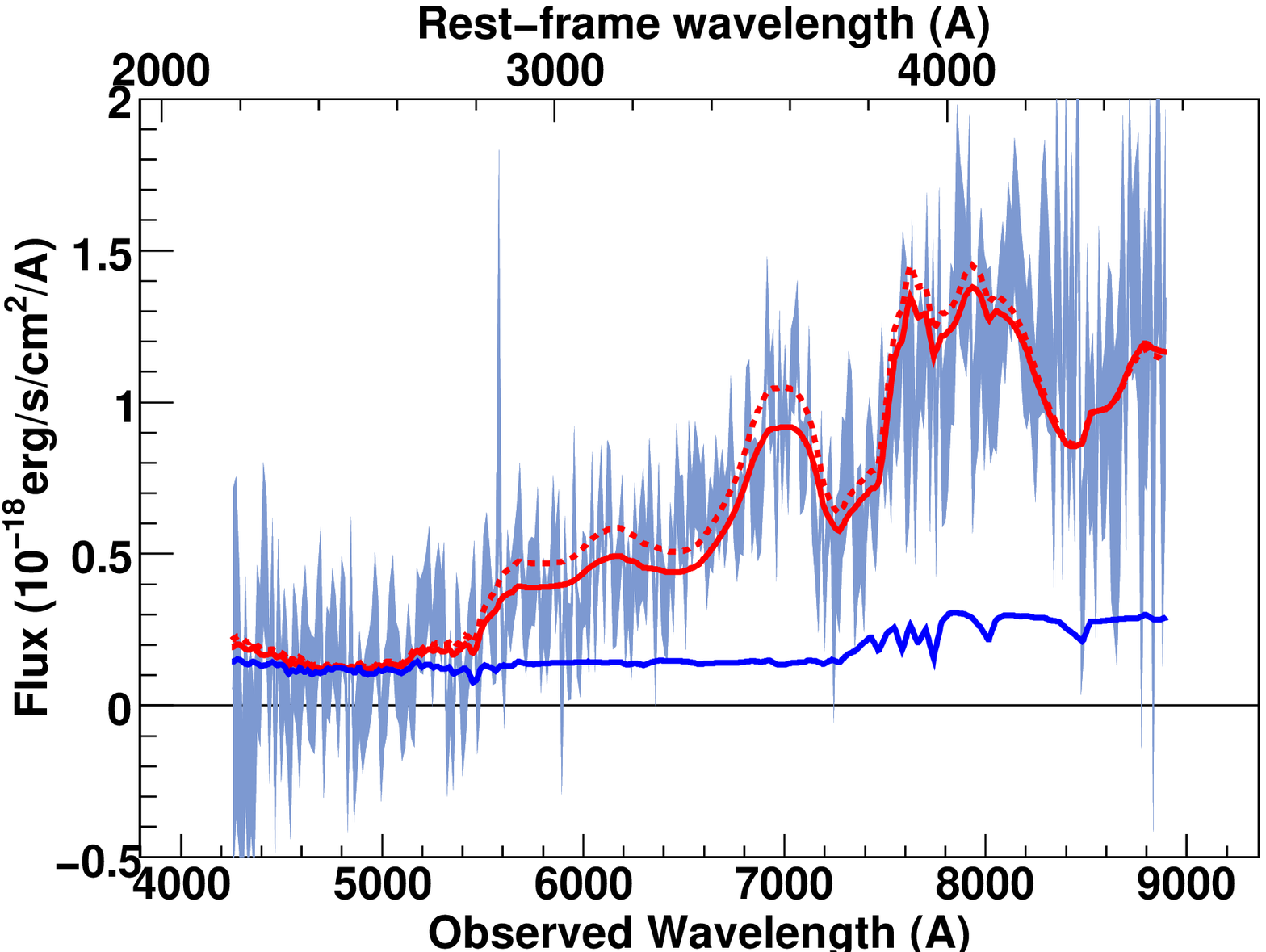}
                    \end{minipage}
                    &
                    \begin{minipage}[b]{10cm}
                    \hspace{-5mm}\includegraphics[width=10cm]{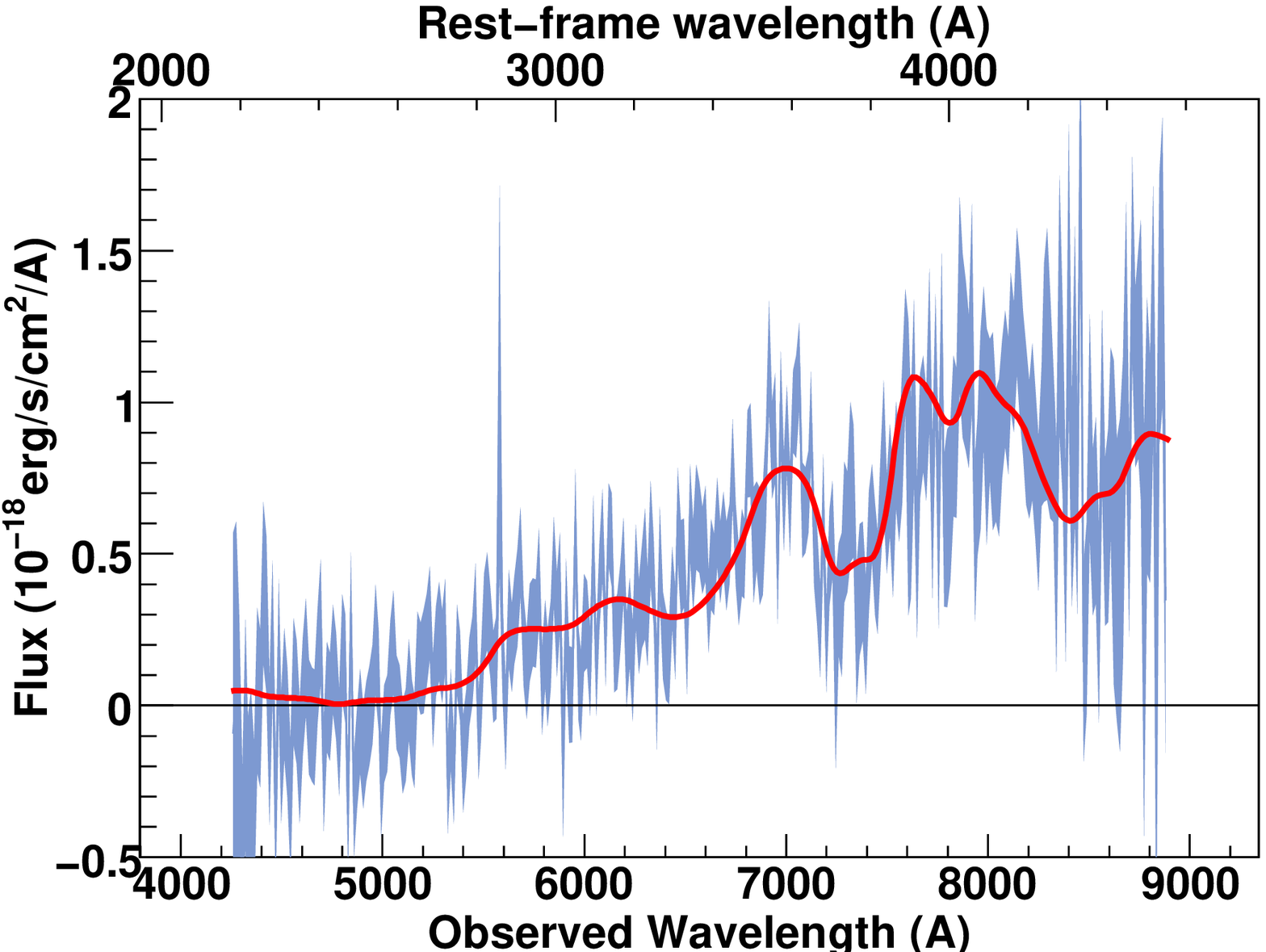}
                    \end{minipage}
                    \\
                    \end{tabular}
                    \caption{\label{fig:04D1pd682} SNLS- 04D1pd682: a
                      SN~Ia supernova at z=0.95. The spectrum phase is
                      2.5. A E(1) host model has been subtracted.}
                    \end{center}
                    \end{figure*}

\begin{figure*}[ht]
                    \begin{center}
                    \begin{tabular}[b]{cc}
                    \hspace{-5mm}
                    \vspace*{1cm}
                    \begin{minipage}[b]{10cm}
                    \centering
                    \includegraphics[width=10cm]{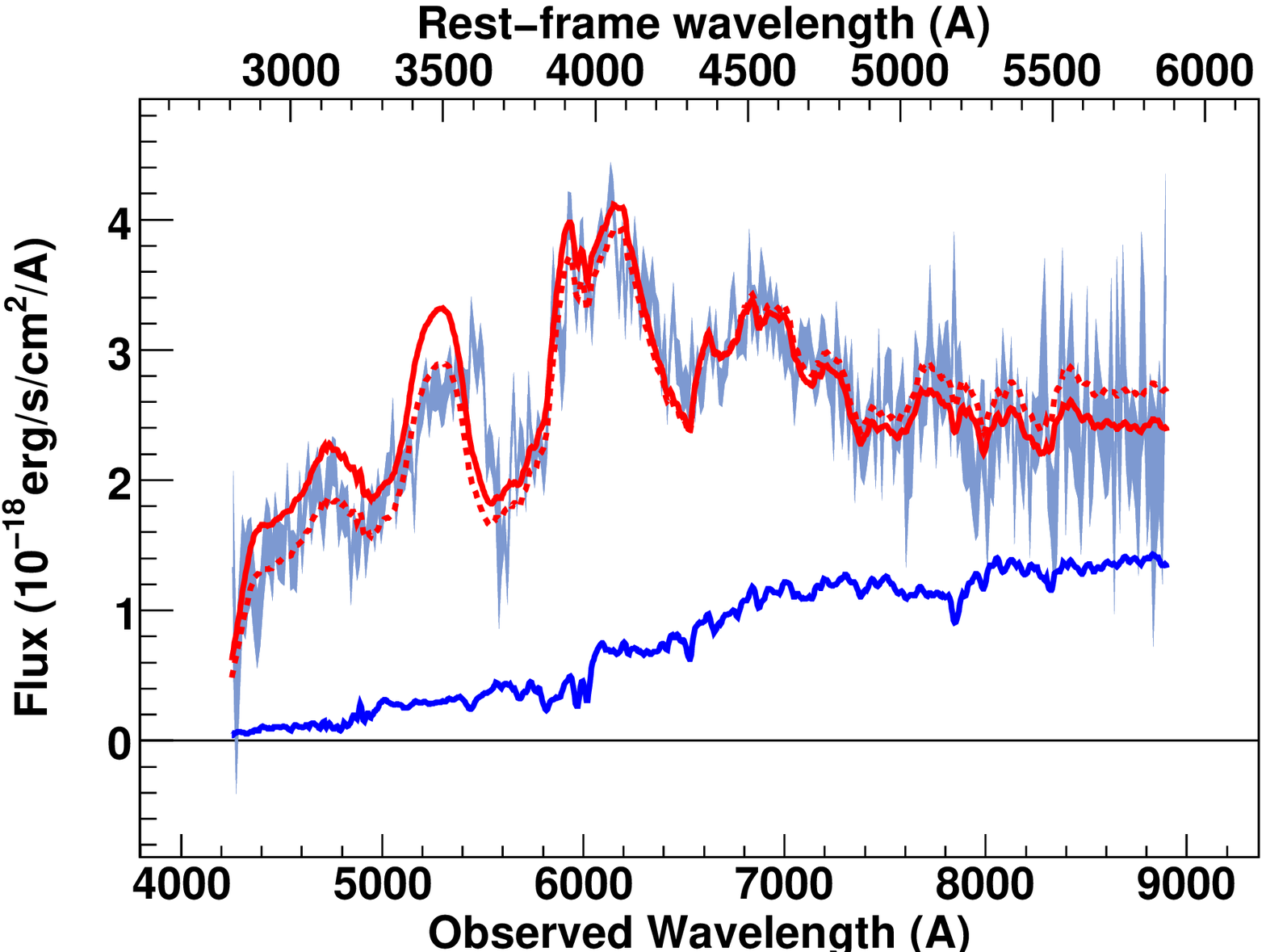}
                    \end{minipage}
                    &
                    \begin{minipage}[b]{10cm}
                    \hspace{-5mm}\includegraphics[width=10cm]{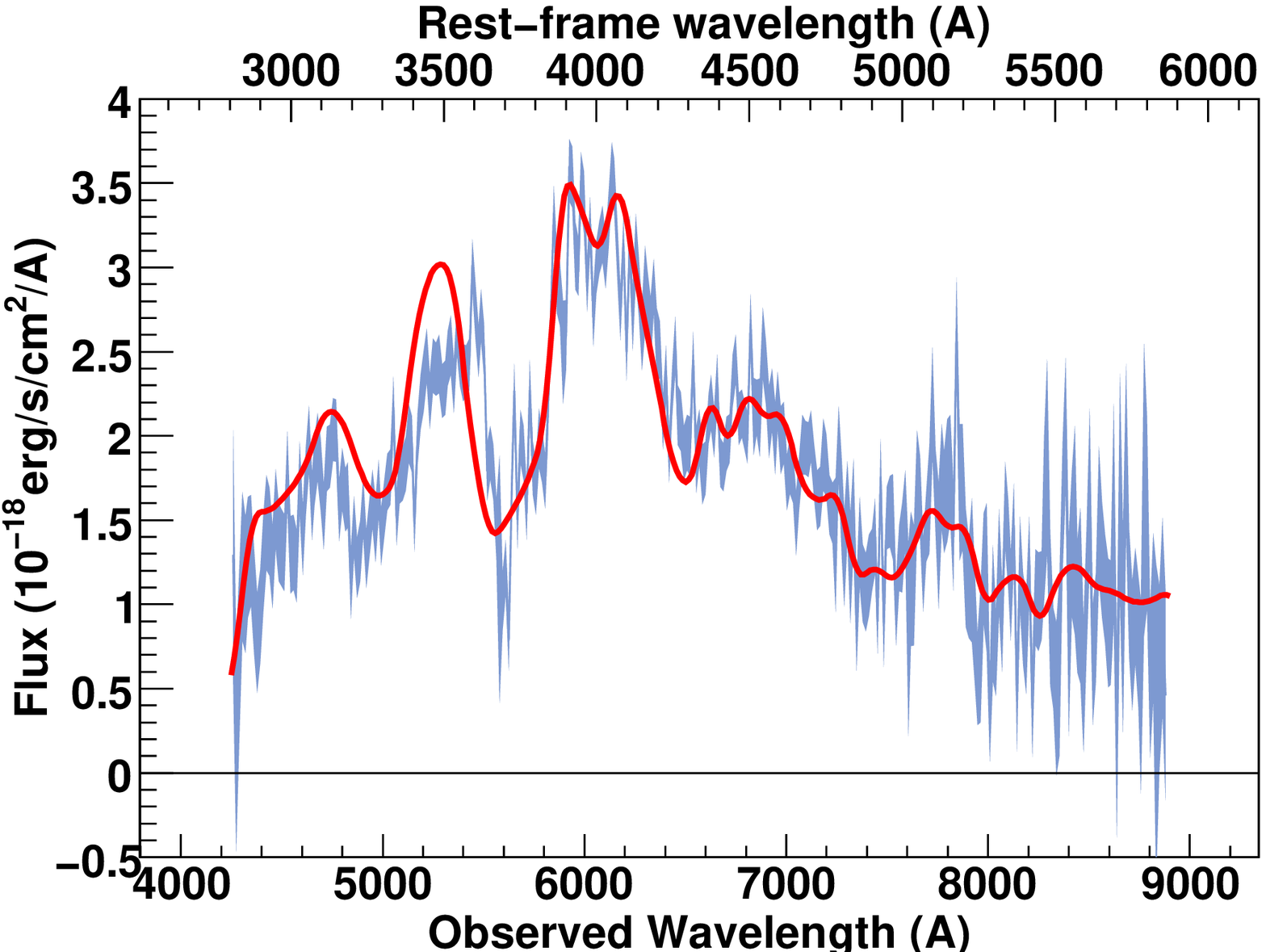}
                    \end{minipage}
                    \\
                    \end{tabular}
                    \caption{\label{fig:04D1pg684} SNLS-  04D1pg684: a SN~Ia supernova at z=0.515. The spectrum phase is -1.3. A         Sa-Sb host model has been subtracted.}
                    \end{center}
                    \end{figure*}

\begin{figure*}[ht]
                    \begin{center}
                    \begin{tabular}[b]{cc}
                    \hspace{-5mm}
                    \vspace*{1cm}
                    \begin{minipage}[b]{10cm}
                    \centering
                    \includegraphics[width=10cm]{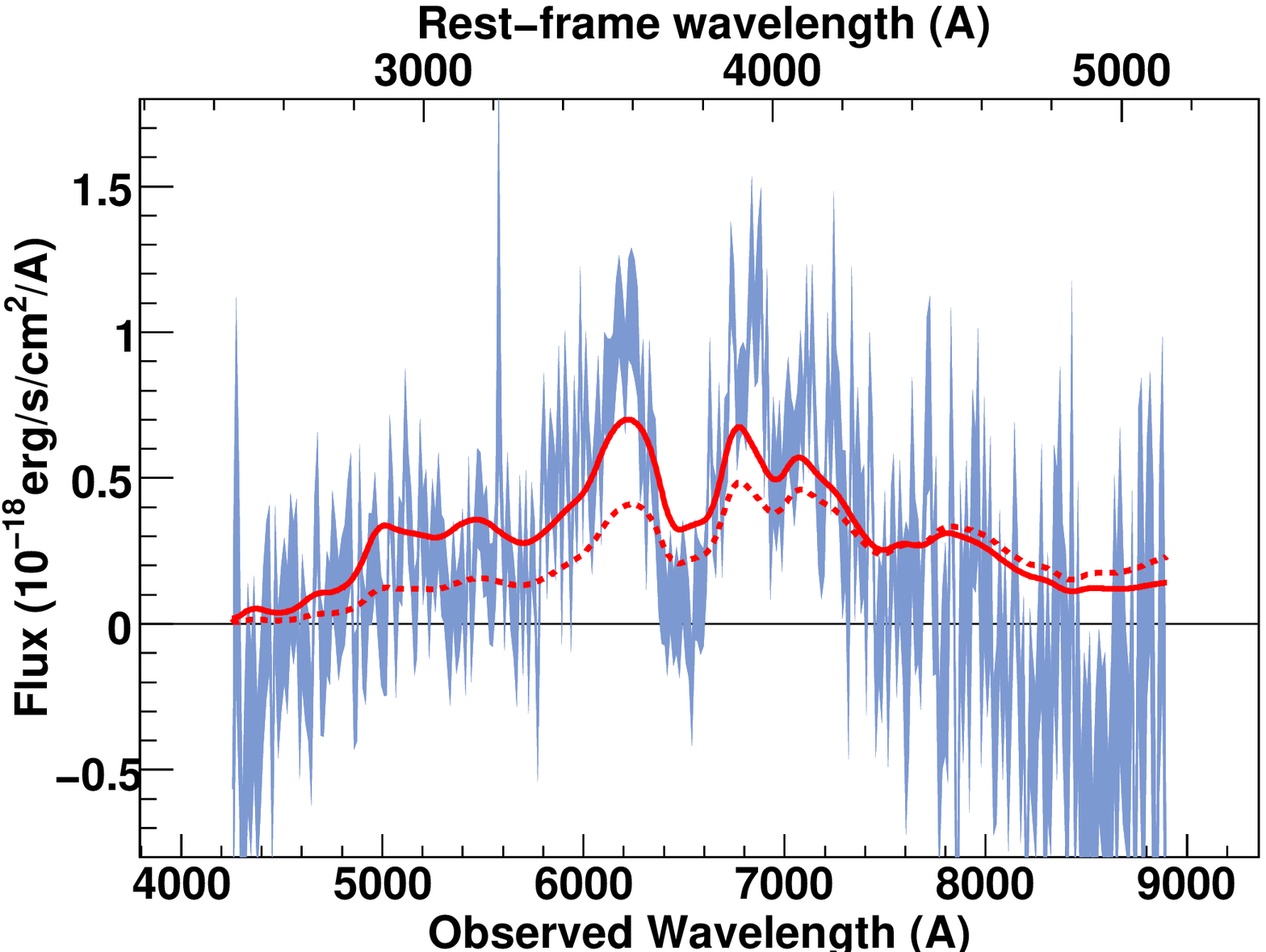}
                    \end{minipage}
                    &
                    \begin{minipage}[b]{10cm}
                    \end{minipage}
                    \\
                    \end{tabular}
                    \caption{\label{fig:04D1pp683} SNLS-  04D1pp683: a SN~Ia supernova at z=0.735. The spectrum phase is 2.2.  Best-fit obtained for a model with no host galaxy component.}
                    \end{center}
                    \end{figure*}

\begin{figure*}[ht]
                    \begin{center}
                    \begin{tabular}[b]{cc}
                    \hspace{-5mm}
                    \vspace*{1cm}
                    \begin{minipage}[b]{10cm}
                    \centering
                    \includegraphics[width=10cm]{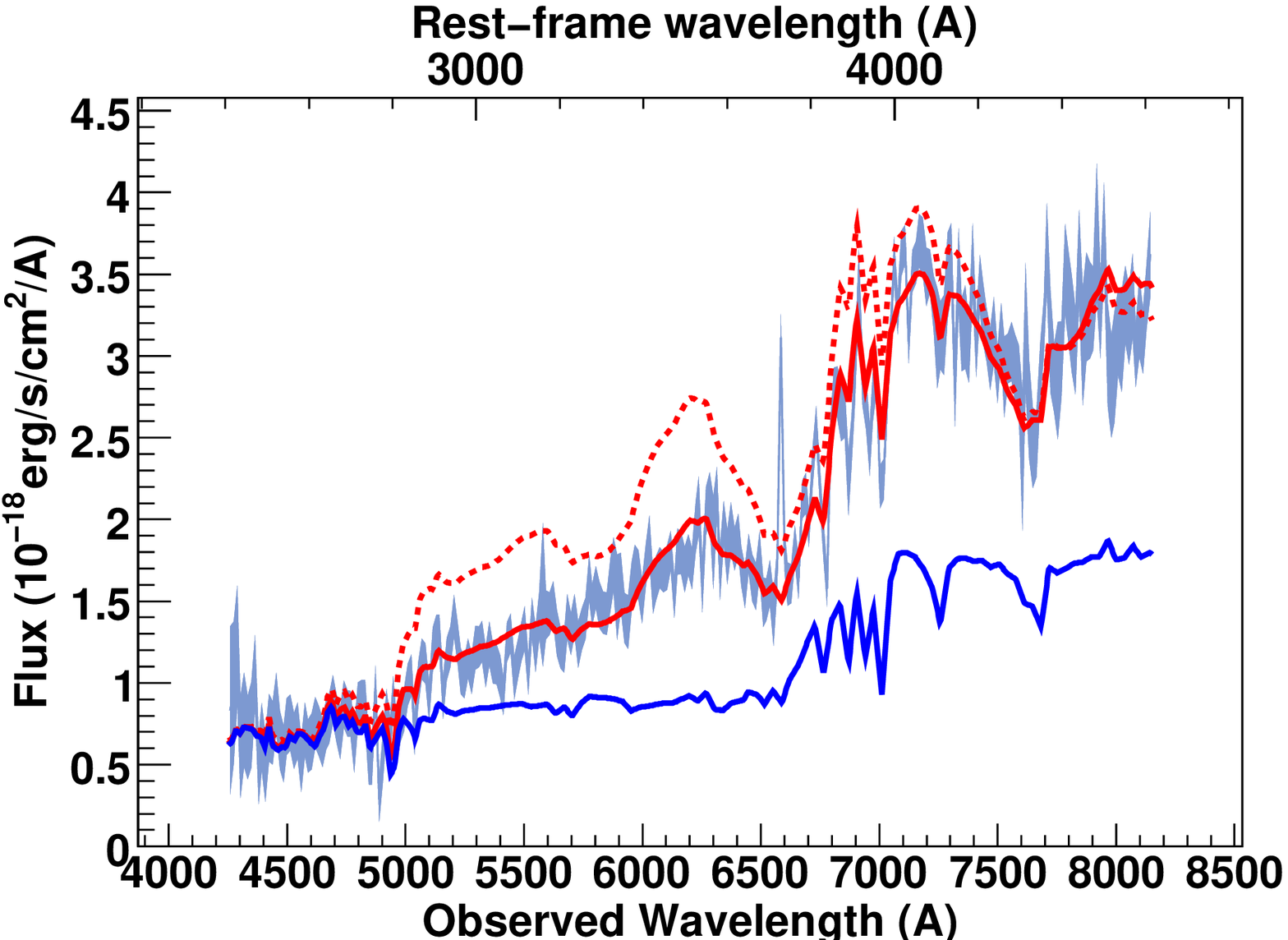}
                    \end{minipage}
                    &
                    \begin{minipage}[b]{10cm}
                    \hspace{-5mm}\includegraphics[width=10cm]{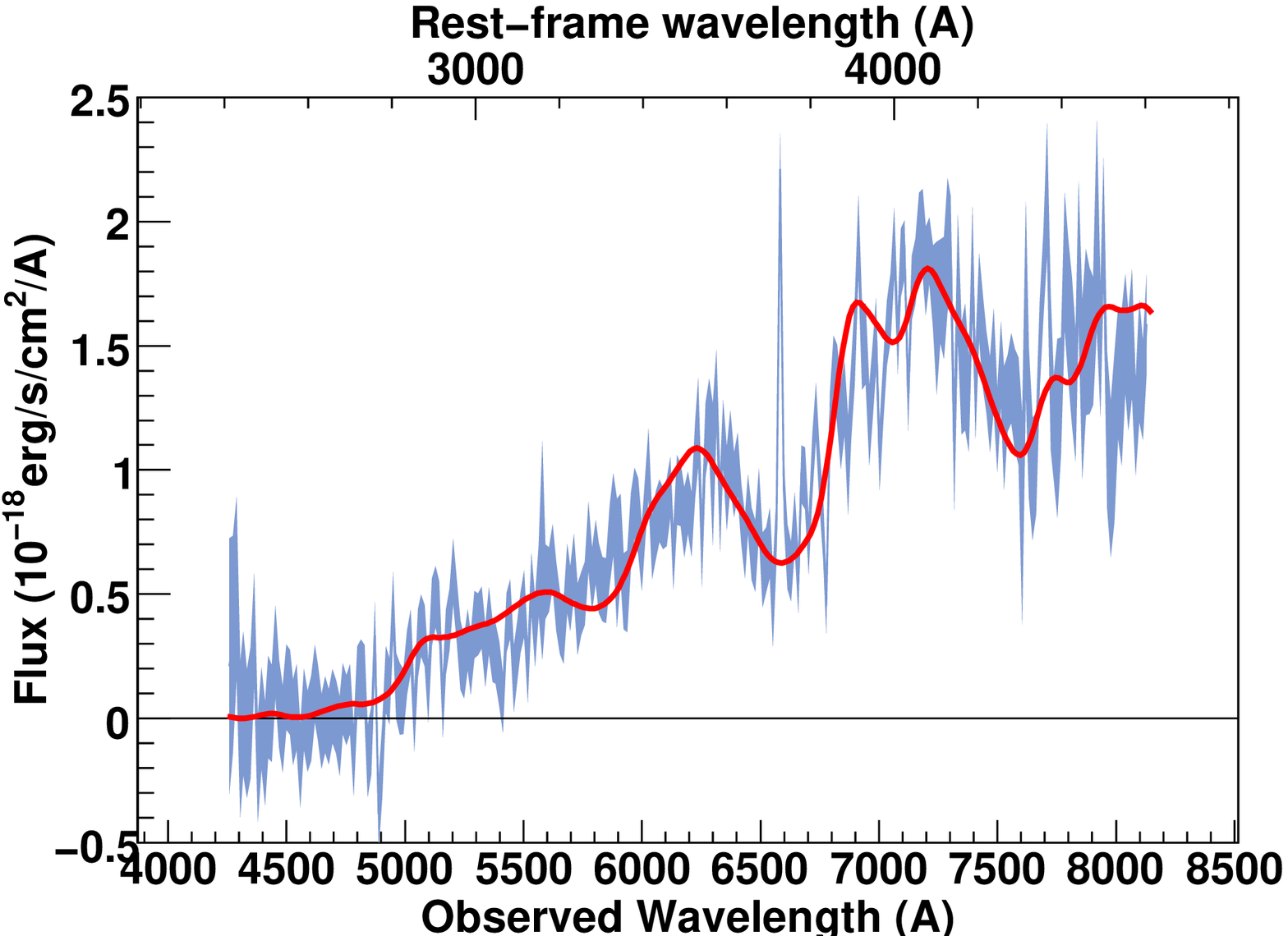}
                    \end{minipage}
                    \\
                    \end{tabular}
                    \caption{\label{fig:04D1qd688} SNLS-  04D1qd688: a SN~Ia$\star$ supernova at z=0.767. The spectrum phase is -0.2. A         E(1) host model has been subtracted.}
                    \end{center}
                    \end{figure*}

\begin{figure*}[ht]
                    \begin{center}
                    \begin{tabular}[b]{cc}
                    \hspace{-5mm}
                    \vspace*{1cm}
                    \begin{minipage}[b]{10cm}
                    \centering
                    \includegraphics[width=10cm]{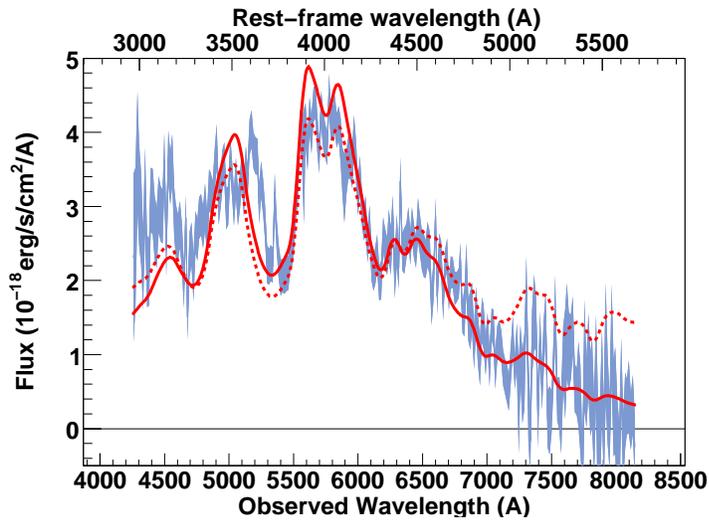}
                    \end{minipage}
                    &
                    \begin{minipage}[b]{10cm}
                    \end{minipage}
                    \\
                    \end{tabular}
                    \caption{\label{fig:04D1rh710} SNLS-  04D1rh710: a SN~Ia supernova at z=0.436. The spectrum phase is 0.0. Best-fit obtained for a model with no host galaxy.}
                    \end{center}
                    \end{figure*}

\begin{figure*}[ht]
                    \begin{center}
                    \begin{tabular}[b]{cc}
                    \hspace{-5mm}
                    \vspace*{1cm}
                    \begin{minipage}[b]{10cm}
                    \centering
                    \includegraphics[width=10cm]{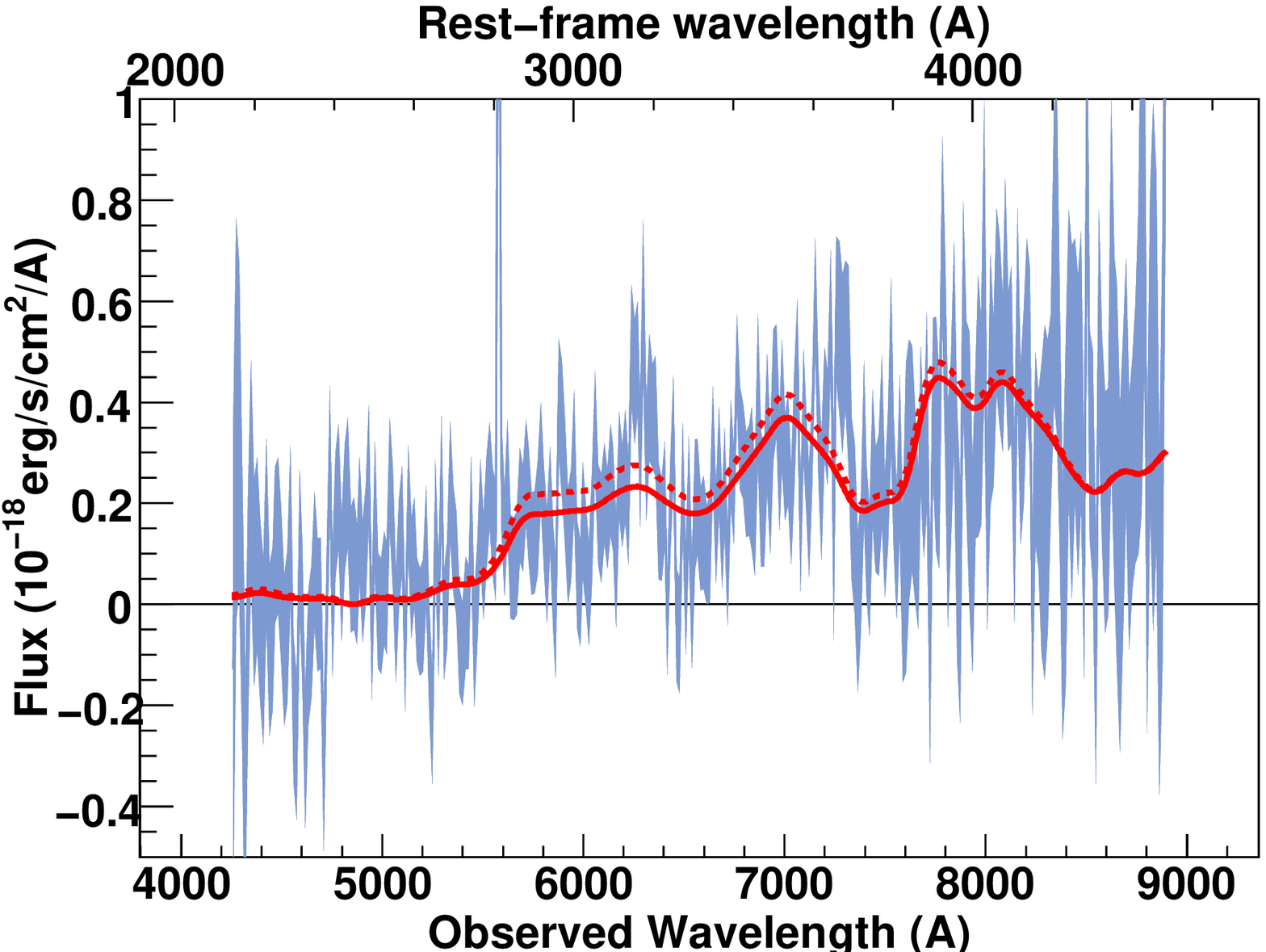}
                    \end{minipage}
                    &
                    \begin{minipage}[b]{10cm}
                    \end{minipage}
                    \\
                    \end{tabular}
                    \caption{\label{fig:04D1rx713} SNLS-  04D1rx713: a SN~Ia$\star$ supernova at z=0.985. The spectrum phase is 0.9. Best-fit obtained for a model with no host galaxy component.}
                    \end{center}
                    \end{figure*}

\begin{figure*}[ht]
                    \begin{center}
                    \begin{tabular}[b]{cc}
                    \hspace{-5mm}
                    \vspace*{1cm}
                    \begin{minipage}[b]{10cm}
                    \centering
                    \includegraphics[width=10cm]{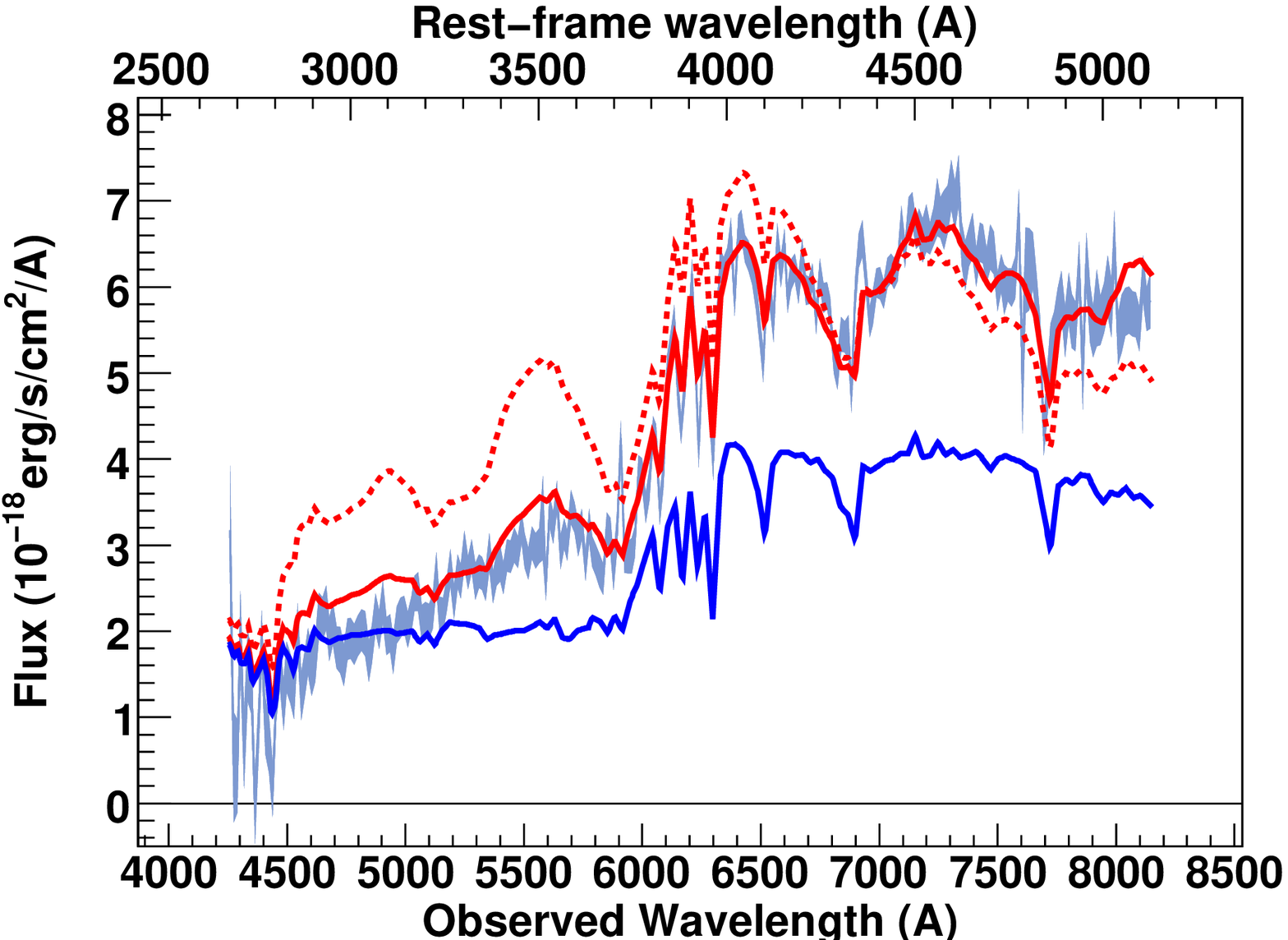}
                    \end{minipage}
                    &
                    \begin{minipage}[b]{10cm}
                    \hspace{-5mm}\includegraphics[width=10cm]{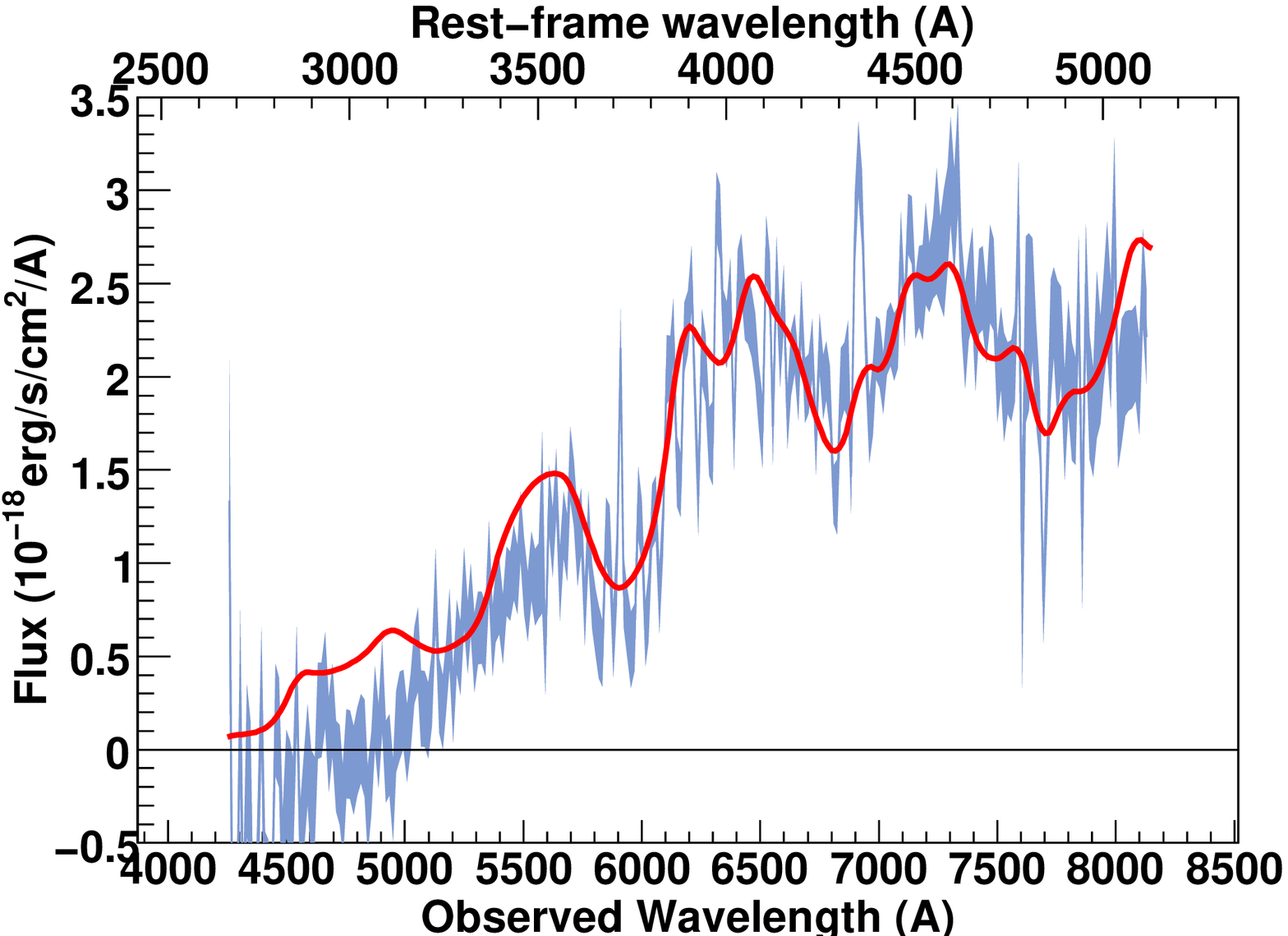}
                    \end{minipage}
                    \\
                    \end{tabular}
                    \caption{\label{fig:04D1sa716} SNLS-  04D1sa716: a SN~Ia supernova at z=0.585. The spectrum phase is -2.6. A         E(1) host model has been subtracted. Note the bad fit in the UV region due to subtracting too blue a host model.}
                    \end{center}
                    \end{figure*}

\begin{figure*}[ht]
                    \begin{center}
                    \begin{tabular}[b]{cc}
                    \hspace{-5mm}
                    \vspace*{1cm}
                    \begin{minipage}[b]{10cm}
                    \centering
                    \includegraphics[width=10cm]{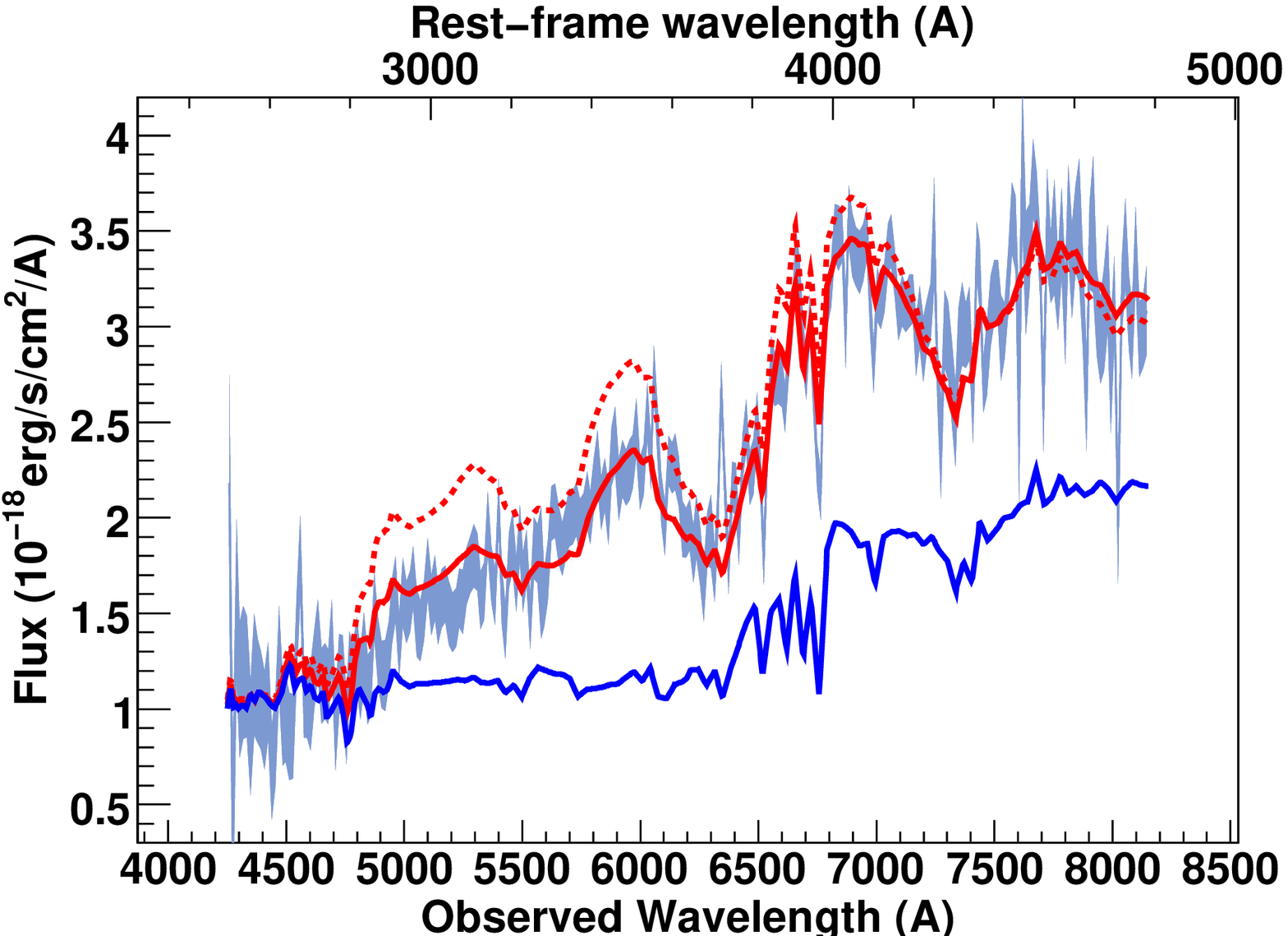}
                    \end{minipage}
                    &
                    \begin{minipage}[b]{10cm}
                    \hspace{-5mm}\includegraphics[width=10cm]{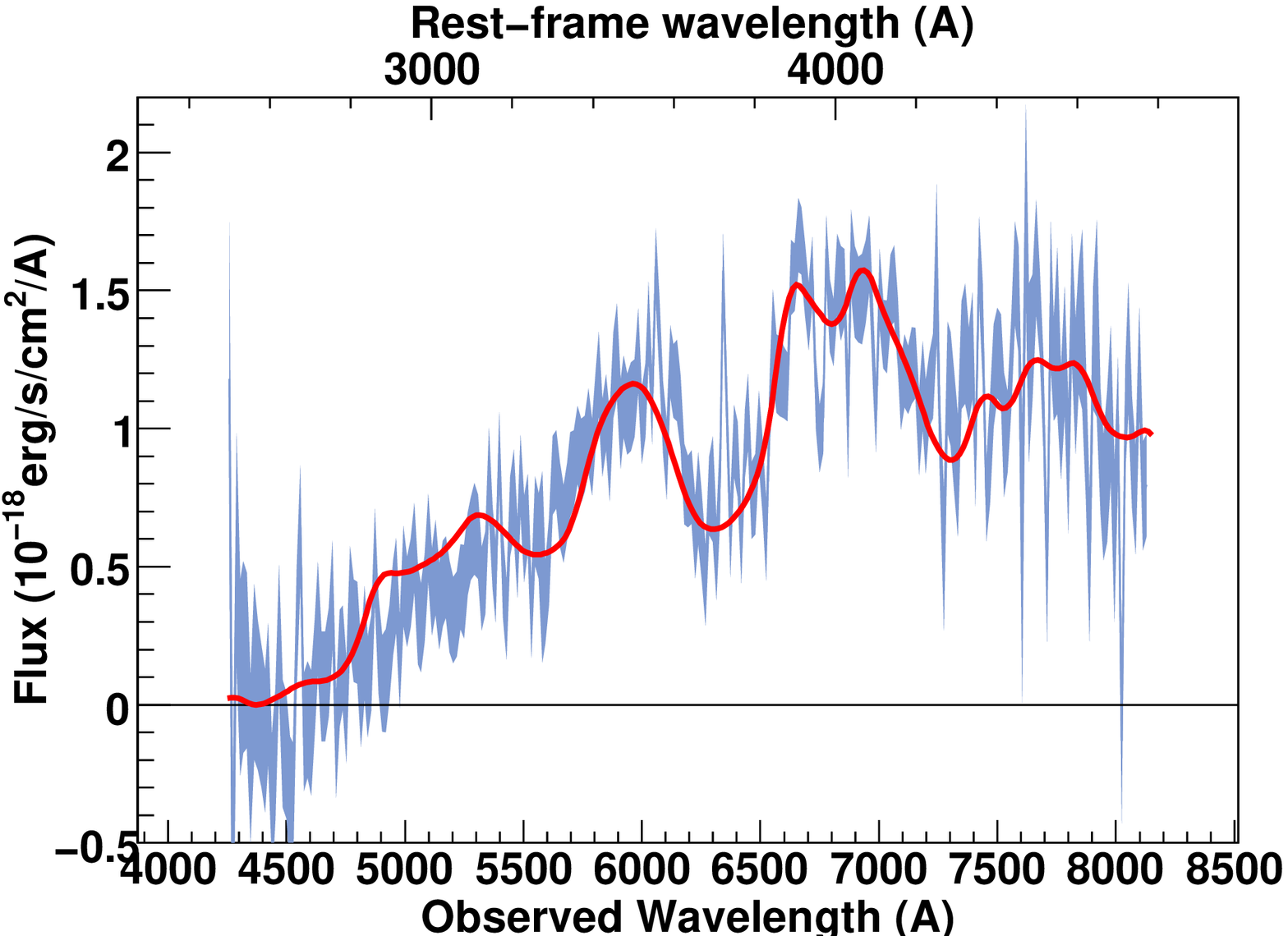}
                    \end{minipage}
                    \\
                    \end{tabular}
                    \caption{\label{fig:04D1si716} SNLS-  04D1si716: a SN~Ia supernova at z=0.702. The spectrum phase is -1.7. A        Sa(9) host model has been subtracted.}
                    \end{center}
                    \end{figure*}

\begin{figure*}[ht]
                    \begin{center}
                    \begin{tabular}[b]{cc}
                    \hspace{-5mm}
                    \vspace*{1cm}
                    \begin{minipage}[b]{10cm}
                    \centering
                    \includegraphics[width=10cm]{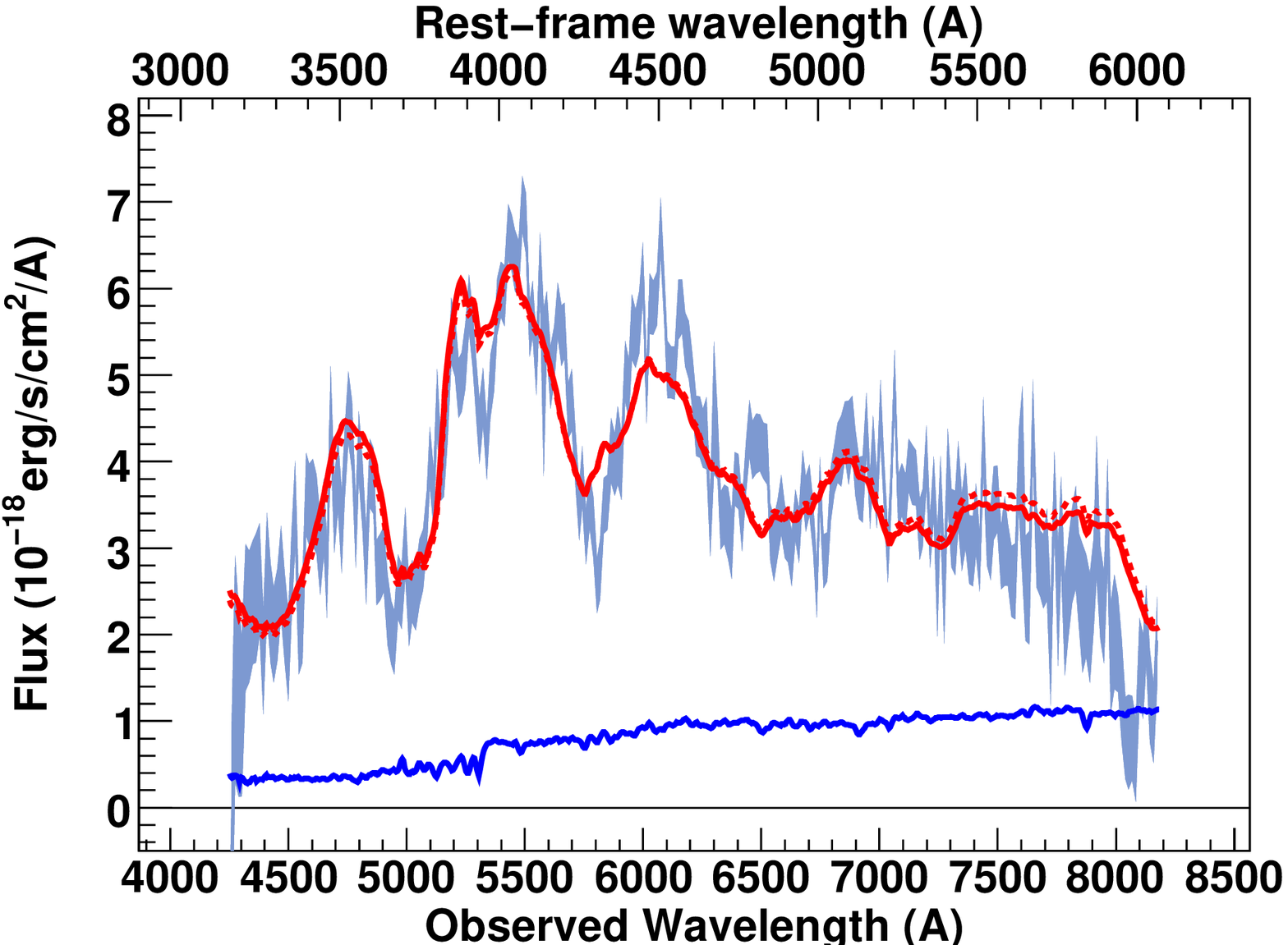}
                    \end{minipage}
                    &
                    \begin{minipage}[b]{10cm}
                    \hspace{-5mm}\includegraphics[width=10cm]{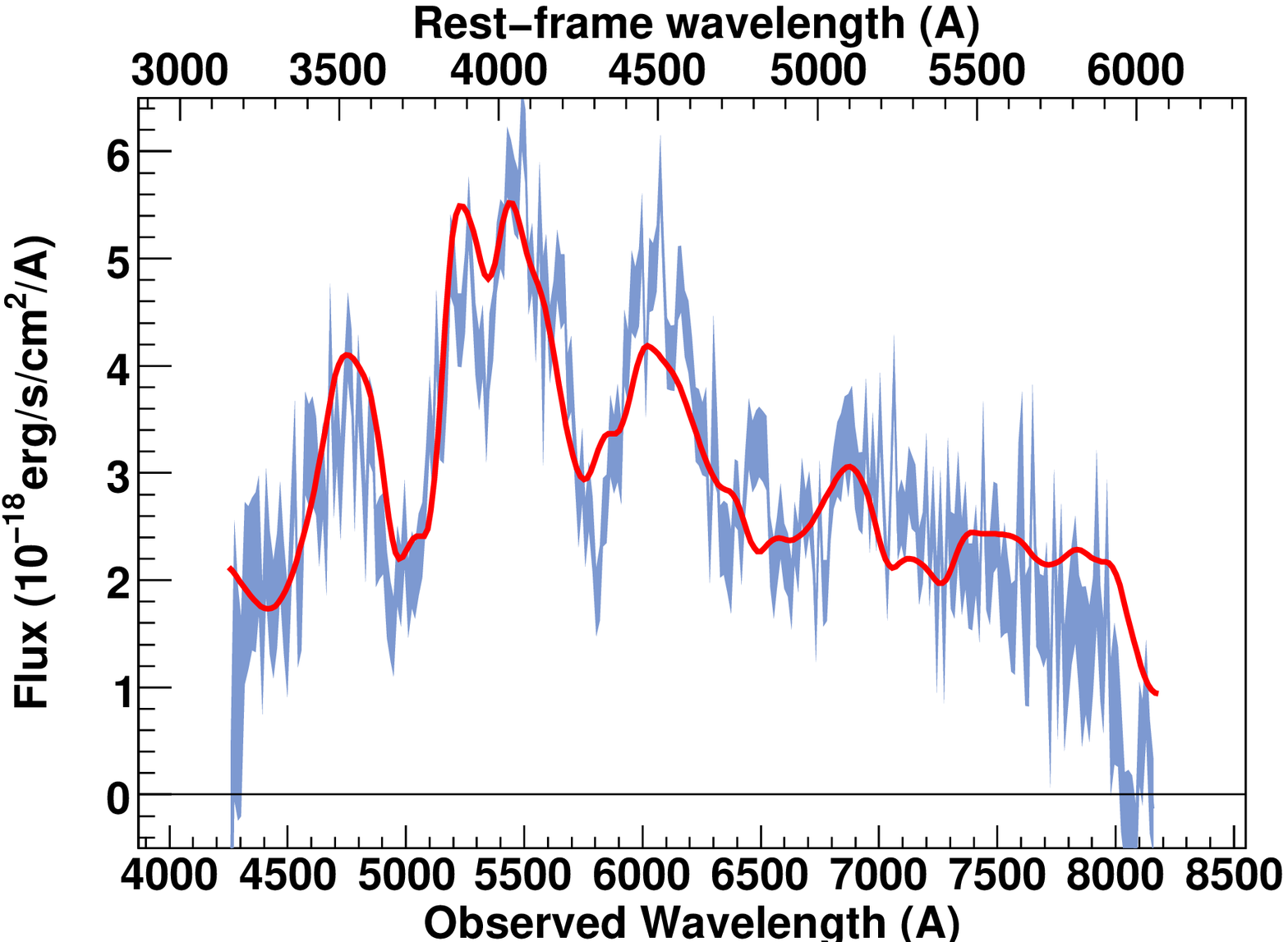}
                    \end{minipage}
                    \\
                    \end{tabular}
                    \caption{\label{fig:04D2ac395} SNLS-  04D2ac395: a SN~Ia supernova at z=0.348. The spectrum phase is 1.3. A         Sa-Sb host model has been subtracted.}
                    \end{center}
                    \end{figure*}

\begin{figure*}[ht]
                    \begin{center}
                    \begin{tabular}[b]{cc}
                    \hspace{-5mm}
                    \vspace*{1cm}
                    \begin{minipage}[b]{10cm}
                    \centering
                    \includegraphics[width=10cm]{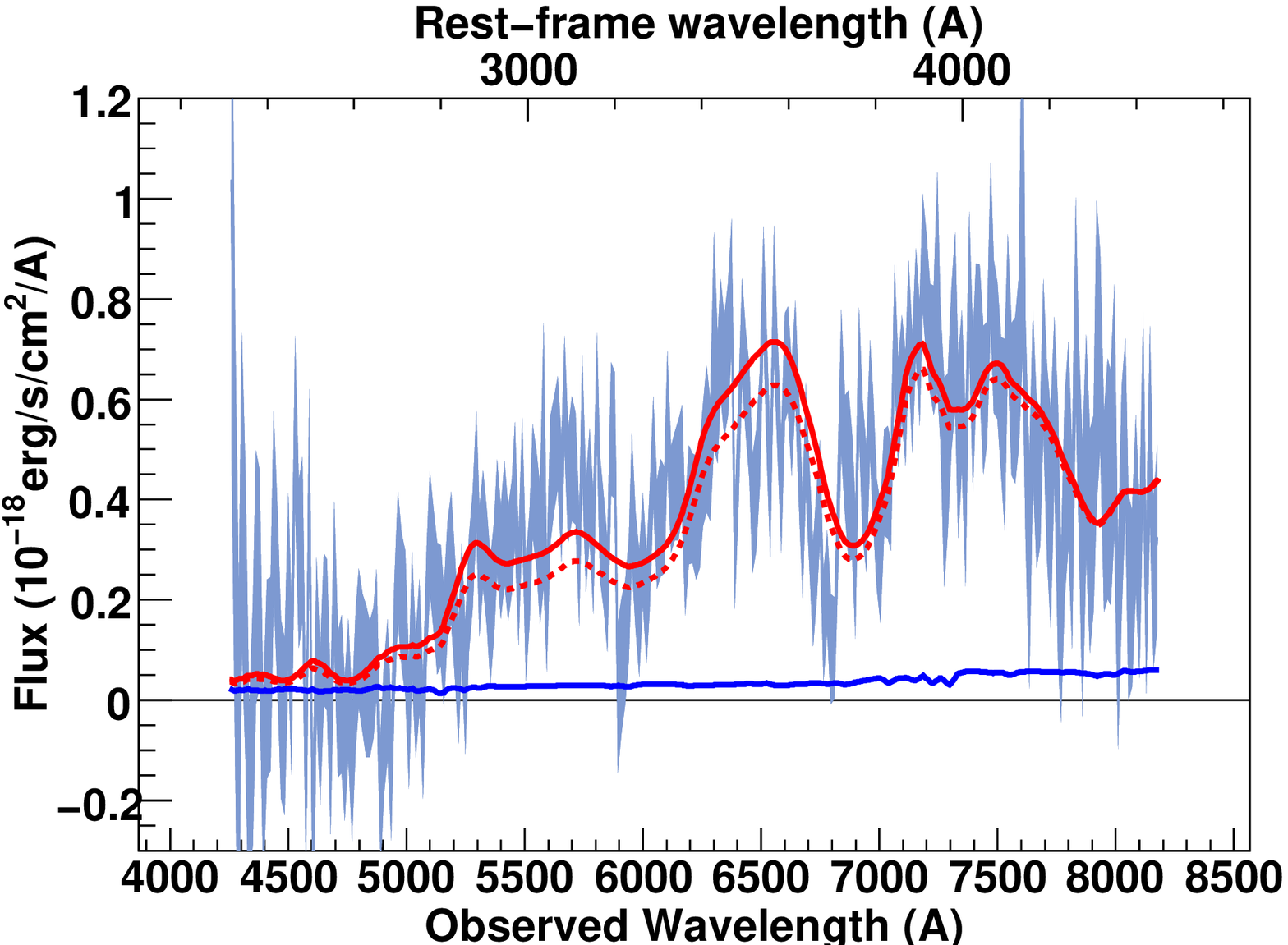}
                    \end{minipage}
                    &
                    \begin{minipage}[b]{10cm}
                    \hspace{-5mm}\includegraphics[width=10cm]{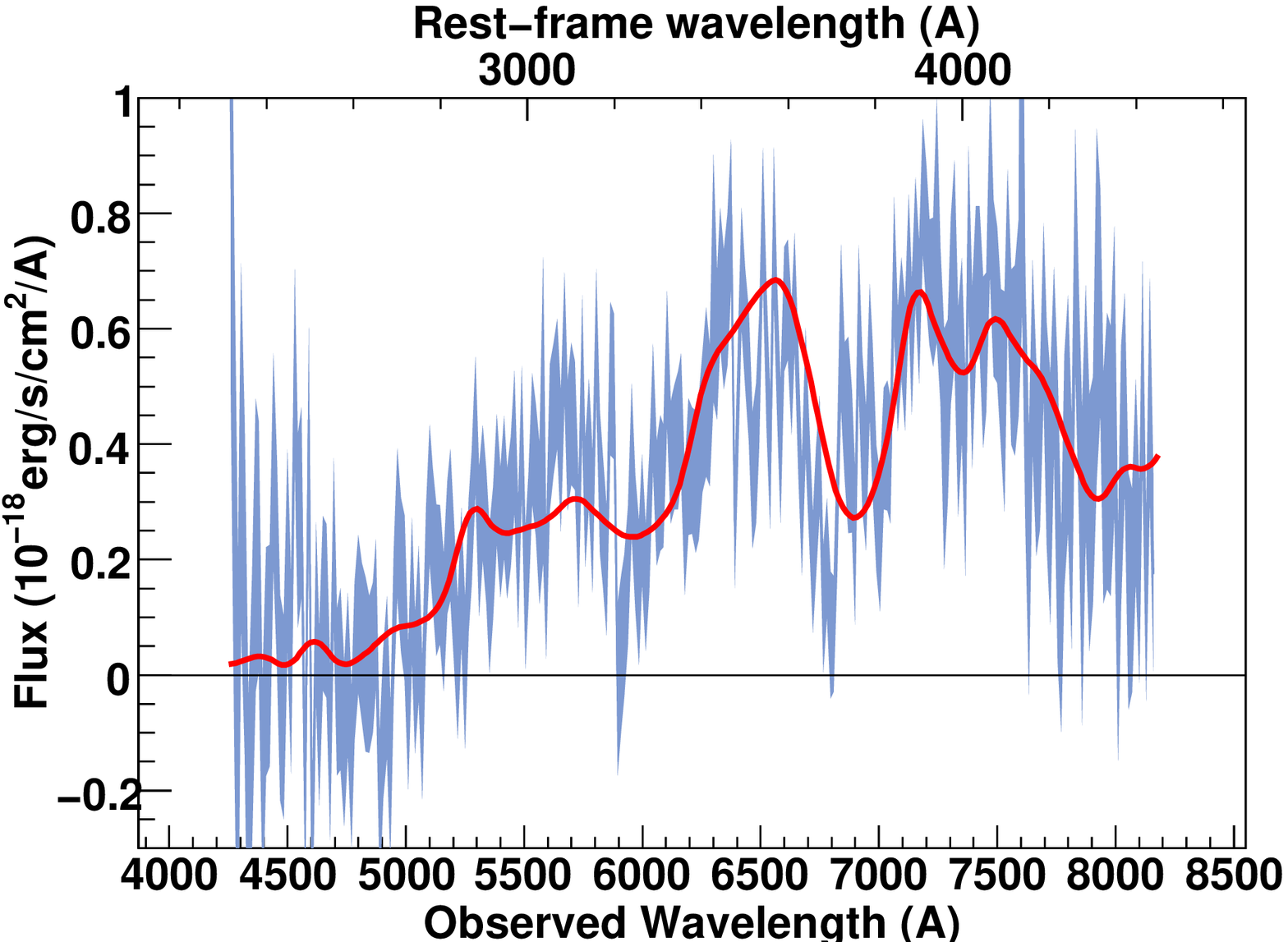}
                    \end{minipage}
                    \\
                    \end{tabular}
                    \caption{\label{fig:04D2al390} SNLS-  04D2al390: a SN~Ia supernova at z=0.836. The spectrum phase is -2.5. A         E(2) host model has been subtracted.}
                    \end{center}
                    \end{figure*}

\begin{figure*}[ht]
                    \begin{center}
                    \begin{tabular}[b]{cc}
                    \hspace{-5mm}
                    \vspace*{1cm}
                    \begin{minipage}[b]{10cm}
                    \centering
                    \includegraphics[width=10cm]{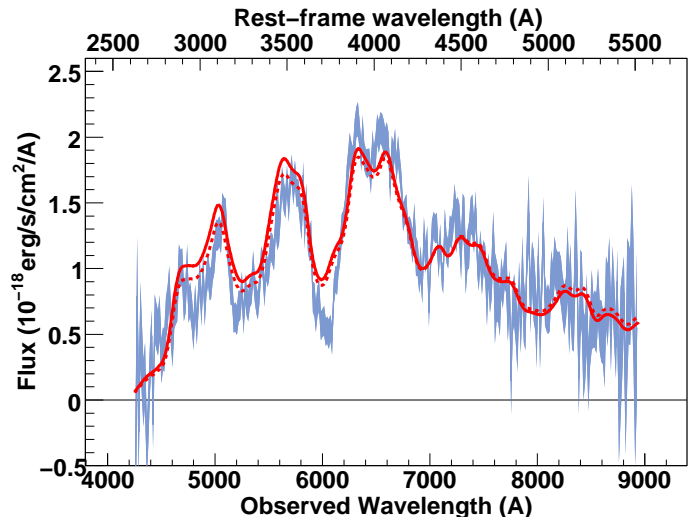}
                    \end{minipage}
                    &
                    \begin{minipage}[b]{10cm}
                    \end{minipage}
                    \\
                    \end{tabular}
                    \caption{\label{fig:04D2an386} SNLS-  04D2an386: a SN~Ia supernova at z=0.62. The spectrum phase is -3.4. Best-fit obtained for a model with no host galaxy component. Note that the \ion{Ca}{\sc ii} feature at rest-frame 3700 $\AA$ is deeper than in the model.}
                    \end{center}
                    \end{figure*}

\begin{figure*}[ht]
                    \begin{center}
                    \begin{tabular}[b]{cc}
                    \hspace{-5mm}
                    \vspace*{1cm}
                    \begin{minipage}[b]{10cm}
                    \centering
                    \includegraphics[width=10cm]{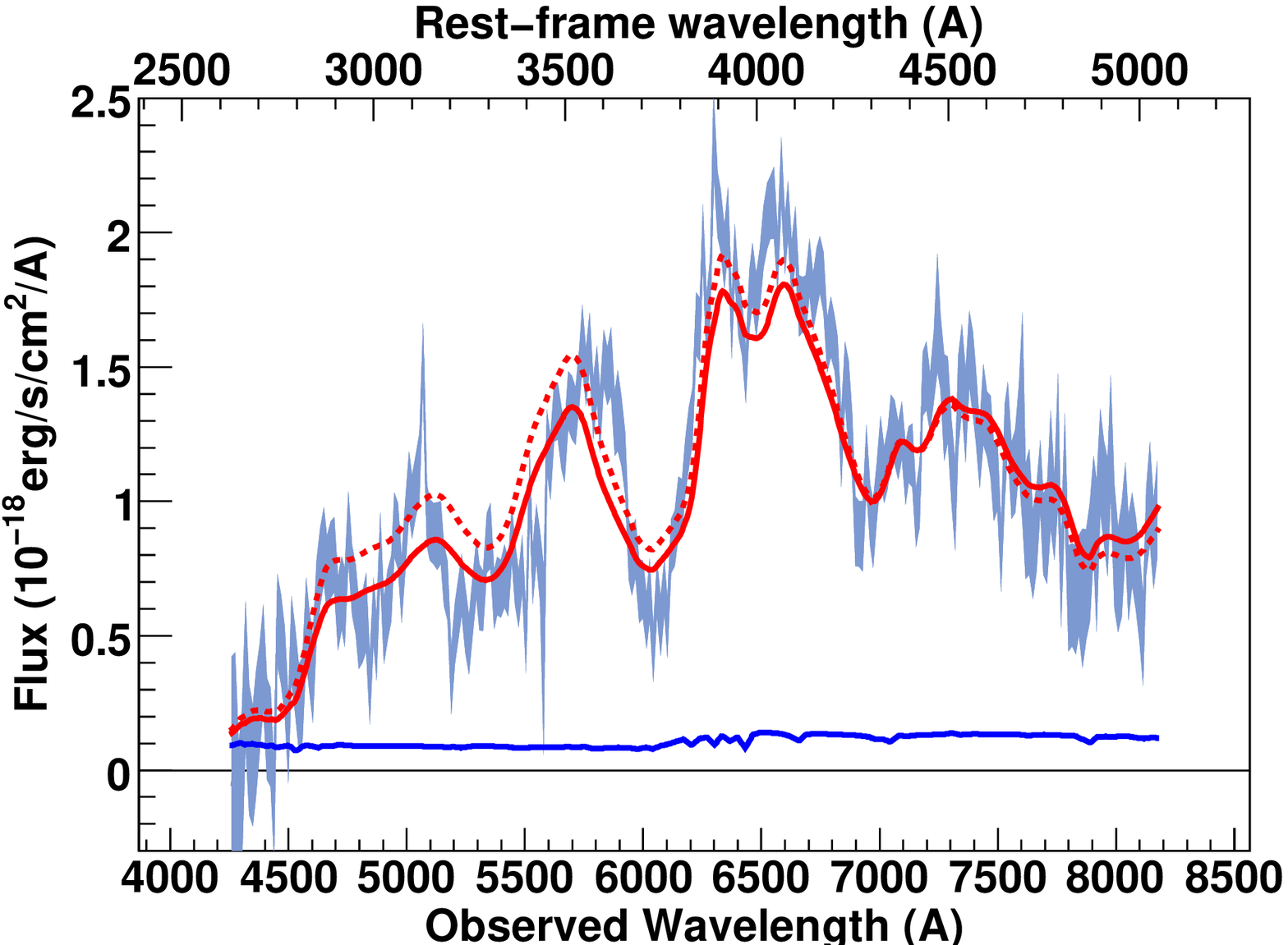}
                    \end{minipage}
                    &
                    \begin{minipage}[b]{10cm}
                    \hspace{-5mm}\includegraphics[width=10cm]{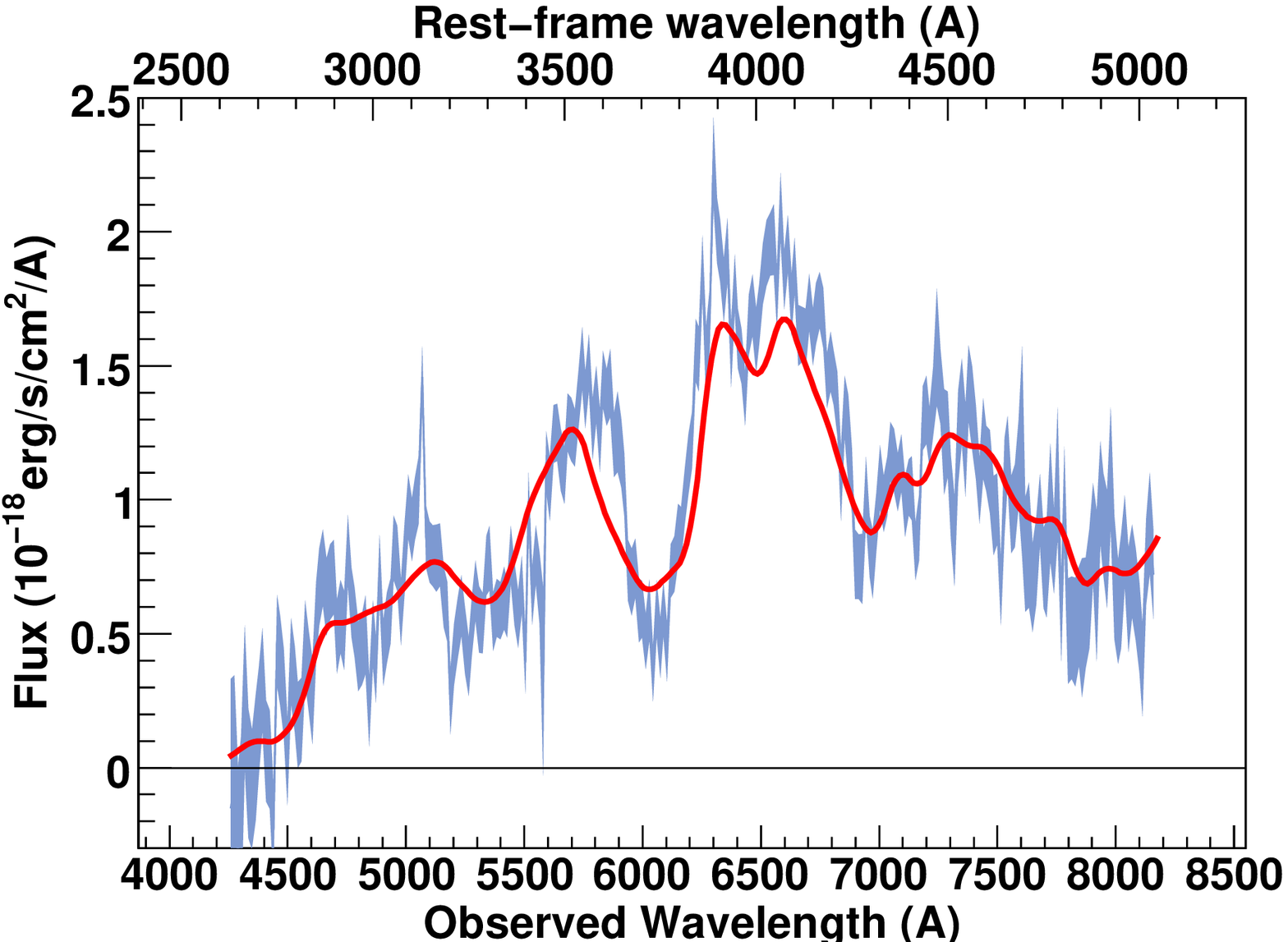}
                    \end{minipage}
                    \\
                    \end{tabular}
                    \caption{\label{fig:04D2an392} SNLS-  04D2an392: a SN~Ia supernova at z=0.62. The spectrum phase is 0.3. A        Sc(9) host model has been subtracted.}
                    \end{center}
                    \end{figure*}

\begin{figure*}[ht]
                    \begin{center}
                    \begin{tabular}[b]{cc}
                    \hspace{-5mm}
                    \vspace*{1cm}
                    \begin{minipage}[b]{10cm}
                    \centering
                    \includegraphics[width=10cm]{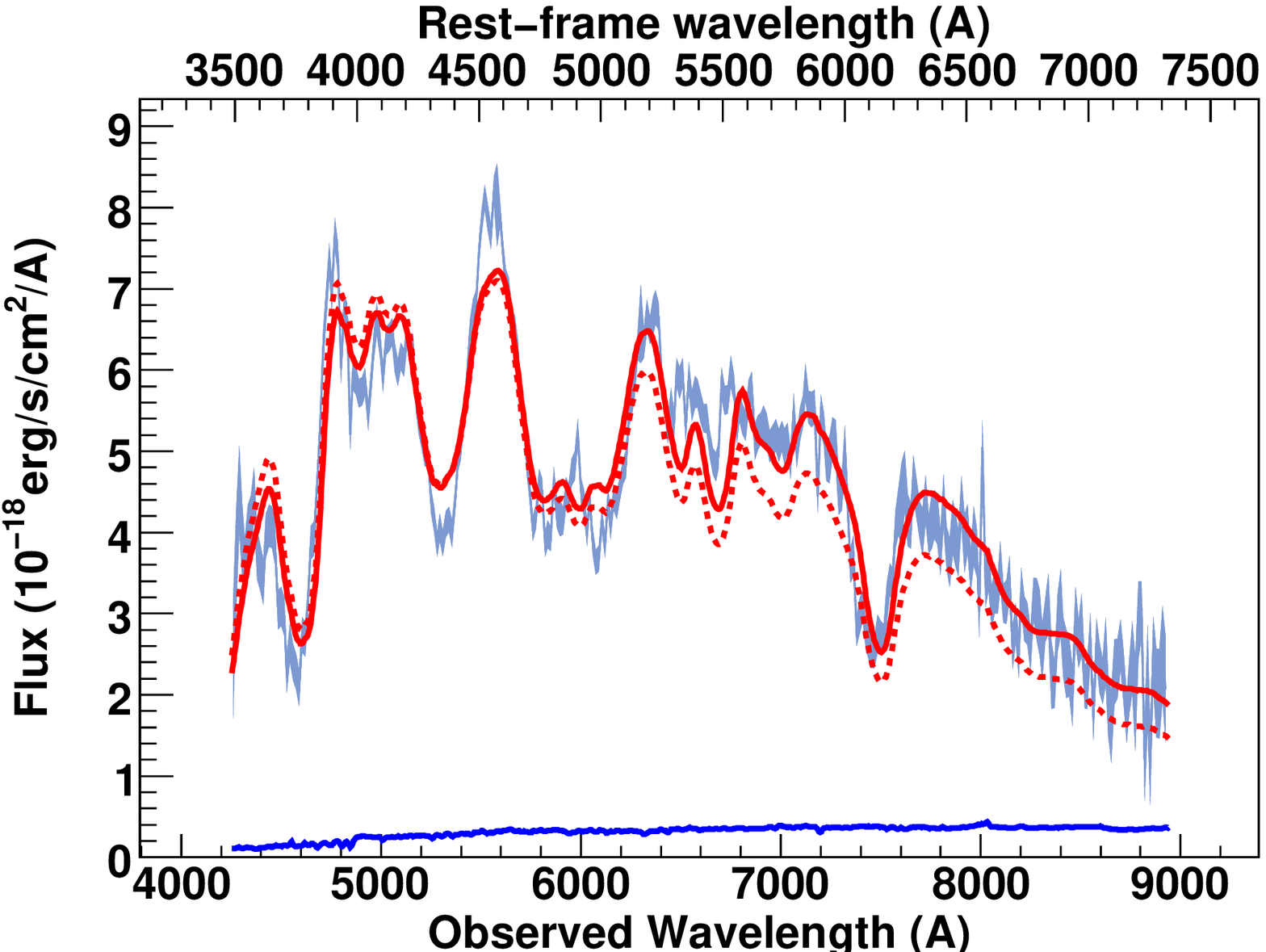}
                    \end{minipage}
                    &
                    \begin{minipage}[b]{10cm}
                    \hspace{-5mm}\includegraphics[width=10cm]{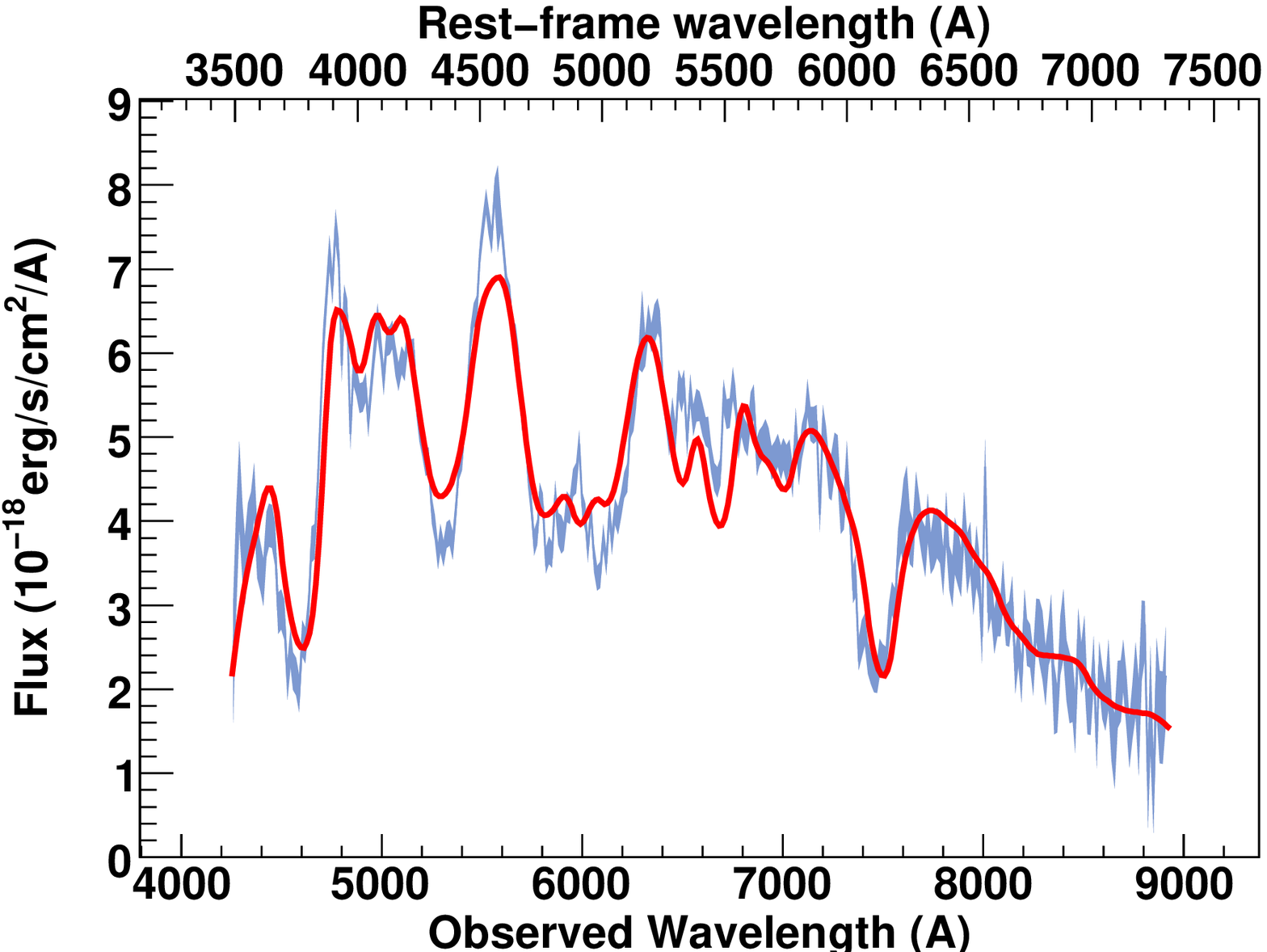}
                    \end{minipage}
                    \\
                    \end{tabular}
                    \caption{\label{fig:04D2bt444} SNLS-  04D2bt444: a SN~Ia supernova at z=0.220. The spectrum phase is 6.6. A         E(6) host model has been subtracted.}
                    \end{center}
                    \end{figure*}

\begin{figure*}[ht]
                    \begin{center}
                    \begin{tabular}[b]{cc}
                    \hspace{-5mm}
                    \vspace*{1cm}
                    \begin{minipage}[b]{10cm}
                    \centering
                    \includegraphics[width=10cm]{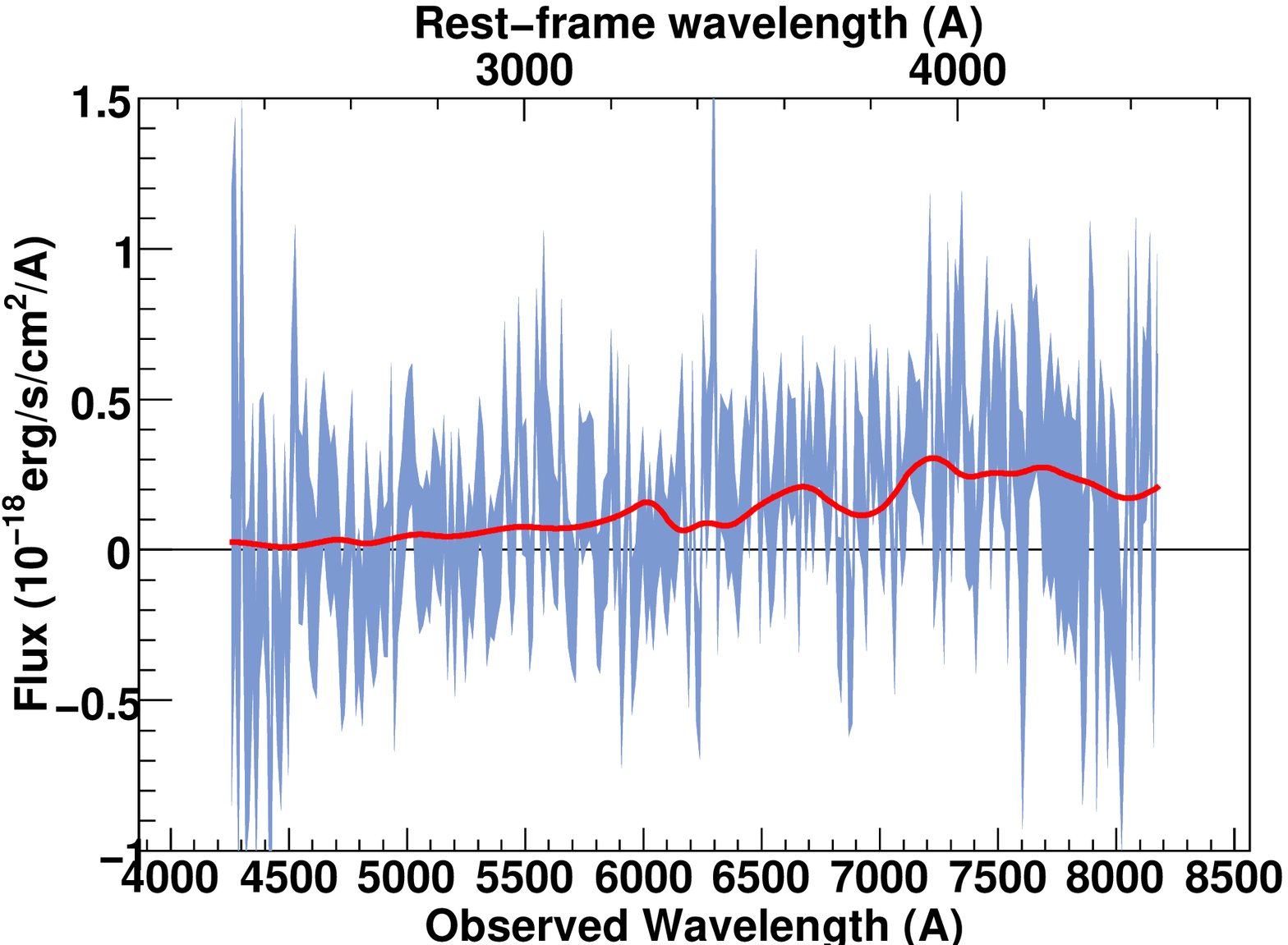}
                    \end{minipage}
                    &
                    \begin{minipage}[b]{10cm}
                    \end{minipage}
                    \\
                    \end{tabular}
                    \caption{\label{fig:04D2ca445} SNLS-  04D2ca445: a SN~Ia$\star$ supernova at z=0.835. The spectrum phase is 12.0. Best-fit obtained for a model with no host galaxy component.}
                    \end{center}
                    \end{figure*}

\begin{figure*}[ht]
                    \begin{center}
                    \begin{tabular}[b]{cc}
                    \hspace{-5mm}
                    \vspace*{1cm}
                    \begin{minipage}[b]{10cm}
                    \centering
                    \includegraphics[width=10cm]{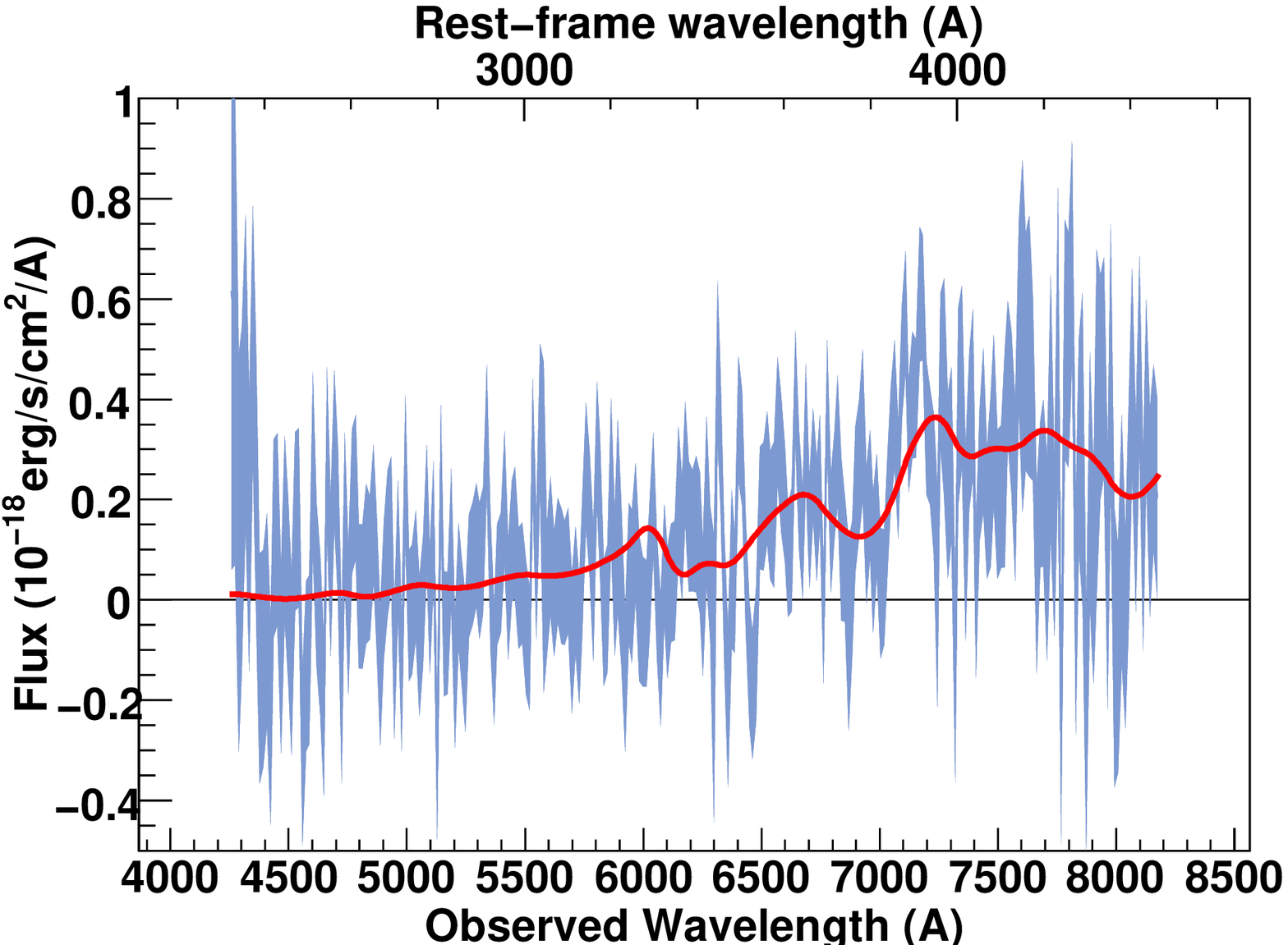}
                    \end{minipage}
                    &
                    \begin{minipage}[b]{10cm}
                    \end{minipage}
                    \\
                    \end{tabular}
                    \caption{\label{fig:04D2ca447} SNLS-  04D2ca447: a SN~Ia$\star$ supernova at z=0.835. The spectrum phase is 13.1. Best-fit obtained for a model with no host galaxy component.}
                    \end{center}
                    \end{figure*}

\begin{figure*}[ht]
                    \begin{center}
                    \begin{tabular}[b]{cc}
                    \hspace{-5mm}
                    \vspace*{1cm}
                    \begin{minipage}[b]{10cm}
                    \centering
                    \includegraphics[width=10cm]{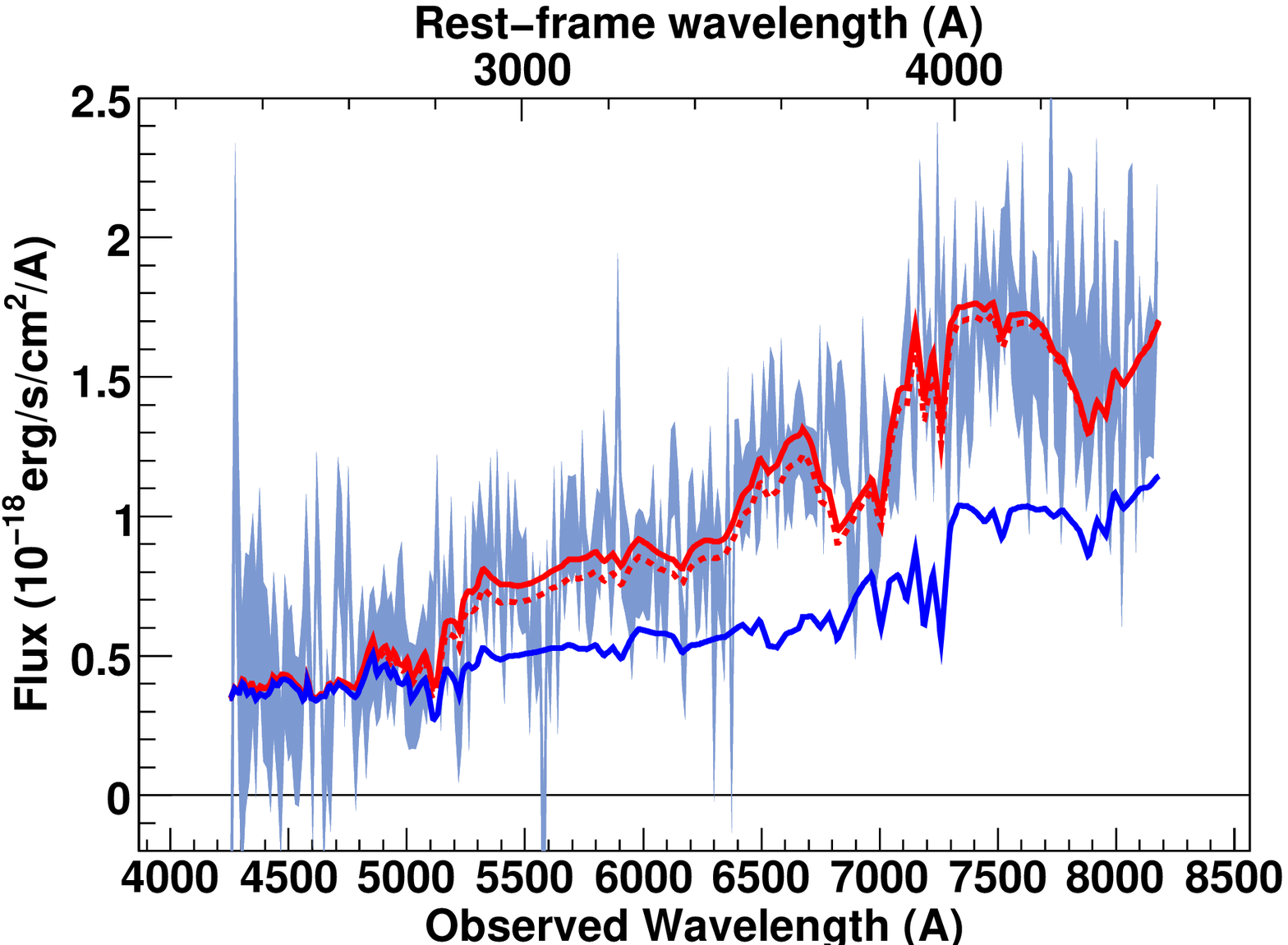}
                    \end{minipage}
                    &
                    \begin{minipage}[b]{10cm}
                    \hspace{-5mm}\includegraphics[width=10cm]{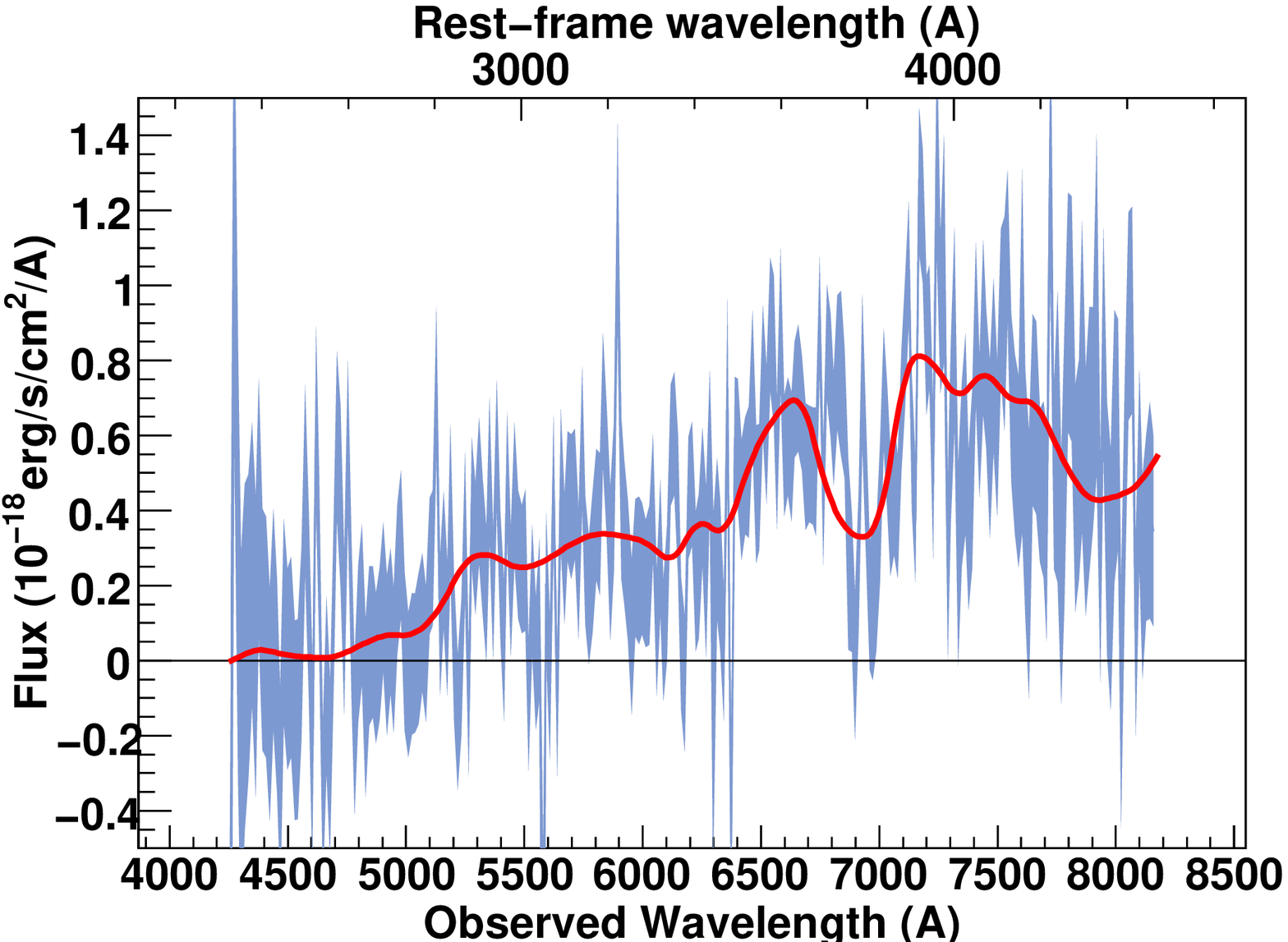}
                    \end{minipage}
                    \\
                    \end{tabular}
                    \caption{\label{fig:04D2cc446} SNLS-  04D2cc446: a SN~Ia$\star$ supernova at z=0.838. The spectrum phase is 6.1. A        S0(3) host model has been subtracted.}
                    \end{center}
                    \end{figure*}

\begin{figure*}[ht]
                    \begin{center}
                    \begin{tabular}[b]{cc}
                    \hspace{-5mm}
                    \vspace*{1cm}
                    \begin{minipage}[b]{10cm}
                    \centering
                    \includegraphics[width=10cm]{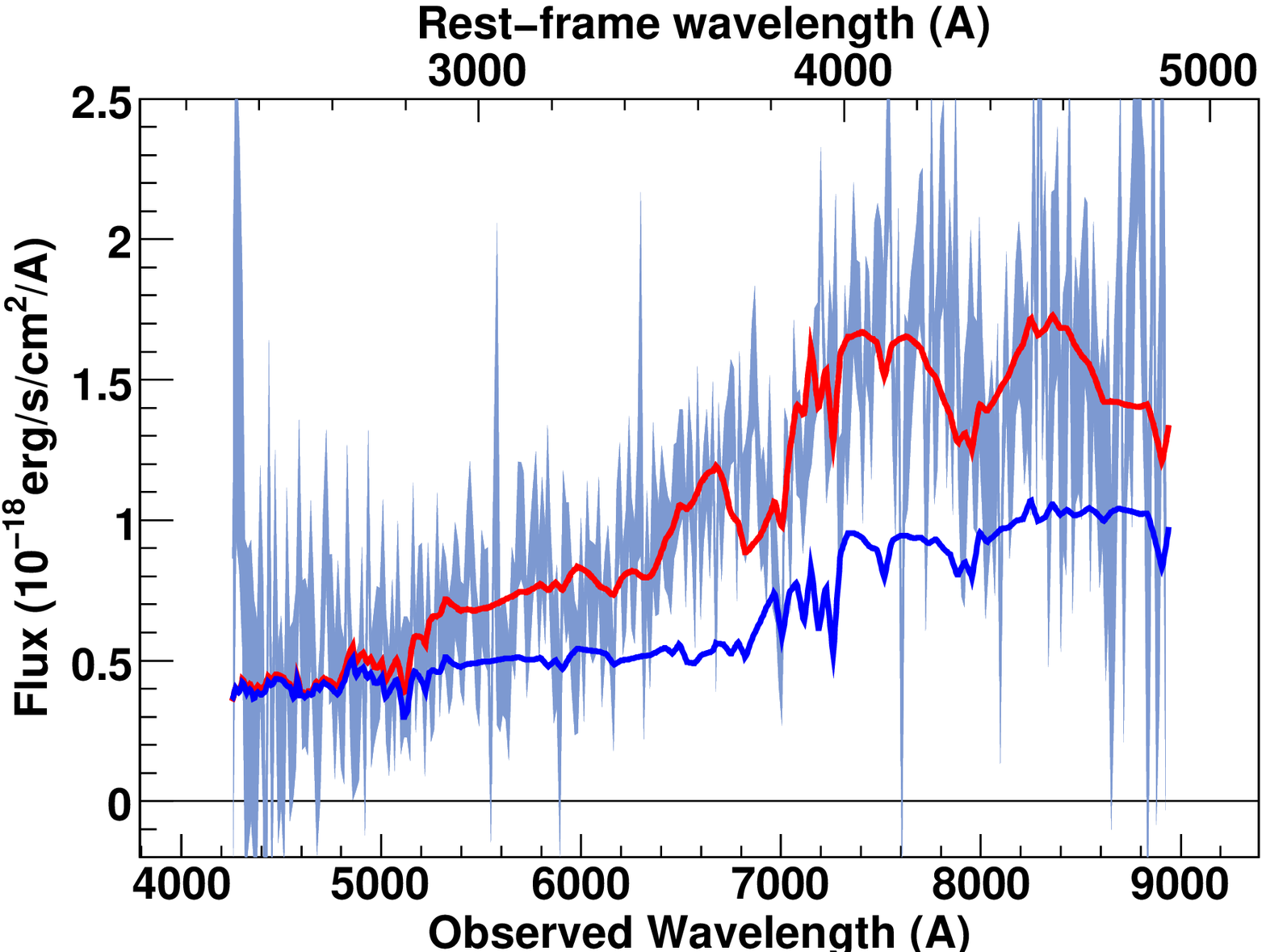}
                    \end{minipage}
                    &
                    \begin{minipage}[b]{10cm}
                    \hspace{-5mm}\includegraphics[width=10cm]{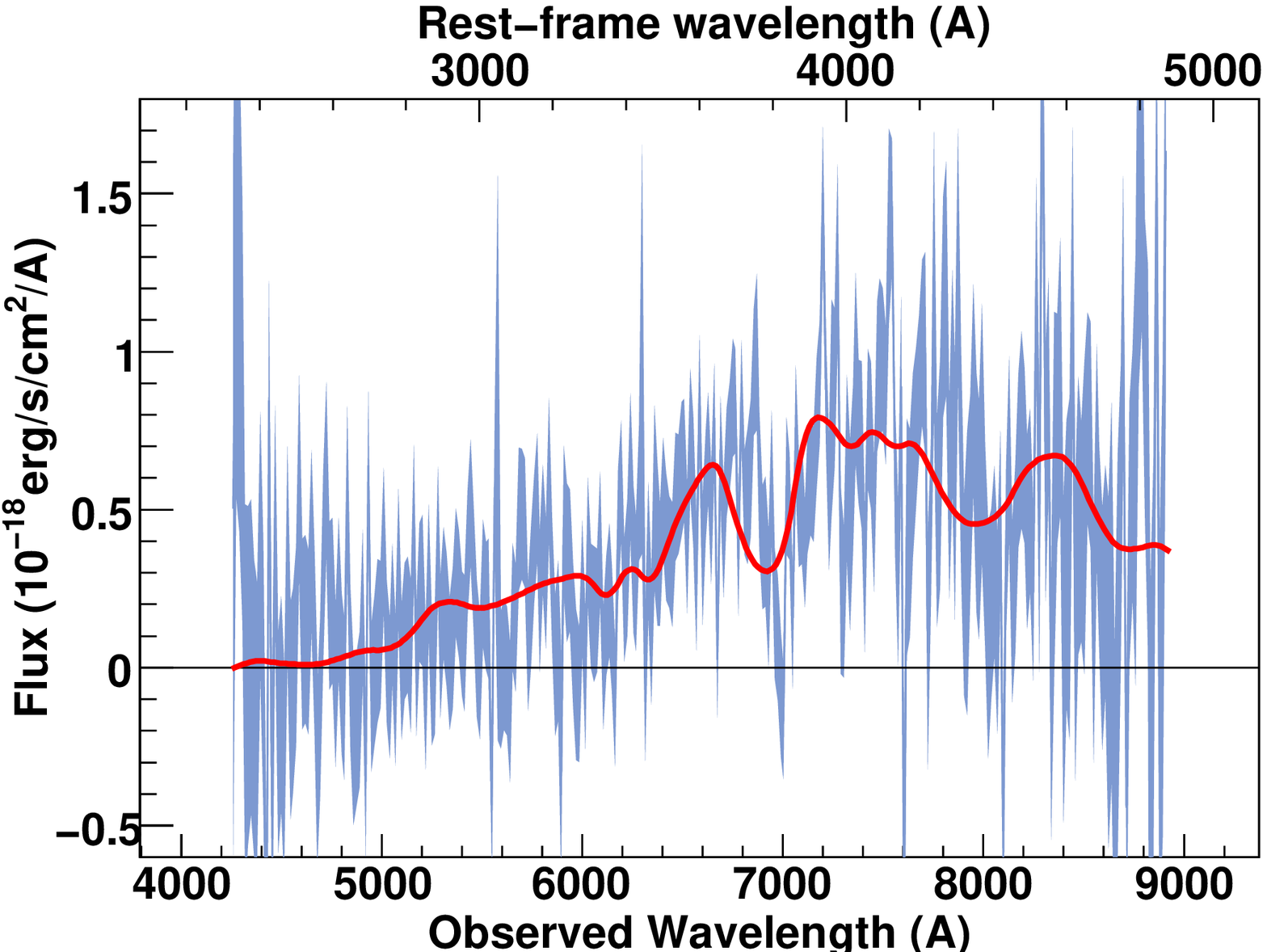}
                    \end{minipage}
                    \\
                    \end{tabular}
                    \caption{\label{fig:04D2cc448} SNLS-  04D2cc448: a SN~Ia$\star$ supernova at z=0.838. The spectrum phase is 7.2. A        S0(2) host model has been subtracted.}
                    \end{center}
                    \end{figure*}

\begin{figure*}[ht]
                    \begin{center}
                    \begin{tabular}[b]{cc}
                    \hspace{-5mm}
                    \vspace*{1cm}
                    \begin{minipage}[b]{10cm}
                    \centering
                    \includegraphics[width=10cm]{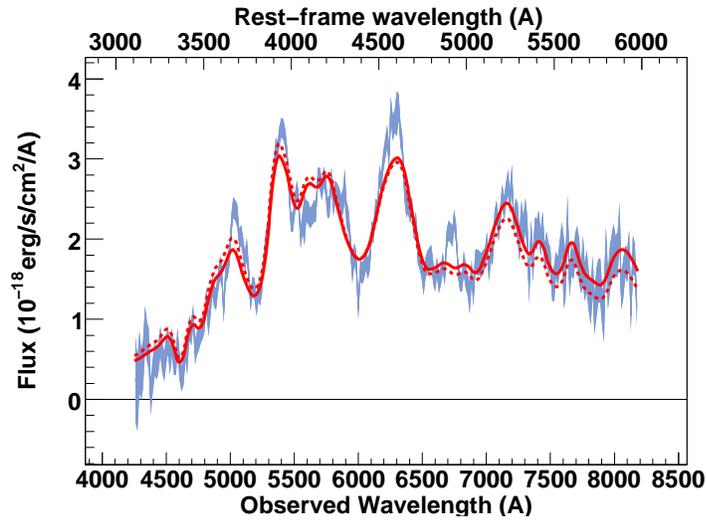}
                    \end{minipage}
                    &
                    \begin{minipage}[b]{10cm}
                    \end{minipage}
                    \\
                    \end{tabular}
                    \caption{\label{fig:04D2cf447} SNLS-  04D2cf447: a SN~Ia supernova at z=0.369. The spectrum phase is 8.5. Best-fit obtained for a model with no host galaxy component. Note that the model is unable to reproduce the feature at rest-frame 4900 $\AA$.}
                    \end{center}
                    \end{figure*}

\begin{figure*}[ht]
                    \begin{center}
                    \begin{tabular}[b]{cc}
                    \hspace{-5mm}
                    \vspace*{1cm}
                    \begin{minipage}[b]{10cm}
                    \centering
                    \includegraphics[width=10cm]{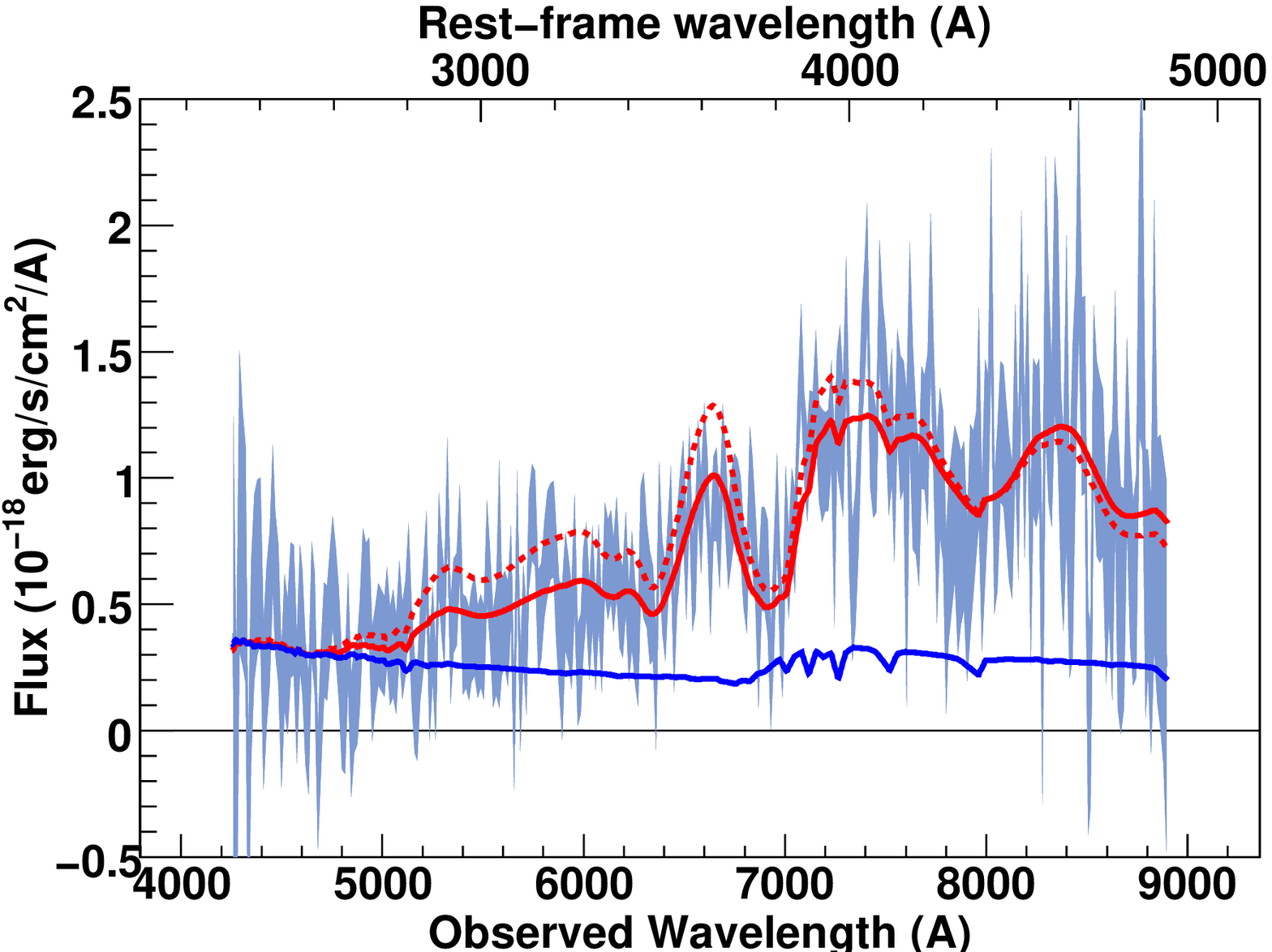}
                    \end{minipage}
                    &
                    \begin{minipage}[b]{10cm}
                    \hspace{-5mm}\includegraphics[width=10cm]{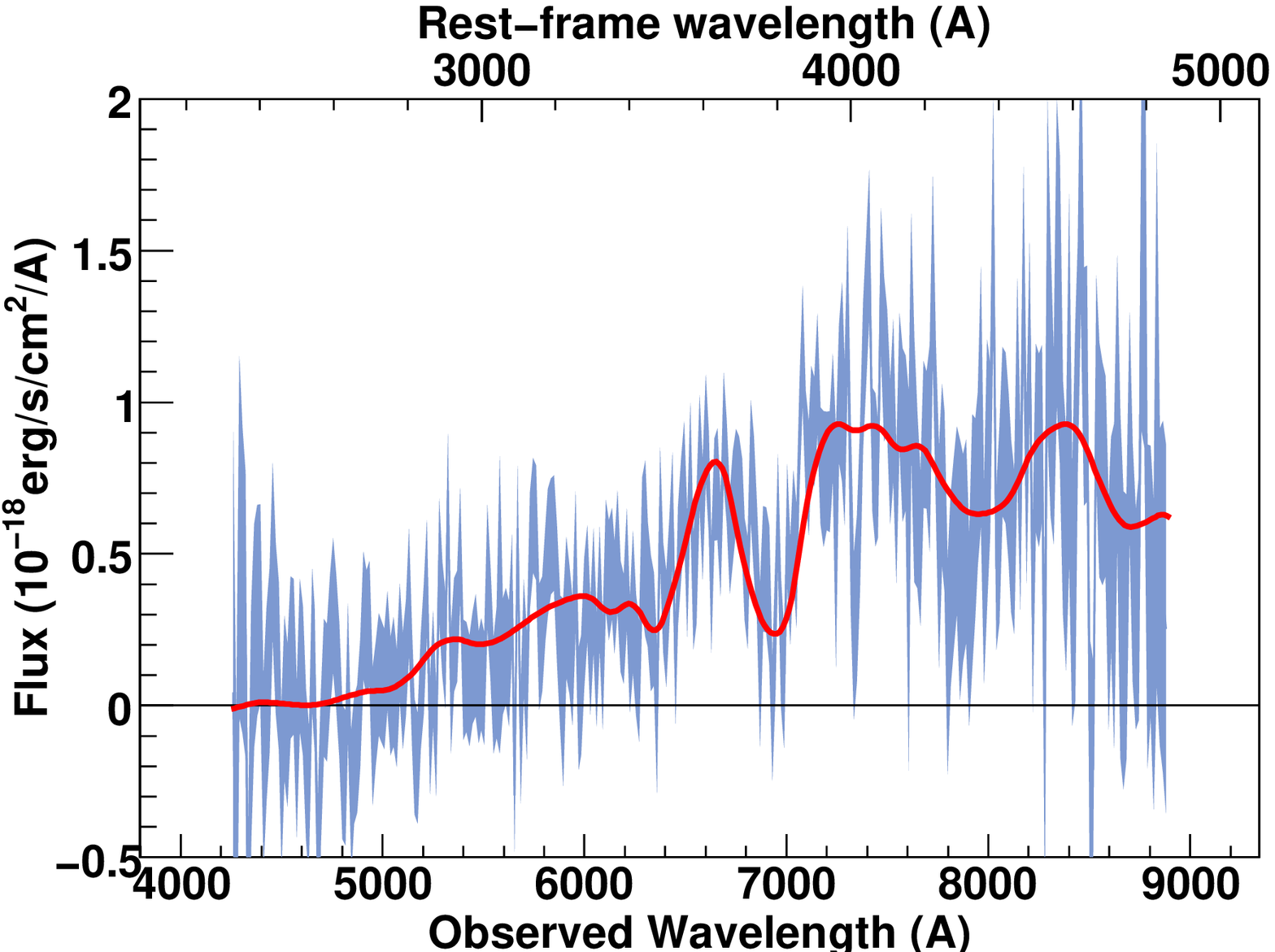}
                    \end{minipage}
                    \\
                    \end{tabular}%
                    \caption{\label{fig:06D4cq1306} SNLS- 06D4cq1306: a SN~Ia supernova at z=0.411. The spectrum phase is -1.4. Best-fit obtained for a model with no host galaxy component.}
                    \end{center}
                    \end{figure*}

Table \ref{tab:distributionIa} gives the number of SNe for each class
(SN~Ia and SN~Ia$\star$) per field and in total.  Figures
\ref{fig:03D1ar270} to \ref{fig:06D4cq1306} show the full (host+SN)
PHASE extracted spectrum (left panel). For each spectrum, the name of
the SN is followed by the date of spectroscopy, in terms of the number
of days elapsed since January 1st, 2003. The red dashed line is the
SALT2 model obtained with no recalibration, the red solid line being
the same model after recalibration. Whenever fitting with a host
galaxy template is necessary, we show the best-fitting host spectrum
model (blue solid line). In these cases, we present in the right panel
the host subtracted SN spectrum obtained by subtracting the host model
(blue solid line in left panel) from the PHASE extracted spectrum. The
best-fit, recalibrated, SN~Ia model (solid red line) is overplotted on
the host-subtracted SN spectrum.

\begin{figure*}[ht]
                    \begin{center}
                    \begin{tabular}[b]{cc}
                    \hspace{-5mm}
                    \vspace*{1cm}
                    \begin{minipage}[b]{10cm}
                    \centering
                    \includegraphics[width=10cm]{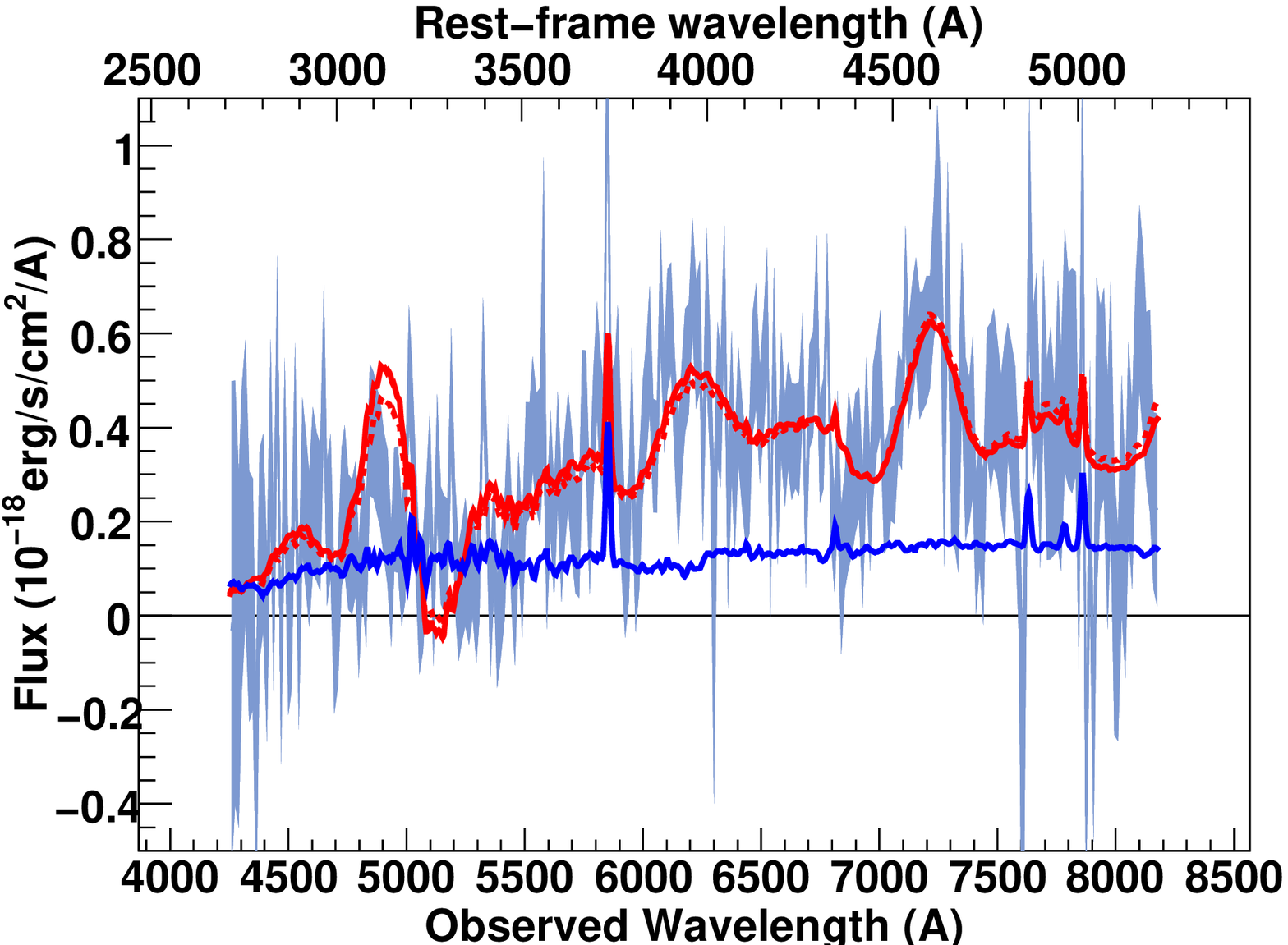}
                    \end{minipage}
                    &
                    \begin{minipage}[b]{10cm}
                    \hspace{-5mm}\includegraphics[width=10cm]{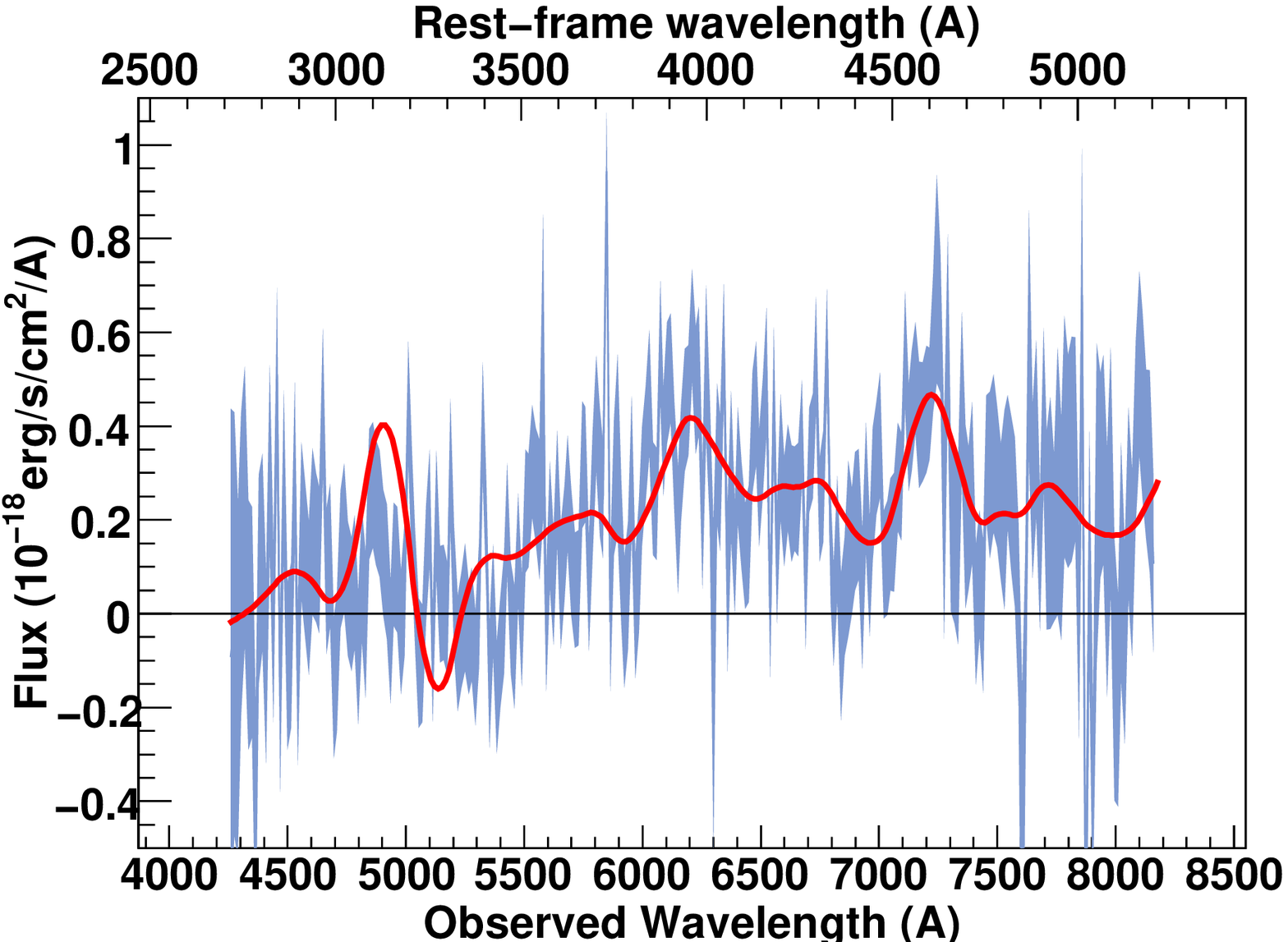}
                    \end{minipage}
                    \\
                    \end{tabular}
                    \caption{\label{fig:04D2cw445} SNLS-  04D2cw445: a SN~Ia$\star$ supernova at z=0.568. The spectrum phase is 19.3. A         Sb-Sc host model has been subtracted. The UV part of the model is poorly constrained due to the lack of UV templates at late phases in the version of the SALT2 training set used in this work.}
                    \end{center}
                    \end{figure*}

As discussed above, host subtraction is a key issue and techniques are
not perfect. In particular, the PEGASE templates we use have no
emission lines. As a consequence a number of our spectra show residual
host lines, e.g., SNLS-04D1sa (Fig.  \ref{fig:04D1sa716}) with
residual [\ion{O}{\sc ii}] emission and \ion{Ca}{\sc ii} H\&K
absorption.  In some cases, the fit is poor in some portion of the
wavelength range. There can be different explanations for this.  In
the case of SNLS-04D1hx (Fig.  \ref{fig:04D1hx628}), there are two
galaxies along the line of sight. The PHASE host model used for the
extraction is not accurate, and the extracted spectrum shows strong
host residuals. In the case of SNLS-03D4gf, SNLS-03D4gg and
SNLS-04D2cw (Figs.  \ref{fig:03D4gf329}, \ref{fig:03D4gg330} and
\ref{fig:04D2cw445}), the SALT2 model is unable to reproduce the UV
wavelength region due to the lack of UV coverage in the SALT2 training
sample used. More specific comments on individual SNe~Ia are given in
the corresponding caption.

\subsection{Average properties of the SN~Ia and SN~Ia$\star$ samples}
\label{sec:averageproperties}

\begin{figure*}
\begin{center} 
\resizebox{12cm}{10cm}{\includegraphics{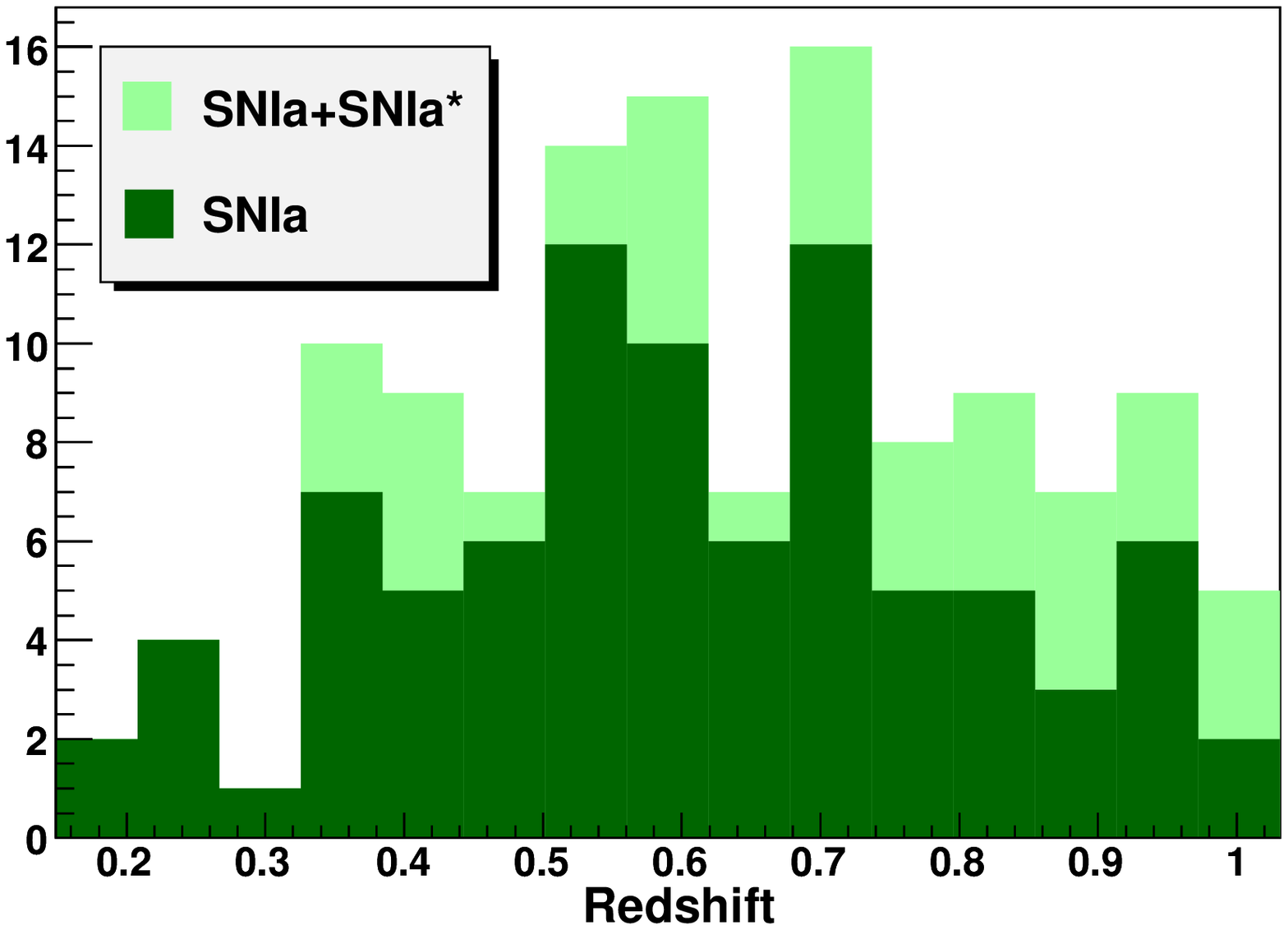}}\\[2mm]
\end{center}
\caption{Redshift distribution of SNLS candidates identified as \Ia\ or
SN~Ia$\star$. In light green, the histogram for all SNe~Ia
(SN~Ia+SN~Ia$\star$). In dark green, the histogram for SNe~Ia
alone. The average redshifts for the SNe~Ia and SNe~Ia$\star$
subsamples are $<z>_{SNIa}=0.60\pm 0.02$ and
$<z>_{SNIa\star}=0.70\pm0.03$ for 86 and 38 SNe, respectively.}
\label{fig:histoz}
\end{figure*}

\begin{figure*}
\begin{center} 
\resizebox{12cm}{10cm}{\includegraphics{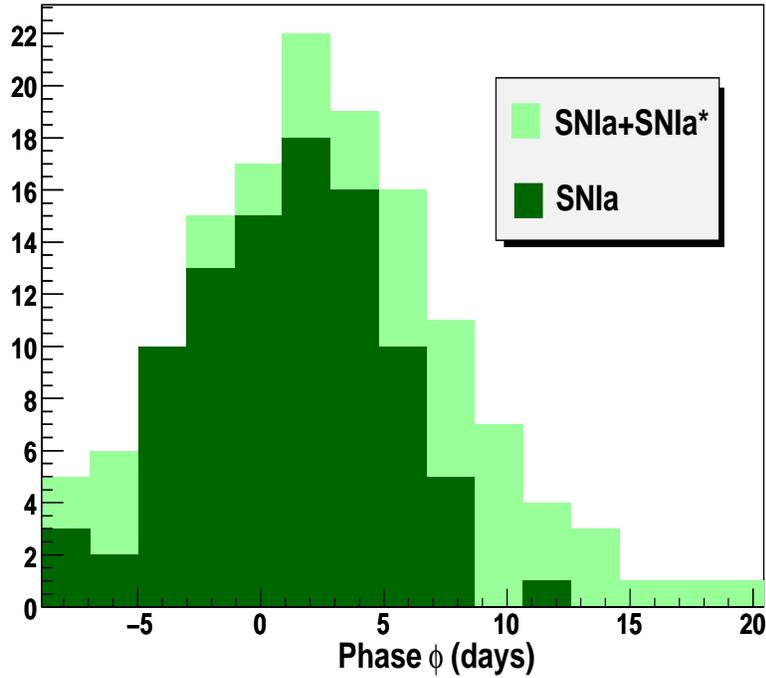}}\\[2mm]

\end{center}
\caption{Phase distribution of SNLS candidates identified as \Ia\ or
SN~Ia$\star$.  In light green, the histogram for all SNe~Ia
(SN~Ia+SN~Ia$\star$). In dark green, the histogram for SNe~Ia
alone. The average phases for the SNe~Ia and SNe~Ia$\star$ subsamples
are $<\phi>_{SN~Ia}=1.1\pm 0.4$ days and $<\phi>_{SN~Ia\star}=5.8\pm
1.1$ days, respectively. }
\label{fig:histophi}
\end{figure*}

The main parameters characterizing the SN~Ia and SN~Ia$\star$
subsamples are given in Table \ref{tab:tabparamsIa} and are discussed
below. Figures \ref{fig:histoz} and \ref{fig:histophi} show the
redshift and phase distributions of both SN~Ia and SN~Ia$\star$
samples. Redshifts range from $0.149$ to $1.031$.  As expected, the
SN~Ia$\star$ subsample has a higher average redshift
($<z>_{SN~Ia\star}=0.70\pm 0.03$) than the SN~Ia subsample
($z_{SN~Ia}=0.60\pm0.02$). The average redshift of the whole sample is
$<z>=0.63$ and the median redshift is 0.62. Below $z\approx 0.3$, all
\Iae\ are identified as certain SN~Ia.

The average phase of the SN~Ia$\star$ subsample is significantly
higher ($<\phi>_{SN~Ia\star}=5.8\pm 1.1$ days) than for the SN~Ia
($<\phi>_{SN~Ia}=1.1\pm 0.4$ days). This is generally caused by
similarities between SN~Ia spectra one week past maximum and SN~Ic
spectra. These cases are labeled SN~Ia$\star$ unless \ion{Si}{\sc ii}
is clearly seen.  Fig.  \ref{fig:histophi} shows that most spectra at
phases later than 10 days are classified as SN~Ia$\star$. This also
reflects the lower-S/N in these (fainter) spectra.  The average S/N
per 5 $\AA$ bin is $\sim 6$ for SN~Ia, $\sim 2$ for SN~Ia$\star$.

We now compare the rest frame $B$-band magnitudes, at maximum light,
for the two samples. For each SN, $m_B^*$, the apparent rest frame
$B$-band magnitude, is determined as part of the light curve fit with
SALT2. In order to assess the significance of any discrepancy between
the SN~Ia and SN~Ia$\star$ samples in terms of physical properties of
the SNe, we compute, for each SN, the ``distance corrected''
magnitude\footnote{Note that $m_B^{*c}$ is not the quantity used to
  constrain the cosmological parameters, as it is not corrected for
  $x_1$ and $c$.}  $m_B^{*c}\equiv m_B^*-5\log (c^{-1}H_0 d_L(z,\bf
\Theta))$, where $d_L$ is the luminosity distance, $c$ is here the
speed of light and ${\bf \Theta}=\{H_0,\Omega_M,\Omega_\Lambda\}$ the
set of cosmological parameters of the underlying cosmology. Here, we
adopt the values ${\bf \Theta}=\{70,0.27,0.73\}$\footnote{$H_0$ is in
  units of km/s/Mpc.} \citep{Astier06}.  We find
$<m_B^{*c}>_{SN~Ia}=23.95\pm 0.04$ and
$<m_B^{*c}>_{SN~Ia\star}=24.12\pm 0.08$. As expected, the SNe we
classify as SN~Ia$\star$ are fainter, on average, than the ones we
classify as SN~Ia (by $<\Delta m_B^{*c}>= 0.17\pm 0.09$ mag). This
reflects the fact that low S/N spectra at a given phase fall
preferentially into the SN~Ia$\star$ category, as they are more
difficult to identify.

\begin{figure*}
\begin{center} 
\resizebox{12cm}{10cm}{\includegraphics{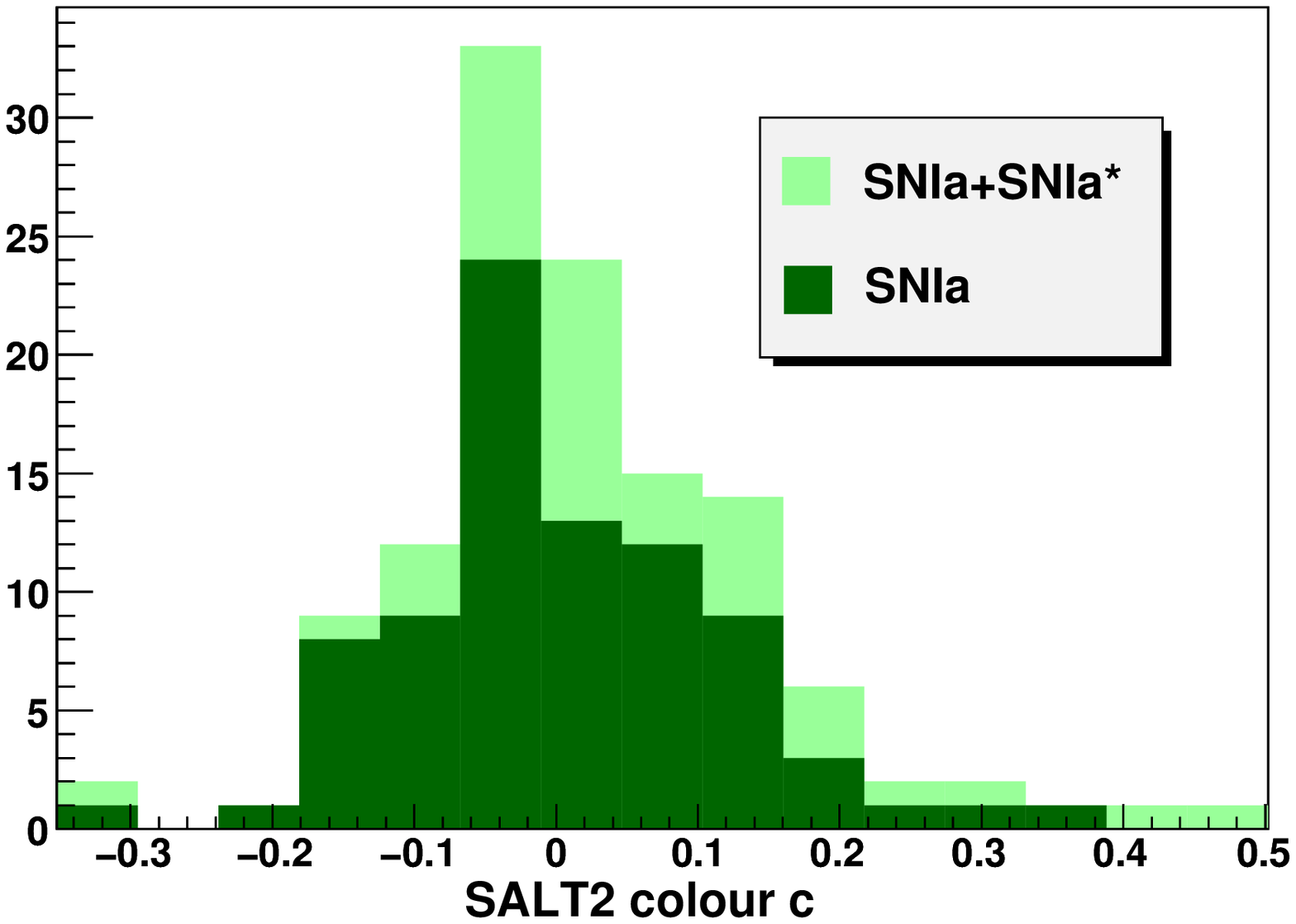}}\\[2mm]
\end{center}
\caption{Colour distribution of SNLS candidates identified as \Ia\ or
SN~Ia$\star$. In light green, the histogram for all SNe~Ia
(SN~Ia+SN~Ia$\star$). In dark green, the histogram for SNe~Ia alone.
The average colours for the SNe~Ia and SNe~Ia$\star$ subsamples are
$<c>_{SNIa}=0.009\pm 0.013$ mag and $<c>_{SNIa\star}=0.05\pm 0.03$
mag, respectively.}
\label{fig:histocol}
\end{figure*}

\begin{figure*}
\begin{center} 
\resizebox{12cm}{10cm}{\includegraphics{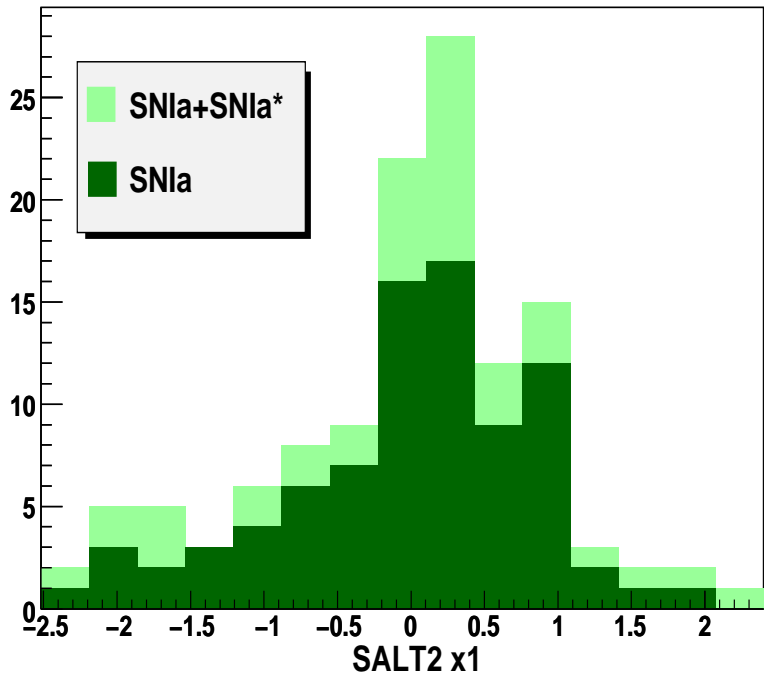}}\\[2mm]
\end{center}
\caption{SALT2 $x_1$ distribution of SNLS candidates identified as \Ia\ or
SN~Ia$\star$. In light green, the histogram for all SNe~Ia
(SN~Ia+SN~Ia$\star$). In dark green, the histogram for SNe~Ia
alone. The average $x_1$ for the SNe~Ia and SNe~Ia$\star$ subsamples
are $<x_1>_{SNIa}=-0.02\pm 0.09$ and $<x_1>_{SNIa\star}=0.004\pm 0.18$,
respectively.}
\label{fig:histox1}
\end{figure*}

It is of interest to account for the difference found in the
$<m_B^{*c}>$ values. One obvious explanation is that it is related to
the well-known ``brighter-slower'' \citep{Phillips93} and
``brighter-bluer'' \citep{Tripp99} relationships observed in SNe~Ia
populations. Figures \ref{fig:histocol} and \ref{fig:histox1} show the
SALT2 colour $c$ and $x_1$ parameter distributions for the two
subsamples. The SN~Ia$\star$ appear to be slightly redder, as their
average colour $<c>_{SN~Ia\star}=0.05\pm 0.03$ is higher than for
SN~Ia: $<c>_{SN~Ia}=0.009\pm 0.013$. Regarding $x_1$, we find
$<x_1>_{SN~Ia}=-0.02\pm 0.09$ and $<x_1>_{SN~Ia\star}=0.004\pm 0.18$,
which are consistent within $1\sigma$ errors (the SN~Ia sample having
faster, narrower light curves than the SN~Ia$\star$ sample, on
average). The ``brighter-slower'' and ``brighter-bluer'' relationships
translate into a magnitude difference between the two samples of
$<\Delta m_B^{*c}>=-\alpha_x<\Delta x_1> + \beta<\Delta c>$. Using
$\alpha_x=0.13$ and $\beta=1.77$ \citep{Guy07}, we find $<\Delta
m_B^{*c}>\approx 0.07\pm 0.06$. The ``observed'' $<m_B^{*c}>$
difference between the two samples is therefore consistent with the
``brighter-slower'' and ``brighter-bluer'' relationships. This gives
us confidence that our SN~Ia$\star$ sample is not significantly
polluted by the inclusion of red, non-\Ia\ objects, such as SNe~Ic.
We can estimate the contamination of the SN~Ia$\star$ sample by
core-collapse SNe by comparing the number of certain SN~Ib/c we have
typed to the number of certain SNe Ia.  In the whole SNLS data, $\sim
3$\% of securely identified SNe are of the SN~Ib/c type. Applying this
ratio to the 38 SN~Ia$\star$ of the VLT sample shows that we should
expect at most 1 contaminant. Note however that the detection
efficiency of the SNLS does not allow us to detect SN~Ib/c SNe beyond
$z<0.4$ \citep{Bazin09}.  As we only have 3 SN~Ia$\star$ with $z<0.4$,
this leads to a more realistic estimate of 0.1 SN~Ia$\star$ that might
be contaminant.

\subsection{Peculiar SN~Ia}
\label{subsec:pec}

\begin{figure*}[htbp]
\begin{center} 
  \begin{minipage}[b]{17cm} \centering
    \includegraphics[width=12cm]{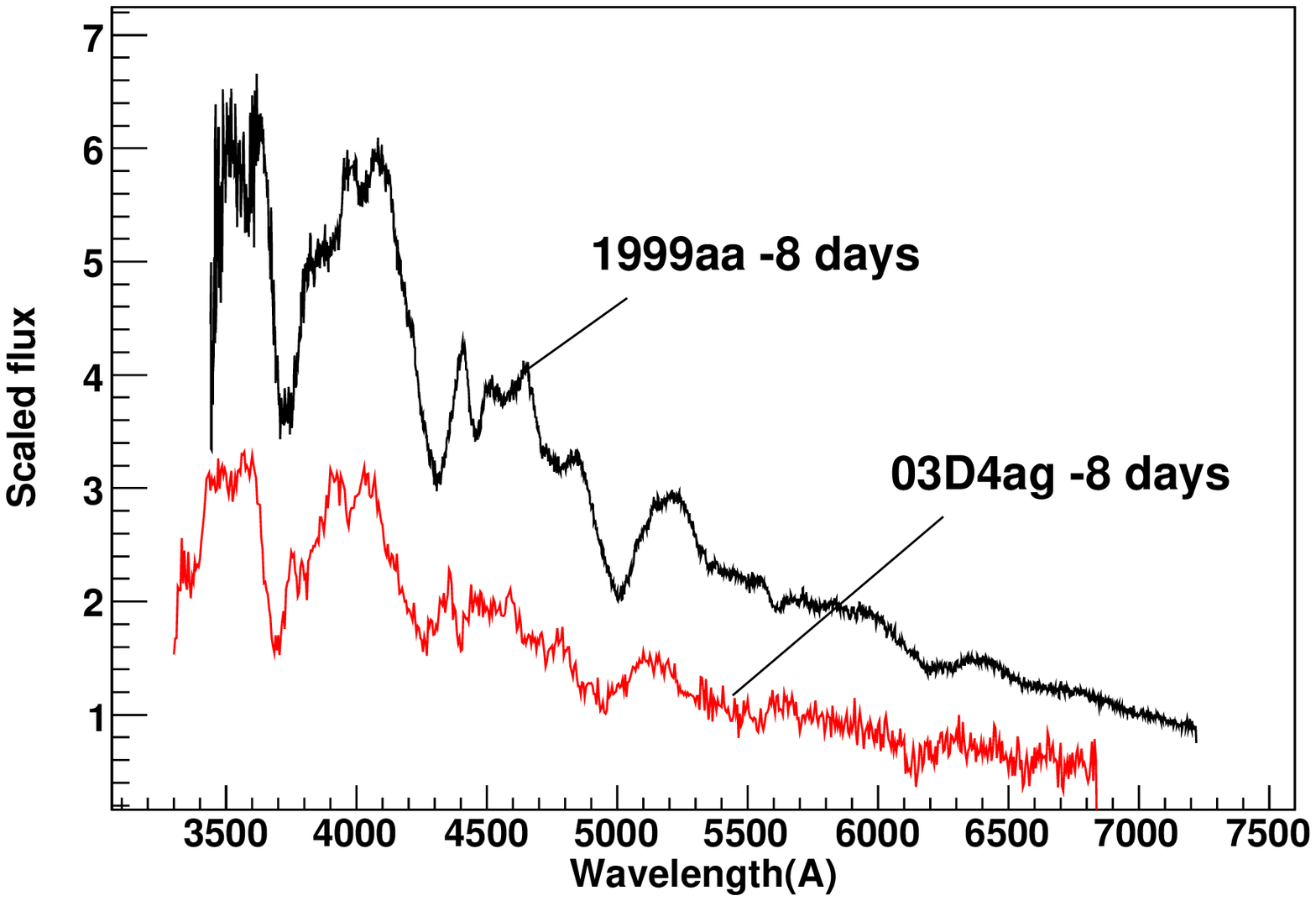}
  \end{minipage} \vspace*{2cm}
  \begin{minipage}[b]{17cm} \centering
    \includegraphics[width=12cm]{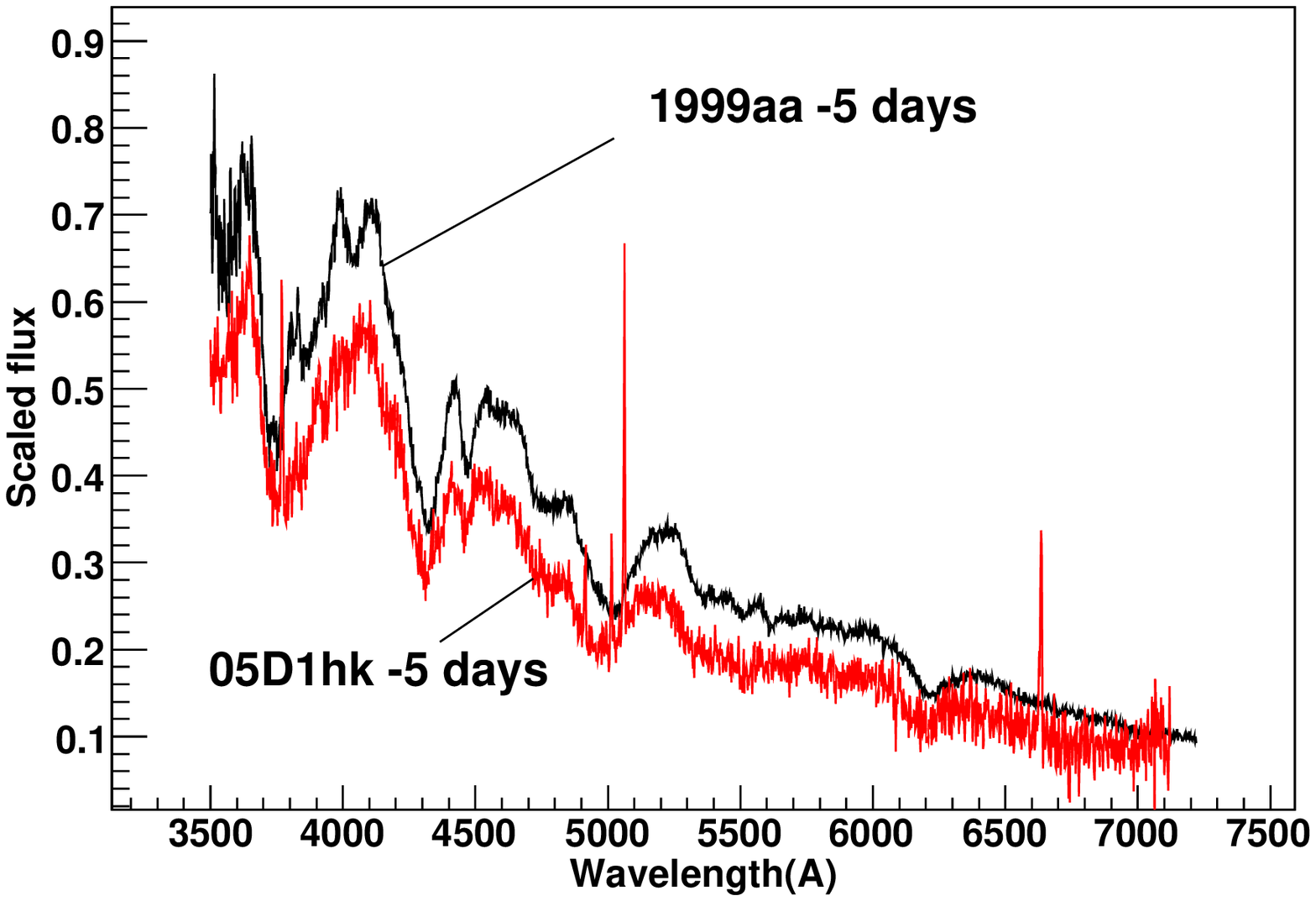}
  \end{minipage}
\end{center}
\caption{\label{fig:figpec} SNLS-03D4ag (top) and SNLS-05D1hk (bottom)
rest frame spectra compared to SN~1999aa templates at -8 and -5 days
respectively (from \citealt{Matheson08}). SN~1999aa spectra are not
de-redshifted. None of the spectra are corrected for peculiar
velocities, which explains the shifts observed in the positions of
absorption features. In both cases, note the very blue spectrum and
shallow \ion{Si}{\sc ii} $\lambda$ 6150.}
\end{figure*}

We have identified two \Ia\ in our sample showing strong similarities
with the spectra of SN~1999aa \citep{Garavini04} (see Fig.
\ref{fig:figpec}). These are 03D4ag and 05D1hk (this latter SN was
already noted as a peculiar \Ia\ by \citealt{Ellis08}).
Table~\ref{tab:tabpec} summarises the properties of these two SNe as
well as the parameters obtained when fitting their spectra with SALT2.
Both SNe are at relatively low redshift ($z=0.285$ and 0.263 for
03D4ag and 05D1hk respectively) and at early phase ($\sim -9$ and
$\sim -5$ days). 05D1hk has a large $x_1$ value, which translates into
a $\Delta m_{15}$ parameter \citep{Phillips93} of $0.89$ using the
formula given in \citet{Guy07}. By comparison, SN~1999aa has $\Delta
m_{15}=0.85$ \citep{Jha06}. For 03D4ag, we find $\Delta m_{15}=0.97$.
Visual inspection of the spectra shows that 03D4ag and 05D1hk have
properties typical of high stretch \Ia: shallow silicon absorption and
a blue spectrum, even for the early phases under consideration.

The possibility of SNLS-03D1co being a 1991bg-like event has been
discussed in \citet{Bronder08}, as there is also a Gemini spectrum of
this event. \citet{Bronder08} found that 03D1co had a large
\ion{Mg}{\sc ii} $\lambda$ 4300 equivalent width (EW), consistent with
the excess absorption measured in under-luminous, low-z SNe spectra.
However, the presence of \ion{Si}{\sc ii} $\lambda$ 4000 in the Gemini
spectrum contradicts this hypothesis, as this feature is not seen in
under-luminous SNe because of extra absorption due to \ion{Ti}{\sc ii}
and \ion{Fe}{\sc ii} in the same wavelength range.  Based on this
observation, \cite{Bronder08} concluded that 03D1co was likely to be a
``normal'' SN~Ia, in agreement with its normal light curves
\citep{Astier06}. The SALT2 fit of the VLT spectrum of 03D1co confirms
this conclusion (see Fig. \ref{fig:03D1co307}).

\subsection{Contamination by non-SN Ia}
\label{sec:nonIa}

An obvious challenge in the identification of distant \Iae\ is to
avoid confusion with other types, such as SNe~Ic, especially one week
after maximum light and beyond.  Template cross-correlation techniques
such as SNID can help \citep{Matheson05}, but still rely on
(unavoidably) incomplete template libraries.  Our SALT2 identification
procedure provides some additional information on non SN~Ia types (or
peculiar SNe~Ia) through the fitted parameter values, which can differ
from those of a typical SN~Ia, e.g., an unusually red colour, and/or a
high $x_1$ parameter, and/or a high recalibration parameter.  This is
not equivalent to a direct identification of non \Ia\ objects, but
does allow a mechanism by which peculiar events can be identified. We
explore these parameters in this section.

We start with the SALT2 ``recalibration parameters'' used to adjust
the SALT2 photometric model to the observed data. We focus on the
first order recalibration parameter $\gamma_1$ (Section
\ref{sec:idsalt2}).  This parameter can be interpreted as the ``tilt''
required to adjust the observed spectrum to the SALT2 colour model. A
large recalibration is a sign that the SALT2 model is not able to
properly model the data, as would be the case if we were trying to fit
a non-SN~Ia spectrum, but could also occur for a SN Ia spectrum whose
properties were very different from those of the training sample.
Large tilts can also be needed when the host subtraction has failed
due to inadequate host modelling or too strong a contamination,
although no strong correlation exists between $\gamma_1$ and the host
fraction.

The average $\gamma_1$ value for the SN~Ia$\star$ is
$<\gamma_1>=0.26\pm 0.29$, with $<\gamma_1>=0.26\pm 0.11$ for the
SN~Ia: the required tilts needed to recalibrate the photometric model
are only moderate.  The dispersions in the $\gamma_1$ values are quite
large for both samples, with a larger variation for the SN~Ia$\star$:
$\sigma_{\gamma_1}^{SN~Ia}=1.1$ and
$\sigma_{\gamma_1}^{SN~Ia\star}=1.9$. This can be explained by the
inclusion in the SN~Ia$\star$ sample of more host-contaminated
spectra. If we select only SNe for which a separate extraction from
their host was not possible, the mean host fractions $f_g$ (i.e. the
contribution of the host model to the full spectrum averaged over the
whole spectral range) are $<f_g>=0.32\pm 0.03$ and $<f_g>=0.47\pm
0.05$ for the SN~Ia and SN~Ia$\star$ subsamples, respectively (see
Table \ref{tab:tabparamsIa}). Clearly, spectra identified as
SN~Ia$\star$ are more host contaminated than SN~Ia spectra.

\begin{figure*}
\begin{center} %\rotatebox{90}{
\resizebox{12cm}{10cm}{\includegraphics{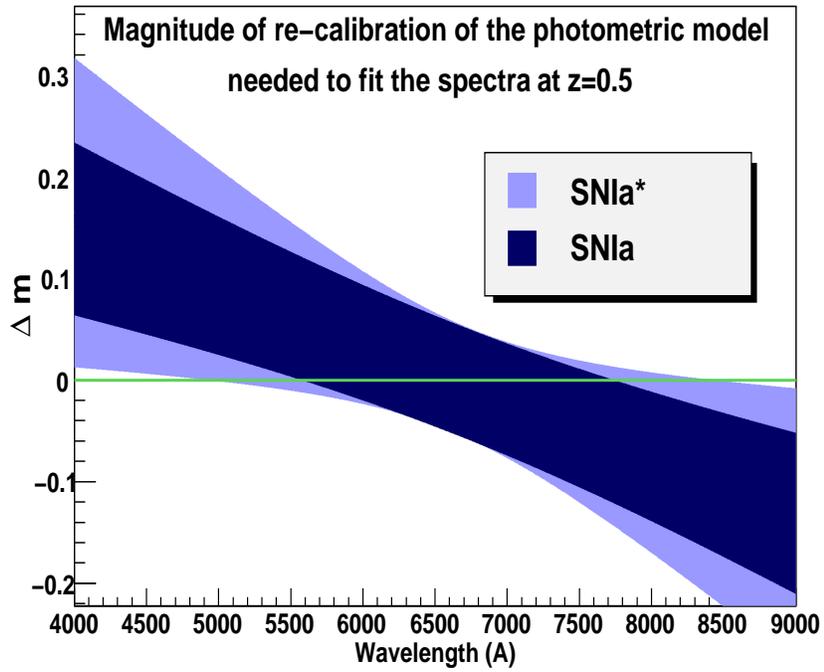}}\\[2mm]
\end{center}
\caption{Illustration of the magnitude $\Delta m$ of the average
recalibration (``tilt") needed for the photometric models to fit the
spectra, as a function of wavelength. For the purpose of this
illustration, we have selected a subset of VLT spectra with redshifts
between $z=0.4$ and $z=0.6$. The average redshift is then $<z>=0.512$
for SN~Ia and $<z>=0.518$ for SN~Ia$\star$. Errors shown in the figure
(in light blue for SN~Ia$\star$ and in dark blue for SN~Ia), are
computed from the average value of the recalibration parameter
$\gamma_1$, estimated from the subset, for both SN~Ia and SN~Ia$\star$
populations. It appears that, if recalibration is about the same, on
average, for SN~Ia and SN~Ia$\star$ at this redshift, the $\gamma_1$
values are more dispersed for the SN~Ia$\star$. The global flux
correction is mild, $\approx$15\% at both ends of the wavelength
scale.}
\label{fig:recalib}
\end{figure*}

We do not find any systematic trend with redshift or phase in the
``tilt'' values needed to accomodate the SALT2 model with the spectra.
In Fig.  \ref{fig:recalib}, we show the magnitude $\Delta m$ of the
recalibration of the photometric model as a function of wavelength,
for a subset of our spectra chosen between $z=0.4$ and $z=0.6$
($<z>^{SN~Ia}=0.512$ and $<z>^{SN~Ia\star}=0.518$ for this subset).
The light blue area is for SNe~Ia$\star$, the dark blue one SNe~Ia. At
this redshift, we find that a $\approx 15$\% recalibration is needed
at both ends of the effective spectral range, for both the SN~Ia and
SN~Ia$\star$ categories.

\begin{figure*}[htbp]
\begin{center} 
  \begin{minipage}[b]{17cm} \centering
    \includegraphics[width=12cm]{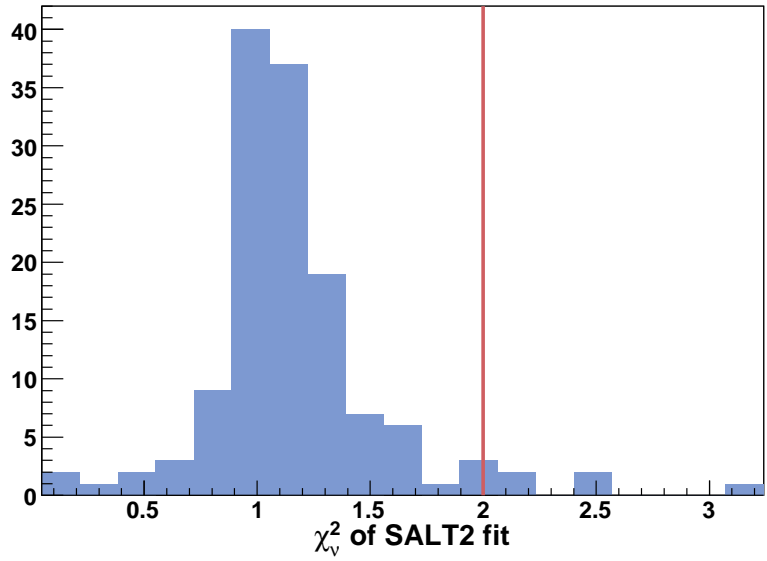}
  \end{minipage}
\end{center}
\caption{\label{fig:chi2distrib} Distribution of SALT2 $\chi^2_\nu$
values for the whole (SN~Ia and SN~Ia$\star$) sample. The vertical
red/grey solid line indicates the cut at $\chi^2_\nu=2$ used to
identify potential non- or peculiar SN~Ia. The two highest
$\chi^2_\nu$ objects are not shown.}
\end{figure*}

A potentially more quantitative criterion is the (reduced)
$\chi^2_\nu$ of the SALT2 spectral fits. We show in Fig.
\ref{fig:chi2distrib} the reduced $\chi^2_\nu$ histogram of all SALT2
fits (both for the SNe~Ia and SNe~Ia$\star$ samples). The bulk of the
$\chi^2_\nu$ values are around 1 (or slightly higher), but with a tail
of 11 objects with $\chi^2_\nu>2$ (Fig.~\ref{fig:chi2distrib}; the two
highest $\chi^2_\nu$ objects are not shown). With these objects, we
find $<\chi^2_\nu>=1.33$, while excluding them yields
$<\chi^2_\nu>=1.11$.  Among the 11 high $\chi^2_\nu$ objects, one
finds 03D4ag and 05D1hk, the over-luminous events identified in
Section~\ref{subsec:pec}. Their best-fit $\chi^2_\nu$ values are 2.48
and 3.12 respectively.  The other objects are (ordered by increasing
$\chi^2_\nu$): SNLS-03D1fc(2.01), SNLS-04D4ht(2.03),
SNLS-05D2ct(2.04), SNLS-05D4cw(2.18), SNLS-04D2bt(2.23),
SNLS-04D2fs(2.42), SNLS-04D1dc(3.24), SNLS-03D4au(4.63) and
SNLS-05D4ff(16.33).

03D1fc (Fig. \ref{fig:03D1fc358}; \Ia\ at $z=0.332$) has a separate
extraction from its host. The spectrum has a high S/N and the SALT2
fit parameters are typical of a normal SN~Ia.  \ion{Si}{\sc ii}
$\lambda$ 6150 is visible but shallow. The spectrum is not as blue as
the one of 05D1hk (which is at about the same phase, i.e. -5 days),
but the photometric model requires three recalibration parameters to
consistently fit the spectrum. Although the stretch is normal
($s=1.006$), this SN might be slightly peculiar.

\begin{figure*}[ht]
                    \begin{center}
                    \begin{tabular}[b]{cc}
                    \hspace{-5mm}
                    \vspace*{1cm}
                    \begin{minipage}[b]{10cm}
                    \centering
                    \includegraphics[width=10cm]{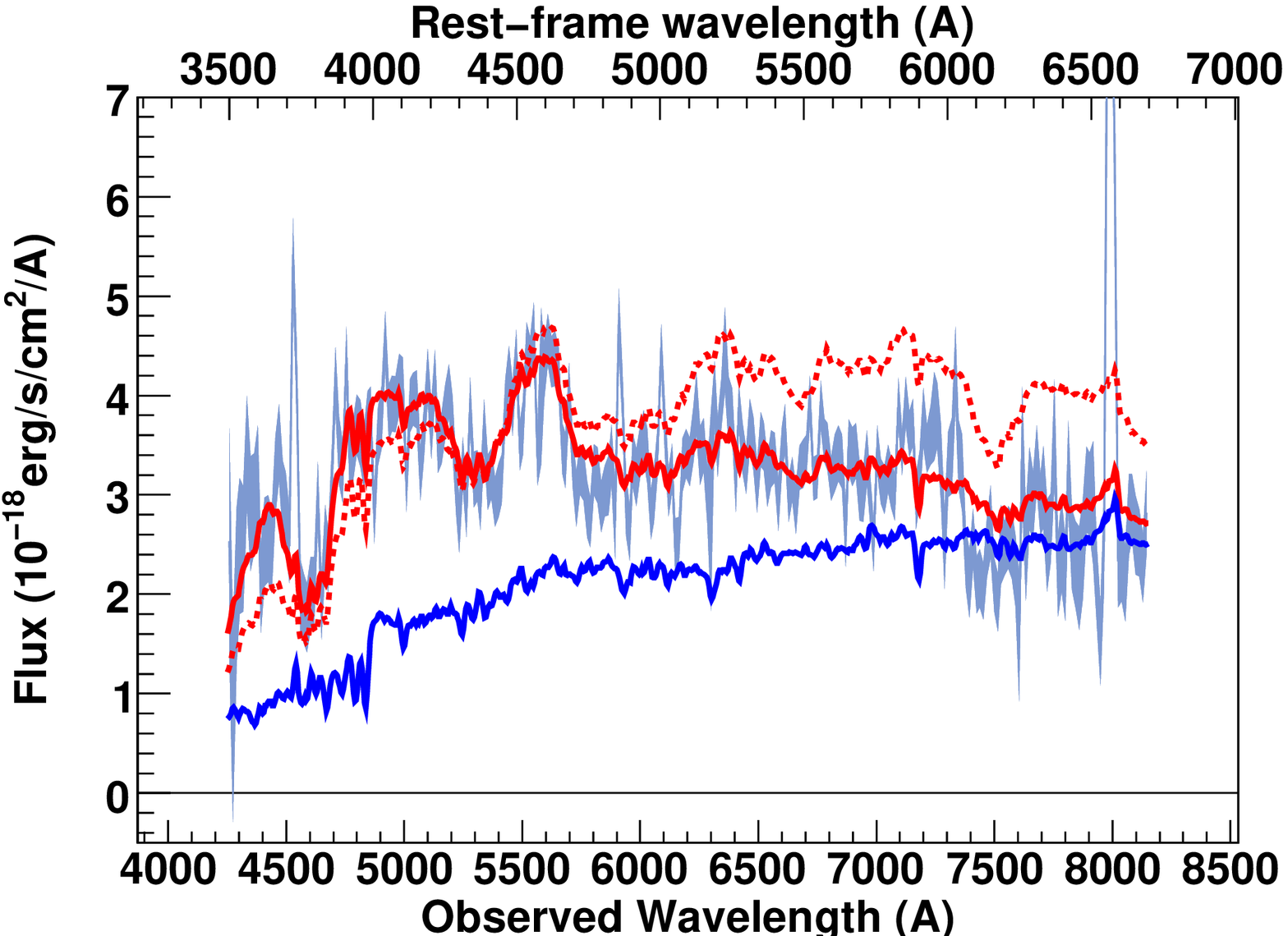}
                    \end{minipage}
                    &
                    \begin{minipage}[b]{10cm}
                    \hspace{-5mm}\includegraphics[width=10cm]{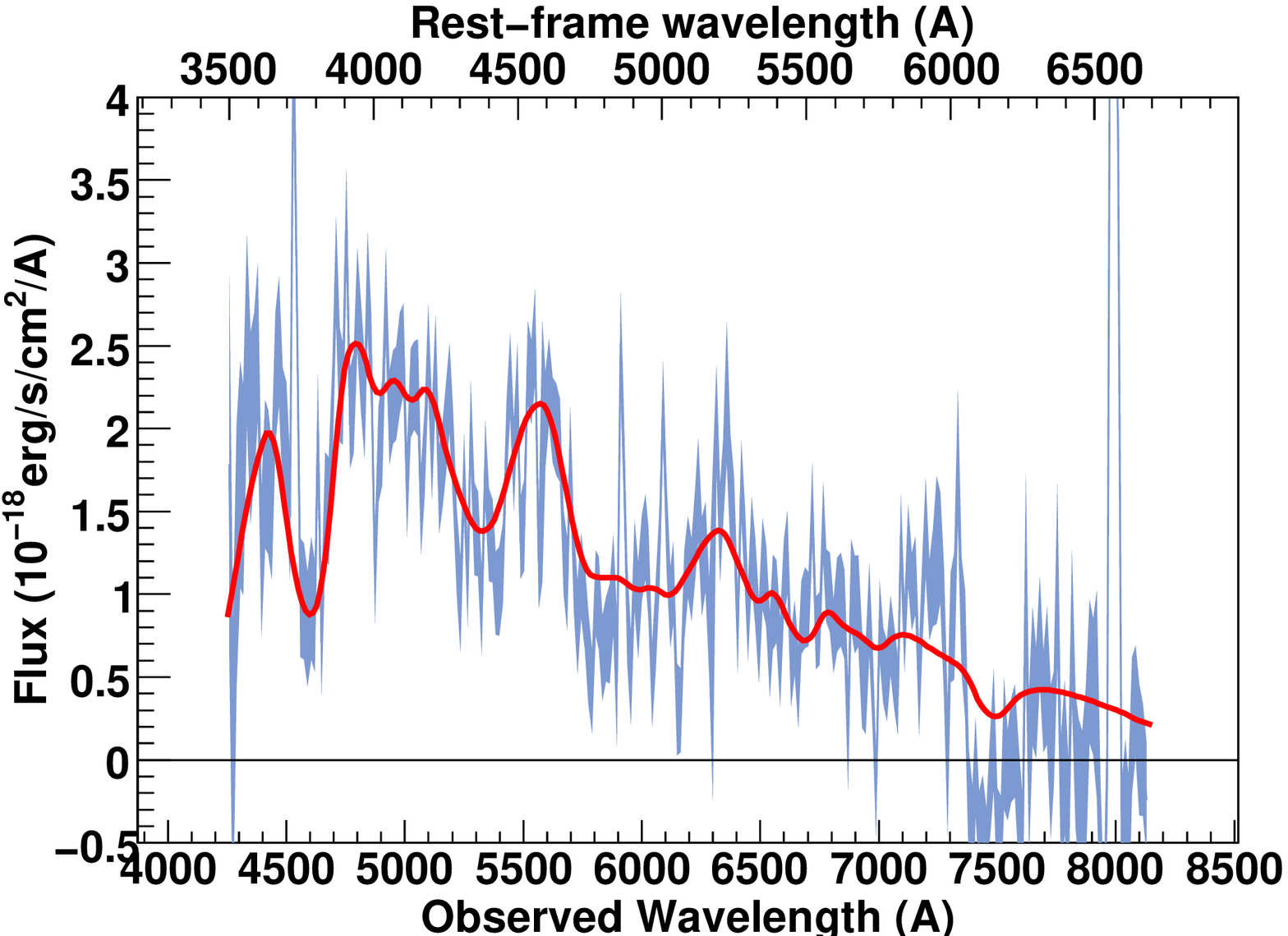}
                    \end{minipage}
                    \\
                    \end{tabular}
                    \caption{\label{fig:04D4ht628} SNLS-  04D4ht628: a SN~Ia supernova at z=0.217. The spectrum phase is 6.6. A         Sa-Sb host model has been subtracted.}
                    \end{center}
                    \end{figure*}

04D4ht (Fig. \ref{fig:04D4ht628}; \Ia\ at $z=0.217$) is heavily host
contaminated ($>60\%$ of host in the best-fitting model). It has an
high colour value ($c=0.5$). 04D4ht is very red but with strong
\ion{Si}{\sc ii} in its spectrum and is identified as a SN~Ia.
However, it is located very near the core of its host (a late type
spiral) and potentially heavily extinguished.

05D2ct (Fig. \ref{fig:05D2ct799}; SN~Ia$\star$ at $z=0.734$) is
another example of a SN close to its host centre (more than half of
the extracted signal is modeled by a Sd galaxy), slightly red
($c=0.14$) but with a normal stretch ($s=0.994$). It is likely to be
extinguished. Due to its fairly high redshift and host contamination,
no \ion{Si}{\sc ii} is visible either at 6150 or at 4000 $\AA$ and it
is identified as a SN~Ia$\star$.

05D4cw (Fig. \ref{fig:05D4cw950}; SN~Ia at $z=0.375$) is a blue
($c=-0.15$) SN. It is heavily host-contaminated, probably an early
type galaxy. As it is close to its host centre, a separate extraction
was not possible. However, \ion{Si}{\sc ii} $\lambda$ 6150 is clearly
visible, and it is classified as a SN~Ia.

04D2bt (Fig. \ref{fig:04D2bt444}; SN~Ia at $z=0.220$) is located in the
bulge of its early type host. It is red ($c=0.18$), likely due to host
extinction. \ion{Si}{\sc ii} and \ion{S}{\sc ii} are clearly visible,
the classification is SN~Ia.

\begin{figure*}[ht]
                    \begin{center}
                    \begin{tabular}[b]{cc}
                    \hspace{-5mm}
                    \vspace*{1cm}
                    \begin{minipage}[b]{10cm}
                    \centering
                    \includegraphics[width=10cm]{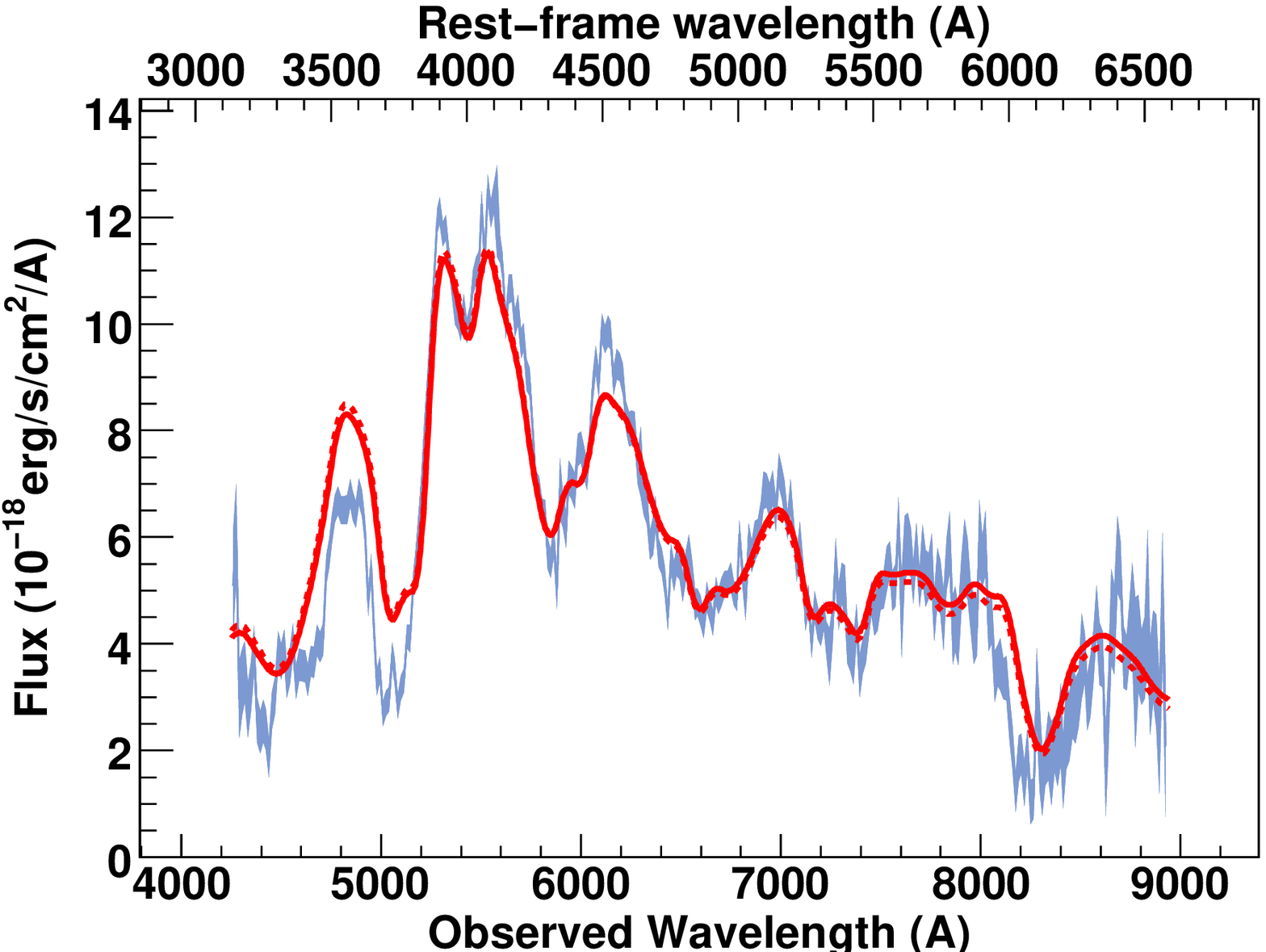}
                    \end{minipage}
                    &
                    \begin{minipage}[b]{10cm}
                    \end{minipage}
                    \\
                    \end{tabular}
                    \caption{\label{fig:04D2fs470} SNLS-  04D2fs470: a SN~Ia supernova at z=0.357. The spectrum phase is 1.7. Best-fit obtained for a model with no host galaxy component.}
                    \end{center}
                    \end{figure*}

04D2fs (Fig. \ref{fig:04D2fs470}; SN~Ia at $z=0.357$) has a large S/N
with \ion{Si}{\sc ii} and \ion{S}{\sc ii} clearly visible.  The high
$\chi^2_\nu$ value can be explained by the high S/N (low noise) of the
spectrum.

04D1dc (Fig. \ref{fig:04D1dc589}; SN~Ia at $z=0.211$) is similar to
04D2fs with a very high S/N. \ion{Si}{\sc ii} and \ion{S}{\sc ii} are
obvious. Again, the high $\chi^2_\nu$ value reflects the small errors
on the spectrum flux.

03D4au (Fig. \ref{fig:03D4au186}; SN~Ia$\star$ at $z=0.468$) is red
($c=0.18$) and host contaminated, a week past maximum light. It is
located right at the centre of its late-type host, probably
extinguished, and with strong emission lines that are difficult to
subtract, explaining the high $\chi^2_\nu$ value.

05D4ff (Fig. \ref{fig:05D4ff1003}; SN~Ia$\star$ at $z=0.402$) is
heavily buried in its late-type host. The presence of strong
[\ion{O}{\sc ii}], [\ion{O}{\sc iii}] and H$\beta$ emission lines, not
present in the host PEGASE template, explains the poor $\chi^2_\nu$.

In conclusion, a cut in $\chi^2_\nu$ can help with the identification
of peculiar (and possibly non-SN~Ia) events, but it is not a
straightforward identification. Several effects can conspire to give a
high $\chi^2_\nu$ value, even in the case of obvious SNe~Ia. The
results of this section suggest that our two subsamples are unlikely
to be significantly contaminated by non-SNe~Ia.

\section{Composite spectra of the SN~Ia sample}

\label{sec:composit}

In this section, we build average VLT spectra in six regions of
phase-redshift space to compare the evolution of their properties with
redshift.

\subsection{Methodology}

We select spectra with phase $\phi\leq -4$ days (pre-maximum spectra);
$-4<\phi<4$ days (maximum spectra); $\phi\geq 4$ days (post-maximum
spectra) for SNe at redshift $z<0.5$ and $z\geq 0.5$, respectively. We
only use SN~Ia spectra, as the SN~Ia$\star$ spectra number in each of
the six regions is small and even zero for $z<0.5$ around maximum
(since the spectra in this phase-space region are of sufficient S/N
and quality to be identified as SN~Ia). The number of spectra in each
phase bin for both $z<0.5$ and $z \geq 0.5$ is summarised in Table
\ref{tab:compositeIa}.  For the SN~Ia spectra, we have 12 pre-maximum
spectra (of which 7 are at $z<0.5$), 57 spectra at maximum light (of
which 12 are at $z<0.5$) and 24 post-maximum spectra (of which 6 are
at $z<0.5$). The average redshift is $<z>=0.36\pm 0.02$ for the
$z<0.5$ sample and $<z>=0.70\pm 0.02$ for the $z>0.5$ sample.  We have
more spectra in the $z\geq 0.5$ bins (68 spectra) than in the $z<0.5$
bins (25), as the median redshift of our sample is 0.62.  The spectra
of 03D4ag and 05D1hk are excluded from the $z<0.5$ pre-maximum light
region, as they have been identified as peculiar SNe~Ia (see Section
\ref{subsec:pec}).

To construct the average spectrum in each region, all individual
spectra (shown in the right panels of Fig. \ref{fig:03D1ar270} to
\ref{fig:06D4cq1306} whenever a host model has been subtracted) are
brought into the rest frame and rebinned to 5 $\AA$. All spectra are
colour-corrected using an up-to-date version of the SALT2 colour law
and the colour value $c$ obtained from the SALT2 fit for each SN.
Using a \citet{Cardelli89} extinction law with $R_V=3.1$ and $E(B-V)$
values obtained for each SN from their light curve fit, yields very
similar results in the wavelength range under consideration. We adopt
the SALT2 colour correction in the following. Fluxes are normalised to
the same integral in the range 4450-4550 $\AA$. For each wavelength
bin, an average weighted flux and its corresponding uncertainty are
computed from all spectra included in this bin.  Because of the
variety of the redshifts involved, the number of spectra entering the
average varies from one bin to the other, as does the average phase in
a given bin. In practice, the number of spectra in each bin decreases
at both ends of the wavelength scale but is approximately constant in
the range 4000-7000 $\AA$.

As a consequence of this averaging procedure, the mean phase of the
average spectra varies from one wavelength bin to the other. In
practice, this variation distorts only very moderately the spectra as
the mean phase is roughly the same from one end to the other of the
wavelength scale.  However, for a given phase range (pre-maximum,
maximum or post-maximum), the mean phases of the average spectrum at
$z<0.5$ can significantly differ from the corresponding one at $z\geq
0.5$, which should be borne in mind when comparing the average
spectrum at $z<0.5$ with that at $z\geq 0.5$. In practice, due to a
similar phase sampling of our spectra with redshift, this difference
is marginal. We find the following average phases (in days):
$\phi^{z<0.5}=-6.1$ and $\phi^{z>0.5}=-5.4$ for pre-maximum spectra,
$\phi^{z<0.5}=0.5$ and $\phi^{z>0.5}=0.5$ for spectra at maximum, and
$\phi^{z<0.5}=7.7$ and $\phi^{z>0.5}=5.5$ for post-maximum spectra.

\subsection{Comparison to Hsiao et al. template spectra}

\begin{figure*}
\begin{center}
\resizebox{18cm}{18cm}{\includegraphics{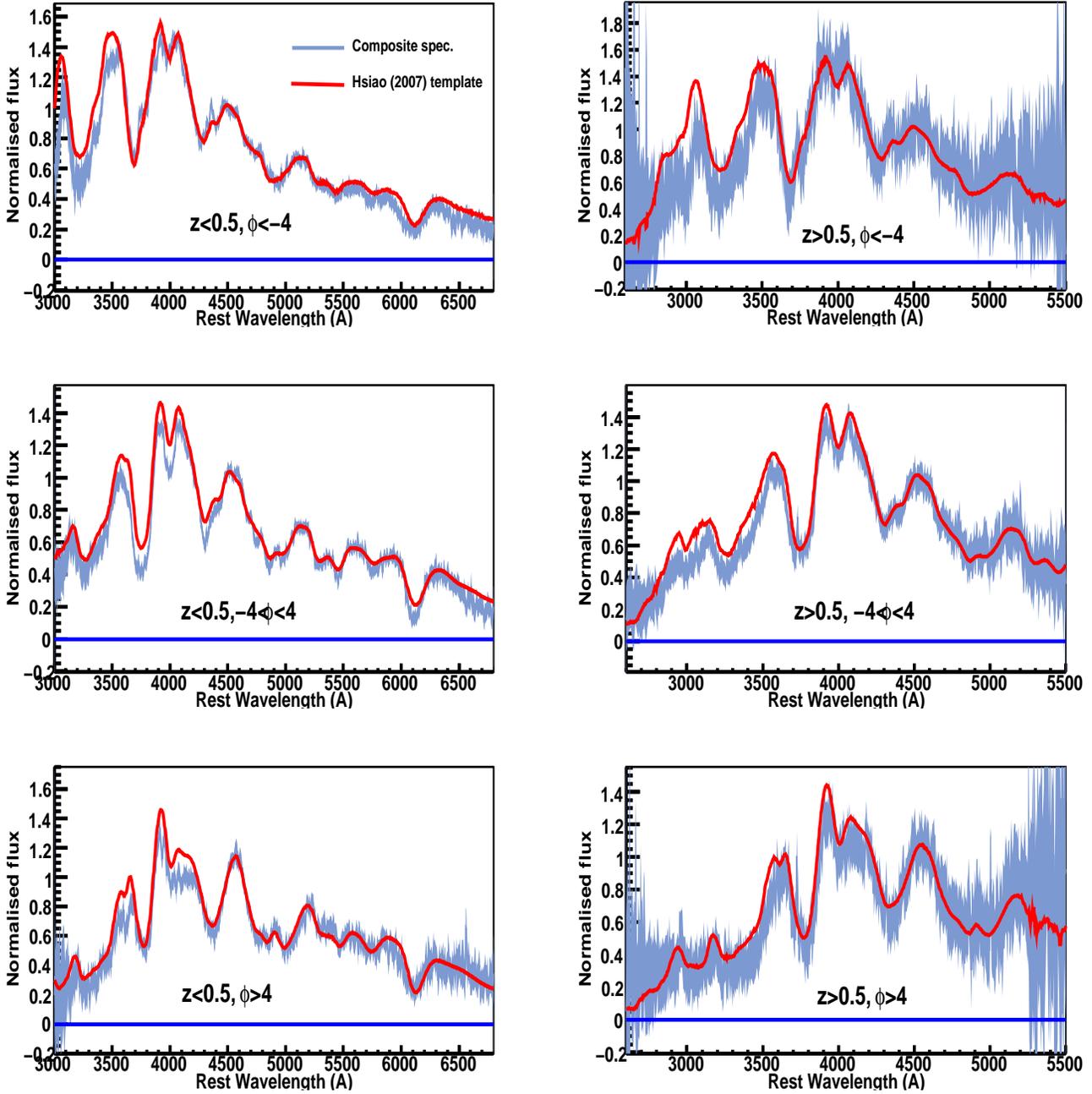}}\\[2mm]

\end{center}
\caption{In grey/blue: Average spectrum built from all VLT SN~Ia
spectra of this study for $z<0.5$ (left column) and $z>0.5$ (right
column), and for pre-maximum ($\Phi<-4$ days, top), around maximum
($-4$ days $< \Phi < +4$ days, middle) and past maximum ($\Phi> +4$
days, bottom). To construct the average spectrum in each panel, all
individual spectra have been brought into the rest frame and rebinned with
5 $\AA$. Spectra have been colour-corrected using the SALT2 colour law
and the colour parameter $c$ for each individual spectrum. Fluxes have
then been normalised to the same integral around 4500 $\AA$. For each
wavelength bin, an average flux $f_\lambda$ and a weighted $1\sigma$ error are
computed from all the spectra with measured flux in this bin. The
number and average phase thus vary at each wavelength step. In red:
an average spectrum built in the same way using this time the
\citet{Hsiao07} template series. For each wavelength bin, the same
number of individual templates and same weights as in the average VLT
spectrum have been used, with the corresponding phases rounded off to
the nearest integer value.  }

\label{fig:compareHsiao}
\end{figure*}

Figure \ref{fig:compareHsiao} shows the result of averaging the SN~Ia
spectra (in blue). The left column is for $z<0.5$ spectra, the right
for $z\geq 0.5$ spectra. From top to bottom, pre-maximum, maximum and
post-maximum spectra are shown. Average spectra built from the
\citet{Hsiao07} template series (red curve) are overlaid on top of the
VLT average spectra. Note that about two-thirds of the spectra used in
constructing the \citet{Hsiao07} template are at low redshift
($z<0.1$) and lack UV coverage, so the UV section of the red curves
come from the remaining high-redshift spectra used in building the
\citet{Hsiao07} template.  In each region of
Fig.~\ref{fig:compareHsiao}, phases in each wavelength bin are the
same for the template and the VLT composite spectrum, weighted in
exactly the same way.

The overall agreement between the VLT average spectra and the
\citet{Hsiao07} template is good in all regions.  Note that no colour
correction has been applied to the \citet{Hsiao07} templates. As this
template is designed for use in light curve fitters that implement
``warping'' techniques \citep[e.g., SiFTO,][]{Conley08}, there is no
{\em a priori} reason that the continuum should agree with our
composite spectra.  However, we find that the agreement is almost
perfect in the optical region of the spectra except around the
\ion{Ca}{\sc ii} $\lambda$ 3700, \ion{Si}{\sc ii} $\lambda$ 4000 and
\ion{Si}{\sc ii} $\lambda$ 6150 features, for the maximum average
spectrum at $z<0.5$. In the UV region, one notices more discrepancies
in the fluxes (a low number of spectra are used in this region).  Part
of this effect might be due to the fact that we normalise the spectra
in the optical region. This spectral region is also the most sensitive
to differential slit losses.  We find a satisfying match of the
positions of the UV $\lambda$ 3000 - 3400 features. Note that the peak
around 3200 $\AA$ decreases from pre- to post-maximum phases in a
proportion that is well reproduced by the \citet{Hsiao07} model, both
at $z<0.5$ and $z\geq 0.5$.

\subsection{Comparison of $z<0.5$ and $z\geq 0.5$ spectra}

We now compare the average spectra at $z<0.5$ and $z\geq 0.5$ for
pre-maximum phases (upper panels of Fig. \ref{fig:comparez}),
maximum-light (middle panels) and post-maximum (lower panels). The
comparison is done in the region of intersection, from the UV up to
the mid-optical wavelengths.  For each panel, the blue curve is the
$z\geq 0.5$ average spectrum, and the black curve is the corresponding
$z<0.5$ spectrum. At the bottom of each panel, we plot the residual
$\Delta^{VLT}(\lambda)=f^{z\geq 0.5}(\lambda)-f^{z<0.5}(\lambda)$
(solid black line) with $\pm 1\sigma$ errors.

The mean spectra can be different for two reasons: they might be
intrinsically different (i.e. evolution in spectral properties), or
the phase distribution of the samples can differ. We therefore compute
the residual $\Delta^{Hsiao}(\lambda)$ (solid thick red curve) for the
\citet{Hsiao07} templates shown in Fig.  \ref{fig:compareHsiao}. As
the underlying assumption in building such spectral templates is that
there is no evolution between low and high redshift,
$\Delta^{Hsaio}(\lambda)$ should in principle be zero over the whole
spectral range.  Any deviation from zero, in a given wavelength range,
can be attributed to a difference in the average phase of the $z<0.5$
and $z\geq 0.5$ composite spectra in this range.  When inspecting
differences in the VLT composite spectra at $z<0.5$ and $z\geq 0.5$,
it is important to refer to $\Delta^{Hsiao}(\lambda)$: if
$\Delta^{VLT}(\lambda)$ follows $\Delta^{Hsiao}(\lambda)$, deviations
can be traced back to differing phase distributions of the $z<0.5$ and
$z\geq 0.5$ composite spectra. In the opposite case ($\Delta^{VLT}$ is
not zero while $\Delta^{Hsiao}(\lambda)$ is zero), any differences
should be real.

To quantitatively assess the significance of the differences, we
define a reduced $\chi^2_\nu$ measure of the agreement of the $z\geq
0.5$ VLT composite spectrum and the $z<0.5$ composite spectrum. The
variance entering the definition of $\chi^2_\nu$ is the sum of the
variances of the two composite spectra. A value $\chi^2_\nu \sim 1$
indicates that the two spectra are consistent with one another. We now
examine each phase bin in turn.

\subsubsection{ Pre-maximum spectra}

For pre-maximum spectra ($\phi\leq -4$), the composite spectra are
typically consistent ($\chi^2_\nu \approx 1.14$ for 588 degrees of
freedom - D.O.F.). The spectra are most discrepant in the UV region
(3300 to 4000 $\AA$).  A possible explanation is that, as the bands
bluer than the rest frame $B$-band are given a lower weight in the
SALT2 light curve fit, the colour correction applied to the spectra
that uses the colour parameter $c$ derived from the light curve fit,
is more efficient in the red than in the blue.  Alternatively, this
might reflect a greater variability in this spectral region
\citep[e.g.,][]{Ellis08}. More variability is expected here due to
variations in metallicities of the progenitors, particularly at
pre-maximum phases \citep{Hoeflich98, Lentz00}.  The residuals in Fig.
\ref{fig:comparez} show the UV region of the pre- and maximum spectra
shows more discrepancy than at post-maximum.

\subsubsection{ Maximum-light spectra}

The largest discrepancy between $z<0.5$ and $z\geq 0.5$ is found at
maximum light ($-4<\phi<4$). This is the phase where we have the
highest number of spectra (12 at $z<0.5$, 45 at $z \geq 0.5$), and the
statistical errors are the smallest. We find $\chi^2_\nu\approx 4.80$
for 601 D.O.F. -- the two average spectra are formally not consistent.
The largest differences are seen around \ion{Ca}{\sc ii} $\lambda$
3700, \ion{Si}{\sc ii} $\lambda$ 4000, \ion{Mg}{\sc ii} $\lambda$ 4300
and \ion{Fe}{\sc ii} $\lambda$ 4800.  Overall, the $z<0.5$ maximum
spectrum has deeper absorptions than its $z\geq 0.5$ counterpart.

In the UV up to 3500 $\AA$, the residuals $\Delta^{VLT}(\lambda)$
correlate with the template residuals $\Delta^{Hsiao}(\lambda)$, and
the discrepancies between $z<0.5$ and $z\geq 0.5$ may be due to
differences in mean phases between $z<0.5$ and $z\geq 0.5$ spectra
(recall that due to our averaging procedure, the mean phase varies
with $\lambda$ and depends on the phase of the spectra used to build
the average spectrum in a given wavelength bin). Around \ion{Ca}{\sc
  ii} $\lambda$ 3700 and beyond, the discrepancies are seen in the
spectra, not in the template, and could be real.

\citet{Bronder08} indicate a possible difference in the EW of
\ion{Mg}{\sc ii} $\lambda$ 4300 between low and high redshift, but
they conclude that it is likely due to differences in the epoch
sampling and number of objects at low- and high-redshift. This
difference is not found by \citet{Foley08} when comparing their
composite ESSENCE spectrum with a Lick low-$z$ composite (note that
they use a simpler method for host subtraction, which may alter the
significance of their comparison).  More recently, \citet{Sullivan09}
studied the possible evolution in the EW of intermediate mass elements
(IMEs) in the $0<z<1.3$ redshift range. They do not have a
\ion{Mg}{\sc ii} measurement in their highest-redshift bin, but the
predicted variation of \ion{Mg}{\sc ii} EW is consistent with zero
over their redshift range.

The \ion{Si}{\sc ii}-\ion{Fe}{\sc ii}-\ion{Fe}{\sc iii} $\lambda$ 4800
blend feature is shallower in our $z\geq 0.5$ spectrum than at
$z<0.5$, in qualitative agreement with \citet{Foley08}, who find that
this feature is much weaker in their ESSENCE spectrum. They interpret
this difference as due to a weaker \ion{Fe}{\sc iii} $\lambda$ 5129
line in the high-$z$ spectrum.

The \ion{Si}{\sc ii} $\lambda$ 4000 feature is shallower at high
redshift, in agreement with the findings of \citet{Sullivan09}.
\citet{Bronder08} do not mention such a difference, but
\citet{Foley08} find the same kind of trend for \ion{Si}{\sc ii}
$\lambda$ 6150 (this feature being shallower in their high-redshift
ESSENCE spectrum than in the low-redshift Lick spectrum), although
they do not mention this for \ion{Si}{\sc ii} $\lambda$ 4000. We do
not find this difference in our pre-maximum spectra, though it may be
present at post-maximum.

\subsubsection{ Post-maximum spectra}

At post-maximum ($\phi\geq4$), we find a good agreement between the
$z\geq 0.5$ average spectrum and its lower $z$ counterpart:
$\chi^2_\nu\approx 1.10$ (547 D.O.F.). Variations in the spectral
residuals closely match those in the template residual and reflect
mean phase variations between the $z<0.5$ and $z\geq 0.5$ average
spectra.  Once again, the level of discrepancy, though small, is
highest in the UV and in the \ion{Si}{\sc ii} $\lambda$ 4000 region.
It is interesting to note that, as for the maximum light spectra, the
flux at the position of this line (which has almost disappeared a week
past maximum) is shallower for $z\geq 0.5$ than for the $z<0.5$
spectrum. For \ion{Mg}{\sc ii} $\lambda$ 4300 and \ion{Fe}{\sc ii}
$\lambda$ 4800, no difference is found with redshift.

Inspection of the spectra shows the presence of a residual
[\ion{O}{\sc ii}] host line in the $z\geq 0.5$ spectrum, for which a
separate extraction of the SN and the host is often difficult.
Moreover, our PEGASE host templates do not have emission lines. Other
residual absorption (e.g. \ion{Ca}{\sc ii} H and K and Balmer lines)
is also found in individual spectra.

Our main conclusion of this redshift comparison is that despite the
overall agreement between low ($z<0.5$) and high ($z\geq 0.5$)
spectral data, some discrepancies are found in characteristic
absorption features. This may indicate evolution with redshift, or a
signature of some selection effect. We discuss this in the next
section.

\section{Discussion}
\label{discussion}

In recent years, various large-scale SN programs have published sets
of \Ia\ spectra at intermediate
\citep{Balland06,Balland07,Zheng08,Foley09} and distant
\citep{Howell05,Lidman05,Matheson08,Bronder08,Ellis08,Foley08}
redshifts.  The SNLS VLT \Ia\ spectral data set presented in this
paper supplements these existing sets with 124 new SNe~Ia, or probable
SNe~Ia. This constitutes the largest high-redshift \Iae\ spectroscopic
sample published so far.

Comparing our data to this literature, we have the highest
SN~Ia$\star$/SN~Ia ratio. \citet{Howell05} publish the 1st year of
SNLS spectroscopy at Gemini. The redshift range targeted is $0.155 < z
< 1.01$, the median redshift being 0.81, higher than our median
redshift 0.62. Attributing an index of spectral quality to the Gemini
spectra, they identify 34 SN~Ia and 7 SN~Ia$\star$ among 64 SN
candidates (recall that \citet{Howell05} index CI=4 and 5 to SN~Ia and
index CI=3 to SN~Ia$\star$).  \cite{Lidman05} find 15 SN~Ia and 5
SN~Ia$\star$ in the range $0.212<z<1.181$ ($<z>\approx 0.5$). In both
cases, SN~Ia$\star$ amount to $\sim $20\% of their total SN~Ia sample.
We find 86 SN~Ia (of which two are SN~Ia\_pec) and 38 SN~Ia$\star$ in
our VLT sample, that is $\sim 30$\% of SN~Ia$\star$. Other surveys
such as SDSS-II \citep{Zheng08} and ESSENCE \citep{Matheson05} both
find about 10\% of probable \Ia\ (equivalent to our SN~Ia$\star$). The
SDSS-II survey targets low to intermediate redshifts ($0.05<z<0.4$,
$<z>\approx 0.22$), which probably makes identification easier. The
average redshift of the ESSENCE SNe~Ia is $<z>\approx 0.4$
\citep{Foley09}, significantly lower than ours.

We identify fewer SN~Ia$\star$ towards the end of the SNLS 3rd year
period. The average spectroscopic MJD of SN~Ia classified spectra is
53393 whereas it is 53272 for SN~Ia$\star$ spectra. As all spectra are
treated on an equal footing from the point of view of extraction and
identification, this can not be due to improvement in data processing
or refinements in classification, but more likely indicates an
improved target selection efficiency and optimisation of spectroscopic
time during the course of the survey.

A key result of this work is the construction of average spectra for
various phases and redshift bins from the homogeneous sample of VLT
SNe~Ia. We have used these high quality average spectra to compare
spectral properties at $z<0.5$ and $z\geq 0.5$.  We find differences
in the depth of some optical absorption features (\ion{Si}{\sc ii} and
\ion{Ca}{\sc ii}) around maximum light.  We find that the absorptions
due to these elements in the $z<0.5$ spectrum have weakened in the
corresponding $z\geq0.5$ spectrum.  Recently, \citet{Sullivan09} have
shown evidence for a similar weakening of singly ionized IME EWs at
higher redshift. Using data from various searches, \citet{Howell07}
find an $\sim $8\% increase in average light curve width for
non-subluminous SNe~Ia in the $0.03<z<1.12$ redshift range and
interpret it as a demographic evolution. High-$z$ SNe~Ia have an
average stretch higher than at low redshift, thus are more luminous
and hotter, and should ionize more IMEs, depleting singly ionized
absorptions in high redshift spectra \citep[e.g.,][]{Ellis08}.

\begin{figure*}[htbp]
\begin{center} 
  \begin{minipage}[b]{15cm} \centering 
    \includegraphics[width=12cm]{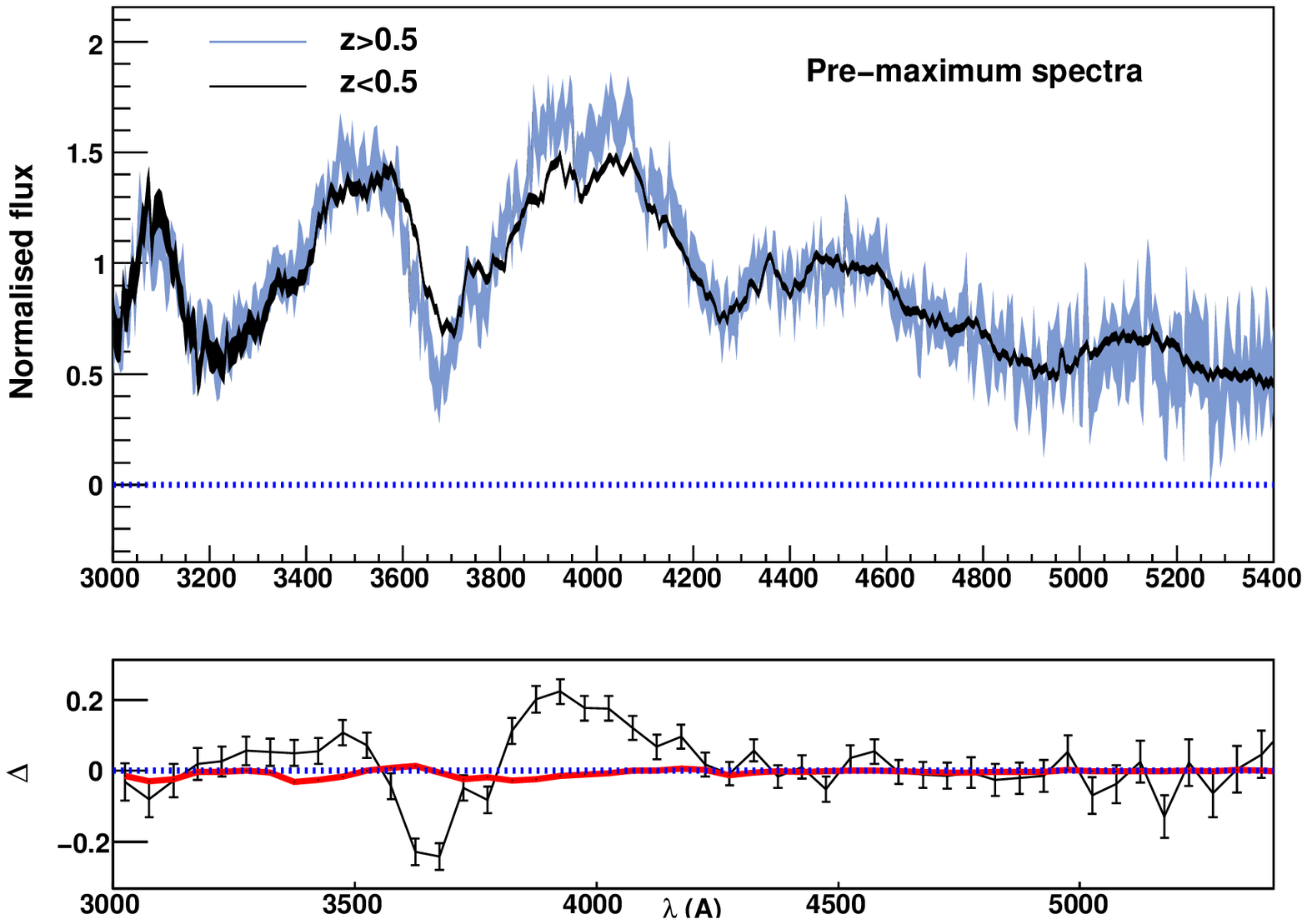}
  \end{minipage} 
  \begin{minipage}[b]{15cm} \centering 
    \includegraphics[width=12cm]{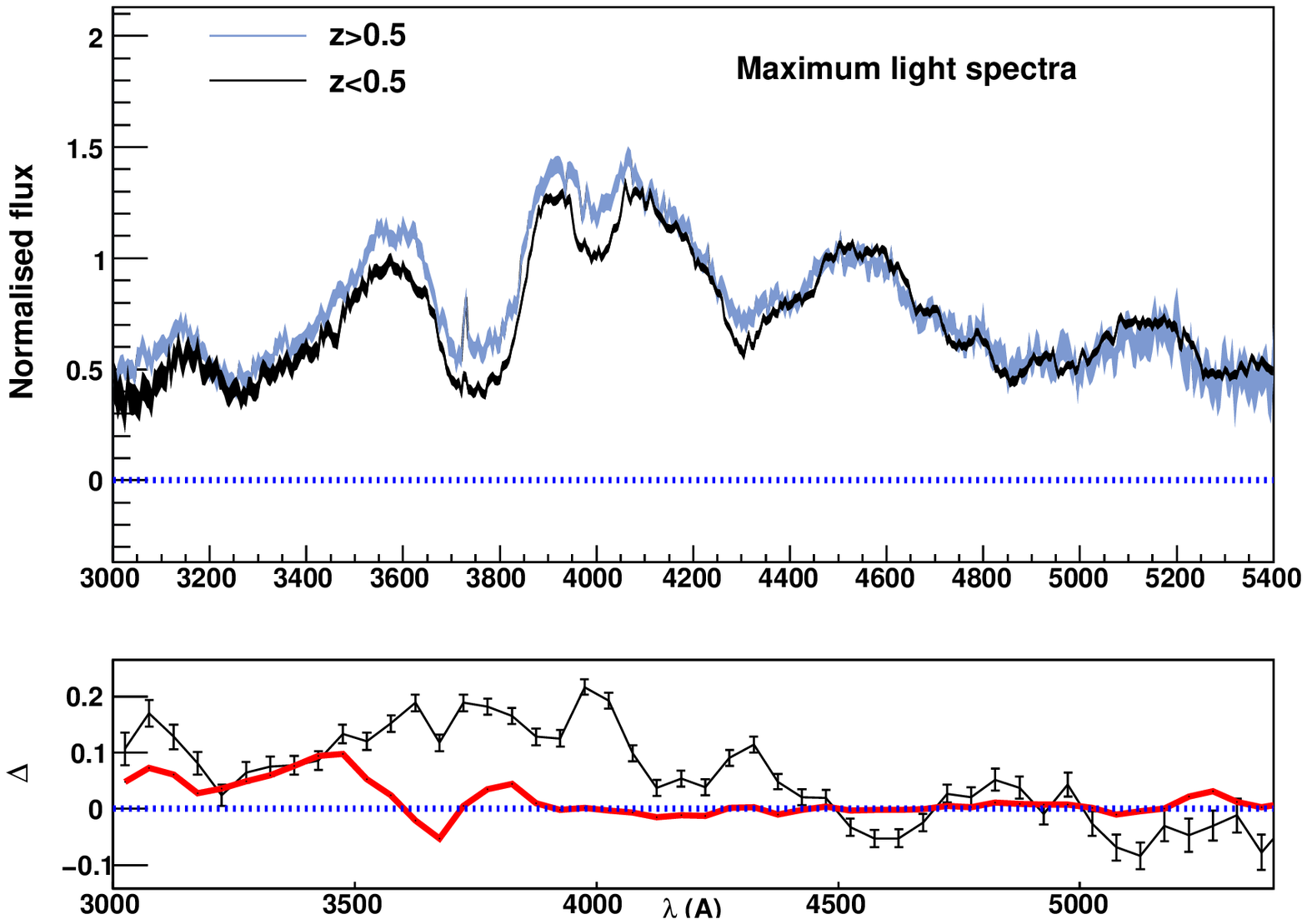}
  \end{minipage} 
  \begin{minipage}[b]{15cm} \centering 
    \includegraphics[width=12cm]{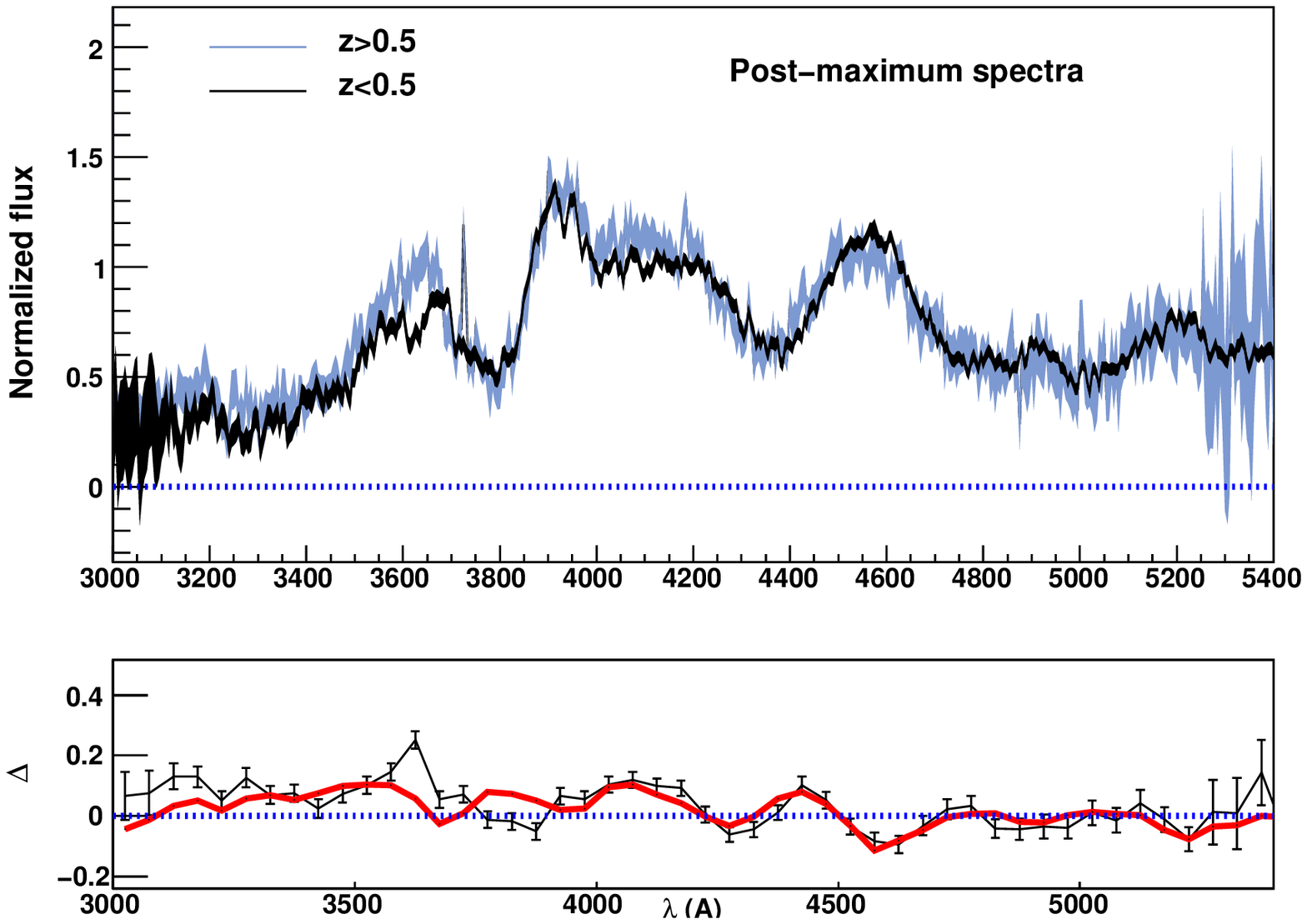}
  \end{minipage}
\end{center}
\caption{\label{fig:comparez} Comparison of $z<0.5$ (in black) and
  $z\geq 0.5$ (in grey/blue) VLT average spectra for pre-maximum
  phases (top), around maximum phases (middle), and post-maximum
  phases (bottom). In each case, the 50 $\AA$ binned residuals $\Delta
  = f^{z\geq 0.5}(\lambda)-f^{z<0.5}(\lambda)$ are shown for the VLT
  spectra (in black) and for the \citet{Hsiao07} templates (in red).
  For the VLT residual, variances are computed as the quadratic sum of
  the variances of the two average spectra ($z<0.5$ and $z\geq 0.5$)
  and $\pm 1\sigma$ errors are shown. Variations in the Hsiao residual
  (red curves) reflect the different phase mixture of the $z<0.5$ and
  $z\geq 0.5$ VLT spectra in each wavelength bins. VLT spectra have been individually colour corrected using the SALT2 colour law, recalibrated at the first order using the $\gamma_1$ 'tilt'
  coefficient and normalised to the same integral in the range 4450-4550\AA.}
\end{figure*}

\begin{figure*}[htbp]
\begin{center} 
  \begin{minipage}[b]{17cm} \centering 
    \includegraphics[width=14cm]{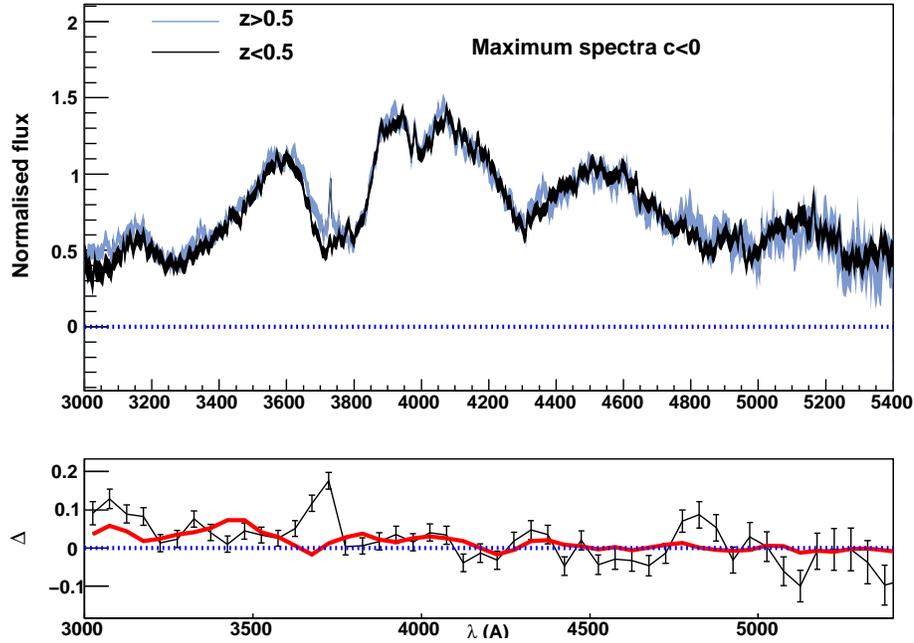}
  \end{minipage}
\end{center}
\caption{\label{fig:cutincolor} Comparison of $z<0.5$ (in black) and
  $z\geq 0.5$ (in grey/blue) VLT average spectra around maximum light
  for bluer than average ($c<0$) subsets (top panel). The 50 $\AA$
  binned residuals $\Delta = f^{z\geq
    0.5}(\lambda)-f^{z<0.5}(\lambda)$ are shown in the bottom panel
  for the VLT spectra (in black) and for the \citet{Hsiao07} templates
  (in red).  For the VLT residual, variances computed as the quadratic
  sum of the variances of the two average spectra ($z<0.5$ and $z\geq
  0.5$) and $\pm 1\sigma$ errors are shown. Compared to the middle
  panel of Fig.  \ref{fig:comparez}, the differences from 3500 to 4500
  $\AA$ between $z<0.5$ and $z\geq 0.5$ have been drastically
  reduced. VLT spectra have been individually colour corrected using the SALT2 colour law, recalibrated and normalised to the same integral in the range 4450-4550\AA.}
\end{figure*}

The average stretches of our two redshift subsamples are
$<s>^{z<0.5}=0.975\pm 0.017$ and $<s>^{z\geq0.5}=0.983\pm 0.009$
respectively. These two values are similar given the uncertainties.
This is not inconsistent with the results of
\citet{Howell05,Sullivan09}, as the predicted change in the spectra
over our redshift range due to a demographic shift is likely to be
very small.  The average colours\footnote{We limit our set to SNe~Ia
  with $c\leq 0.3$ to reject red, heavily extinguished objects. This
  cuts three SNe~Ia out of the sample.} are $<c>^{z<0.5}=0.06\pm 0.03$
and $<c>^{z\geq0.5}=-0.01\pm 0.01$. The average rest frame distance
corrected magnitudes (using ${\bf
  \Theta}=\{H_0,\Omega_M,\Omega_\Lambda\}= \{70,0.27,0.73\}$
\citealt{Astier06}) are $<m_B^{*c}>_{z<0.5}=24.10\pm 0.10$
($\sigma=0.52$) and $<m_B^{*c}>_{z\geq 0.5}=23.88\pm 0.03$
($\sigma=0.26$), the $z\geq 0.5$ subsample having brighter supernovae,
on average, in qualitative agreement with what is expected from
Malmquist bias, as brighter supernovae, preferentially selected at
higher redshift, tend to be bluer.  In order to assess the
significance of the apparent spectral differences in our maximum light
spectra, we have built two new subsamples (one for each redshift bin)
by selecting spectra with phase in the range $-4d < \phi < 4d$ and
bluer than average (i.e. spectra of SN~Ia with colour $c<0$). Using
these cuts, we end up with only 6 spectra for the $z<0.5$ bin, and 26
spectra for the $z\geq 0.5$ bin. As we expect to observe bluer objects
at higher redshift due to Malquist bias and to the 'brighter-bluer'
correlation, this should select two subsamples with roughly comparable
photometric properties. The average distance corrected magnitudes and
dispersions of these new subsamples are $<m_B^{*c}>_{z<0.5}=23.77\pm
0.04$ ($\sigma=0.11$) and $<m_B^{*c}>_{z\geq 0.5}=23.75\pm 0.03$
($\sigma=0.17$).  The two first moments of the $m_B^{*c}$
distributions are thus similar.  We build new average spectra at
maximum light in the same way as in Section \ref{sec:composit}. The
results are shown in Fig.  \ref{fig:cutincolor} and must be compared
to the middle panel of Fig.  \ref{fig:comparez}. Most of the
differences have diseappeared, except around 3700 $\AA$ and 3900
$\AA$. We find $\chi^2_\nu=0.97$ (544 D.O.F.).  The difference at 3700
$\AA$ is due to a residual [\ion{O}{\sc ii}] line in the $z\geq 0.5$.
If the differences seen in Fig. \ref{fig:comparez} were due to
imperfect calibration, slit losses or imperfect host subtraction, we
would expect to see comparable differences when selecting the two
'blue' populations shown in Fig. \ref{fig:cutincolor}. The fact that
the composite spectra nicely overlap for these two populations is in
itself an indication that the differences with redshift in our maximum
light spectra are real. We have checked that the differences remain
when selecting a subset of spectra with a host galaxy fraction below
20\%. They also remain when we select a subset of spectra with little
recalibration ($|\gamma_1|<$0.5). Thus it is unlikely that the
differences are due to slit losses or a poorer host subtraction at
higher redshift.  These differences more likely result from the
selection of brighter and bluer supernovae at higher redshift.

\begin{figure*}[htbp]
\begin{center} % \vspace*{1cm}
  \begin{minipage}[b]{17cm} \centering 
    \includegraphics[width=14cm]{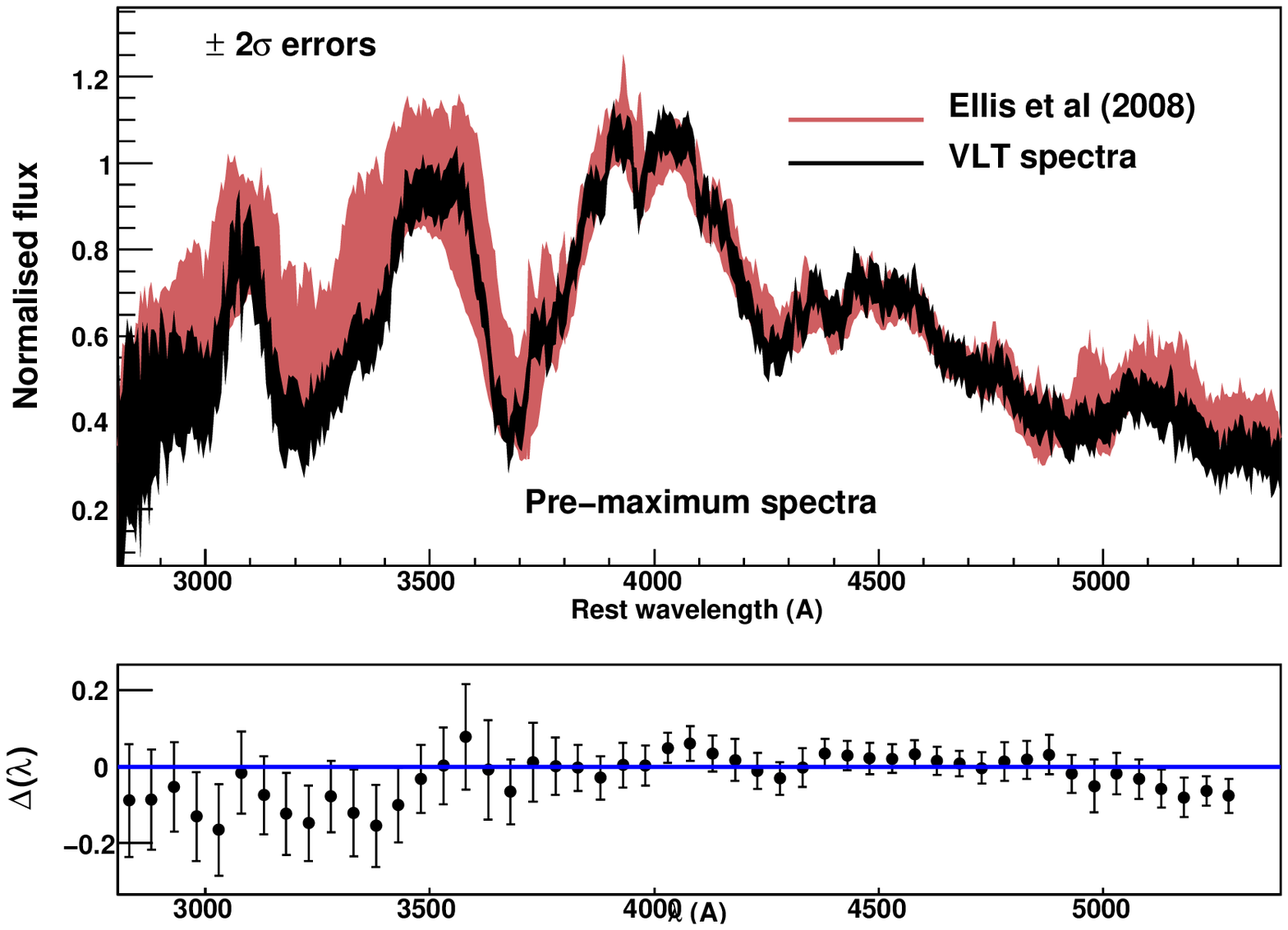}
  \end{minipage} \vspace*{2cm}
  \begin{minipage}[b]{17cm} \centering 
    \includegraphics[width=14cm]{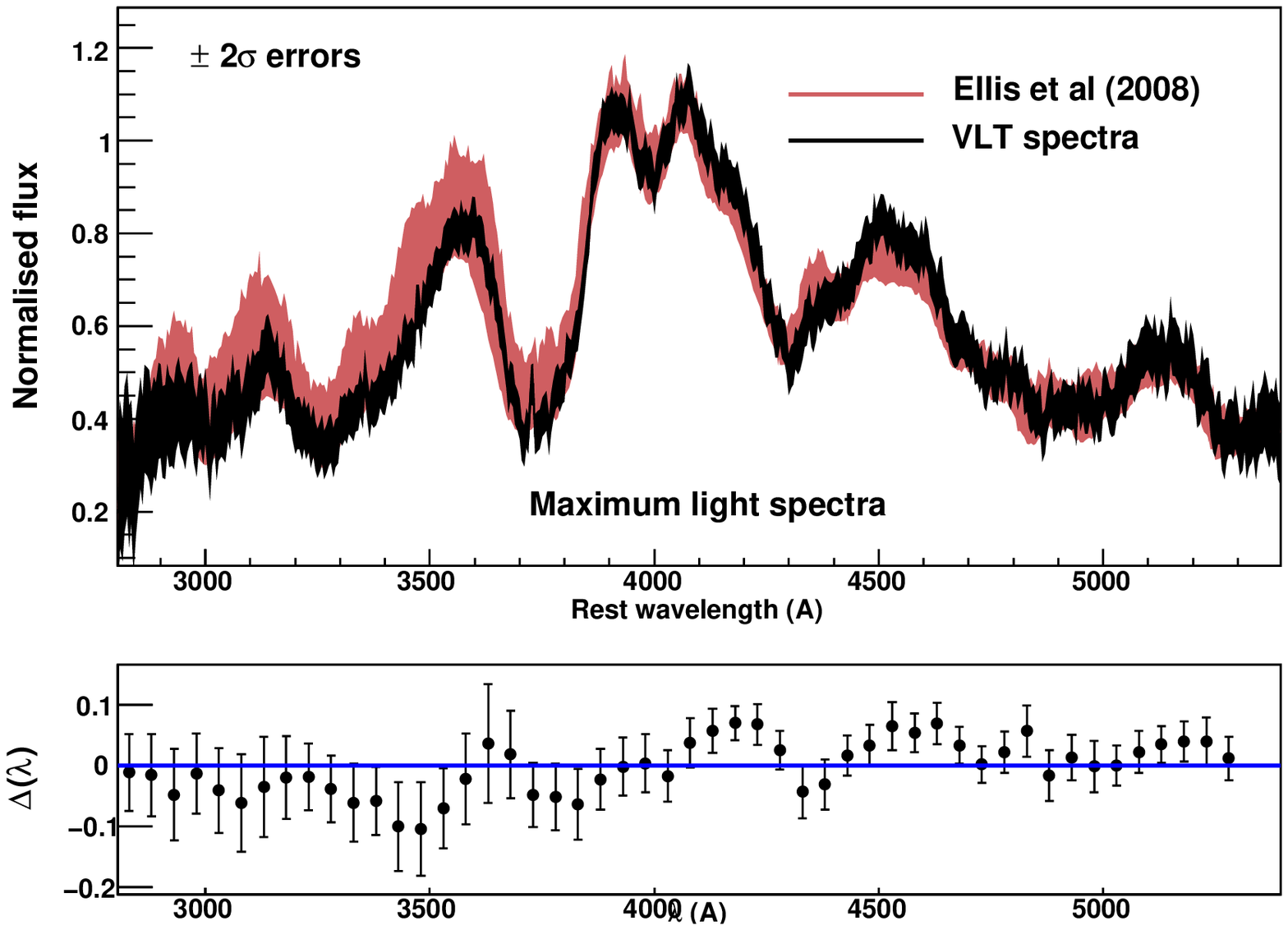}
  \end{minipage}
\end{center}
\caption{\label{fig:compuv} Comparison of the VLT average spectra
  obtained from a subset of 51 spectra in the range $z=0.35-0.7$
  (black), with the Keck average spectra from \citet{Ellis08} (in
  grey/red), for pre-maximum phases (top), and at maximum phases (bottom).
  VLT spectra have been individually colour corrected using the SALT2 colour law
  and recalibrated at the first order using the $\gamma_1$ 'tilt'
  coefficient.  All spectra have been normalised to unity at 3900 $\AA$. For
  each phase range, the residuals
  $\Delta(\lambda)=\Delta^{VLT}(\lambda)-\Delta^{Keck}(\lambda)$ are
  shown. Variances are computed as the quadratic sum of the variances
  of the two average spectra (VLT and Keck). Errors are $\pm
  2\sigma$.}
\end{figure*}

We now compare our average spectra to the ones of \citet{Ellis08}.
These were built from spectra obtained at Keck as part of detailed
studies programs using SNLS SNe. Some SNe~Ia are common to both
samples.  We find a qualitative agreement with \citet{Ellis08}
regarding variations in the UV part of the spectra.  Figure
\ref{fig:compuv} shows the comparison between a subset of 51 VLT
average spectra between $z=0.35$ and $z=0.7$ (with $<z>\sim 0.5$ for
this subset), for pre- and maximum phases (in black) with the average
spectra obtained from 23 high S/N Keck spectra, with $<z>\sim 0.5$ (in
red, from Tables 3 and 4 of their paper). Both VLT and Keck average
spectra are shown with lower and upper 2$\sigma$ errors and have been
normalised to unity in the same wavelength region (at 3900 $\AA$).
VLT spectra have been recalibrated using the SALT2 recalibration
function at first order (the ``tilt'' coefficient $\gamma_1$). The
average phases of the Keck spectra are comparable to the VLT ones.

The two sets of pre-maximum spectra are consistent: $\chi^2_\nu\approx
1.00$ (500 D.O.F.). For around maximum spectra, we find
$\chi^2_\nu\approx 1.17$ (500 D.O.F.). The VLT average spectra are on
the lower limit of the Keck spectra in the UV region for both phase
ranges. In the optical region, the average spectra are in remarkable
agreement for pre-maximum spectra. For maximum-light spectra, the
agreement is fair in the optical, with some structure seen in the
residuals. It is in the UV region that differences are most
noticeable. There could be several explainations: this region is more
sensitive to differential slit losses, and we have normalised the
spectra at 3900 $\AA$ (had we chosen a shorter wavelength, the UV
discrepancies would have been less apparent, but the optical
differences greater). Also, the SALT2 colour law has been used to
build the average VLT spectra, whereas \citet{Ellis08} use the SALT1
colour law, though this has a negligible effect on the average
spectra.  We find that the \ion{Ca}{\sc ii} $\lambda$ 3700 and the
\ion{Si}{\sc ii} $\lambda$ 4000 absorptions are consistent, both in
pre-maximum and at maximum spectra.

Recently, \citet{Foley08} have published composite spectra from
ESSENCE data and they compare them in various bins of redshift and
phase, to the Lick $z<0.1$ counterpart composite spectrum. Comparing
the ESSENCE and VLT spectra in a similar fashion to our comparison
with the Ellis spectra, we find substantial differences at all phases.
This may be due to the fact that \citet{Foley08} subtract a single
average host spectrum, derived from a PCA analysis, from all of their
SNe. Regarding specific spectral features, they find that the most
important differences at maximum light between low- and high-redshift
are 1) \ion{Fe}{\sc ii} $\lambda$ 5129, that yields a smaller EW of
the \{\ion{Fe}{\sc ii} $\lambda$ 4800\} blend in the ESSENCE spectra
than in the Lick composite, and 2) the lack of absorption at 3000
$\AA$ in the highest redshift bins of their ESSENCE composite spectra.
We qualitatively agree with their analysis of \ion{Fe}{\sc ii}
$\lambda$ 4800. We however find the presence of absorption at 3000
$\AA$ in our average spectra, for all phases at $z\geq 0.5$, as seen
in Fig. \ref{fig:compareHsiao}. For $z<0.5$, the absorption is at the
limit of our effective spectral range (see Fig.
\ref{fig:compareHsiao}), but the absorption seems present, at least
for pre-maximum and at maximum spectra (it is very noisy for
post-maximum spectra). Following \citet{Foley08}, this difference
could arise from a selection bias in the ESSENCE sample that is not
present in the VLT sample.

\section{Conclusion}
\label{conclusion}

We have presented 139 spectra of 124 SNe~Ia (SN~Ia type) or probable
SNe~Ia (SN~Ia$\star$ type) at $0.149<z<1.031$ observed with the FORS1
instrument at the VLT during the first three years of the SNLS survey.
This is the largest SN Ia spectral dataset in this redshift range. We
have developed a dedicated pipeline, PHASE, for extracting clean SN
spectra free from host contamination. Our approach takes advantage of
the rolling search mode of the SNLS by using deep stacked reference
images to estimate the host spatial profile, used during the SN
extraction.  We have also developed an identification technique based
on the simultaneous fit of light curve and spectral data with SALT2.
We have obtained two sets of certain and probable SNe~Ia (the SN~Ia
and SN~Ia$\star$ categories), whose statistical properties have been
studied in detail. We find that:

\begin{itemize}
\item The statistical properties of our SN~Ia and SN~Ia$\star$ samples
  are similar. The SALT2 colour $c$ and $x_1$ parameters are
  consistent for both subsamples, as is the ``tilt'' parameter,
  $\gamma_1$. These are indications that our samples are not strongly
  contaminated by non SNe~Ia, validating the inclusion of SN~Ia$\star$
  in the Hubble diagram.
\item The average redshift and phase for SN~Ia$\star$ are higher than
  for SN~Ia. In particular, the average phase of SN~Ia$\star$ is 5.8
  days, a phase at which the spectral S/N has decreased with respect
  to maximum and \Iae\ spectra closely resemble SNe~Ic,
\item The uncorrected peak SN~Ia$\star$ magnitudes appear fainter than
  the SN~Ia, consistent with the SN Ia ``brighter-bluer''
  relationship,
\item On average, spectra of SNe identified as SN~Ia$\star$ have a
  higher host contamination than SN~Ia spectra,
\item We find two peculiar SNe whose spectra and properties resemble
  those of the over-luminous SN~1999aa supernova.
\end{itemize}

In an attempt to make a first characterisation of the physical
properties of our spectral data, we have built average spectra for
pre-maximum, maximum-light, and post-maximum phases, using only SN~Ia
spectra, in two redshift bins ($z<0.5$ and $z\geq0.5$).  A consistent
comparison of these spectra to \citet{Hsiao07} template spectra shows
a remarkable concordance in the positions and relative flux of the
spectral features. Of particular interest is the UV region of the
spectra.  Though more differences are found in this region than in the
optical region, the overall agreement is excellent.

We have also internally compared the average SN~Ia spectra for $z<0.5$
and $z\geq 0.5$. We find evidence for differences in the intermediate
mass element absorptions in the average optical spectrum of SNe~Ia
between $z\sim 0.35$ and $z\sim 0.7$ (the average redshifts of the
$z<0.5$ and $z\geq0.5$ bins).  The brighter SNe spectra have weaker
absorptions of singly ionized IMEs. Some discrepancy arises in the UV
region of the spectra. As far as our spectroscopic sample is
concerned, we have shown that these spectral differences can be
essentially accounted for by the Malmquist bias.  The use of SNe~Ia as
``calibrated candles'' for cosmological purposes is thus fully
justified.

Finally, we have compared our VLT composite spectra with the
\citet{Ellis08} ones, both for pre- and at maximum phases. We find a
good agreement between the two sets of spectra, which, given their
fully independent processing, gives us confidence in the quality of
both spectral sets.

The VLT spectra presented in this paper for which sufficient
photometric information exists, once merged with the Gemini and Keck
SNLS spectra \citep{Howell05, Bronder08, Ellis08}, will constitute the
spectroscopic sample for the SNLS 3rd year cosmological analyses.

\begin{acknowledgements}
  We gratefully acknowledge the assistance of the VLT Queue Scheduling
  Observing Team. We thank the anonymous referee for helpful comments
  on the manuscript. French authors acknowledge support from CNRS/IN2P3,
  CNRS/INSU and PNC. CENTRA members were supported by Funda\c{c}\~ao
  para a Ci\^encia e Tecnologia (FCT), Portugal under
  POCTI/CTE-AST/57664/2004. VA acknowledges support from
  FCT under grant no SFRH/BD/11119/2002 and SF grant no
  SFRH/BPD/31817/2006. MS acknowledges support from the Royal Society.

\end{acknowledgements}

\bibliographystyle{aa}
\bibliography{bibi}

\onecolumn

\begin{longtable}[b]{lcccccccc}
\caption[]{\label{tab:tab2} List of SNLS 3rd year ESO-VLT SN~Ia and SN~Ia$\star$ spectra$^a$.}\\[2mm]
\hline\hline\\
SN name & RA (J2000) & DEC (J2000) & Spectrum Date & Exp. Time (s)  & Seeing ('') & Dsep ('')$^b$ & Air Mass  & Mag $i_{M}$ \\[2mm]
\hline
\\
03D1ar & +02:27:14.67 & -04:19:04.74 & 2003-09-28 & 3$\times$750 & 0.79 & 1.05 & 1.21 & 22.9 \\
03D1bf & +02:24:02.39 & -04:55:57.11 & 2003-09-27 & 3$\times$750 & 0.96 & 0.79 & 1.09 & 23.5 \\
03D1bm & +02:24:36.07 & -04:12:44.51 & 2003-09-30 & 3$\times$750 & 0.55 & 0.29 & 1.08 & 23.4 \\
03D1bp & +02:26:37.74 & -04:50:19.61 & 2003-09-28 & 3$\times$750 & 0.82 & 1.15 & 1.10 & 21.9 \\
03D1co & +02:26:16.26 & -04:56:06.18 & 2003-11-04 & 5$\times$750 & 0.70 & * & 1.22 & 23.2 \\
03D1dt & +02:26:31.19 & -04:03:08.43 & 2003-11-30 & 3$\times$750 & 0.69 & * & 1.17 & 22.4 \\
03D1fc & +02:25:43.60 & -04:08:38.61 & 2003-12-25 & 5$\times$750 & 0.74 & 0.44 & 1.32 & 21.4 \\
03D1fl & +02:25:58.30 & -04:07:44.03 & 2003-12-22 & 5$\times$750 & 0.71 & 0.38 & 1.24 & 22.6 \\
03D1gt & +02:24:56.01 & -04:07:36.81 & 2004-01-21 & 5$\times$750 & 0.77 & 0.48 & 1.36 & 23.3 \\
03D4ag & +22:14:45.83 & -17:44:22.76 & 2003-06-30 & 2$\times$750 & 0.56 & 1.09 & 1.03 & 21.0 \\
03D4at & +22:14:24.04 & -17:46:36.22 & 2003-07-06 & 3$\times$750 & 0.75 & 0.15 & 1.09 & 22.8 \\
03D4au & +22:16:09.93 & -18:04:39.32 & 2003-07-06 & 3$\times$750 & 0.69 & * & 1.02 & 23.0 \\
03D4cx & +22:14:33.75 & -17:35:15.04 & 2003-09-03 & 3$\times$750 & 0.54 & 0.28 & 1.03 & 23.5 \\
03D4cx$^c$ & +22:14:33.75 & -17:35:15.04 & 2003-09-06 & 3$\times$750 & 0.74 & * & 2.15 & 23.5 \\
03D4cy$^d$ & +22:13:40.42 & -17:40:53.91 & 2003-09-27 & 3$\times$750 & 0.88 & 1.15 & 1.01 & 23.6 \\
03D4di & +22:14:10.24 & -17:30:23.82 & 2003-09-03 & 3$\times$750 & 0.59 & 0.52 & 1.11 & 23.4 \\
03D4di$^c$ & +22:14:10.24 & -17:30:23.82 & 2003-09-07 & 3$\times$750 & 0.55 & 0.46 & 1.04 & 23.4 \\
03D4dy & +22:14:50.53 & -17:57:23.32 & 2003-09-30 & 3$\times$750 & 0.76 & * & 1.02 & 22.4 \\
03D4gf & +22:14:22.93 & -17:44:02.52 & 2003-11-25 & 5$\times$750 & 0.70 & 1.31 & 1.29 & 22.5 \\
03D4gg & +22:16:40.19 & -18:09:52.05 & 2003-11-27 & 5$\times$750 & 0.51 & 1.10 & 1.40 & 22.5 \\
04D1ag & +02:24:41.10 & -04:17:19.50 & 2004-01-19 & 3$\times$750 & 0.83 & 0.68 & 1.26 & 22.3 \\
04D1aj & +02:25:53.96 & -04:59:40.60 & 2004-01-17 & 5$\times$750 & 0.98 & * & 1.35 & 22.9 \\
04D1aj & +02:25:53.96 & -04:59:40.60 & 2004-01-20 & 5$\times$750 & 0.79 & * & 1.34 & 22.9 \\
04D1ak & +02:27:33.36 & -04:19:38.78 & 2004-01-24 & 5$\times$750 & 0.77 & 0.60 & 1.38 & 23.1 \\
04D1dc & +02:26:18.46 & -04:18:43.09 & 2004-08-12 & 2$\times$750 & 0.75 & 1.18 & 1.23 & 20.7 \\
04D1ff & +02:25:38.62 & -04:54:09.47 & 2004-09-11 & 5$\times$750 & 0.83 & 0.05 & 1.07 & 23.2 \\
04D1hx$^e$ & +02:24:42.50 & -04:47:25.17 & 2004-09-20 & 5$\times$750 & 0.77 & 0.52 & 1.08 & 22.8 \\
04D1iv & +02:24:48.05 & -04:09:09.21 & 2004-09-22 & 5$\times$750 & 0.88 & 0.69 & 1.07 & 23.5 \\
04D1jd & +02:27:50.84 & -04:56:37.64 & 2004-09-22 & 5$\times$750 & 0.86 & * & 1.17 & 23.4 \\
04D1kj & +02:27:52.66 & -04:10:48.91 & 2004-10-12 & 3$\times$750 & 0.54 & 0.08 & 1.09 & 22.5 \\
04D1ks & +02:24:09.52 & -04:58:43.40 & 2004-10-11 & 3$\times$750 & 0.75 & * & 1.16 & 23.0 \\
04D1ow & +02:26:42.70 & -04:18:22.28 & 2004-11-14 & 4$\times$900 & 0.65 & * & 1.08 & 23.2 \\
04D1pc & +02:26:24.88 & -04:23:57.84 & 2004-11-15 & 4$\times$900 & 0.54 & 0.47 & 1.08 & 23.5 \\
04D1pd & +02:27:39.75 & -04:37:52.22 & 2004-11-13 & 4$\times$900 & 0.66 & 0.22 & 1.07 & 23.7 \\
04D1pg & +02:27:04.16 & -04:10:31.03 & 2004-11-15 & 3$\times$750 & 0.56 & * & 1.19 & 22.8 \\
04D1pp & +02:25:12.48 & -04:42:05.70 & 2004-11-14 & 4$\times$900 & 0.56 & 0.20 & 1.11 & 23.0 \\
04D1qd & +02:26:33.08 & -04:06:25.84 & 2004-11-19 & 4$\times$900 & 0.87 & * & 1.37 & 23.1 \\
04D1rh & +02:27:47.15 & -04:15:13.56 & 2004-12-11 & 3$\times$750 & 0.77 & 0.39 & 1.66 & 21.9 \\
04D1rx & +02:25:11.03 & -04:37:06.06 & 2004-12-14 & 4$\times$900 & 0.74 & 1.60 & 1.07 & 23.7 \\
04D1sa & +02:27:56.16 & -04:10:33.85 & 2004-12-17 & 3$\times$750 & 0.97 & * & 1.12 & 22.2 \\
04D1si & +02:24:48.80 & -04:34:08.22 & 2004-12-16 & 4$\times$900 & 1.17 & * & 1.07 & 24.0 \\
04D2ac & +10:00:18.88 & +02:41:21.34 & 2004-01-31 & 3$\times$750 & 0.86 & 0.29 & 1.13 & 21.6 \\
04D2al & +10:01:52.48 & +02:09:50.78 & 2004-01-26 & 6$\times$750 & 0.83 & * & 1.26 & 23.2 \\
04D2an & +10:00:52.31 & +02:02:28.42 & 2004-01-22 & 5$\times$750 & 0.65 & * & 1.37 & 22.7 \\
04D2an & +10:00:52.31 & +02:02:28.42 & 2004-01-28 & 5$\times$750 & 0.70 & * & 1.14 & 22.7 \\
04D2bt & +09:59:32.69 & +02:14:52.99 & 2004-03-20 & 5$\times$750 & 0.62 & 0.35 & 1.19 & 21.0 \\
04D2ca & +10:01:20.49 & +02:20:21.67 & 2004-03-23 & 5$\times$750 & 0.72 & 0.98 & 1.32 & 23.9 \\
04D2ca & +10:01:20.49 & +02:20:21.67 & 2004-03-21 & 2$\times$750 & 0.80 & 0.98 & 1.48 & 23.9 \\
04D2cc & +09:58:44.69 & +02:16:14.53 & 2004-03-22 & 5$\times$750 & 0.82 & * & 1.40 & 23.3 \\
04D2cc & +09:58:44.69 & +02:16:14.53 & 2004-03-24 & 5$\times$750 & 0.74 & * & 2.15 & 23.3 \\
%\hline\\
\newpage{Table 1} (cont'd)\\[2mm]
\hline\hline\\
SN name & RA (J2000) & DEC (J2000) & Spectrum Date & Exp. Time (s) & Seeing ('') & Dsep ('')$^b$ & Air Mass  & Mag $i_{M}$ \\[2mm]
\hline\\
04D2cf & +10:01:56.09 & +01:52:46.12 & 2004-03-23 & 5$\times$750 & 0.79 & 2.07 & 1.13 & 21.8 \\
04D2cw & +10:01:22.78 & +02:11:54.81 & 2004-03-21 & 5$\times$750 & 0.70 & * & 1.13 & 23.7 \\
04D2fp & +09:59:28.12 & +02:19:15.65 & 2004-04-15 & 3$\times$750 & 0.66 & 1.26 & 1.23 & 21.9 \\
04D2fs & +10:00:22.09 & +01:45:55.43 & 2004-04-15 & 3$\times$750 & 0.60 & * & 1.13 & 21.7\\
04D2gc & +10:01:39.27 & +01:52:58.99 & 2004-04-15 & 5$\times$750 & 0.62 & 0.25 & 1.13 & 22.6 \\
04D2gp & +09:59:20.37 & +02:30:31.96 & 2004-04-20 & 5$\times$750 & 0.77 & 1.16 & 1.20 & 23.3 \\
04D2iu & +10:01:13.18 & +02:24:54.18 & 2004-05-12 & 4$\times$750 & 0.92 & * & 1.48 & 23.3 \\
04D2iu & +10:01:13.18 & +02:24:54.18 & 2004-05-12 & 4$\times$750 & 0.78 & * & 1.14 & 23.3 \\
04D2ja & +09:58:48.50 & +01:46:18.25 & 2004-05-13 & 3$\times$750 & 0.55 & 0.31 & 1.15 & 23.1 \\
04D2ja & +09:58:48.50 & +01:46:18.25 & 2004-05-14 & 1$\times$750 & 0.72 & 0.31 & 1.12 & 23.1 \\
04D2mc & +10:00:35.56 & +01:47:03.05 & 2005-01-06 & 3$\times$750 & 0.84 & 0.50 & 1.12 & 22.0 \\
04D4an & +22:15:57.10 & -17:41:43.81 & 2004-07-12 & 5$\times$750 & 0.87 & 0.29 & 1.01 & 23.3 \\
04D4bk & +22:15:07.70 & -18:03:36.99 & 2004-07-12 & 5$\times$750 & 1.01 & * & 1.15 & 23.3 \\
04D4bq & +22:14:49.42 & -17:49:39.34 & 2004-07-12 & 3$\times$750 & 1.36 & 1.41 & 1.05 & 22.5 \\
04D4bq & +22:14:49.42 & -17:49:39.34 & 2004-07-16 & 5$\times$750 & 0.83 & 1.41 & 1.08 & 22.5 \\
04D4dw & +22:16:44.67 & -17:50:02.66 & 2004-07-20 & 5$\times$750 & 0.65 & * & 1.06 & 23.6 \\
04D4fx & +22:16:38.14 & -18:03:58.89 & 2004-08-12 & 5$\times$750 & 0.70 & * & 1.08 & 22.7 \\
04D4gz & +22:16:59.01 & -17:37:18.85 & 2004-09-12 & 6$\times$750 & 1.06 & 0.27 & 1.09 & 22.7 \\
04D4gz & +22:16:59.01 & -17:37:18.85 & 2004-08-21 & 5$\times$750 & 1.00 & 0.27 & 1.32 & 22.7 \\
04D4hf & +22:16:57.90 & -17:41:13.29 & 2004-08-21 & 5$\times$750 & 0.93 & * & 1.08 & 23.8 \\
04D4ht & +22:14:33.28 & -17:21:31.21 & 2004-09-20 & 5$\times$750 & 0.85 & 0.30 & 1.46 & 21.5 \\
04D4ib & +22:16:41.71 & -18:06:18.32 & 2004-09-20 & 5$\times$750 & 0.84 & * & 1.24 & 22.7 \\
04D4id & +22:16:21.40 & -17:13:44.50 & 2004-09-20 & 5$\times$750 & 0.85 & 1.17 & 1.11 & 23.0 \\
04D4jr & +22:14:14.32 & -17:21:00.78 & 2004-10-12 & 3$\times$750 & 0.64 & * & 1.01 & 22.0 \\
04D4ju & +22:17:02.73 & -17:19:58.13 & 2004-10-12 & 3$\times$750 & 0.64 & 0.34 & 1.04 & 23.0 \\
04D4jw & +22:17:18.89 & -17:39:55.78 & 2004-10-12 & 5$\times$900 & 0.91 & * & 1.04 & 23.7 \\
05D1cb & +02:26:57.06 & -04:07:03.24 & 2005-09-12 & 4$\times$900 & 0.87 & 1.31 & 1.07 & 22.8 \\
05D1ck & +02:24:24.85 & -04:42:46.08 & 2005-09-11 & 4$\times$900 & 1.04 & 0.32 & 1.13 & 23.3\\
05D1cl & +02:25:05.66 & -04:13:14.95 & 2005-09-27 & 6$\times$900 & 0.83 & * & 1.15 & 23.2 \\
05D1dn & +02:24:26.65 & -04:59:29.61 & 2005-09-29 & 4$\times$900 & 0.81 & 1.11 & 1.11 & 22.5 \\
05D1hk & +02:24:39.18 & -04:38:03.31 & 2005-12-09 & 3$\times$750  & 0.68 & *  & 1.07  & 20.9\\
05D1hn & +02:24:36.24 & -04:10:54.78 & 2005-12-09 & 3$\times$750 & 0.85 & 0.17 & 1.09 & 20.6\\
05D1iz & +02:26:12.77 & -04:46:24.13 & 2005-12-05 & 4$\times$900 & 0.71 & * & 1.08 & 23.6\\
05D1ke & +02:26:46.29 & -04:51:44.99 & 2005-12-30 & 3$\times$750 & 0.65 & 0.42 & 1.09 & 22.7\\
05D2ac & +09:58:59.24 & +02:29:22.21 & 2005-01-08 & 3$\times$750 & 0.56 & 0.47 & 1.16 & 22.0 \\
05D2ay & +10:01:08.32 & +02:07:49.16 & 2005-01-16 & 4$\times$900 & 0.70 & 2.53 & 1.14 & 23.7 \\
05D2bt & +10:01:40.24 & +02:33:58.18 & 2005-02-03 & 3$\times$500 & 0.64 & 0.22 & 1.13 & 22.8 \\
05D2bv & +10:02:17.00 & +02:14:26.16 & 2005-02-03 & 3$\times$750 & 0.71 & 0.39 & 1.14 & 22.2 \\
05D2bw & +09:58:30.56 & +02:32:11.04 & 2005-02-03 & 4$\times$900 & 0.69 & * & 1.27 & 23.5 \\
05D2by & +10:00:28.03 & +02:31:17.54 & 2005-02-14 & 4$\times$900 & 0.95 & * & 1.66 & 23.5 \\
05D2cb & +09:59:24.58 & +02:19:41.45 & 2005-02-14 & 2$\times$700 & 1.02 & * & 2.56 & 22.6 \\
05D2ci & +10:00:02.04 & +02:12:57.07 & 2005-03-07 & 2$\times$900 & 0.40 & 0.52 & 1.31 & 23.0 \\
05D2ci & +10:00:02.04 & +02:12:57.07 & 2005-03-10 & 2$\times$900 & 0.72 & 0.52 & 1.12 & 23.0\\
05D2ci & +10:00:02.04 & +02:12:57.07 & 2005-03-06 & 2$\times$900 & 1.01 & 0.52 & 1.24 & 23.0 \\
%05D2cl & +09:58:48.60 & +02:36:00.37 & 2005-03-06 & 9$\times$900 & 0.90 & * & 1.18 & 23.4 \\
05D2ct & +10:01:42.18 & +02:07:25.66 & 2005-03-10 & 8$\times$900 & 0.74 & * & 1.23 & 23.6 \\
05D2dt & +10:01:23.93 & +01:51:27.87 & 2005-03-18 & 4$\times$900 & 0.82 & * & 1.32 & 22.9 \\
05D2dw & +09:58:32.07 & +02:01:56.07 & 2005-03-18 & 4$\times$900 & 0.66 & 1.18 & 1.13 & 21.8 \\
05D2dy & +10:00:58.10 & +02:10:59.16 & 2005-03-16 & 4$\times$900 & 0.75 & 1.29 & 1.19 & 22.2 \\
05D2eb & +10:00:14.56 & +02:24:26.78 & 2005-03-16 & 4$\times$900 & 0.77 & 0.78 & 1.13 & 22.4 \\
05D2ec & +09:59:26.18 & +02:00:49.06 & 2005-03-16 & 4$\times$900 & 0.78 & 0.40 & 1.23 & 22.8 \\
%\hline\\
\newpage{Table 1} (cont'd)\\[2mm]
\hline\hline\\
SN name & RA (J2000) & DEC (J2000) & Spectrum Date & Exp. Time (s) & Seeing ('') & Dsep ('')$^b$ & Air Mass  & Mag $i_{M}$ \\[2mm]
\hline\\
05D2ei & +10:01:39.12 & +01:49:11.83 & 2005-04-13 & 3$\times$750 & 0.70 & 0.63 & 1.95 & 21.3 \\
05D2fq & +09:59:08.50 & +02:36:07.20 & 2005-04-07 & 4$\times$900 & 0.77 & 1.26 & 1.14 & 23.0 \\
05D2he & +10:01:26.51 & +02:19:03.82 & 2005-04-13 & 4$\times$900 & 0.83 & 0.34 & 1.16 & 23.0 \\
05D2ie & +10:01:02.90 & +02:39:29.11 & 2005-04-09 & 4$\times$900 & 0.79 & 0.28 & 1.33 & 22.2 \\
05D2nn & +09:58:49.75 & +02:42:36.95 & 2005-12-07 & 4$\times$900 & 0.81 & * & 1.30 & 23.5\\
05D4af & +22:16:33.14 & -18:00:17.72 & 2005-07-03 & 4$\times$900 & 0.60 & 1.49 & 1.02 & 22.4 \\
05D4ag & +22:13:47.43 & -18:09:54.95 & 2005-07-11 & 4$\times$900 & 0.58 & * & 1.16 & 23.1 \\
05D4ay & +22:14:33.19 & -17:46:03.62 & 2005-07-03 & 3$\times$900 & 0.69 & * & 1.12 & 23.3 \\
05D4ay & +22:14:33.19 & -17:46:03.62 & 2005-07-05 & 1$\times$900 & 0.74 & * & 1.18 & 23.3 \\
05D4be & +22:16:53.39 & -17:14:09.96 & 2005-07-07 & 3$\times$750 & 0.56 & 0.18 & 1.03 & 22.1 \\
05D4bi & +22:15:56.55 & -17:59:09.85 & 2005-07-07 & 4$\times$900 & 0.64 & * & 1.11 & 22.9 \\
05D4bj & +22:14:59.66 & -18:08:00.26 & 2005-07-10 & 4$\times$900 & 0.65 & * & 1.02 & 23.0 \\
05D4cn & +22:13:31.45 & -17:17:19.84 & 2005-08-10 & 4$\times$900 & 0.74 & 1.00 & 1.10 & 23.0 \\
05D4cq & +22:14:09.65 & -18:13:36.17 & 2005-08-04 & 3$\times$750 & 0.57 & 1.29 & 1.01 & 22.8 \\
05D4cs & +22:17:17.82 & -17:52:48.54 & 2005-08-04 & 3$\times$750 & 0.90 & * & 1.02 & 22.8 \\
05D4cw & +22:14:50.08 & -17:44:19.56 & 2005-08-08 & 3$\times$750 & 0.64 & * & 1.03 & 21.6 \\
05D4dw & +22:16:55.88 & -18:03:05.26 & 2005-09-10 & 6$\times$900 & 0.83 & 0.28 & 1.05 & 23.4 \\
05D4ef & +22:13:59.31 & -18:12:44.44 & 2005-09-24 & 6$\times$900 & 1.53 & * & 1.47 & 23.0 \\
05D4ej & +22:15:52.52 & -18:11:44.59 & 2005-09-27 & 4$\times$900 & 0.65 & 0.03 & 1.05 & 22.9 \\
05D4ek & +22:16:27.52 & -17:44:10.74 & 2005-09-26 & 4$\times$900 & 0.77 & 1.19 & 1.06 & 22.6 \\
%05D4em & +22:14:45.62 & -17:32:29.03 & 2005-09-12 & 6$\times$900 & 0.74 & 1.08 & 1.20 & 23.5 \\
05D4ev & +22:13:30.42 & -17:42:02.83 & 2005-09-12 & 6$\times$900 & 0.49 & 2.87 & 1.02 & 23.4 \\
05D4fe & +22:15:40.27 & -17:31:42.38 & 2005-09-27 & 4$\times$900 & 0.60 & * & 1.01 & 23.9 \\
05D4ff & +22:16:20.16 & -18:02:33.20 & 2005-09-29 & 4$\times$900 & 0.79 & * & 1.12 & 22.0 \\
05D4fg & +22:16:41.35 & -17:35:43.66 & 2005-09-29 & 4$\times$900 & 0.74 & 0.56 & 1.09 & 23.0 \\
06D1ab & +02:24:59.03 & -04:40:51.20 & 2006-01-06 & 3$\times$750 & 1.05 & 0.53 & 1.15 & 21.5\\
06D2bk & +09:58:42.87 & +02:10:19.14 & 2006-02-09 & 3$\times$750 & 0.72 & 0.41 & 1.18 & 22.6\\
06D2ca & +09:59:09.56 & +02:18:09.06 & 2006-02-27 & 4$\times$900 & 0.91 & 0.33 & 1.27 & 22.5\\
06D2ca & +09:59:09.56 & +02:18:09.06 & 2006-02-28 & 1$\times$900 & 0.90 & 0.33 & 1.32 & 22.5\\
%new
06D2cb$^f$ & +10:01:23.49 & +02:27:22.36 & 2006-03-02 & 6$\times$900 & 0.88 & * & 1.13 & 24.1\\
06D2cc & +09:59:26.89 & +02:27:04.88 & 2006-02-28 & 4$\times$900 & 0.80 & 1.42 & 1.18 & 22.6\\
%new
06D2cd$^{c,f}$ & +10:00:57.96 & +02:32:02.54 & 2006-03-04 & 6$\times$900 & 0.60 & * & 1.24 & 24.0\\
%new
06D2ce$^{c,f}$ & +10:01:44.22 & +02:42:47.93 & 2006-03-03 & 5$\times$900 & 0.76 & * & 1.26 & 23.3\\
06D2ck & +10:01:28.02 & +01:51:49.07 & 2006-03-08 & 4$\times$900 & 1.07 & 0.34 & 1.30 & 22.7\\
06D2ga & +09:59:08.11 & +01:56:18.58 & 2006-05-25 & 4$\times$900 & 0.66 & 3.60 & 1.27 & 23.4\\
06D4ce & +22:17:17.25 & -17:14:09.37 & 2006-07-19 & 4$\times$900 & 0.86 & * & 1.23 & 23.1\\
06D4ce & +22:17:17.25 & -17:14:09.37 & 2006-07-20 & 4$\times$900 & 0.79 & * & 1.04 & 23.2\\
06D4cl & +22:16:57.05 & -17:27:02.28 & 2006-07-21 & 8$\times$900 & 0.70 &  4.34 & 1.03 & 23.4\\
06D4co & +22:15:26.52 & -17:52:09.07 & 2006-07-30 & 3$\times$750 & 1.08 & * & 1.01 & 21.8\\
06D4cq & +22:16:55.52 & -17:42:43.13 & 2006-07-30 & 4$\times$750 & 1.07 & 0.84 & 1.10 & 21.8\\[2mm]
\hline
\end{longtable}

\noindent
$^a$ Except for 8 \Ia\ taken in MOS mode.

\noindent
$^b$ Host-supernova distance (from PHASE model) in arcsec. An asterisk 
either denotes cases for which a separate extraction of the supernova and the host component was not possible with PHASE or the absence of a detectable 
host.

\noindent
$^c$ Observed with Grism 300I.

\noindent
$^d$ SNLS-03D4cy is identified as a SN~Ia on the basis of a Gemini 
spectrum
published in \citet{Howell05}. The VLT spectrum is shown here for 
reference.

\noindent
$^e$ There are two underlying galaxies in the line-of-sight of 
SNLS-04D1hx.

\noindent 
$^f$ FORS2 spectrum
\begin{longtable}[b]{llccclcc}
\caption[]{\label{tab:tab1}
Results of identifications of the SNLS 3rd year ESO-VLT SN~Ia and SN~Ia$\star$ spectra.}\\[2mm]
\hline\hline\\
SN name & Type & $z$ & $z$ source & $\Phi$ & Host model$^a$ & Host fraction & $<S/N>^b$\\[2mm]
\hline
\\
03D1ar & Ia$\star$ & 0.408$\pm0.001$ & H & 5.3 & E(7) & 0.16 &       2.3            \\ 
03D1bf & Ia$\star$ & 0.703$\pm0.001$ & H & -2.7 & E(3) & 0.16 &     1.4              \\ 
03D1bm & Ia$\star$ & 0.575$\pm0.001$ & H & -5.1 & E(3) & 0.07 &     2.1              \\ 
03D1bp & Ia$\star$ & 0.347$\pm0.001$ & H & -7.5 & E(2) & 0.86 &     2.2              \\ 
03D1co & Ia & 0.679$\pm0.001$ & H & -4.1 & E(8) & 0.10 &      2.0             \\ 
03D1dt & Ia & 0.612$\pm0.001$ & H & 5.1 & Sc(4) & 0.65 &      3.3             \\ 
03D1fc & Ia & 0.332$\pm0.001$ & H & -4.4 & E(3) & 0.33 &         21.7          \\ 
03D1fl & Ia & 0.687$\pm0.001$ & H & 0.5 & NoGalaxy & - &    3.0               \\ 
03D1gt & Ia & 0.560$\pm0.001$ & H & 7.0 & E(4) & 0.50 &       2.4            \\ 
03D4ag & Ia$_\mathrm{pec}$ & 0.285$\pm0.001$ & H & -8.6 & E(2) & 0.27 & 29.5                  \\ 
03D4at & Ia & 0.634$\pm0.001$ & H & 5.5 & Sb-Sc & 0.42 &  3.7                 \\ 
03D4au & Ia$\star$ & 0.468$\pm0.001$ & H & 6.5 & Sb-Sc & 0.43 &   5.2                \\ 
03D4cx & Ia & 0.949$\pm0.014$ & S & 1.2 & NoGalaxy & - &          0.6         \\ 
03D4cx & Ia & 0.949$\pm0.014$ & S & 2.7 & S0(4) & 0.95 &        1.5           \\ 
03D4cy & Ia & 0.927$\pm0.001$ & H & 4.6 & Sb-Sc & 0.16 &          0.5         \\ 
03D4di & Ia$\star$ & 0.899$\pm0.001$ & H & -8.6 & Sb-Sc & 0.11 &         1.8          \\ 
03D4di & Ia$\star$ & 0.899$\pm0.001$ & H & -6.5 & NoGalaxy & - &      1.8             \\ 
03D4dy & Ia & 0.61$\pm0.01$ & S & 4.8 & NoGalaxy & - &       5.5            \\ 
03D4gf & Ia$\star$ & 0.580$\pm0.001$ & H & 20.4 & S0(1) & 0.46 &       1.5            \\ 
03D4gg & Ia$\star$ & 0.592$\pm0.001$ & H & 16.4 & Sc(10) & 0.73 &      1.7             \\ 
04D1ag & Ia & 0.557$\pm0.001$ & H & 4.3 & E(12) & 0.08 &            5.4       \\ 
04D1aj & Ia$\star$ & 0.721$\pm0.011$ & S & 11.6 & NoGalaxy & - &     1.0              \\ 
04D1aj & Ia$\star$ & 0.721$\pm0.011$ & S & 13.4 & NoGalaxy & - &     1.2              \\ 
04D1ak & Ia$\star$ & 0.526$\pm0.001$ & H & 11.8 & NoGalaxy & - &     0.6              \\ 
04D1dc & Ia & 0.211$\pm0.001$ & H & -0.4 & NoGalaxy & - &      33.4             \\ 
04D1ff & Ia & 0.86$\pm0.01$ & S & 4.6 & Sc(2) & 0.24 &         2.0          \\ 
04D1hx & Ia & 0.560$\pm0.001$ & H & 5.9 & NoGalaxy & - &      1.9             \\ 
04D1iv & Ia & 0.998$\pm0.001$ & H & 3.0 & NoGalaxy & - &       1.3            \\ 
04D1jd & Ia$\star$ & 0.778$\pm0.001$ & H & 6.8 & E(2) & 0.58 &         1.9          \\ 
04D1kj & Ia & 0.585$\pm0.001$ & H & -3.7 & NoGalaxy & - &         6.7          \\ 
04D1ks & Ia & 0.798$\pm0.012$ & S & -1.0 & NoGalaxy & - &      2.8             \\ 
04D1ow & Ia & 0.915$\pm0.001$& H & 6.4 & NoGalaxy & - &       1.6            \\ 
04D1pc & Ia & 0.770$\pm0.001$ & H & 0.1 & E(3) & 0.22 &          3.0         \\ 
04D1pd & Ia & 0.95$\pm0.01$ & S & 2.5 & E(1) & 0.31 &          2.1         \\ 
04D1pg & Ia & 0.515$\pm0.001$ & H & -1.3 & Sa-Sb & 0.61 &         8.9          \\ 
04D1pp & Ia & 0.735$\pm0.001$ & H & 2.2 & NoGalaxy & - &       1.2            \\ 
04D1qd & Ia$\star$ & 0.767$\pm0.001$ & H & -0.2 & E(1) & 0.71 &        4.8           \\ 
04D1rh & Ia & 0.436$\pm0.001$ & H & 0.0 & NoGalaxy & - &          7.8         \\ 
04D1rx & Ia$\star$ & 0.985$\pm0.001$ & H & 0.9 & NoGalaxy & - &      1.4             \\ 
04D1sa & Ia & 0.585$\pm0.001$ & H & -2.6 & E(1) & 0.72 &         6.4          \\ 
04D1si & Ia & 0.702$\pm0.001$ & H & -1.7 & Sa(9) & 0.79 &        7.6           \\ 
04D2ac & Ia & 0.348$\pm0.001$ & H & 1.3 & Sa-Sb & 0.20 &          7.6         \\ 
04D2al & Ia & 0.836$\pm0.001$ & H & -2.5 & E(2) & 0.29 &         2.5         \\ 
04D2an & Ia & 0.62$\pm 0.01$ & S & -3.4 & NoGalaxy & - &      7.7             \\ 
04D2an & Ia & 0.62$\pm 0.01$ & S & 0.3 & Sc(9) & 0.06 &         7.4          \\ 
04D2bt & Ia & 0.220$\pm0.001$ & H & 6.6 & E(6) & 0.01 &          9.2         \\ 
04D2ca & Ia$\star$ & 0.835$\pm0.001$ & H & 12.0 & NoGalaxy & - &     0.5              \\ 
04D2ca & Ia$\star$ & 0.835$\pm0.001$ & H & 13.1 & NoGalaxy & - &      0.9             \\ 
04D2cc & Ia$\star$ & 0.838$\pm0.001$ & H & 6.1 & S0(3) & 0.66 &        1.8           \\ 
04D2cc & Ia$\star$ & 0.838$\pm0.001$ & H & 7.2 & S0(2) & 0.67 &        1.5           \\ 
04D2cf & Ia & 0.369$\pm0.001$ & H & 8.5 & NoGalaxy & - &       12.1            \\ 
04D2cw & Ia$\star$ & 0.568$\pm0.001$ & H & 19.3 & Sb-Sc & 0.58 &        1.4           \\ 
%\hline
%\end{longtable}
\newpage
%\begin{longtable}[b]{ccccccc}
Table 2 (cont'd)                    \\[2mm]
\hline\hline\\
SN name & Type & $z$ & $z$ source & $\Phi$ & Host model$^a$ & Host fraction & $<S/N>^b$                  \\[2mm]
\hline
\\
04D2fp & Ia & 0.415$\pm0.001$ & H & 1.8 & NoGalaxy & - &     11.5                   \\ 
04D2fs & Ia & 0.357$\pm0.001$ & H & 1.7 & NoGalaxy & - &     17.8              \\ 
04D2gc & Ia & 0.521$\pm0.001$ & H & -4.9 & E(1) & 0.19 &       5.2            \\ 
04D2gp & Ia & 0.732$\pm0.001$ & H & 2.7 & NoGalaxy & - &     2.5              \\ 
04D2iu & Ia$\star$ & 0.70$\pm0.01$ & S & 9.8 & NoGalaxy & - &    0.7               \\ 
04D2iu & Ia$\star$ & 0.70$\pm0.01$ & S & 10.4 & E(1) & 0.09 &      1.2             \\ 
04D2ja & Ia$\star$ & 0.740$\pm0.001$ & H & 9.2 & E(3) & 0.44 &       0.8            \\ 
04D2ja & Ia$\star$ & 0.740$\pm0.001$ & H & 9.8 & E(3) & 0.19 &       0.5            \\ 
04D2mc & Ia & 0.348$\pm0.001$ & H & 6.5 & Sa-Sb & 0.30 &      10.3             \\ 
04D4an & Ia & 0.613$\pm0.001$ & H & 8.1 & NoGalaxy & - &     0.8              \\ 
04D4bk & Ia$\star$ & 0.88$\pm0.01$ & S & 3.1 & E(1) & 0.20 &        1.6           \\ 
04D4bq & Ia & 0.55$\pm0.01$ & S & 2.6 & NoGalaxy & - &     2.6              \\ 
04D4bq & Ia & 0.55$\pm0.01$ & S & 5.1 & NoGalaxy & - &     6.5              \\ 
04D4dw & Ia$\star$ & 1.031$\pm0.001$ & H & 2.1 & Sa-Sb & 0.68 &       1.5            \\ 
04D4fx & Ia & 0.629$\pm0.001$ & H & -8.4 & NoGalaxy & - &    5.8               \\ 
04D4gz & Ia$\star$ & 0.375$\pm0.001$ & H & -5.8 & Sa-Sb & 0.33 &      2.2             \\ 
04D4gz & Ia$\star$ & 0.375$\pm0.001$ & H & 10.2 & NoGalaxy & - &   1.8                \\ 
04D4hf & Ia$\star$ & 0.936$\pm0.014$ & S & -0.5 & NoGalaxy & - &   0.8                \\ 
04D4ht & Ia & 0.217$\pm0.001$ & H & 6.6 & Sa-Sb & 0.62 &        3.9           \\ 
04D4ib & Ia & 0.699$\pm0.001$ & H & 0.7 & S0-Sa & 0.71 &   4.2                \\ 
04D4id & Ia & 0.769$\pm0.001$ & H & 3.0 & NoGalaxy & - &        2.3           \\ 
04D4jr & Ia & 0.47$\pm0.01$ & S & -5.9 & NoGalaxy & - &       16.5            \\ 
04D4ju & Ia & 0.472$\pm0.001$ & H & -2.2 & S0(1) & 0.44 &         3.5          \\ 
04D4jw & Ia$\star$ & 0.961$\pm0.001$ & H & 2.2 & S0-Sa & 0.69 &  2.3                 \\ 
05D1cb & Ia & 0.632$\pm0.001$ & H & 4.3 & Sb-Sc & 0.28 &      4.5             \\ 
05D1ck & Ia & 0.617$\pm0.001$ & H & -2.7 & Sb-Sc & 0.39 & 0.7                  \\ 
05D1cl & Ia$\star$ & 0.83$\pm0.01$ & S & 8.7 & Sd(4) & 0.36 &            1.8       \\ 
05D1dn & Ia & 0.566$\pm0.001$ & H & -4.7 & NoGalaxy & - &          7.3         \\ 
%05D1dx & Ia & 0.580 & S & 12.41 & Sa-Sb & 0.13 &                   \\ 
05D1hk & Ia$_\mathrm{pec}$ & 0.263$\pm0.001$ & H & -5.0 & Sc(5) & 0.39 & 21.4                 \\ 
05D1hn & Ia & 0.149$\pm0.001$ & H & -1.0 & NoGalaxy & - &  13.0                 \\ 
05D1iz & Ia$\star$ & 0.86$\pm0.01$ & S & 7.8 & Sd(1) & 0.24 &    2.1               \\ 
05D1ke & Ia & 0.69$\pm0.01$ & S & 2.2 & NoGalaxy & - &   4.9                \\ 
05D2ac & Ia & 0.479$\pm0.001$ & H & 2.0 & NoGalaxy & - &   10.9                \\ 
05D2ay & Ia$\star$ & 0.92$\pm0.01$ & S & 5.8 & S0(1) & 0.07 &    1.1               \\ 
05D2bt & Ia & 0.68$\pm0.01$ & S & 1.4 & NoGalaxy & - &   2.1                \\ 
05D2bv & Ia & 0.474$\pm0.001$ & H & -0.1 & Sb-Sc & 0.26 &     8.3              \\ 
05D2bw & Ia & 0.92$\pm0.01$ & S & 1.9 & Sb-Sc & 0.25 &     2.1              \\ 
05D2by & Ia$\star$ & 0.891$\pm0.001$ & H & 0.9 & Sc(11) & 0.83 &   1.3                \\ 
05D2cb & Ia$\star$ & 0.427$\pm0.001$ & H & -5.8 & Sc(11) & 0.57 &  2.9                 \\ 
05D2ci & Ia & 0.630$\pm0.001$ & H & 3.6 & NoGalaxy & - &  0.8                 \\ 
05D2ci & Ia & 0.630$\pm0.001$ & H & 4.9 & NoGalaxy & - &  0.7                 \\ 
05D2ci & Ia & 0.630$\pm0.001$ & H & 6.7 & NoGalaxy & - &  1.4                 \\ 
%05D2cl & Ia$\star$ & 0.857$\pm0.001$ & H & 6.8 & E(2) & 0.86 & 1.8                  \\ 
05D2ct & Ia$\star$ & 0.734$\pm0.001$ & H & 7.8 & Sd(11) & 0.63 &    4.6               \\ 
05D2dt & Ia & 0.574$\pm0.001$ & H & -1.7 & E(3) & 0.43 &    6.0               \\ 
05D2dw & Ia & 0.417$\pm0.001$ & H & -5.3 & S0(1) & 0.12 &   10.3                \\ 
05D2dy & Ia & 0.51$\pm 0.01$ & S & 1.1 & Sb-Sc & 0.02 &     4.8              \\ 
05D2eb & Ia & 0.534$\pm0.001$ & H & -4.7 & S0(1) & 0.04 &   6.2                \\ 
05D2ec & Ia & 0.640$\pm0.001$ & H & 2.7 & NoGalaxy & - &  2.4                 \\ 
05D2ei & Ia$\star$ & 0.366$\pm0.001$ & H & 16.9 & S0(4) & 0.55 &  2.5                 \\ 
05D2fq & Ia & 0.733$\pm0.001$ & H & 2.1 & NoGalaxy & - &  1.9                 \\ 
05D2he & Ia & 0.608$\pm0.001$ & H & 3.0 & E(1) & 0.26 &     3.1              \\ 
%\hline
%\end{longtable}
\newpage
%\begin{longtable}[b]{ccccccc}
Table 2 (cont'd)\\[2mm]                    \\
\hline\hline\\
SN name & Type & $z$ & $z$ source & $\Phi$ & Host model$^a$ & Host fraction & $<S/N>^b$                  \\[2mm]
\hline\\
05D2ie & Ia & 0.348$\pm0.001$ & H & -8.9 & Sc(4) & 0.45 &      4.4             \\ 
%05D2le & Ia & 0.700 & H & 5.30 & NoGalaxy & - &                   \\ 
05D2nn & Ia & 0.87$\pm0.01$ & S & -1.3 & Sa(8) & 0.26 &      1.9             \\ 
05D4af & Ia & 0.499$\pm0.001$ & H & 11.3 & NoGalaxy & - &    8.8               \\ 
05D4ag & Ia$\star$ & 0.64$\pm0.01$ & S & 14.0 & Sb-Sc & 0.10 &      2.0             \\ 
05D4ay & Ia$\star$ & 0.408$\pm0.001$ & H & 8.2 & Sa-Sb & 0.80 & 2.55
\\
05D4ay & Ia$\star$ & 0.408$\pm0.001$ & H & 9.6 & Sa-Sb & 0.80 & 1.8
\\
05D4be & Ia & 0.537$\pm0.001$ & H & 3.9 & E(1) & 0.31 &        5.5           \\ 
05D4bi & Ia & 0.775$\pm0.001$ & H & -1.5 & Sc(1) & 0.40 &      3.7             \\ 
05D4bj & Ia & 0.701$\pm0.001$ & H & -2.0 & Sc(9) & 0.68 &      2.8             \\ 
05D4cn & Ia & 0.763$\pm0.001$ & H & 4.7 & S0(1) & 0.27 &       3.0            \\ 
05D4cq & Ia & 0.701$\pm0.001$ & H & -0.1 & NoGalaxy & - &    4.6               \\ 
05D4cs & Ia & 0.79$\pm0.01$ & S & -1.7 & Sc(12) & 0.16 &     4.3              \\ 
05D4cw & Ia & 0.375$\pm0.001$ & H & 7.0 & S0(6) & 0.57 &       16.4            \\ 
05D4dw & Ia & 0.855$\pm0.001$ & H & 4.7 & Sc(12) & 0.19 &      0.9             \\ 
05D4ef & Ia$\star$ & 0.605$\pm0.001$ & H & 3.5 & E(3) & 0.78 &       1.6            \\ 
05D4ej & Ia & 0.585$\pm0.001$ & H & 7.6 & NoGalaxy & - &     3.5              \\ 
05D4ek & Ia & 0.536$\pm0.001$ & H & 2.1 & NoGalaxy & - &     8.2              \\ 
%05D4em & Ia$\star$ & 0.97$\pm 0.01$ & S & -0.5 & NoGalaxy & - &   2.1                \\ 
05D4ev & Ia & 0.722$\pm0.001$ & H & -3.6 & NoGalaxy & - &       3.4            \\ 
05D4fe & Ia$\star$ & 0.984$\pm0.001$ & H & -2.0 & E(3) & 0.65 &      1.1             \\ 
05D4ff & Ia$\star$ & 0.402$\pm0.001$ & H & 5.1 & Sc(6) & 0.83 &      4.9             \\ 
05D4fg & Ia & 0.839$\pm0.001$ & H & -0.3 & S0(5) & 0.32 &      2.3             \\ 
06D1ab & Ia & 0.182$\pm0.001$ & H & -4.4 & E(7) & 0.11 &       13.9            \\ 
%06D2ag & Ia & 0.310 & H & 3.80 & NoGalaxy & - &     2.0              \\ 
06D2bk & Ia & 0.499$\pm0.001$ & H & 0.9 & Sb-Sc & 0.15 &   3.8                \\ 
06D2ca & Ia & 0.531$\pm0.001$ & H & 0.0 & Sd(5) & 0.27 &             3.2      \\ 
06D2ca & Ia & 0.531$\pm0.001$ & H & 0.7 & Sd(3) & 0.18 &          2.5         \\ 
%new
06D2cb & Ia$\star$ & 1.00$\pm0.01$ & S & 3.0 & NoGalaxy & - & 1.7 \\
06D2cc & Ia & 0.532$\pm0.001$ & H & 3.3 & NoGalaxy & - &        6.2           \\
%new
06D2cd & Ia & 0.930$\pm0.001$ & H & 4.2 & NoGalaxy & - & 5.8 \\
%new
06D2ce & Ia & 0.82$\pm0.01$ & S & 0.0 & NoGalaxy & - & 7.3 \\ 
06D2ck & Ia$\star$ & 0.552$\pm0.001$ & H & 7.7 & Sa-Sb & 0.24 &  1.6                 \\ 
06D2ga & Ia$\star$ & 0.84$\pm 0.01$ & S & 5.4 & Sb-Sc & 0.19 &  2.0                 \\ 
%06D4ba & Ia    &       &   &      &             &     &                  \\
%06D4bw & Ia    &       &   &       &            &     &                 \\
06D4ce & Ia & 0.85$\pm0.01$ & S & 2.7 & NoGalaxy & - &          2.3         \\ 
06D4ce & Ia & 0.85$\pm0.01$ & S & 3.3 & NoGalaxy & - &          2.7         \\ 
06D4cl & Ia & 1.00$\pm 0.01$ & S & -2.5 & E(1) & 0.01 &            2.5       \\ 
06D4co & Ia & 0.437$\pm0.001$ & H & 3.5 & E-S0 & 0.58 &      12.8             \\ 
06D4cq & Ia & 0.411$\pm0.001$ & H & -1.4 & NoGalaxy & - &         8.2          \\[2mm] 
\hline
\end{longtable}

\noindent
$^a$ Host type of the best fitting model. The number in parenthesis is the age
of the best fitting PEGASE template (in Gyrs). Two host types separated by a dash indicate that
the best fitting host model is an interpolation between two \citet{Kinney96} templates. The label
'NoGalaxy' denotes that the best fitting model is obtained with no host model.

\noindent
$^b$ Average signal-to-noise ratio per 5 \AA $\ $ pixel.

\newpage

\begin{longtable}[b]{lccccc}
\caption[]{\label{tab:distributionIa}
Distribution of types per SNLS Deep field and in total}\\[2mm]
\hline\hline
\\
& D1 & D2 & D3 &  D4 & Total number of SNe (spectra)\\[2mm]
\hline
\\
SN~Ia & 27 & 29 & 0 & 30 & 86 (93)\\
SN~Ia$\star$ & 11  & 13  & 0 & 14 & 38 (46)\\[2mm]
\hline
\end{longtable}

\newpage

\begin{longtable}[b]{lcccc}
\caption[]{\label{tab:tabparamsIa}
Average properties of the SN~Ia and SN~Ia$\star$ subsamples}\\[2mm]
\hline\hline
\\
%\multicolumn{6}{c}{SN~Ia} \\
%\hline
& $<z> (\sigma_z)$ & $<\phi>(\sigma_\phi)$ & $<c>(\sigma_c)$ & $<x_1>(\sigma_{x_1})$ \\[2mm]
\hline
\\
SN~Ia & 0.60$\pm 0.02$ (0.20) & 1.1$\pm0.4$ (4.1) & 0.009$\pm 0.013$ (0.12) & -0.02$\pm 0.09$ (0.9)\\
SN~Ia$\star$ & 0.70$\pm 0.03$ (0.21) & 5.8$\pm 1.1$ (7.1) & 0.05$\pm 0.03$ (0.16) & 0.004$\pm 0.18$ (1.1)\\[2mm]
\hline
&&&&\\
&&&&\\
&&&&\\

\hline\hline
\\
$<\gamma_1> (\sigma_{\gamma_1})$ & $m^c_B(\sigma_{m_B})$ & $f_{gal}(\sigma_{f_{gal}})$ & $<S/N>^a$ & $\chi^2_\nu (\sigma_{\chi^2_\nu})$\\[2mm]
\hline
\\
 0.26$\pm 0.11$ (1.1) & 23.95$\pm 0.04$ (0.36)$^b$ & 0.32$\pm 0.03$ 
(0.23) & 5.8$\pm 0.6$ & 1.13 (0.30)$^c$\\
 0.26$\pm 0.29$ (1.9)  & 24.12$\pm 0.08$ (0.50)$^b$ & 0.47$\pm 0.05$ (0.28) & 1.8$\pm 0.2$ & 1.03 (0.16)$^d$\\[2mm]
\hline
\end{longtable}

\noindent
%$^a$ These values are obtained when SN~04D2al is excluded from the SN~Ia subsample. With this supernova included, one finds $<\gamma_1>=0.21\pm 0.13$ and
%$\sigma_{\gamma_1}=1.28$.

\noindent
$^a$ Average signal-to-noise ratio per 5 $\AA$ pixel.

\noindent
$^b$ Distance corrected magnitudes in the B band, using $\Omega_0=0.27$, 
$\Omega_\Lambda=0.73$ and $H_0=70$ km/s/Mpc.

\noindent
$^c$ Excluding $\chi^2_\nu>2$ events (1.24 (0.49) with all events).

\noindent
$^d$ Excluding $\chi^2_\nu>2$ events (1.46 (2.30) with all events).

\newpage

\begin{longtable}[b]{lccccc}
\caption[]{\label{tab:tabpec}
Properties of SNLS-03D4ag and SNLS-05D1hk VLT spectra}\\[2mm]
\hline\hline
\\
SN name & $z$ & $\phi^a$ & $c$ & $x_1(\Delta m_{15}^b)$ & $\gamma_1$\\[2mm]
\hline
\\
03D4ag & 0.285 & -9 & -0.057 &  0.808(0.89) & -0.02\\
05D1hk & 0.263 & -5 & -0.01 & 1.367(0.97) & -1.25\\[2mm]
\hline
\end{longtable}

\noindent
$^a$ Phase in days with respect to B-band maximum light.

\noindent
$^b$ $\Delta m_{15}$ value computed from $x_1$ using \citet{Guy07} formula.\\

\newpage

\begin{longtable}[b]{lcccc}
\caption[]{\label{tab:compositeIa}
Number of SN~Ia spectra in each phase bin (in days) used to create VLT composite spectra.}\\[2mm]
\hline\hline\\
& $\phi \leq  -4$ & -4$<\phi<$4  & $\phi \geq 4$ & Total\\[2mm]
\hline
\\
$z<0.5$     & 7  & 12 & 6 & 25\\
$z\geq 0.5$ & 5  & 45 & 18 & 68 \\
Total       & 12 & 57 & 24 & 93\\[2mm]
\hline
\end{longtable}

\newpage

\newpage

\newpage

\newpage

\clearpage

\newpage

\clearpage

\newpage

\newpage

\clearpage

\newpage

\clearpage

\newpage

\clearpage

\newpage
\clearpage
\Online
\appendix
\section{Spectra}
%%% VLT spectra figures start here
                    
                    \newpage
                    \clearpage
                    
                    \newpage
                    \clearpage
                    
                    \newpage
                    \clearpage
                    
                    \newpage
                    \clearpage
                    
                    \newpage
                    \clearpage
                    
                    \newpage
                    \clearpage
                    
                    \newpage
                    \clearpage
                    
                    \newpage
                    \clearpage
                    
                    \newpage
                    \clearpage
                    
                    \newpage
                    \clearpage
                    
                    \newpage
                    \clearpage
                    
                    \newpage
                    \clearpage
                    
                    \newpage
                    \clearpage
                    
                    \newpage
                    \clearpage
                    
                    \newpage
                    \clearpage
                    
                    \newpage
                    \clearpage
                    
                    \newpage
                    \clearpage
                    
                    \newpage
                    \clearpage
                    
                    \newpage
                    \clearpage
                    
                    \newpage
                    \clearpage
                    
                    \newpage
                    \clearpage
                    
                    \newpage
                    \clearpage
                    
                    \newpage
                    \clearpage
                    
                    \newpage
                    \clearpage
                    
                    \newpage
                    \clearpage
                    
                    \newpage
                    \clearpage
                    
                    \begin{figure*}[ht]
                    \begin{center}
                    \begin{tabular}[b]{cc}
                    \hspace{-5mm}
                    \vspace*{1cm}
                    \begin{minipage}[b]{10cm}
                    \centering
                    \includegraphics[width=10cm]{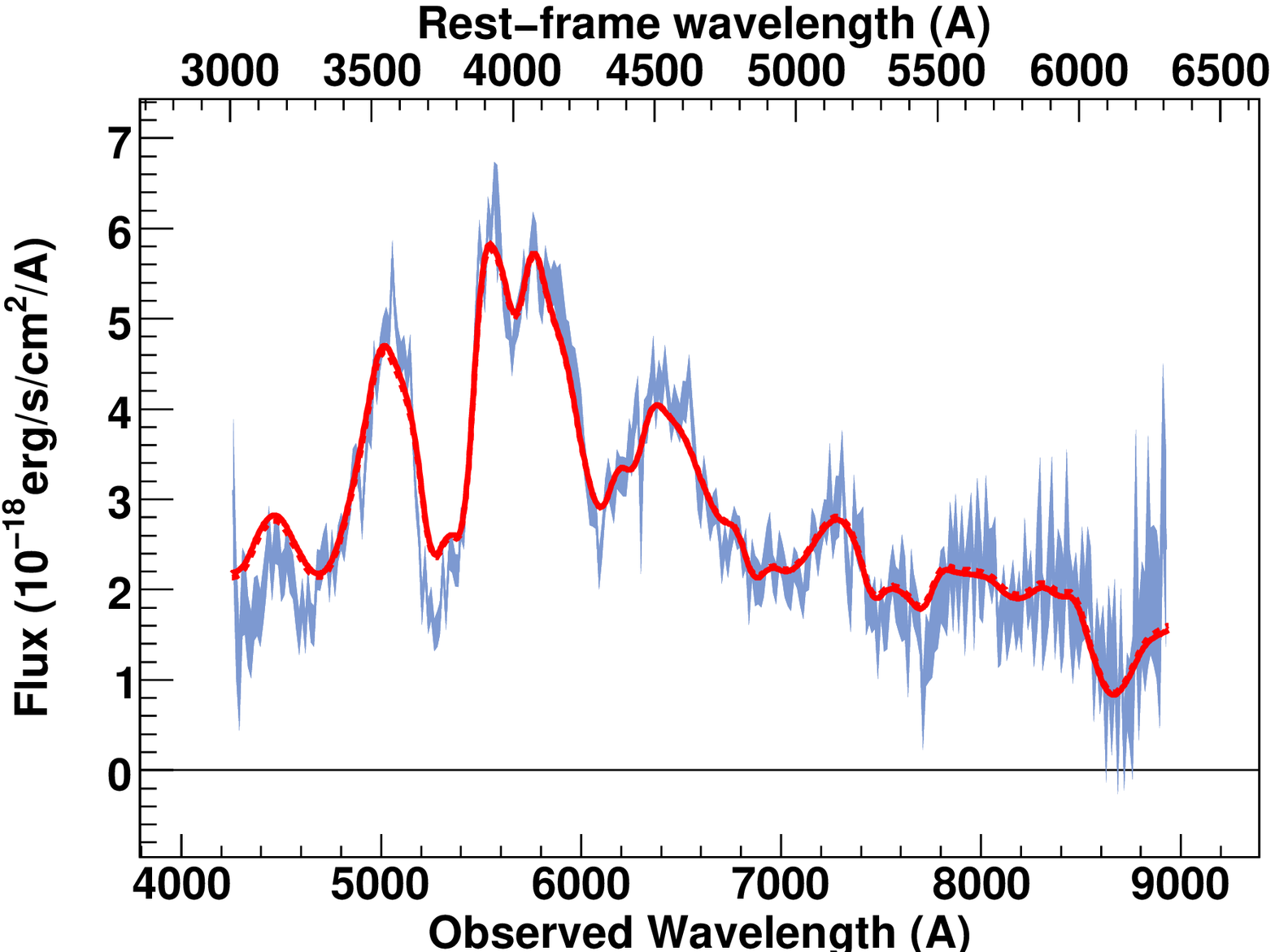}
                    \end{minipage}
                    &
                    \begin{minipage}[b]{10cm}
                    \end{minipage}
                    \\
                    \end{tabular}
                    \caption{\label{fig:04D2fp470} SNLS-  04D2fp470: a SN~Ia supernova at z=0.415. The spectrum phase is 1.8. Best-fit obtained for a model with no host galaxy component.}
                    \end{center}
                    \end{figure*}
                    
                    \newpage
                    \clearpage
                    
                    \begin{figure*}[ht]
                    \begin{center}
                    \begin{tabular}[b]{cc}
                    \hspace{-5mm}
                    \vspace*{1cm}
                    \begin{minipage}[b]{10cm}
                    \centering
                    \includegraphics[width=10cm]{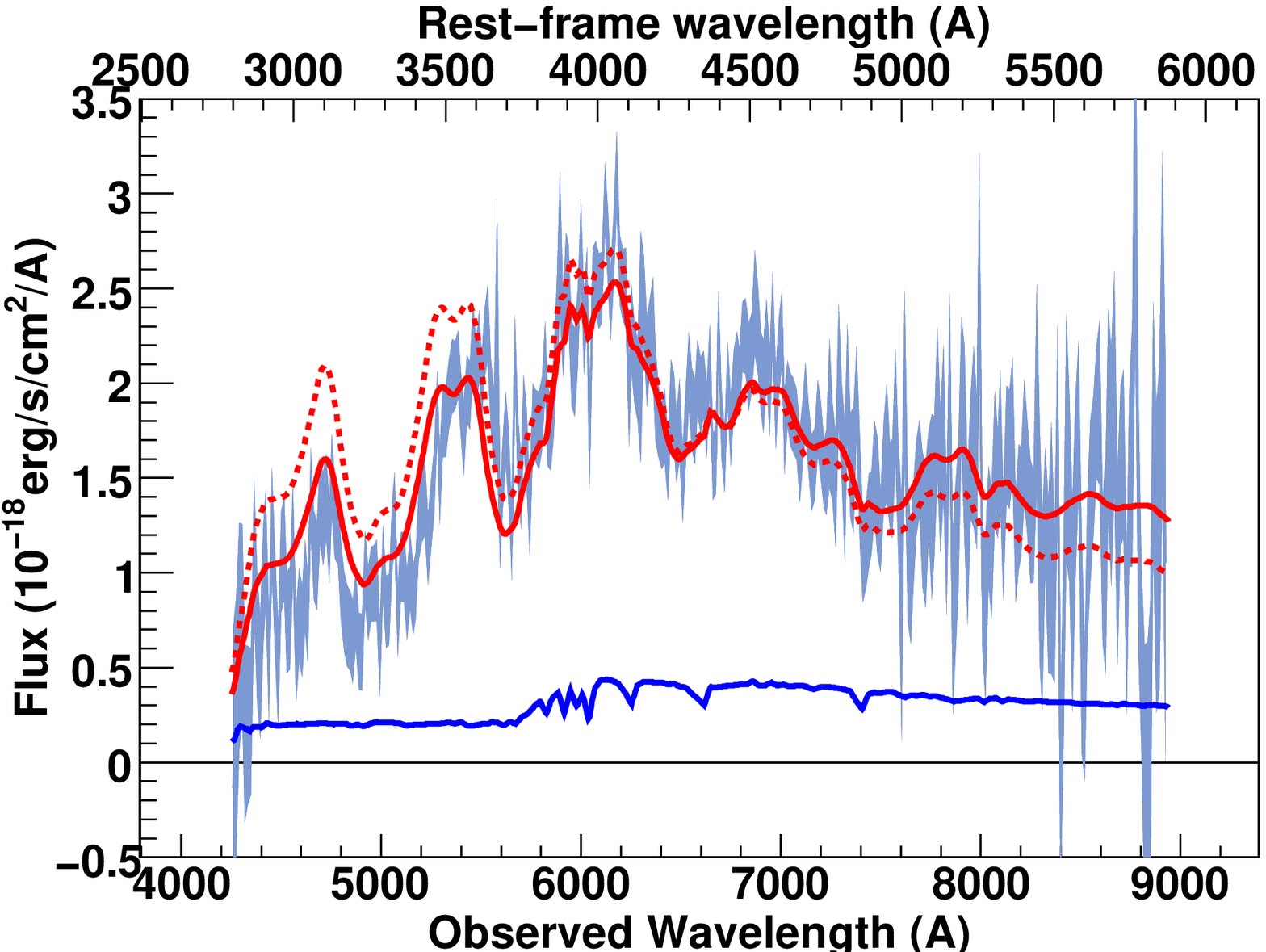}
                    \end{minipage}
                    &
                    \begin{minipage}[b]{10cm}
                    \hspace{-5mm}\includegraphics[width=10cm]{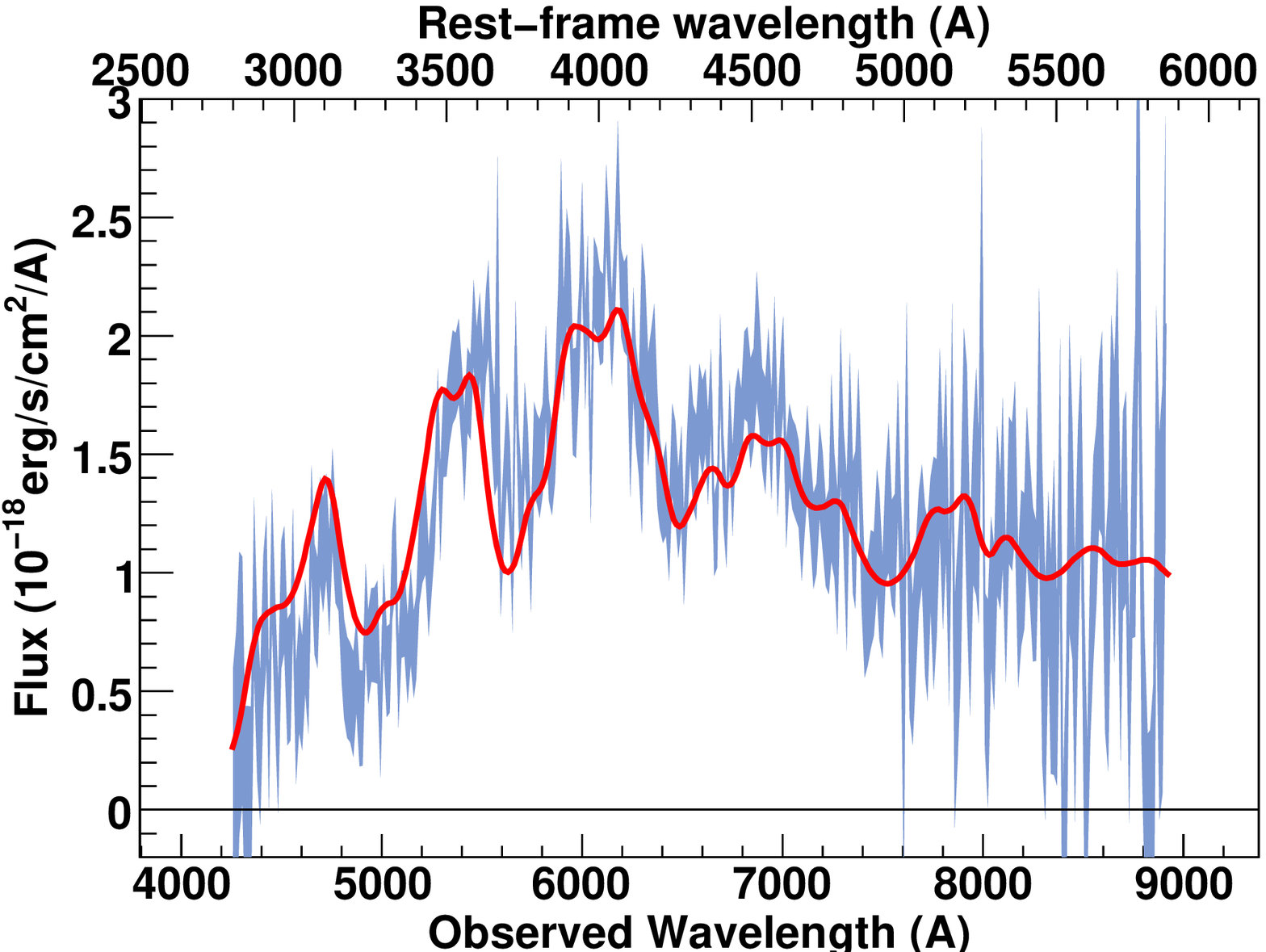}
                    \end{minipage}
                    \\
                    \end{tabular}
                    \caption{\label{fig:04D2gc471} SNLS-  04D2gc471: a SN~Ia supernova at z=0.521. The spectrum phase is -4.9. A         E(1) host model has been subtracted.}
                    \end{center}
                    \end{figure*}
                    
                    \begin{figure*}[ht]
                    \begin{center}
                    \begin{tabular}[b]{cc}
                    \hspace{-5mm}
                    \vspace*{1cm}
                    \begin{minipage}[b]{10cm}
                    \centering
                    \includegraphics[width=10cm]{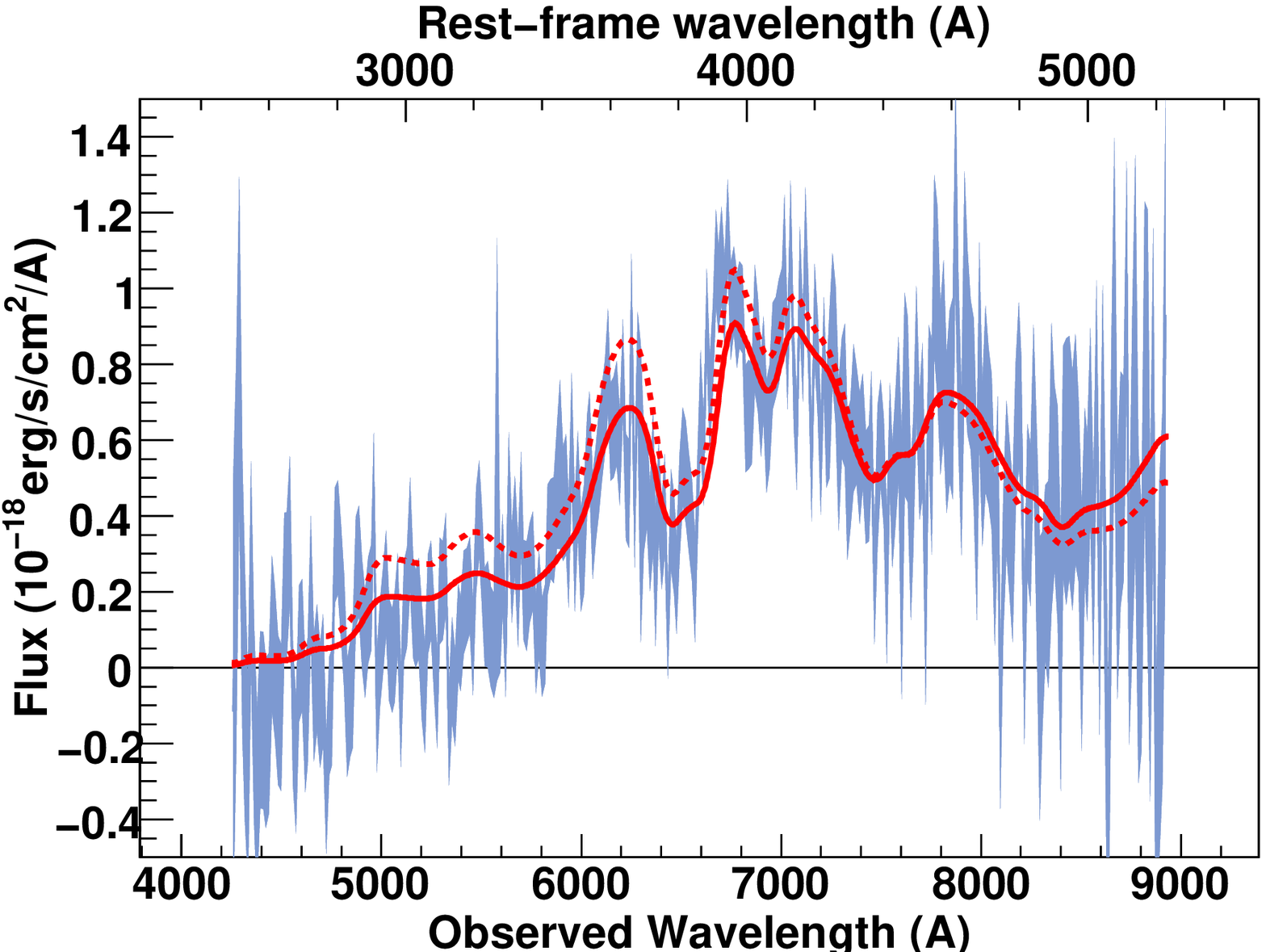}
                    \end{minipage}
                    &
                    \begin{minipage}[b]{10cm}
                    \end{minipage}
                    \\
                    \end{tabular}
                    \caption{\label{fig:04D2gp475} SNLS-  04D2gp475: a SN~Ia supernova at z=0.732. The spectrum phase is 2.7. Best-fit obtained for a model with no host galaxy component.}
                    \end{center}
                    \end{figure*}

                    \newpage
                    \clearpage
                    
                    \begin{figure*}[ht]
                    \begin{center}
                    \begin{tabular}[b]{cc}
                    \hspace{-5mm}
                    \vspace*{1cm}
                    \begin{minipage}[b]{10cm}
                    \centering
                    \includegraphics[width=10cm]{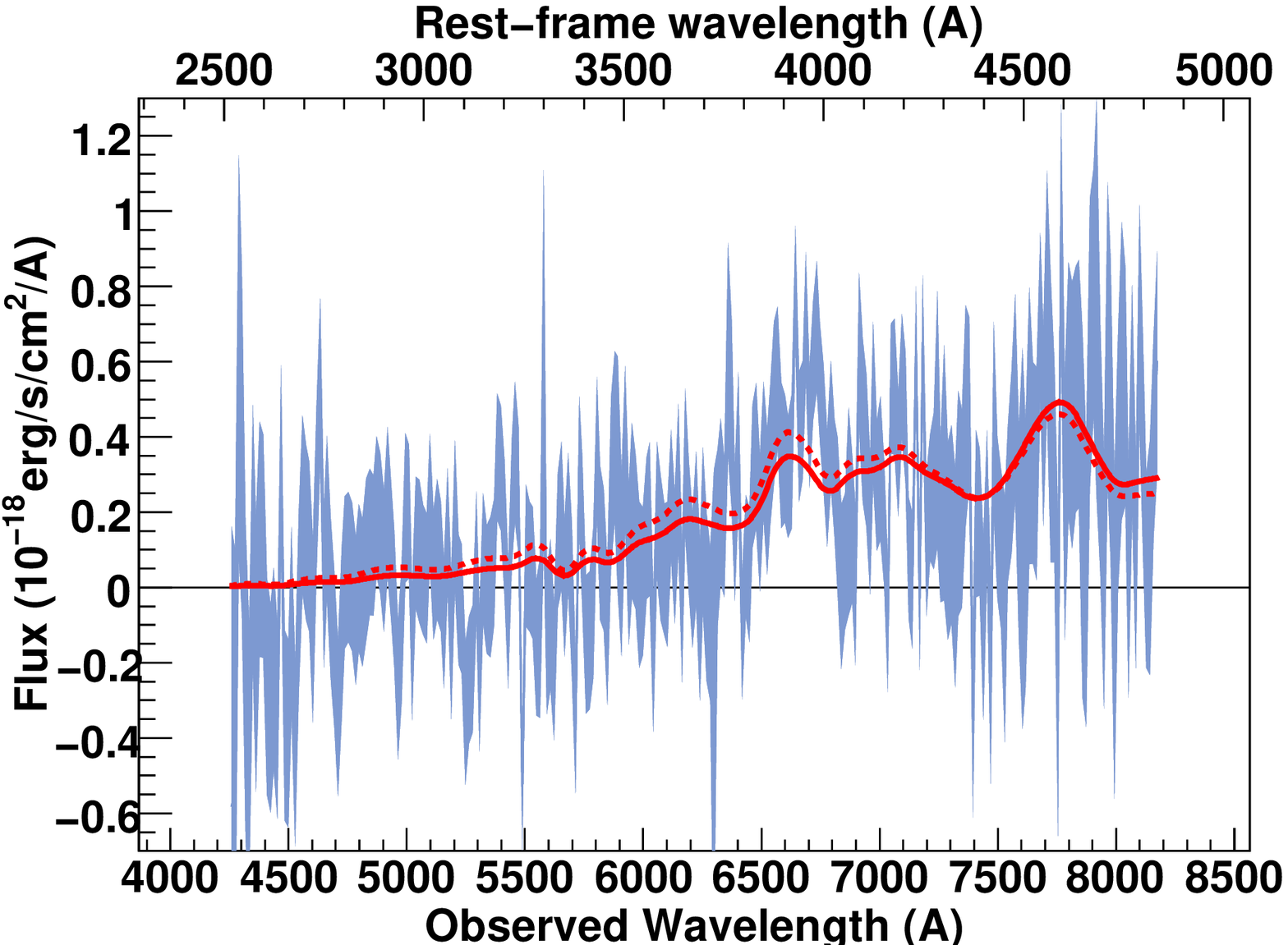}
                    \end{minipage}
                    &
                    \begin{minipage}[b]{10cm}
                    \end{minipage}
                    \\
                    \end{tabular}
                    \caption{\label{fig:04D2iu497} SNLS-  04D2iu497: a SN~Ia$\star$ supernova at z=0.70. The spectrum phase is 9.8. Best-fit obtained for a model with no host galaxy component.}
                    \end{center}
                    \end{figure*}
                    
                    \begin{figure*}[ht]
                    \begin{center}
                    \begin{tabular}[b]{cc}
                    \hspace{-5mm}
                    \vspace*{1cm}
                    \begin{minipage}[b]{10cm}
                    \centering
                    \includegraphics[width=10cm]{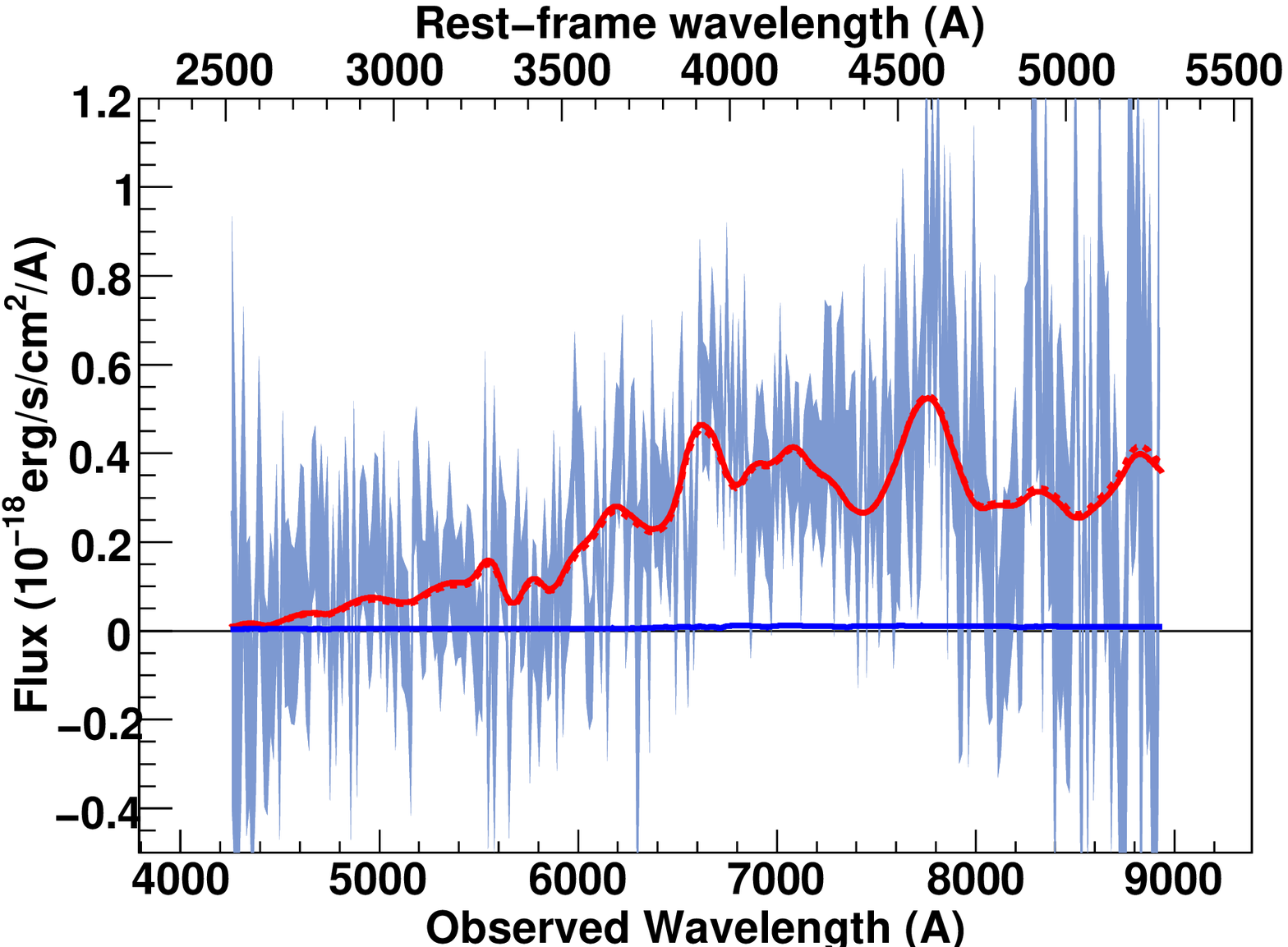}
                    \end{minipage}
                    &
                    \begin{minipage}[b]{10cm}
                    \hspace{-5mm}\includegraphics[width=10cm]{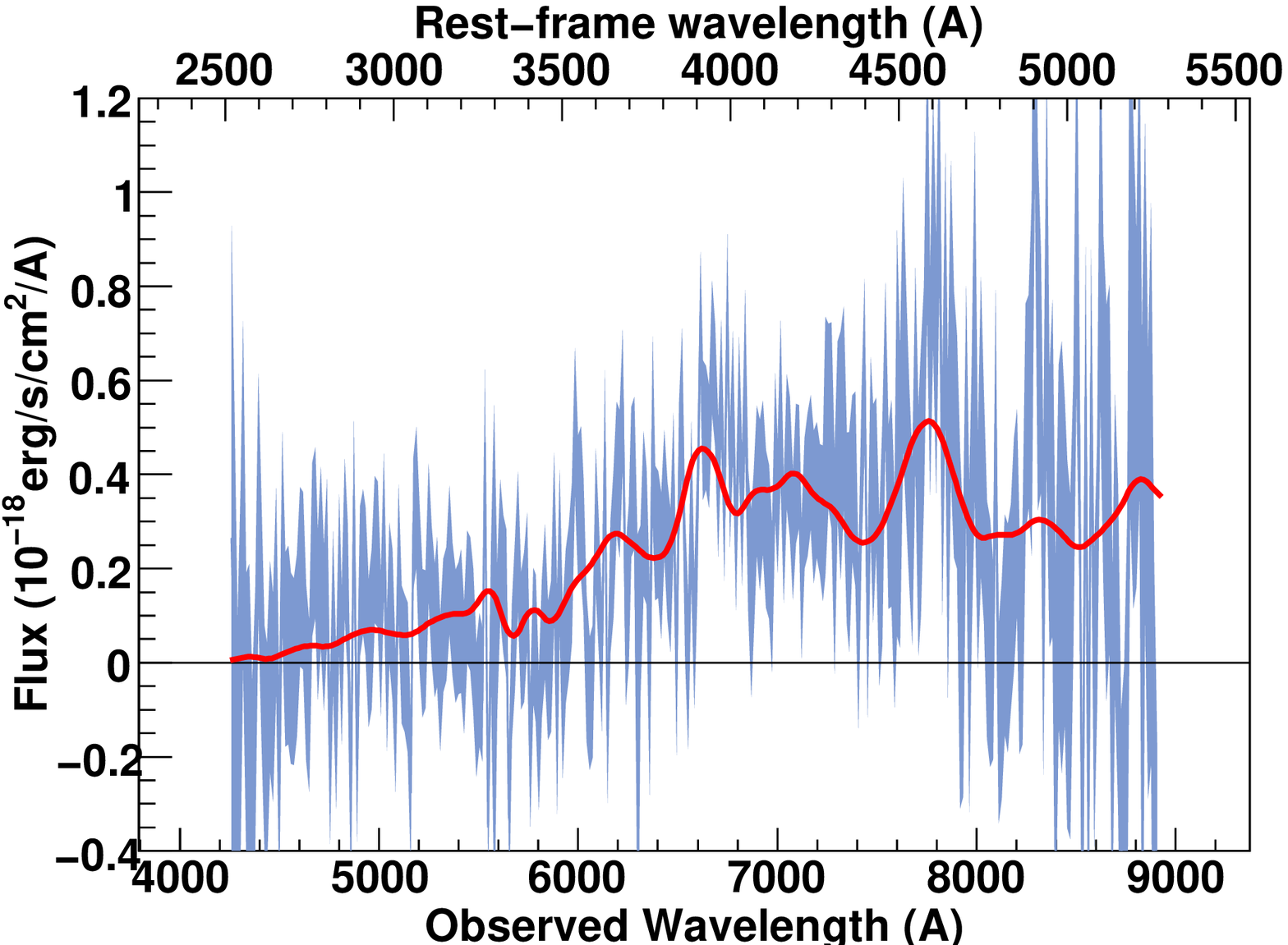}
                    \end{minipage}
                    \\
                    \end{tabular}
                    \caption{\label{fig:04D2iu498} SNLS-  04D2iu498: a SN~Ia$\star$ supernova at z=0.70. The spectrum phase is 10.4. A         E(1) host model has been subtracted.}
                    \end{center}
                    \end{figure*}

                    \newpage
                    \clearpage
                    
                    \begin{figure*}[ht]
                    \begin{center}
                    \begin{tabular}[b]{cc}
                    \hspace{-5mm}
                    \vspace*{1cm}
                    \begin{minipage}[b]{10cm}
                    \centering
                    \includegraphics[width=10cm]{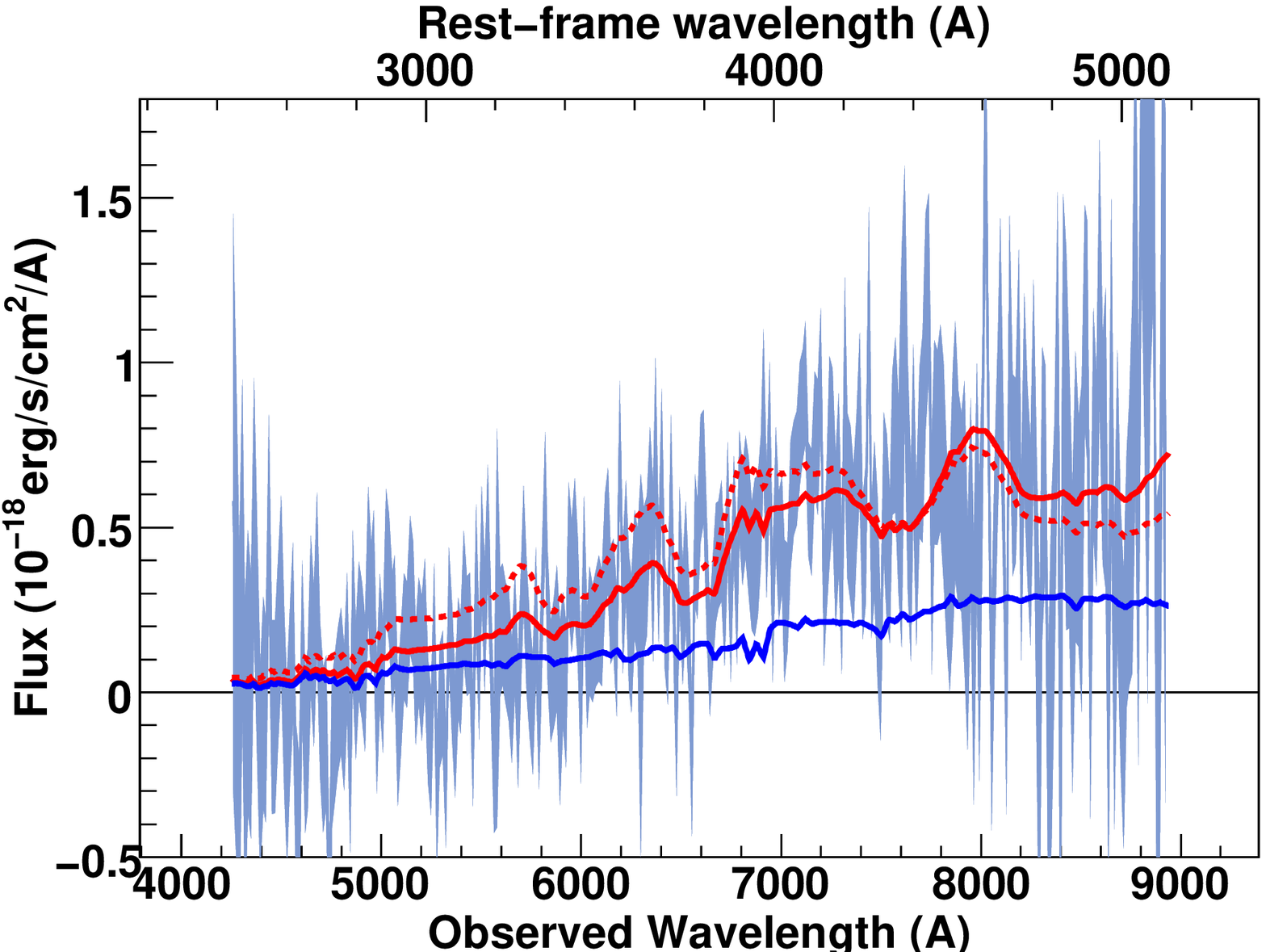}
                    \end{minipage}
                    &
                    \begin{minipage}[b]{10cm}
                    \hspace{-5mm}\includegraphics[width=10cm]{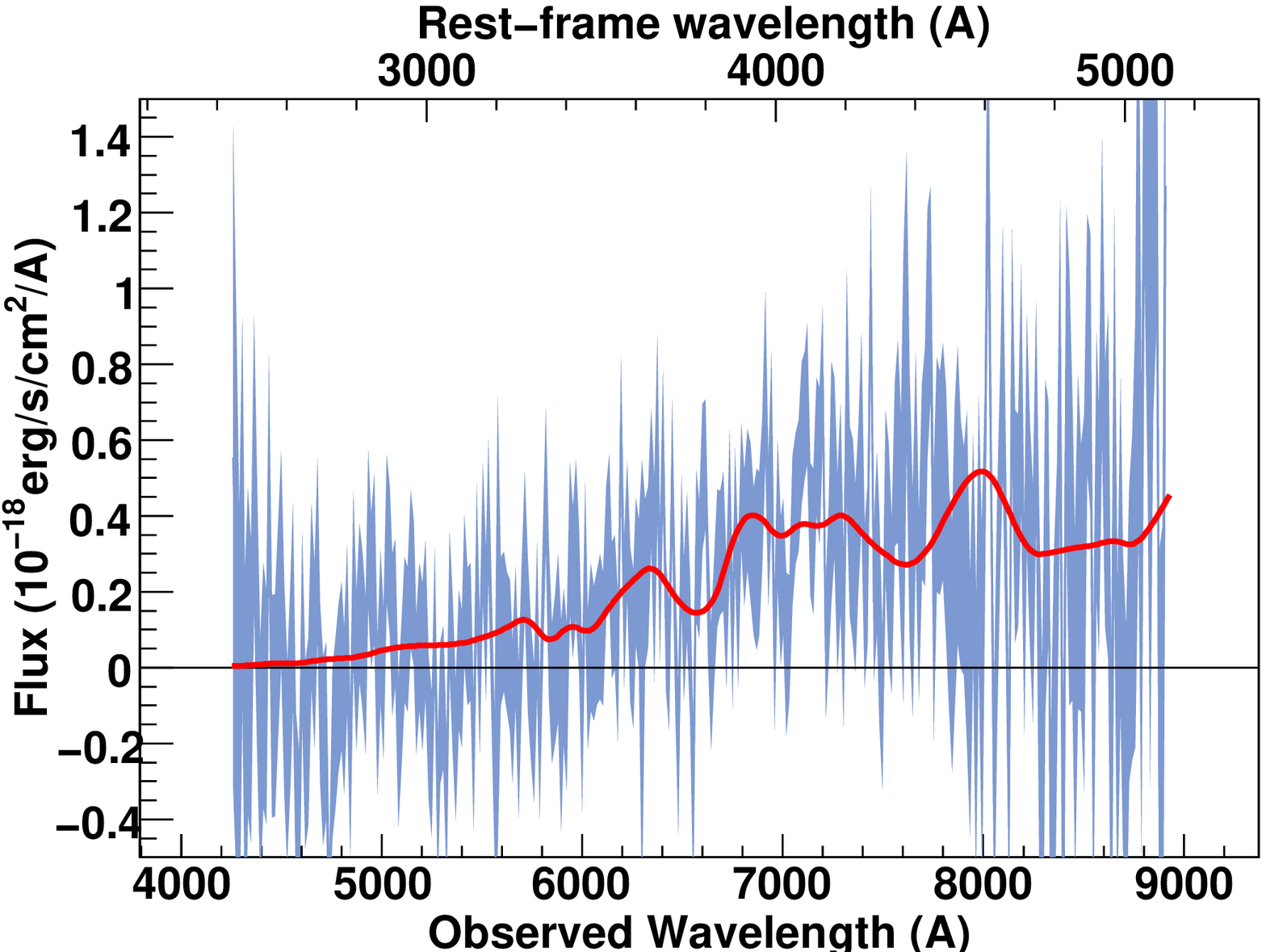}
                    \end{minipage}
                    \\
                    \end{tabular}
                    \caption{\label{fig:04D2ja499} SNLS-  04D2ja499: a SN~Ia$\star$ supernova at z=0.740. The spectrum phase is 9.2. A         E(3) host model has been subtracted. Note the large feature at rest-frame 4800 $\AA$ not reproduced by the model.}
                    \end{center}
                    \end{figure*}
                    
                    \begin{figure*}[ht]
                    \begin{center}
                    \begin{tabular}[b]{cc}
                    \hspace{-5mm}
                    \vspace*{1cm}
                    \begin{minipage}[b]{10cm}
                    \centering
                    \includegraphics[width=10cm]{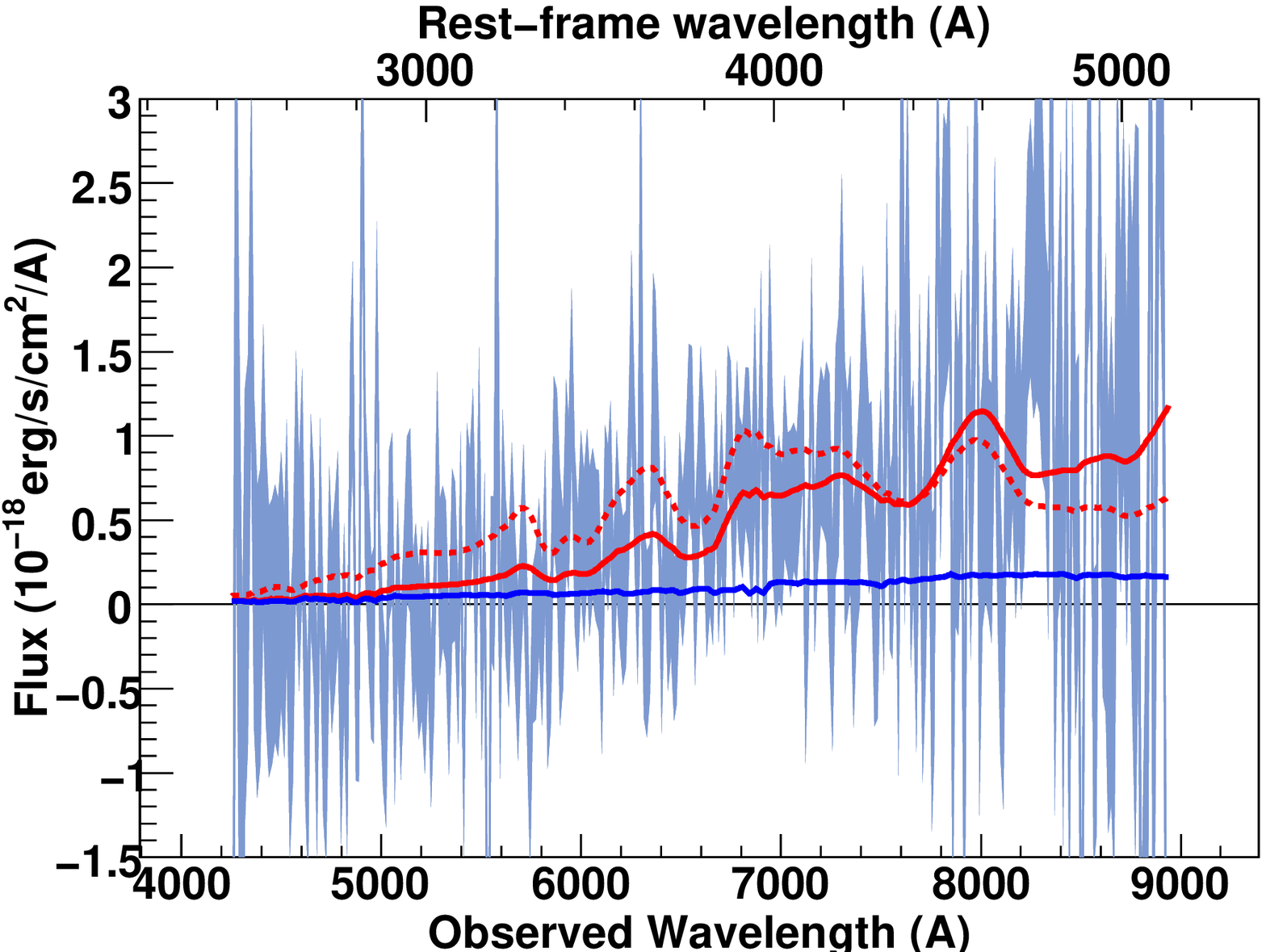}
                    \end{minipage}
                    &
                    \begin{minipage}[b]{10cm}
                    \hspace{-5mm}\includegraphics[width=10cm]{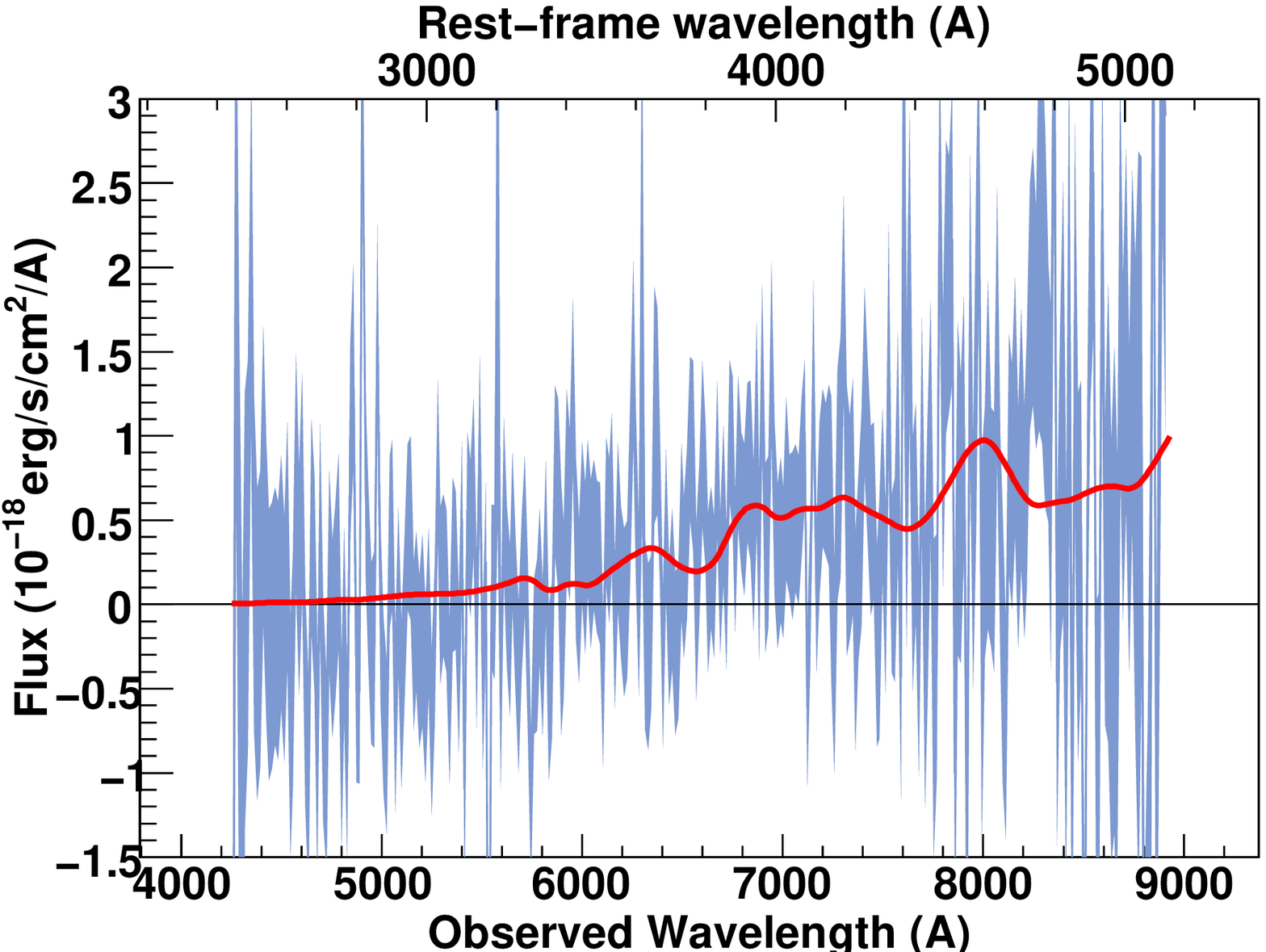}
                    \end{minipage}
                    \\
                    \end{tabular}
                    \caption{\label{fig:04D2ja500} SNLS-  04D2ja500: a SN~Ia$\star$ supernova at z=0.740. The spectrum phase is 9.8. A         E(3) host model has been subtracted.}
                    \end{center}
                    \end{figure*}

                    \newpage
                    \clearpage
                    
                    \begin{figure*}[ht]
                    \begin{center}
                    \begin{tabular}[b]{cc}
                    \hspace{-5mm}
                    \vspace*{1cm}
                    \begin{minipage}[b]{10cm}
                    \centering
                    \includegraphics[width=10cm]{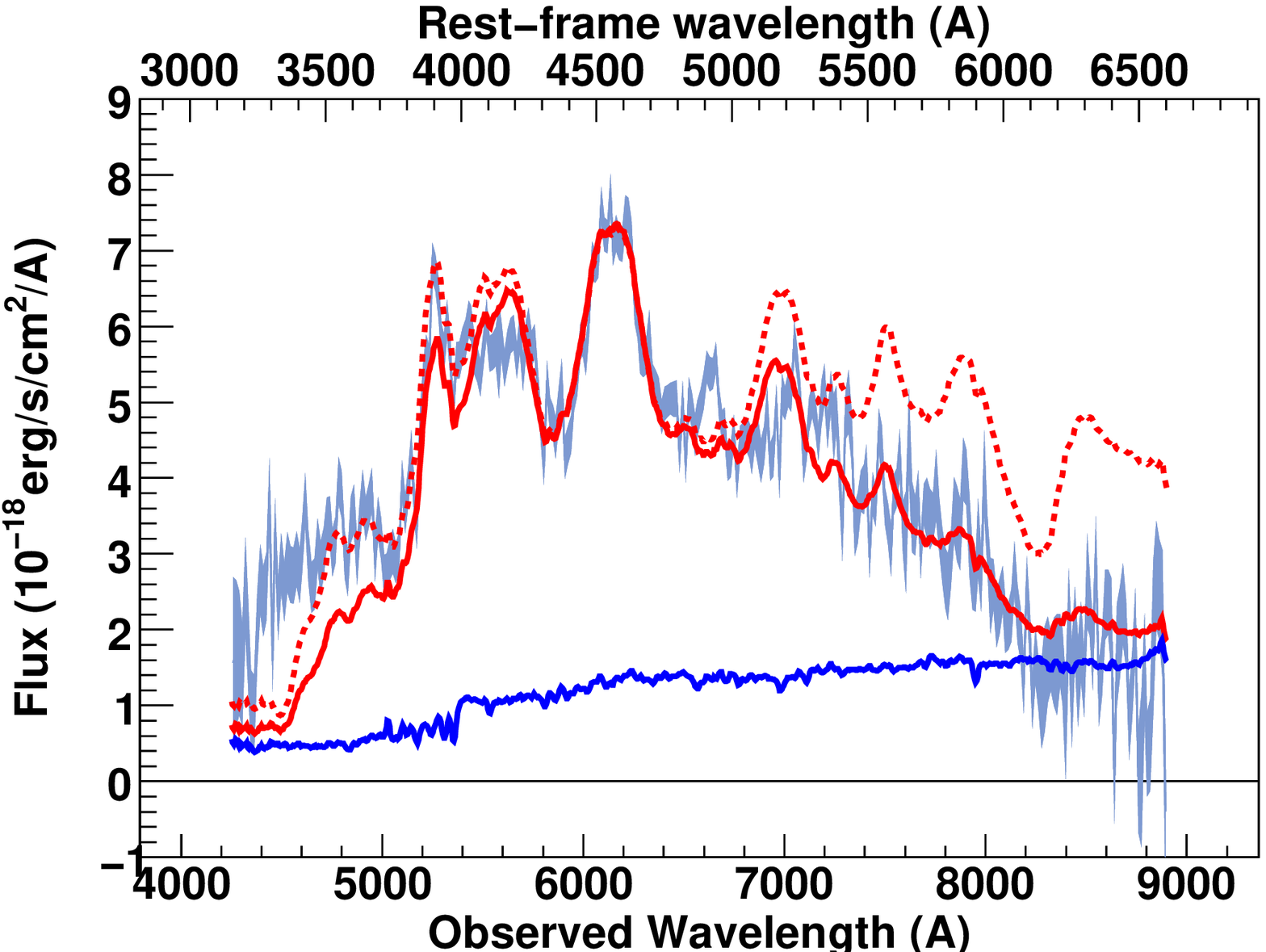}
                    \end{minipage}
                    &
                    \begin{minipage}[b]{10cm}
                    \hspace{-5mm}\includegraphics[width=10cm]{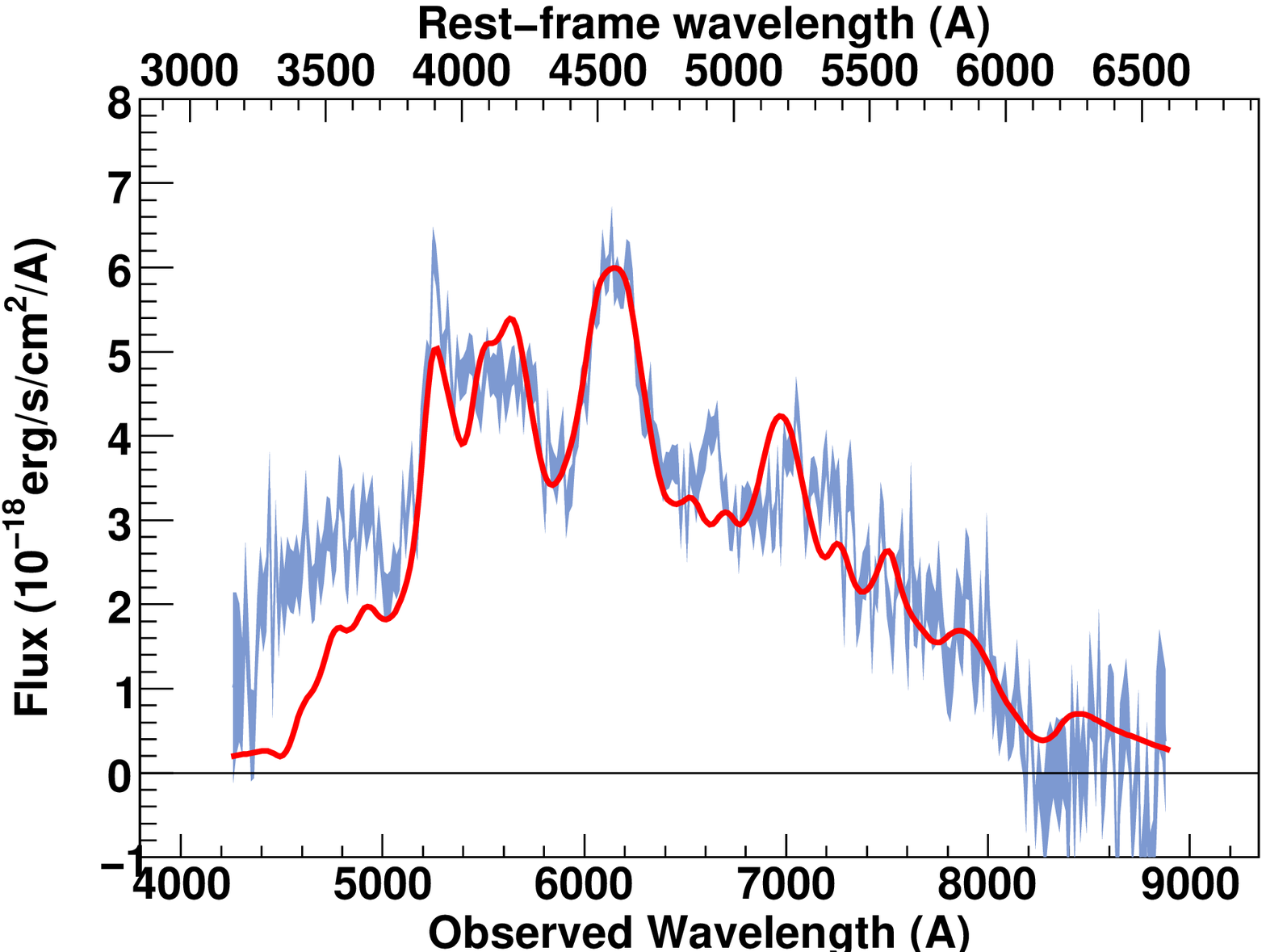}
                    \end{minipage}
                    \\
                    \end{tabular}
                    \caption{\label{fig:04D2mc736} SNLS-  04D2mc736: a SN~Ia supernova at z=0.348. The spectrum phase is 6.5. A       Sa-Sb host model has been subtracted. Three re-calibration parameters have been used to fit this spectrum.}
                    \end{center}
                    \end{figure*}
                    
                    \begin{figure*}[ht]
                    \begin{center}
                    \begin{tabular}[b]{cc}
                    \hspace{-5mm}
                    \vspace*{1cm}
                    \begin{minipage}[b]{10cm}
                    \centering
                    \includegraphics[width=10cm]{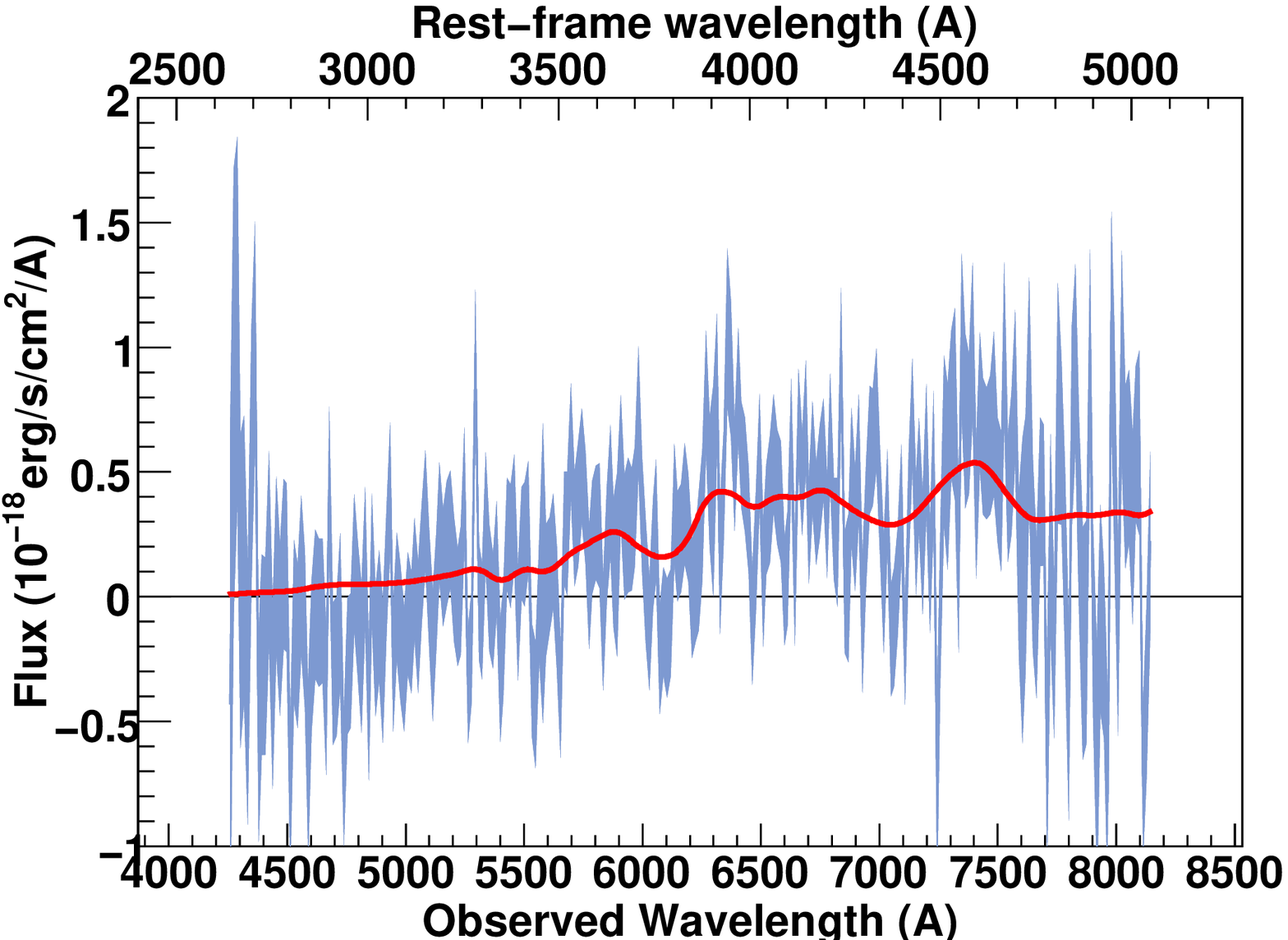}
                    \end{minipage}
                    &
                    \begin{minipage}[b]{10cm}
                    \end{minipage}
                    \\
                    \end{tabular}
                    \caption{\label{fig:04D4an558} SNLS-  04D4an558: a SN~Ia supernova at z=0.613. The spectrum phase is 8.1. Best-fit obtained for a model with no host galaxy component.}
                    \end{center}
                    \end{figure*}

                    \newpage
                    \clearpage
                    
                    \begin{figure*}[ht]
                    \begin{center}
                    \begin{tabular}[b]{cc}
                    \hspace{-5mm}
                    \vspace*{1cm}
                    \begin{minipage}[b]{10cm}
                    \centering
                    \includegraphics[width=10cm]{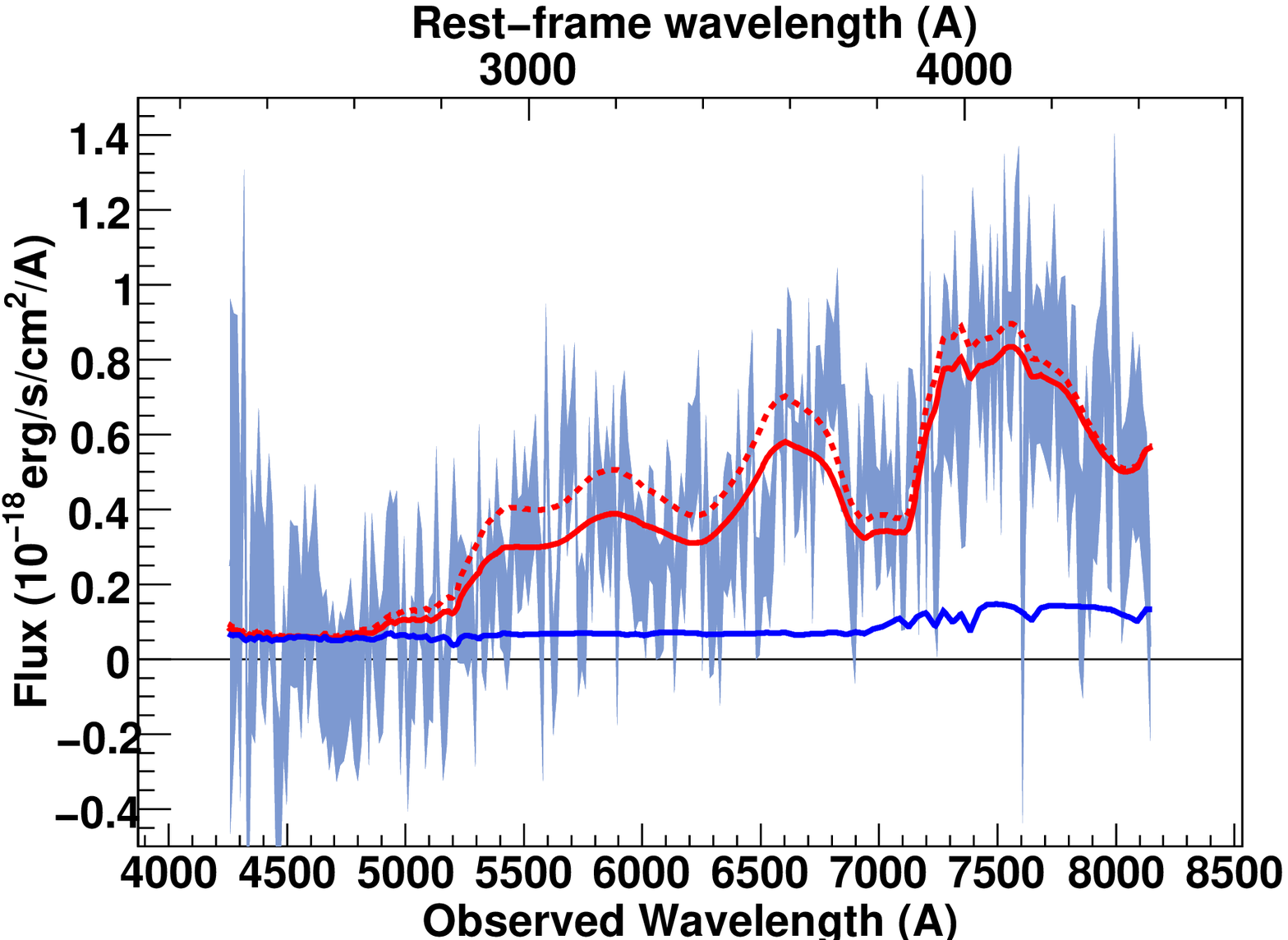}
                    \end{minipage}
                    &
                    \begin{minipage}[b]{10cm}
                    \hspace{-5mm}\includegraphics[width=10cm]{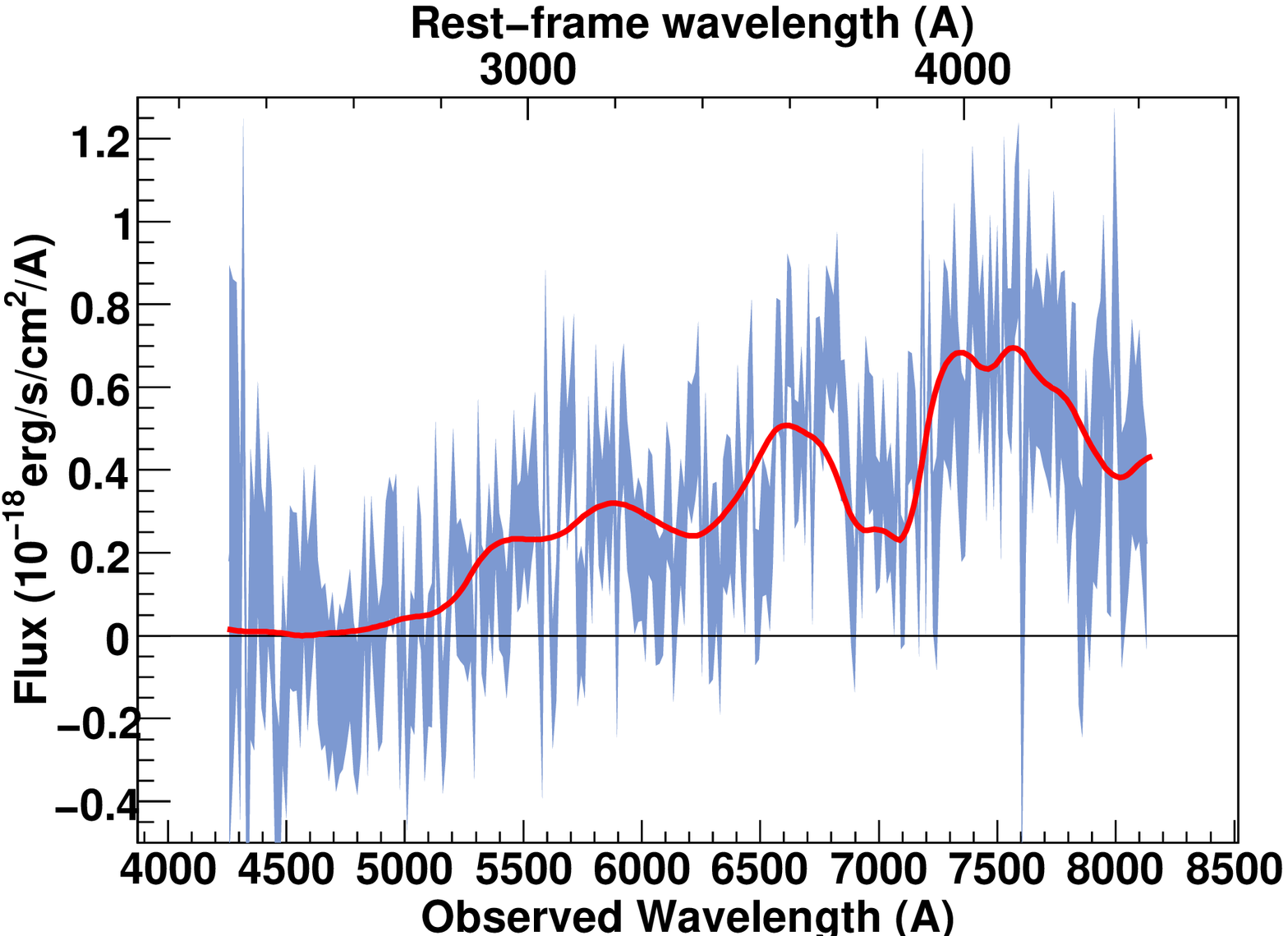}
                    \end{minipage}
                    \\
                    \end{tabular}
                    \caption{\label{fig:04D4bk558} SNLS-  04D4bk558: a SN~Ia$\star$ supernova at z=0.88. The spectrum phase is 3.1. A         E(1) host model has been subtracted.}
                    \end{center}
                    \end{figure*}
                    
                    \begin{figure*}[ht]
                    \begin{center}
                    \begin{tabular}[b]{cc}
                    \hspace{-5mm}
                    \vspace*{1cm}
                    \begin{minipage}[b]{10cm}
                    \centering
                    \includegraphics[width=10cm]{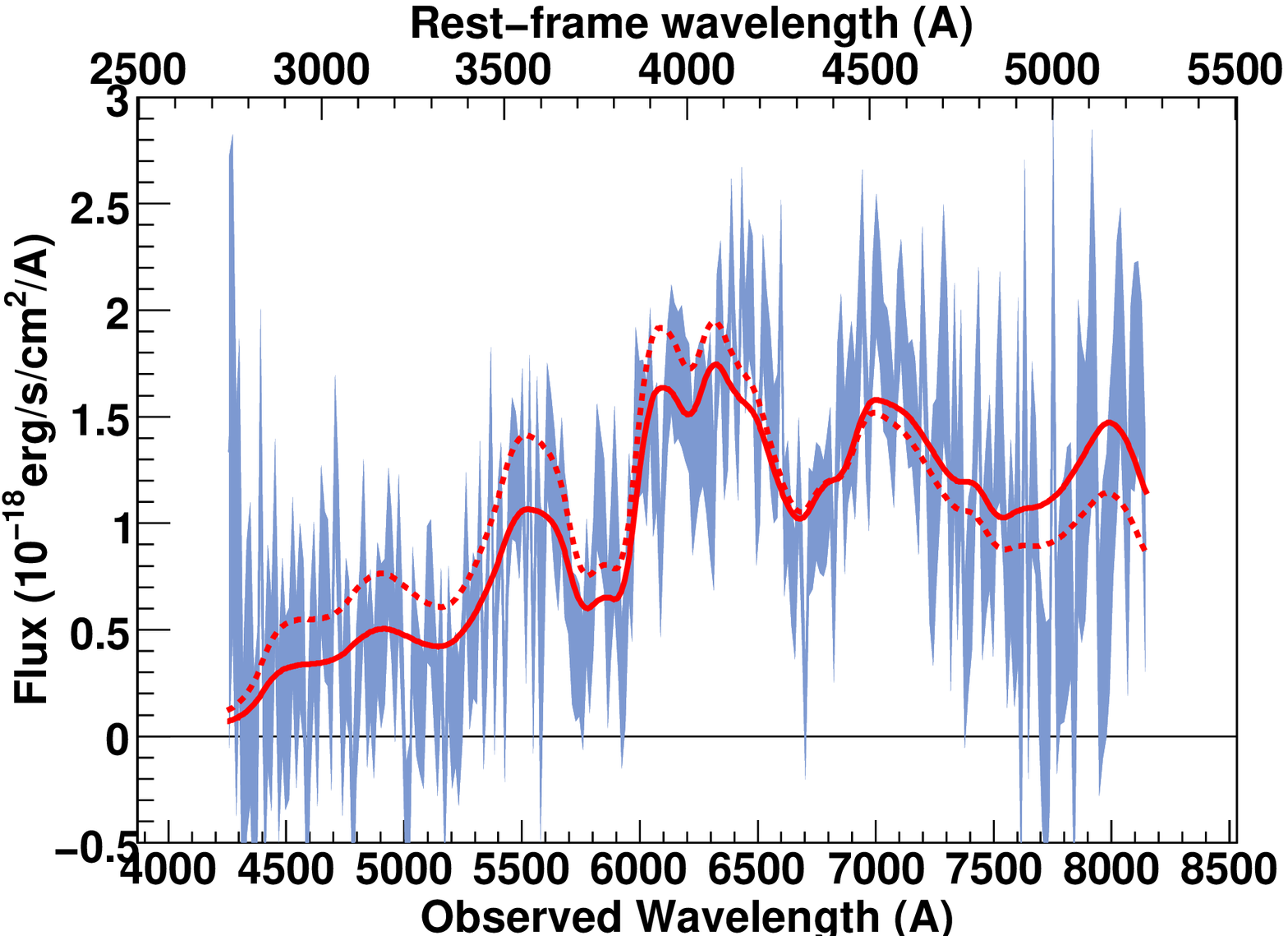}
                    \end{minipage}
                    &
                    \begin{minipage}[b]{10cm}
                    \end{minipage}
                    \\
                    \end{tabular}
                    \caption{\label{fig:04D4bq558} SNLS-  04D4bq558: a SN~Ia supernova at z=0.55. The spectrum phase is 2.6. Best-fit obtained for a model with no host galaxy component.}
                    \end{center}
                    \end{figure*}

                    \newpage
                    \clearpage
                    
                    \begin{figure*}[ht]
                    \begin{center}
                    \begin{tabular}[b]{cc}
                    \hspace{-5mm}
                    \vspace*{1cm}
                    \begin{minipage}[b]{10cm}
                    \centering
                    \includegraphics[width=10cm]{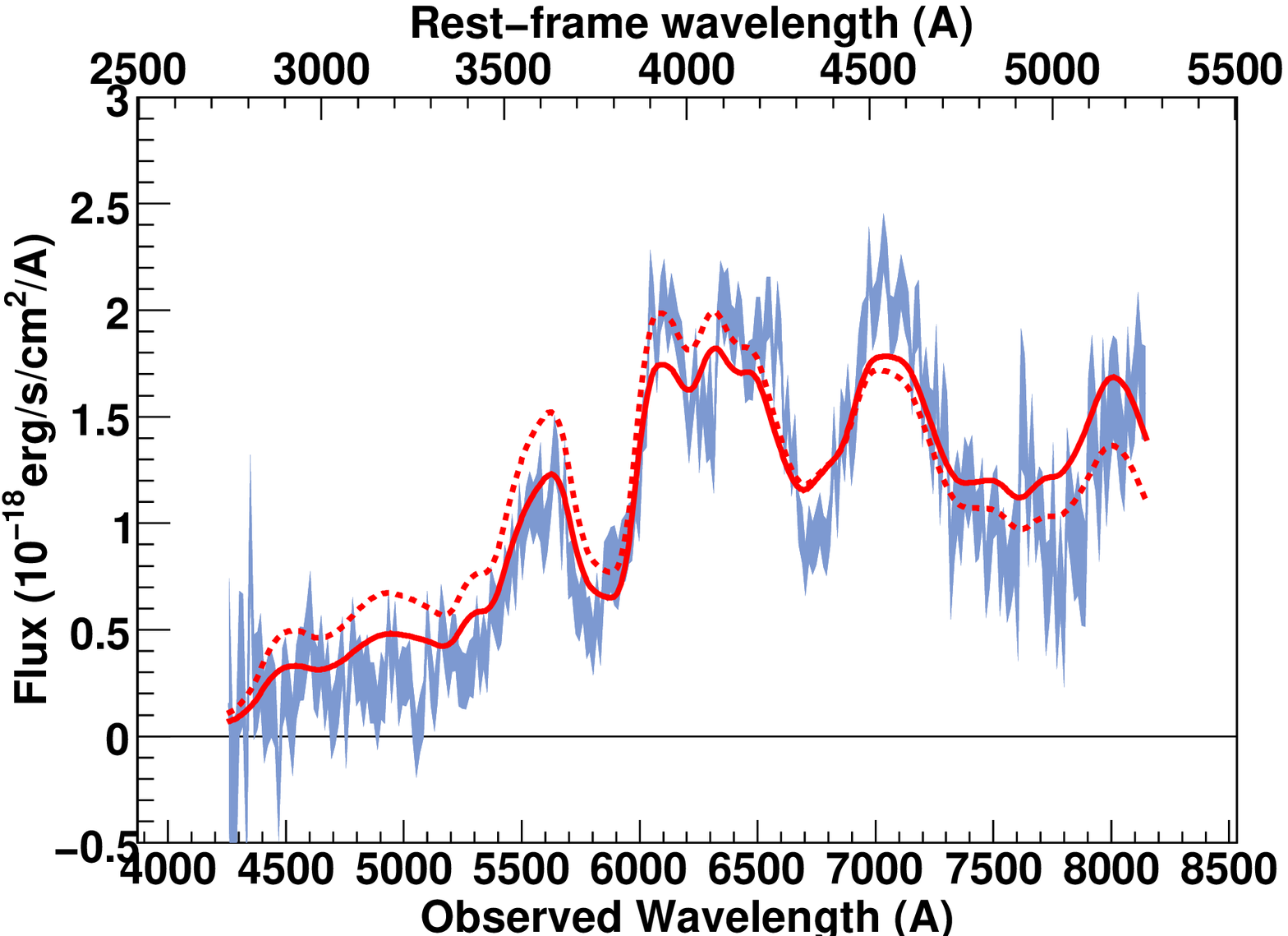}
                    \end{minipage}
                    &
                    \begin{minipage}[b]{10cm}
                    \end{minipage}
                    \\
                    \end{tabular}
                    \caption{\label{fig:04D4bq562} SNLS-  04D4bq562: a SN~Ia supernova at z=0.55. The spectrum phase is 5.1. Best-fit obtained for a model with no host galaxy component.}
                    \end{center}
                    \end{figure*}
                    
                    \begin{figure*}[ht]
                    \begin{center}
                    \begin{tabular}[b]{cc}
                    \hspace{-5mm}
                    \vspace*{1cm}
                    \begin{minipage}[b]{10cm}
                    \centering
                    \includegraphics[width=10cm]{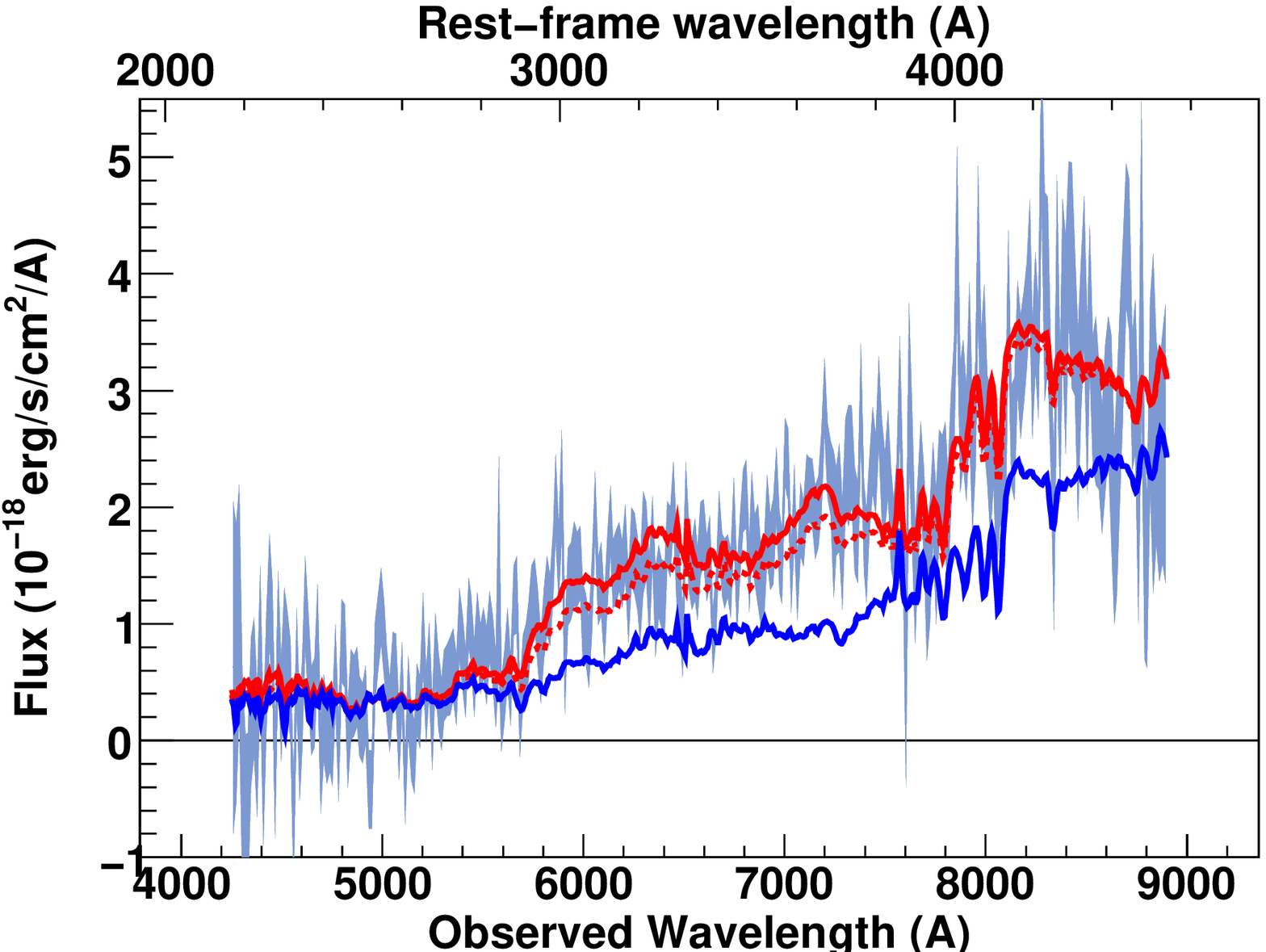}
                    \end{minipage}
                    &
                    \begin{minipage}[b]{10cm}
                    \hspace{-5mm}\includegraphics[width=10cm]{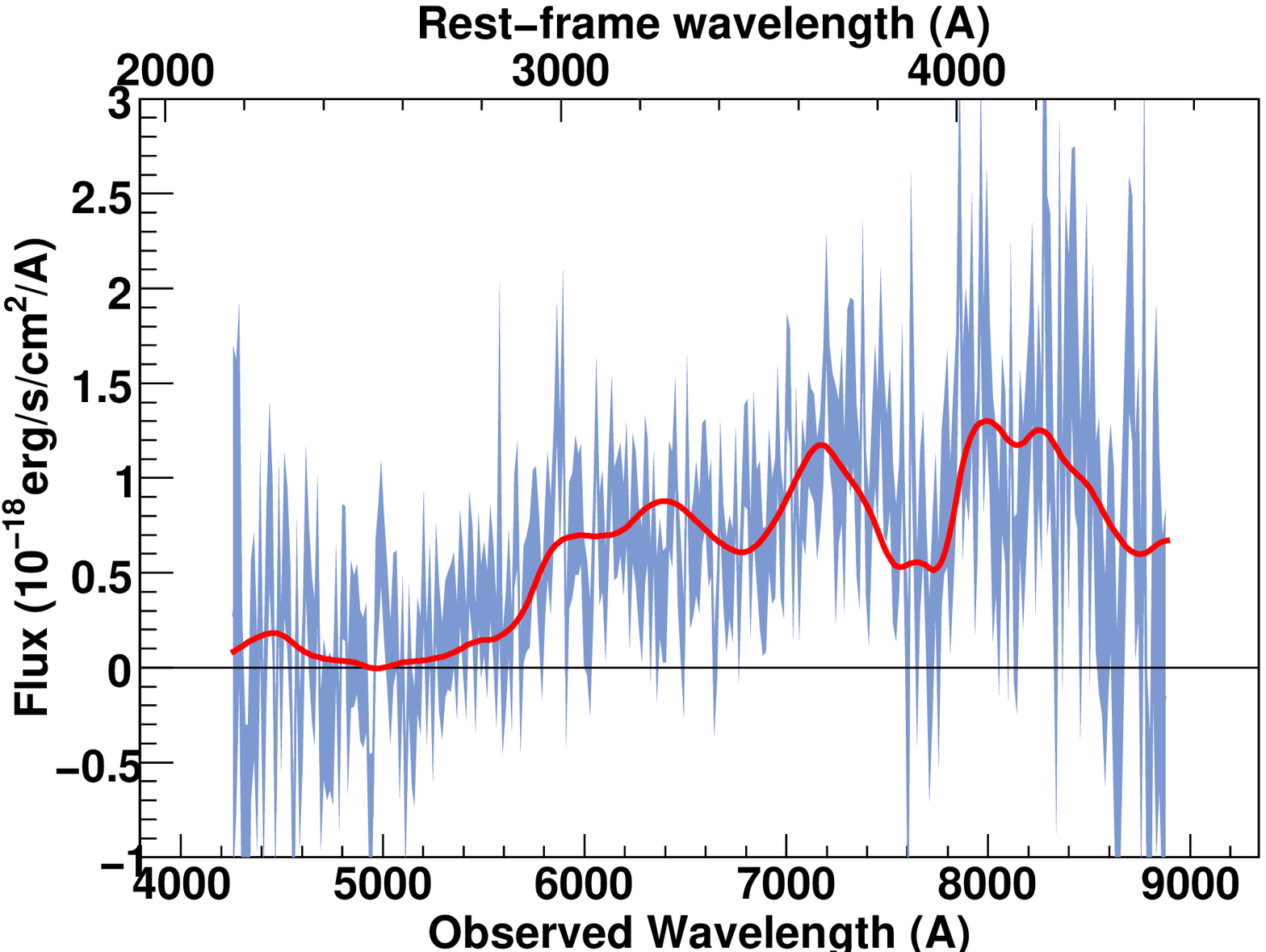}
                    \end{minipage}
                    \\
                    \end{tabular}
                    \caption{\label{fig:04D4dw566} SNLS-  04D4dw566: a SN~Ia$\star$ supernova at z=1.031. The spectrum phase is 2.1. A         Sa-Sb host model has been subtracted.}
                    \end{center}
                    \end{figure*}

                    \newpage
                    \clearpage
                    
                    \begin{figure*}[ht]
                    \begin{center}
                    \begin{tabular}[b]{cc}
                    \hspace{-5mm}
                    \vspace*{1cm}
                    \begin{minipage}[b]{10cm}
                    \centering
                    \includegraphics[width=10cm]{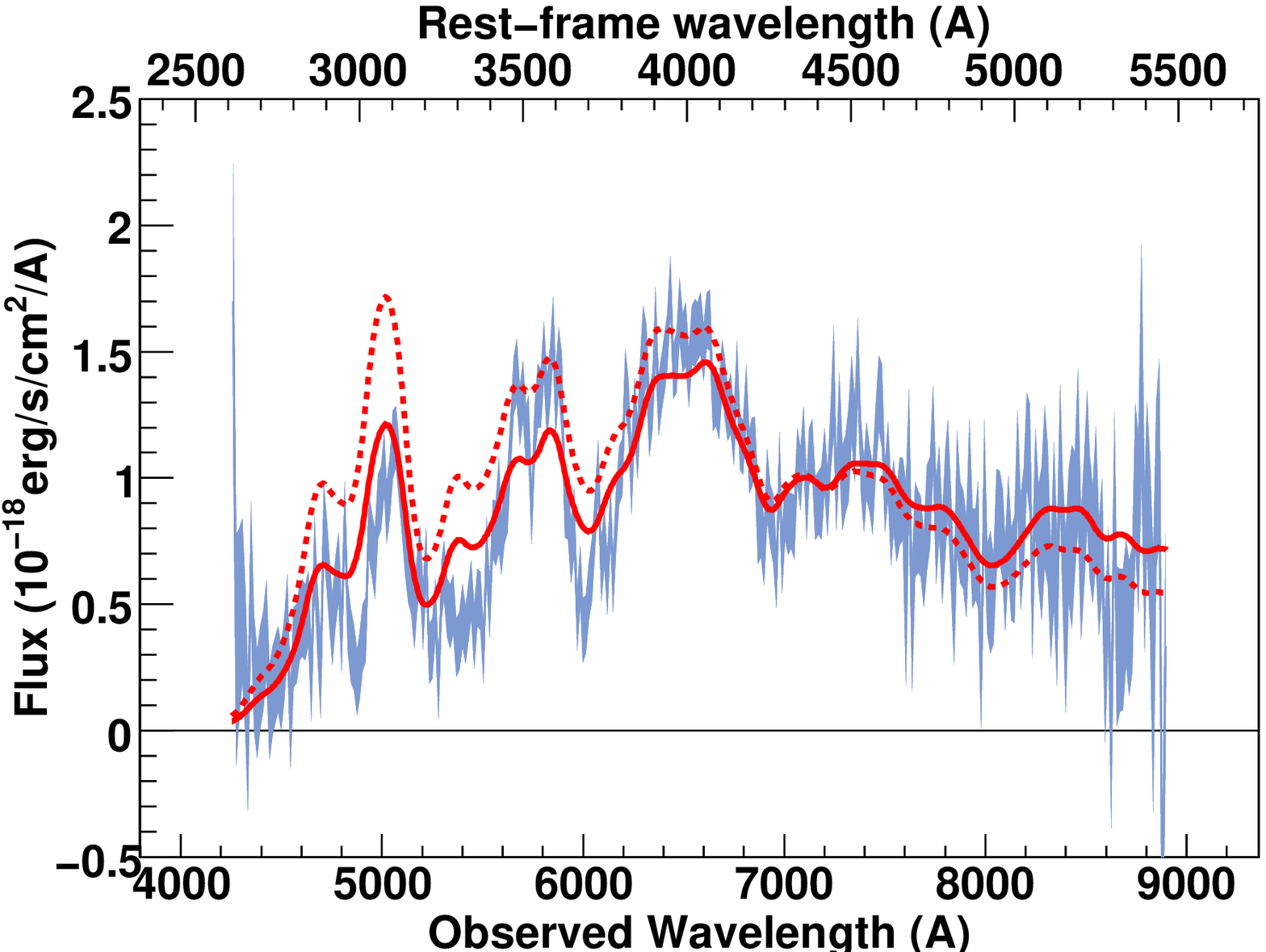}
                    \end{minipage}
                    &
                    \begin{minipage}[b]{10cm}
                    \end{minipage}
                    \\
                    \end{tabular}
                    \caption{\label{fig:04D4fx589} SNLS-  04D4fx589: a SN~Ia supernova at z=0.629. The spectrum phase is -8.4. Best-fit obtained for a model with no host galaxy component.}
                    \end{center}
                    \end{figure*}
                    
                    \begin{figure*}[ht]
                    \begin{center}
                    \begin{tabular}[b]{cc}
                    \hspace{-5mm}
                    \vspace*{1cm}
                    \begin{minipage}[b]{10cm}
                    \centering
                    \includegraphics[width=10cm]{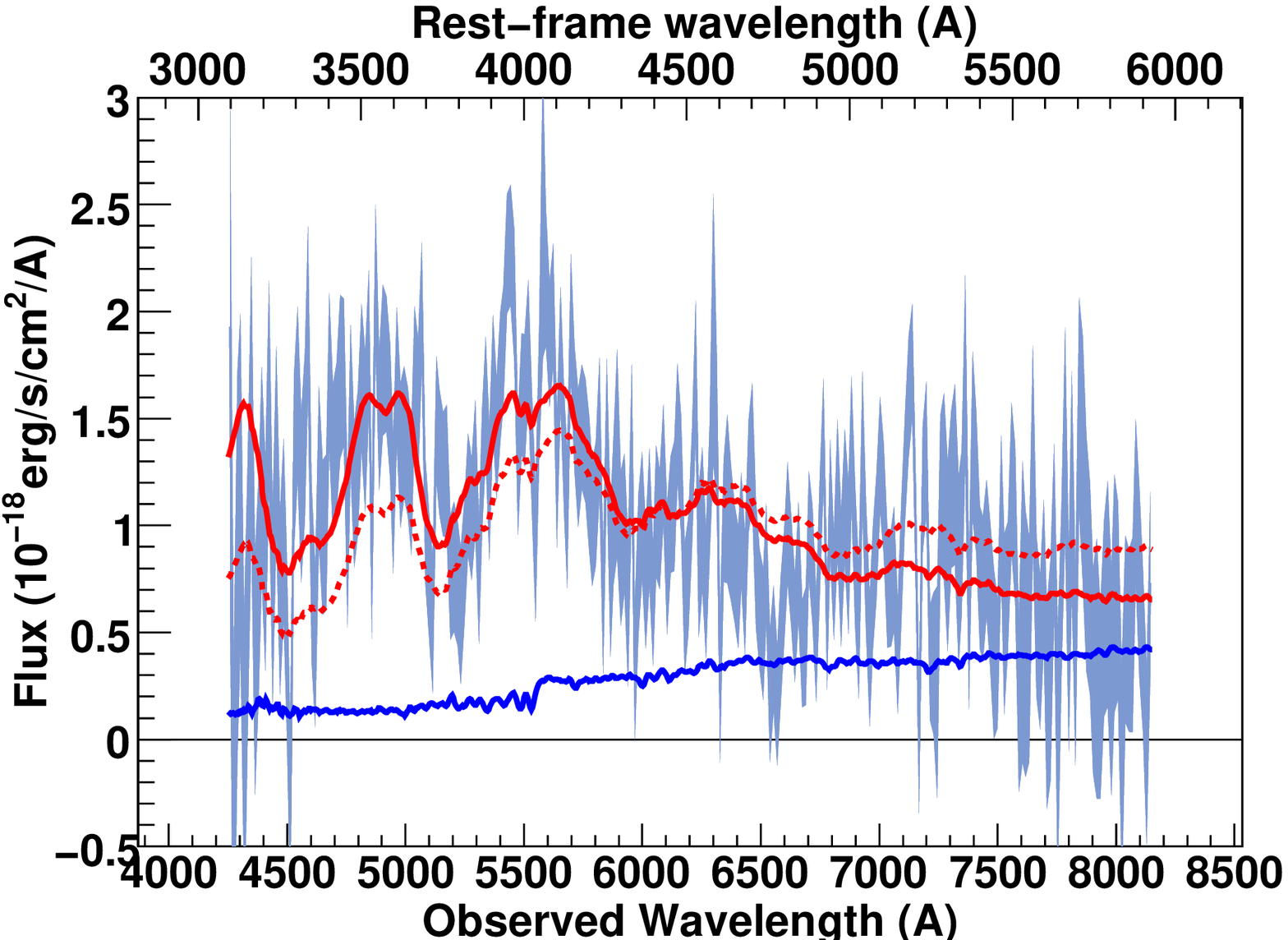}
                    \end{minipage}
                    &
                    \begin{minipage}[b]{10cm}
                    \hspace{-5mm}\includegraphics[width=10cm]{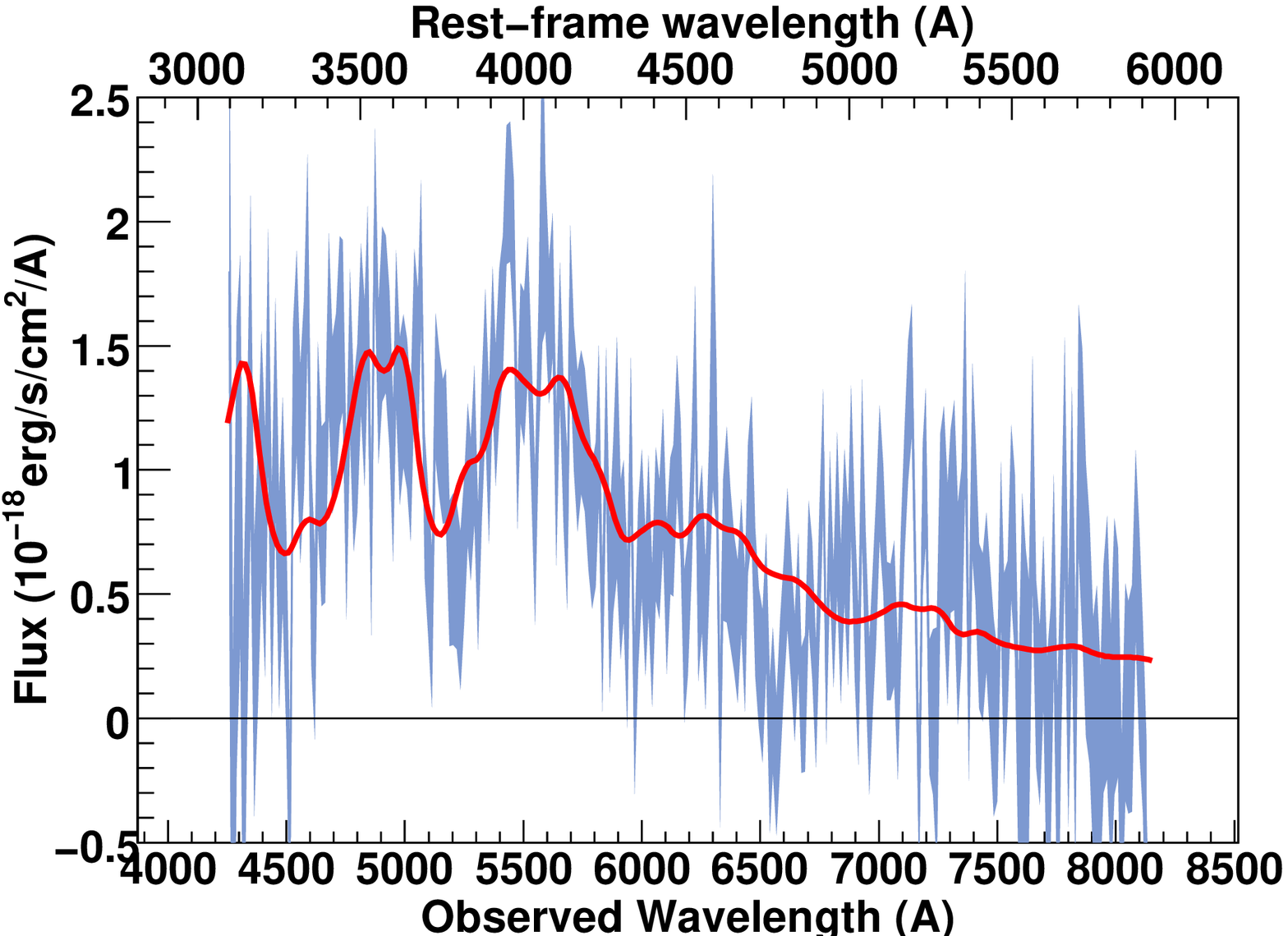}
                    \end{minipage}
                    \\
                    \end{tabular}
                    \caption{\label{fig:04D4gz598} SNLS-  04D4gz598: a SN~Ia$\star$ supernova at z=0.375. The spectrum phase is -5.8. A         Sa-Sb host model has been subtracted.}
                    \end{center}
                    \end{figure*}

                    \newpage
                    \clearpage
                    
                    \begin{figure*}[ht]
                    \begin{center}
                    \begin{tabular}[b]{cc}
                    \hspace{-5mm}
                    \vspace*{1cm}
                    \begin{minipage}[b]{10cm}
                    \centering
                    \includegraphics[width=10cm]{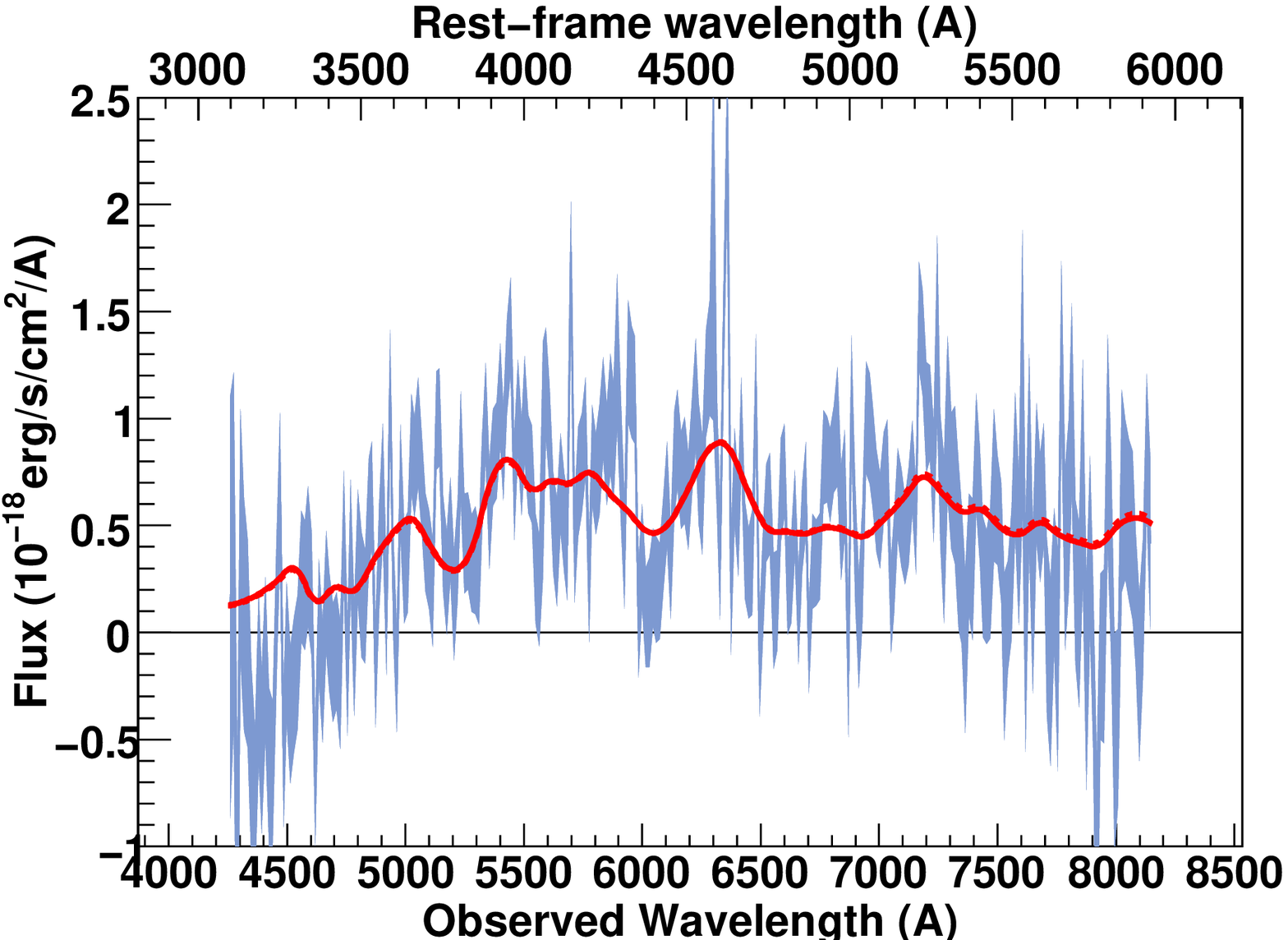}
                    \end{minipage}
                    &
                    \begin{minipage}[b]{10cm}
                    \end{minipage}
                    \\
                    \end{tabular}
                    \caption{\label{fig:04D4gz620} SNLS-  04D4gz620: a SN~Ia$\star$ supernova at z=0.375. The spectrum phase is 10.2. Best-fit obtained for a model with no host galaxy component.}
                    \end{center}
                    \end{figure*}

                    \begin{figure*}[ht]
                    \begin{center}
                    \begin{tabular}[b]{cc}
                    \hspace{-5mm}
                    \vspace*{1cm}
                    \begin{minipage}[b]{10cm}
                    \centering
                    \includegraphics[width=10cm]{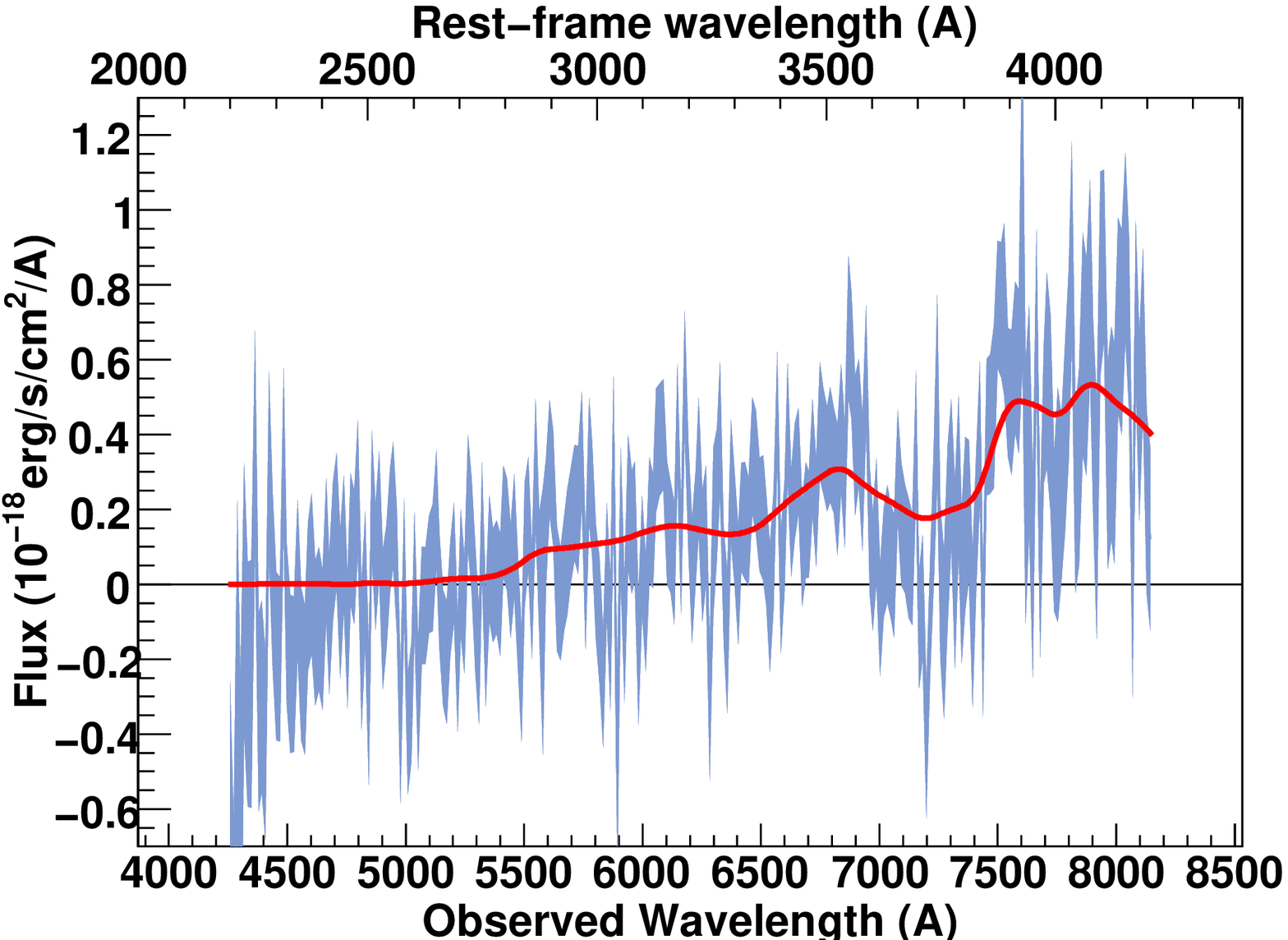}
                    \end{minipage}
                    &
                    \begin{minipage}[b]{10cm}
                    \end{minipage}
                    \\
                    \end{tabular}
                    \caption{\label{fig:04D4hf598} SNLS-  04D4hf598: a SN~Ia$\star$ supernova at z=0.94. The spectrum phase is -0.5. Best-fit obtained for a model with no host galaxy component.}
                    \end{center}
                    \end{figure*}

                    \newpage
                    \clearpage
                    
                    \begin{figure*}[ht]
                    \begin{center}
                    \begin{tabular}[b]{cc}
                    \hspace{-5mm}
                    \vspace*{1cm}
                    \begin{minipage}[b]{10cm}
                    \centering
                    \includegraphics[width=10cm]{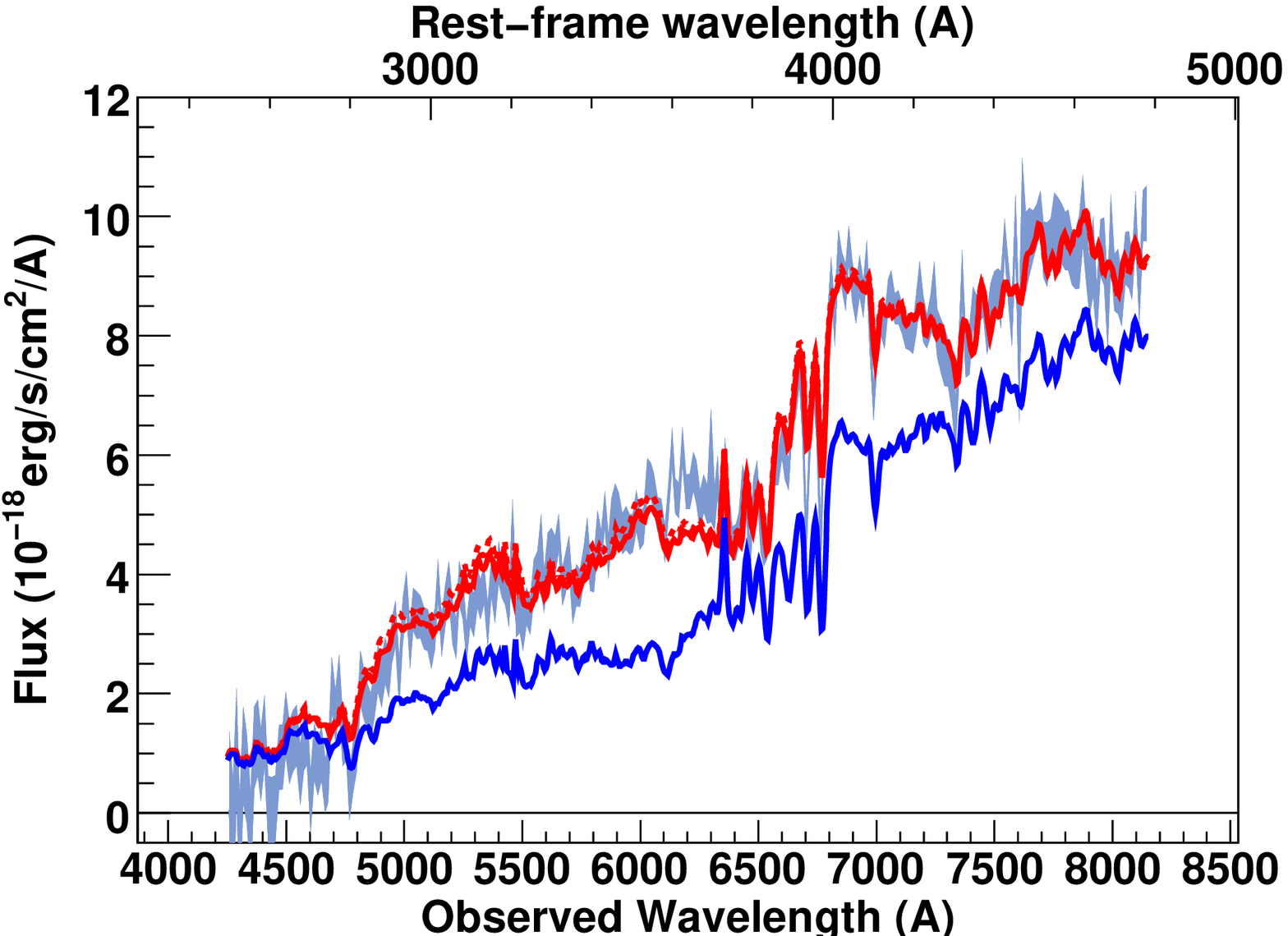}
                    \end{minipage}
                    &
                    \begin{minipage}[b]{10cm}
                    \hspace{-5mm}\includegraphics[width=10cm]{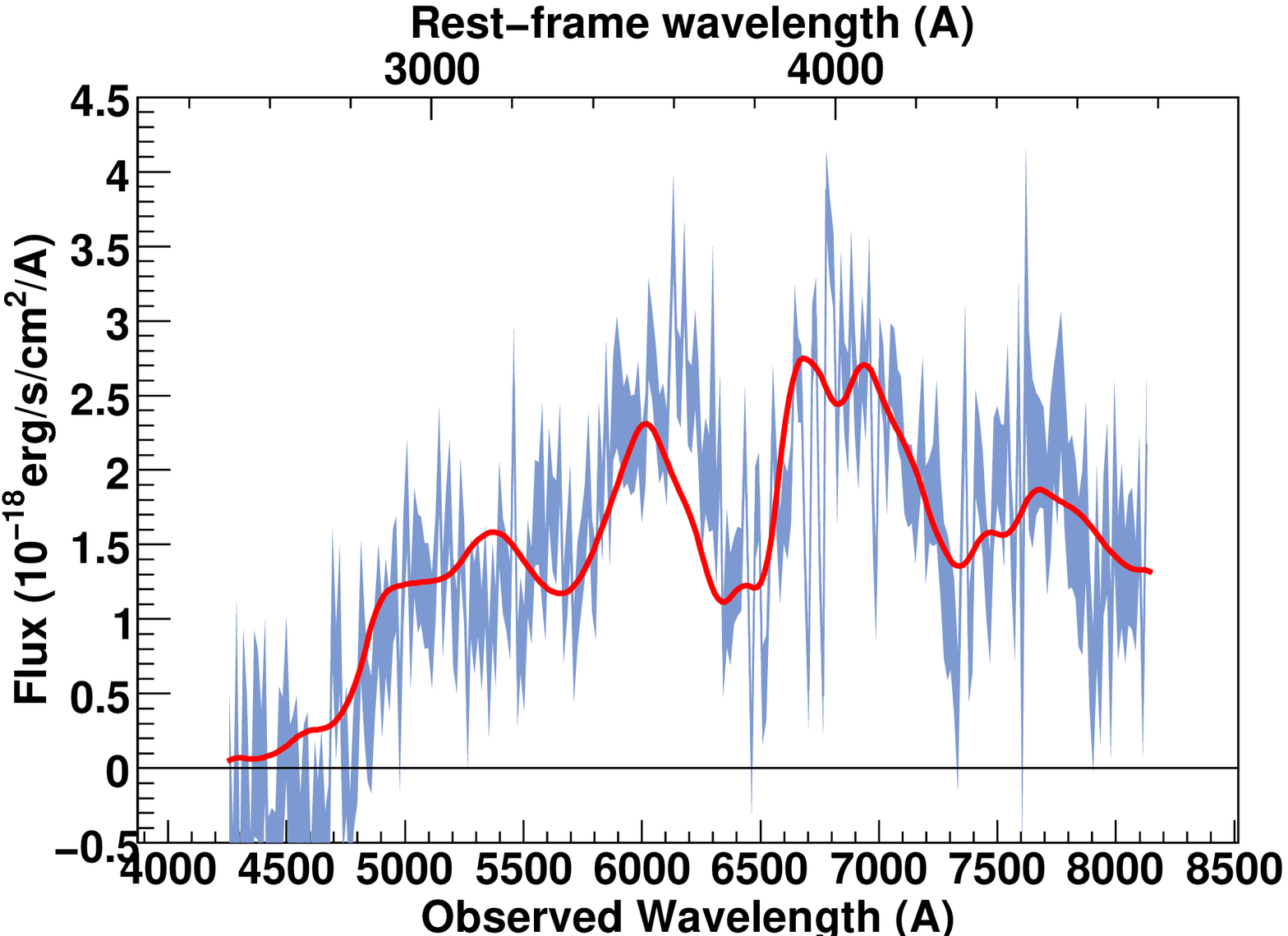}
                    \end{minipage}
                    \\
                    \end{tabular}
                    \caption{\label{fig:04D4ib629} SNLS-  04D4ib629: a SN~Ia supernova at z=0.699. The spectrum phase is 0.7. A S0-Sa host model has been subtracted.}
                    \end{center}
                    \end{figure*}

                    \newpage
                    \clearpage
                    
                    \begin{figure*}[ht]
                    \begin{center}
                    \begin{tabular}[b]{cc}
                    \hspace{-5mm}
                    \vspace*{1cm}
                    \begin{minipage}[b]{10cm}
                    \centering
                    \includegraphics[width=10cm]{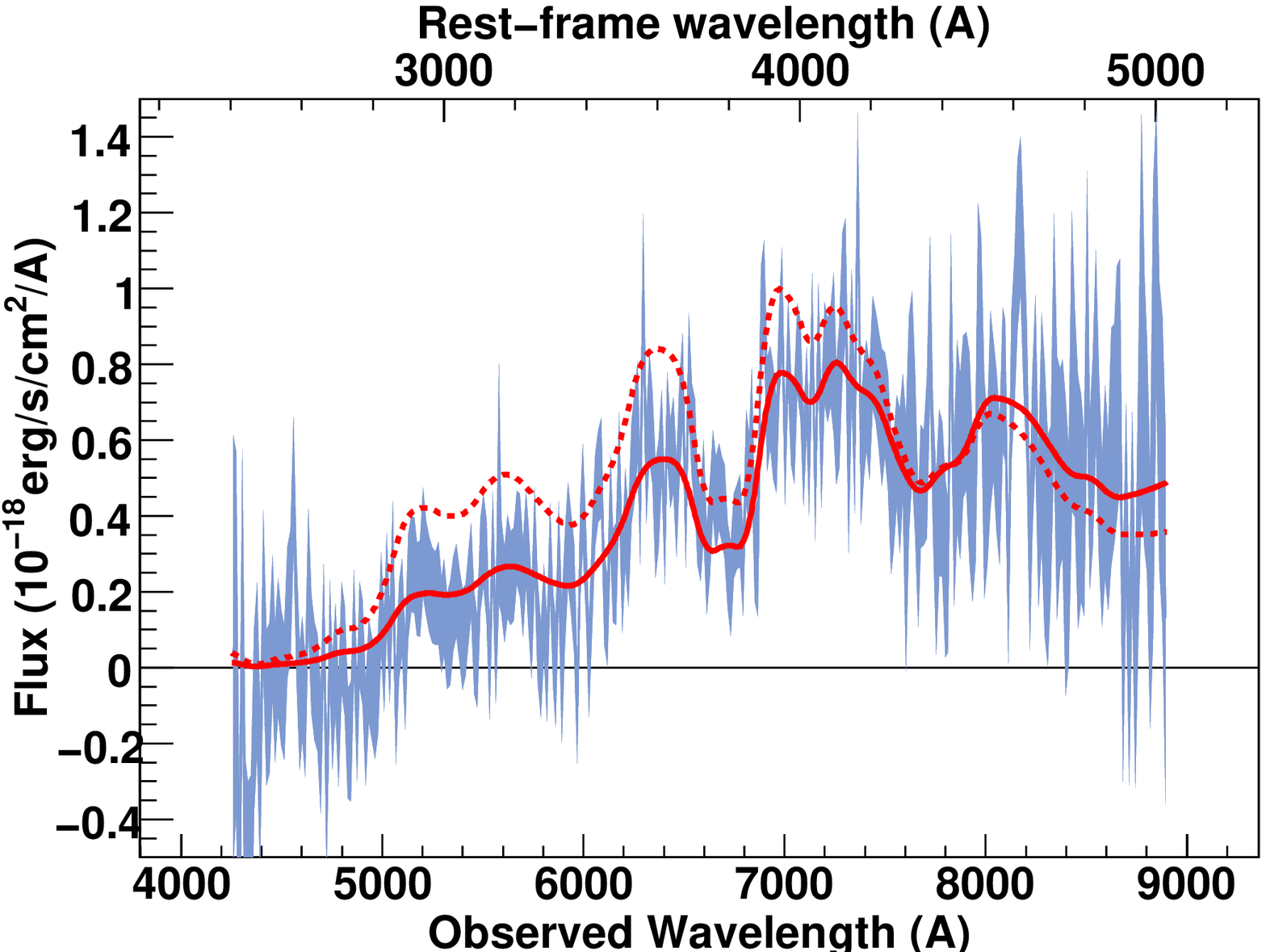}
                    \end{minipage}
                    &
                    \begin{minipage}[b]{10cm}
                    \end{minipage}
                    \\
                    \end{tabular}
                    \caption{\label{fig:04D4id628} SNLS-  04D4id628: a SN~Ia supernova at z=0.769. The spectrum phase is 3.0. Best-fit obtained for a model with no host galaxy component.}
                    \end{center}
                    \end{figure*}
                    
                    \begin{figure*}[ht]
                    \begin{center}
                    \begin{tabular}[b]{cc}
                    \hspace{-5mm}
                    \vspace*{1cm}
                    \begin{minipage}[b]{10cm}
                    \centering
                    \includegraphics[width=10cm]{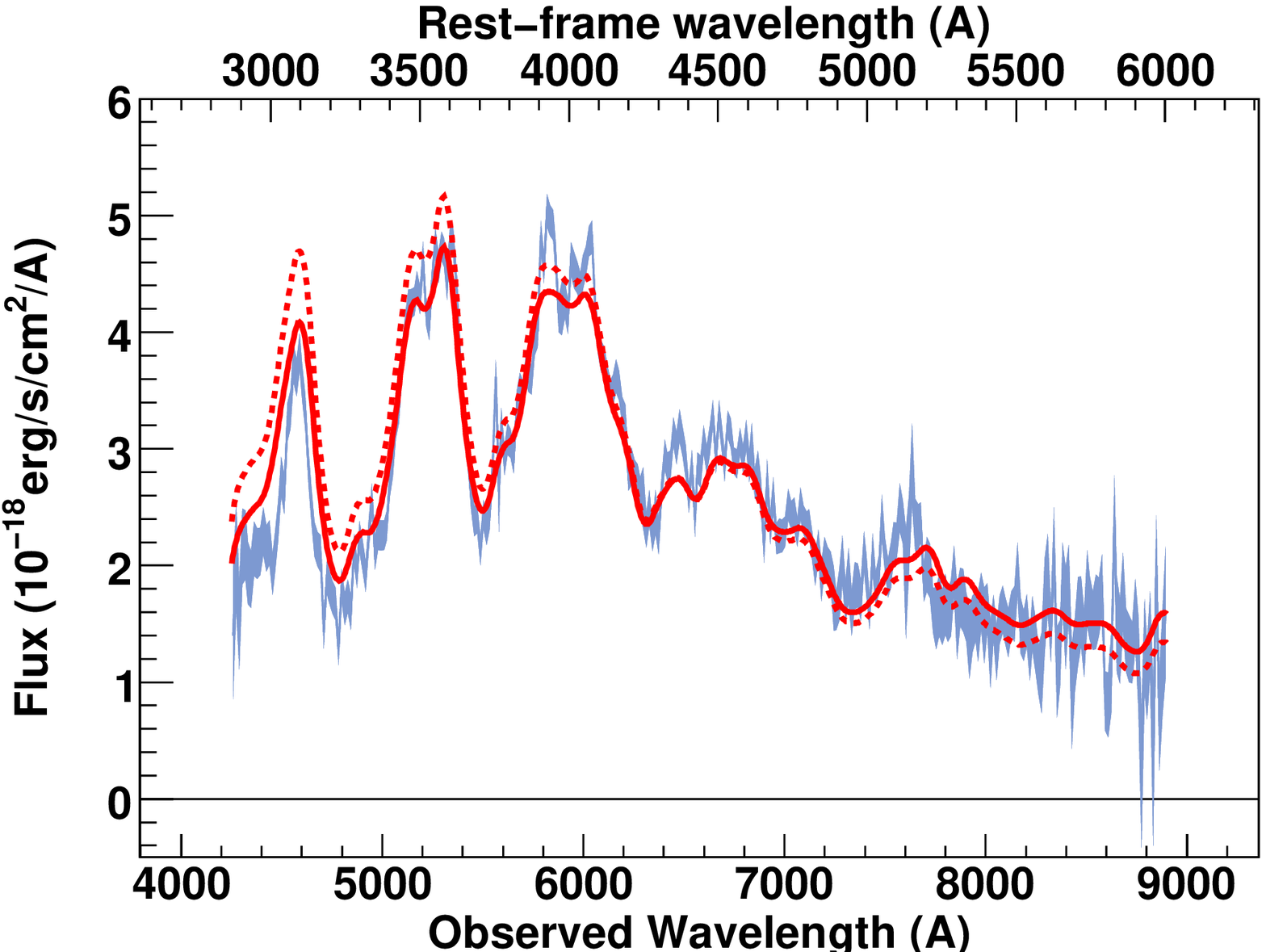}
                    \end{minipage}
                    &
                    \begin{minipage}[b]{10cm}
                    \end{minipage}
                    \\
                    \end{tabular}
                    \caption{\label{fig:04D4jr650} SNLS-  04D4jr650: a SN~Ia supernova at z=0.47. The spectrum phase is -5.9. Best-fit obtained for a model with no host galaxy component.}
                    \end{center}
                    \end{figure*}

                    \newpage
                    \clearpage
                    
                    \begin{figure*}[ht]
                    \begin{center}
                    \begin{tabular}[b]{cc}
                    \hspace{-5mm}
                    \vspace*{1cm}
                    \begin{minipage}[b]{10cm}
                    \centering
                    \includegraphics[width=10cm]{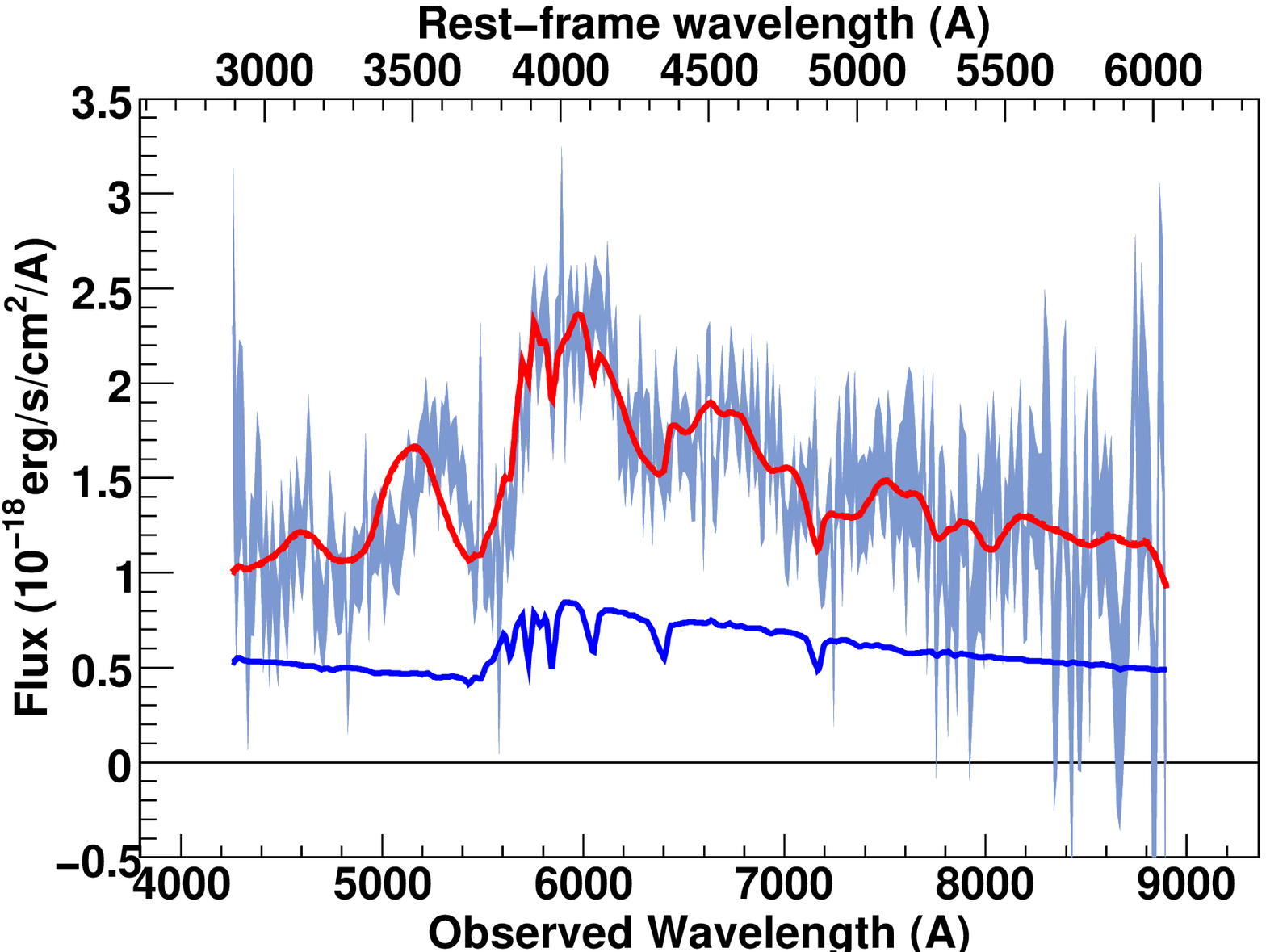}
                    \end{minipage}
                    &
                    \begin{minipage}[b]{10cm}
                    \hspace{-5mm}\includegraphics[width=10cm]{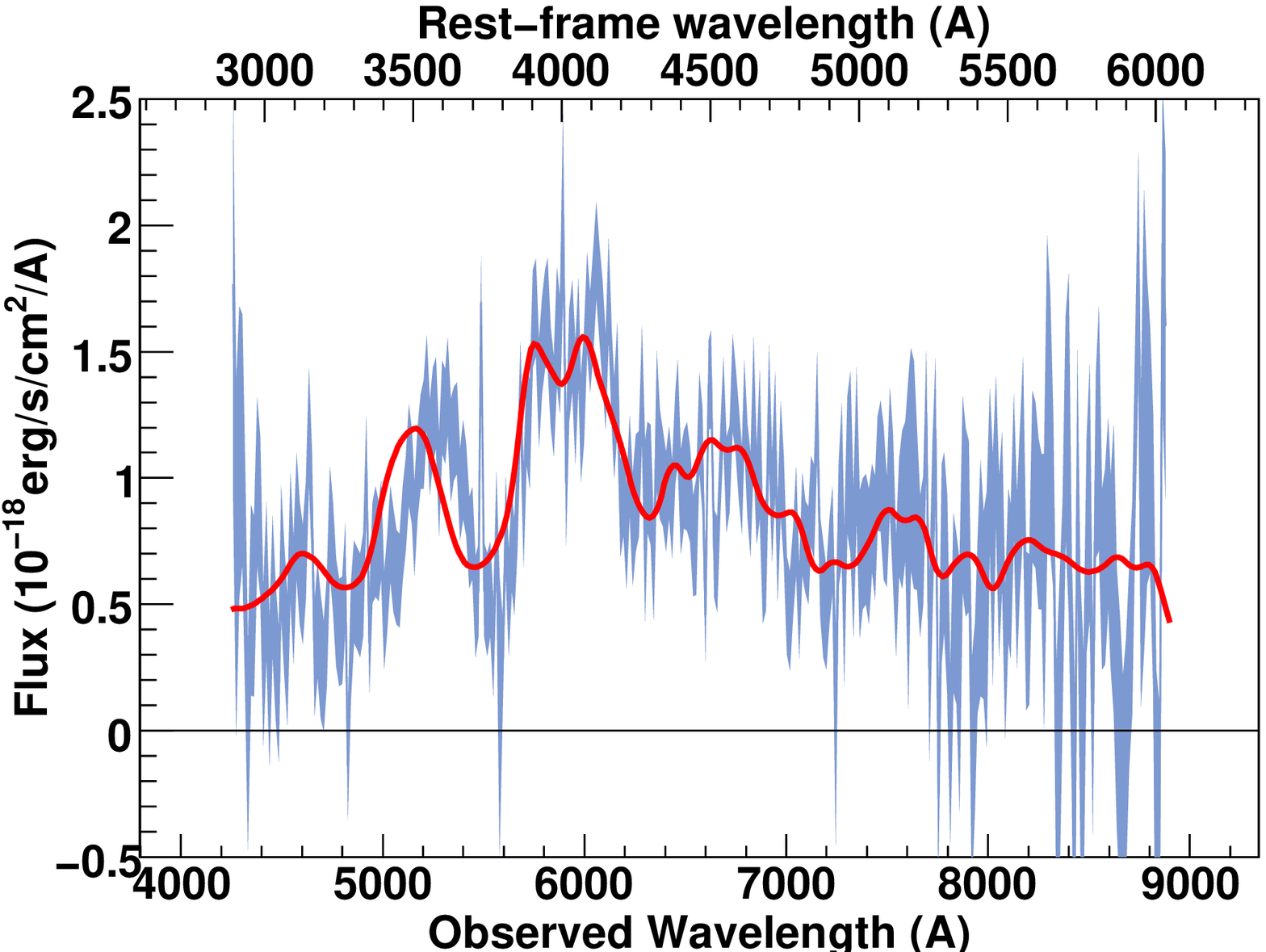}
                    \end{minipage}
                    \\
                    \end{tabular}
                    \caption{\label{fig:04D4ju650} SNLS-  04D4ju650: a SN~Ia supernova at z=0.472. The spectrum phase is -2.2. A        S0(1) host model has been subtracted.}
                    \end{center}
                    \end{figure*}
                    
                    \begin{figure*}[ht]
                    \begin{center}
                    \begin{tabular}[b]{cc}
                    \hspace{-5mm}
                    \vspace*{1cm}
                    \begin{minipage}[b]{10cm}
                    \centering
                    \includegraphics[width=10cm]{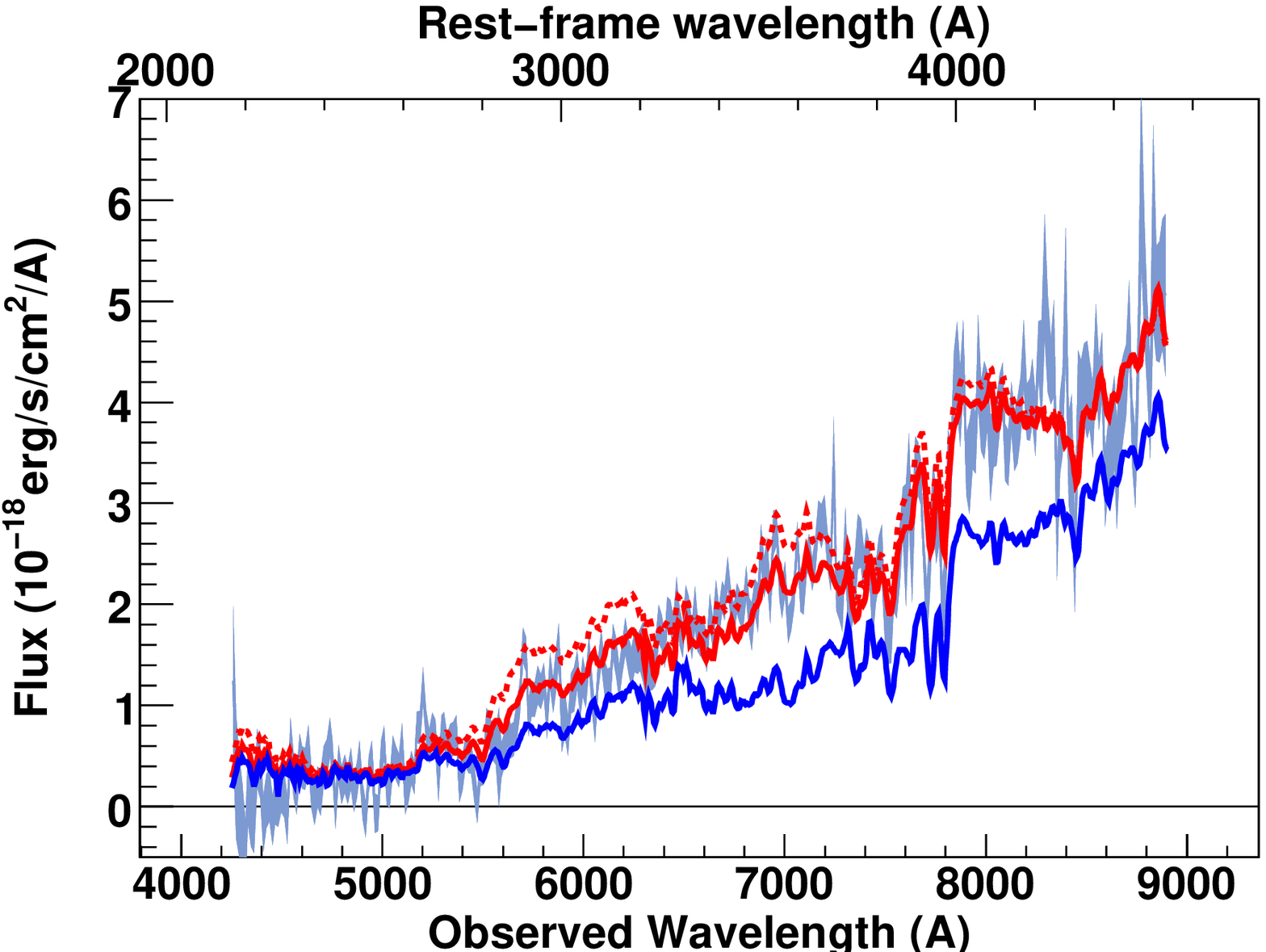}
                    \end{minipage}
                    &
                    \begin{minipage}[b]{10cm}
                    \hspace{-5mm}\includegraphics[width=10cm]{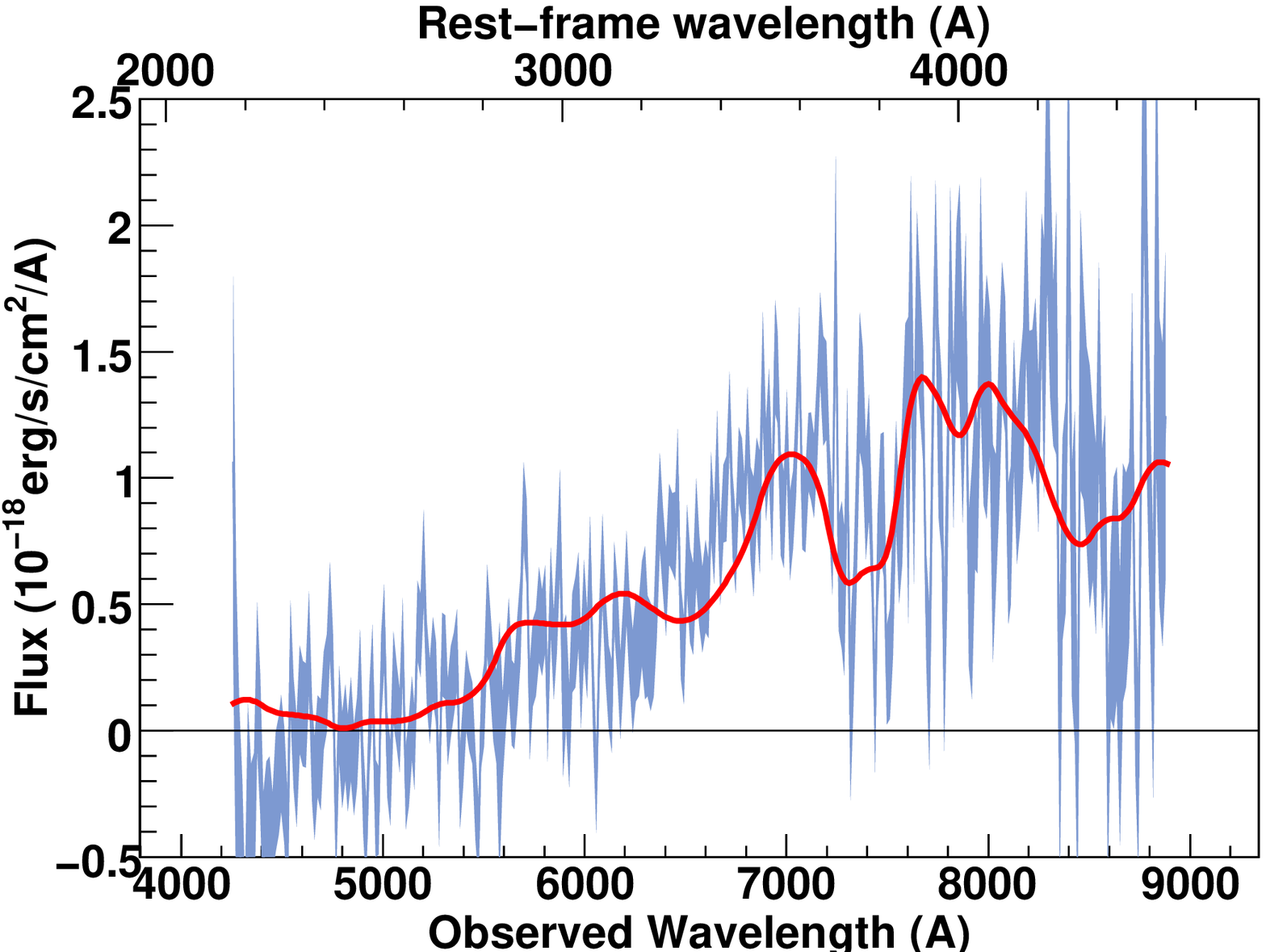}
                    \end{minipage}
                    \\
                    \end{tabular}
                    \caption{\label{fig:04D4jw651} SNLS-  04D4jw651: a SN~Ia$\star$ supernova at z=0.961. The spectrum phase is 2.2. A S0-Sa host model has been subtracted.}
                    \end{center}
                    \end{figure*}

                    \newpage
                    \clearpage
                    
                    \begin{figure*}[ht]
                    \begin{center}
                    \begin{tabular}[b]{cc}
                    \hspace{-5mm}
                    \vspace*{1cm}
                    \begin{minipage}[b]{10cm}
                    \centering
                    \includegraphics[width=10cm]{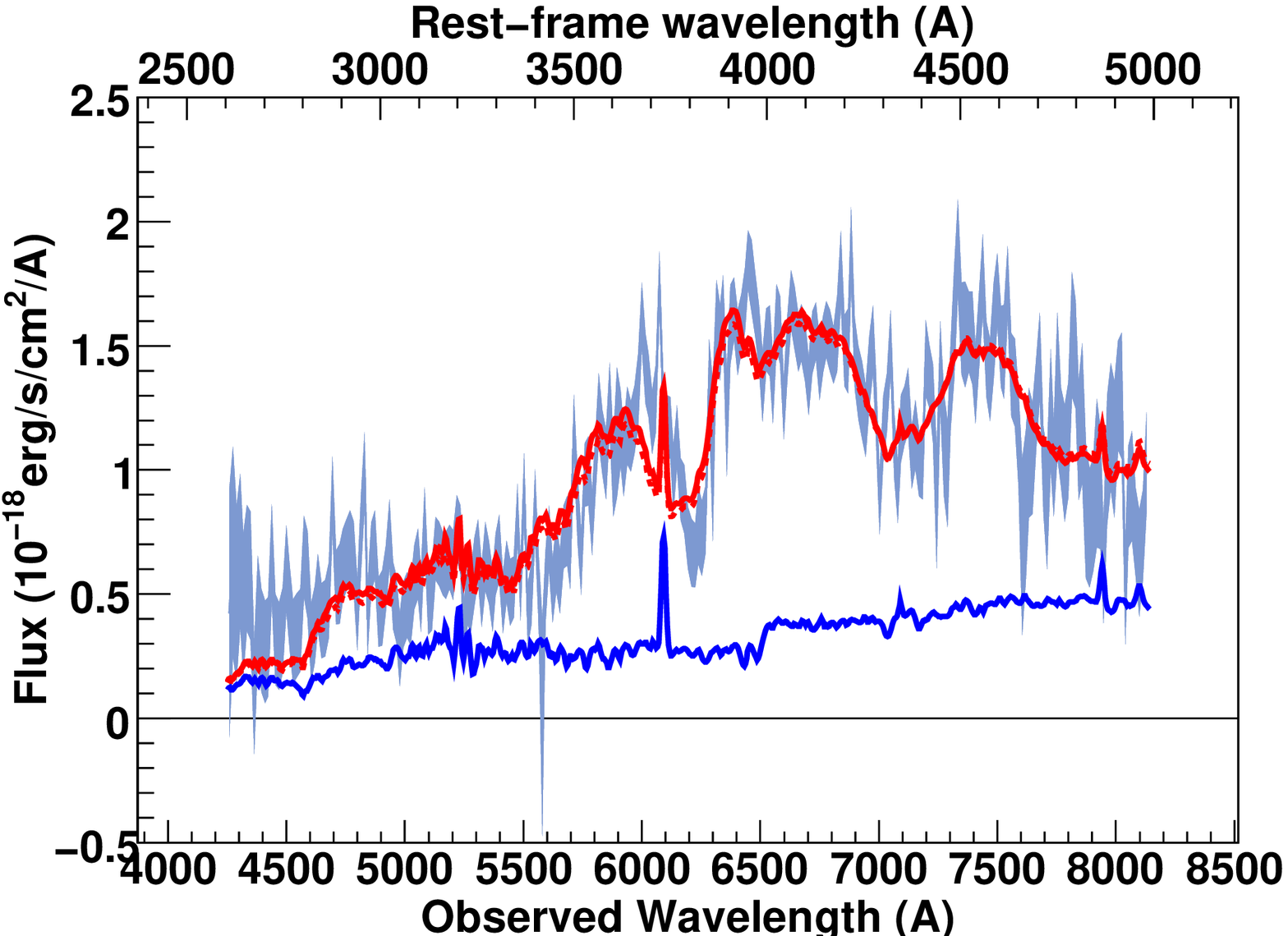}
                    \end{minipage}
                    &
                    \begin{minipage}[b]{10cm}
                    \hspace{-5mm}\includegraphics[width=10cm]{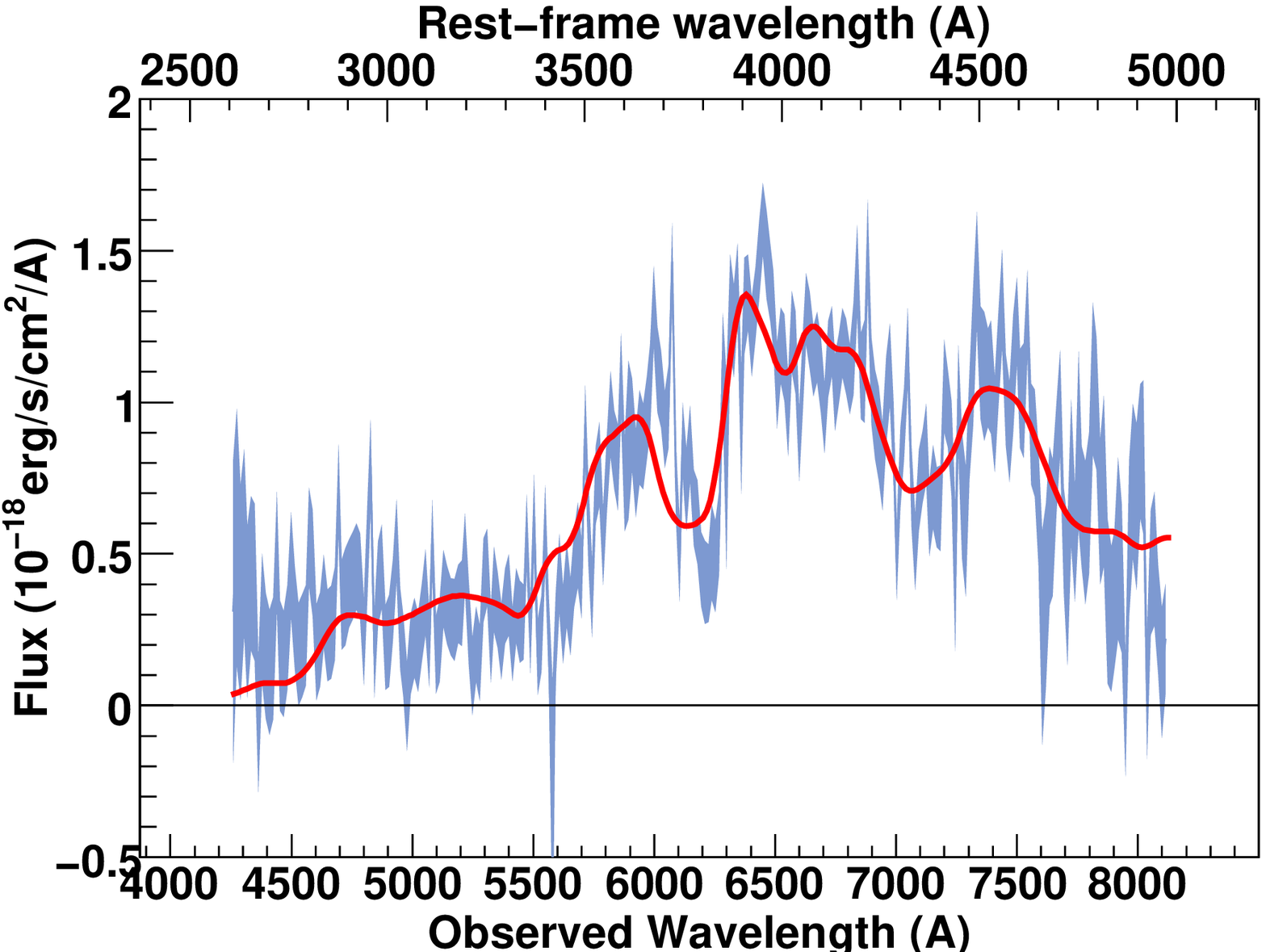}
                    \end{minipage}
                    \\
                    \end{tabular}
                    \caption{\label{fig:05D1cb984} SNLS-  05D1cb984: a SN~Ia supernova at z=0.632. The spectrum phase is 4.3. A Sb-Sc host model has been subtracted.}
                    \end{center}
                    \end{figure*}
                    
                    \begin{figure*}[ht]
                    \begin{center}
                    \begin{tabular}[b]{cc}
                    \hspace{-5mm}
                    \vspace*{1cm}
                    \begin{minipage}[b]{10cm}
                    \centering
                    \includegraphics[width=10cm]{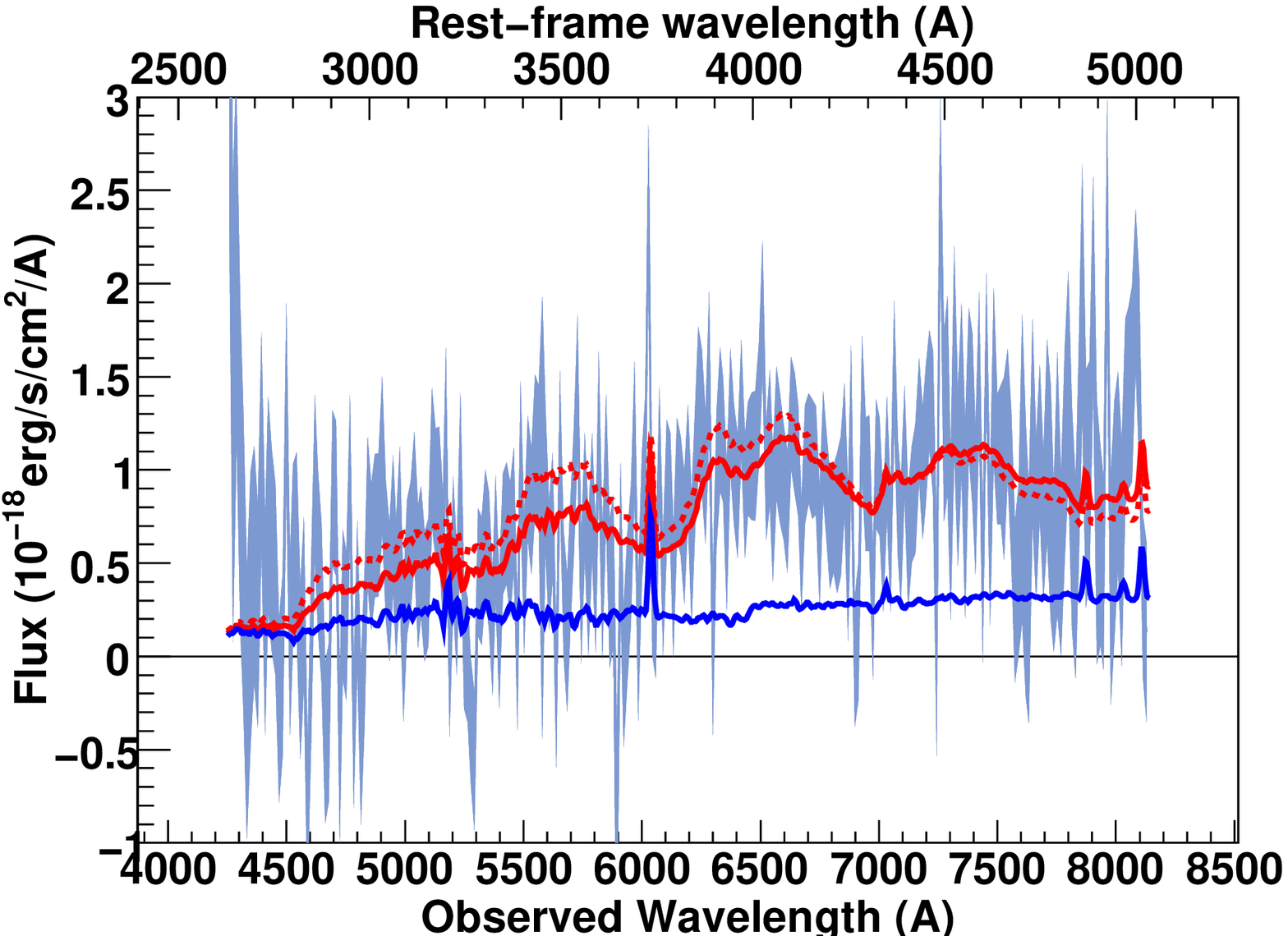}
                    \end{minipage}
                    &
                    \begin{minipage}[b]{10cm}
                    \hspace{-5mm}\includegraphics[width=10cm]{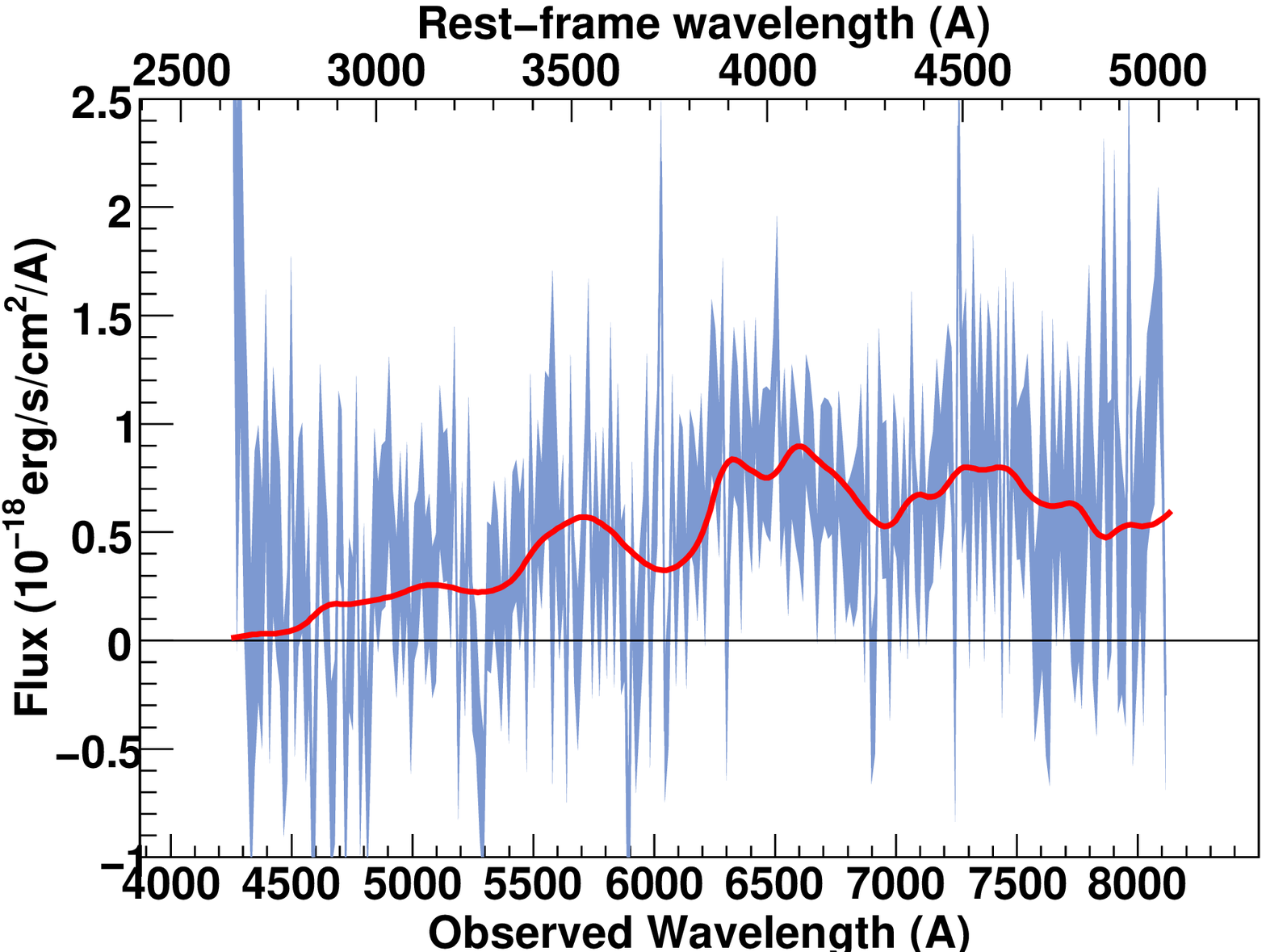}
                    \end{minipage}
                    \\
                    \end{tabular}
                    \caption{\label{fig:05D1ck983} SNLS-  05D1ck983: a SN~Ia supernova at z=0.617. The spectrum phase is -2.7. A         Sb-Sc host model has been subtracted.}
                    \end{center}
                    \end{figure*}

                    \newpage
                    \clearpage
                    
                    \newpage
                    \clearpage
                    
\end{document}